\documentclass[12pt]{article}
\pdfoutput=1
\usepackage[a4paper]{geometry}
\usepackage{jheppub, amsmath,amssymb,amsfonts,amsxtra, mathrsfs, makeidx,graphics,graphicx,amsthm,epsfig, ytableau,bm,longtable,float, color,tikz,mathtools,xfrac,footnote,rotating, lscape, pgfplots}
\usetikzlibrary{positioning}
\usetikzlibrary{chains}
\usetikzlibrary{arrows,fit,decorations.pathreplacing}
\tikzstyle{every picture}+=[remember picture]
\tikzstyle{na} = [baseline=-.5ex]

\usetikzlibrary{arrows, decorations.markings, calc, fadings, decorations.pathreplacing, patterns, decorations.pathmorphing, positioning}

\usepackage{bbm}

\newcommand{\Vol}{\mathrm{Vol}}

\newcommand{\eg}{\textit{e.g.}}

\numberwithin{equation}{section}

\newcommand{\nn}{\nonumber}

\newcommand{\be}{\begin{equation}} \newcommand{\ee}{\end{equation}}
\newcommand{\bea}{\begin{equation} \begin{aligned}} \newcommand{\eea}{\end{aligned} \end{equation}}

\def\tilde{\widetilde}
\def\t{\tilde}
\def\hat{\widehat}
\def\h{\hat}
\def\bar{\overline}

\def\rt2{\sqrt{2}}

\def\det{\mathop{\rm det}}

\def\Tr{\mathop{\rm Tr}}
\def\tr{\mathop{\rm tr}}

\def\CH{{\cal H}}

\def\CN{{\cal N}}

\def\1{{\ds 1}}

\newcommand{\cC}{\mathcal{C}}

\newcommand{\cH}{\mathcal{H}}

\newcommand{\cM}{\mathcal{M}}
\newcommand{\cN}{\mathcal{N}}
\newcommand{\cO}{\mathcal{O}}

\newcommand{\cQ}{\mathcal{Q}}
\newcommand{\cR}{\mathcal{R}}

\newcommand{\cW}{\mathcal{W}}

\newcommand{\cY}{\mathcal{Y}}

\newcommand{\bC}{\mathbb{C}}

\newcommand{\bH}{\mathbb{H}}

\newcommand{\bP}{\mathbb{P}}

\newcommand{\bR}{\mathbb{R}}
\newcommand{\bZ}{\mathbb{Z}}

\def\SU{\mathrm{SU}}

\def\repa{\raise4pt\hbox{$\square$}\mkern-14mu\raise-4pt\hbox{$\square$}}
\def\repab{\overline{\raise4pt\hbox{$\square$}\mkern-14mu\raise-4pt\hbox{$\square$}\mkern-1mu}}

\def\smileface{\ensuremath{\hbox{\large$\bigcirc$}\mkern-15mu\raise-1pt\hbox{\scriptsize$\smallsmile$}%
\mkern-10mu\raise4pt\hbox{..}\mkern4mu}}
\def\frownface{\ensuremath{\hbox{\large$\bigcirc$}\mkern-15mu\raise-1pt\hbox{\scriptsize$\smallfrown$}%
\mkern-10mu\raise4pt\hbox{..}\mkern4mu}}

\DeclareMathOperator{\sign}{sign}

\def\node#1#2{\overset{#1}{\underset{#2}{\circ}}}

\newcommand{\ba}{\begin{array}}
\newcommand{\ea}{\end{array}}
\newcommand{\bi}{\begin{itemize}}
\newcommand{\ei}{\end{itemize}}
\def\vec#1{\bm{#1}}
\def\bea#1\eea{\allowdisplaybreaks \begin{align}#1\end{align}}
 \newcommand{\ben}{\begin{enumerate}}
\newcommand{\een}{\end{enumerate}}
\newcommand{\bean}{\begin{eqnarray*}}
\newcommand{\eean}{\end{eqnarray*}}
\newcommand{\eref}[1]{(\ref{#1})}

\newcommand{\PE}{\mathop{\rm PE}}
\newcommand{\PL}{\mathop{\rm PL}}

\newcommand{\tQ}{\widetilde{Q}}

\newcommand{\BC}{\mathbb{C}}
\newcommand{\BR}{\mathbb{R}}

\newcommand{\BZ}{\mathbb{Z}}

\newcommand{\comment}[1]{}

\newcommand{\Sym}{\mathrm{Sym}}

\newcommand{\fflat}{\mathcal{F}^\flat}

\newcommand{\blue}{\color{blue}}
\newcommand{\gray}{\color{gray}}
\newcommand{\red}{\color{red}}

\title{The moduli spaces of $3d$ $\CN\ge 2$ Chern-Simons gauge theories and their Hilbert series}

\author[a]{Stefano Cremonesi,}
\author[b]{Noppadol Mekareeya}
\author[c,d]{and Alberto Zaffaroni}

\affiliation[a]{Department of Mathematics, KingÕs College London, \\The Strand, London WC2R 2LS, United Kingdom}
\affiliation[b]{Theory Department, CERN, \\CH-1211, Geneva 23, Switzerland}
\affiliation[c]{Dipartimento di Fisica, Universit\`a di Milano-Bicocca, \\ Piazza della Scienza 3, I-20126 Milano, Italy}
\affiliation[d]{INFN, sezione di Milano-Bicocca, I-20126 Milano, Italy}

\emailAdd{stefano.cremonesi@kcl.ac.uk}
\emailAdd{noppadol.mekareeya@cern.ch}
\emailAdd{Alberto.Zaffaroni@mib.infn.it}

\preprint{CERN-TH-2016-167}

\abstract{We present a  formula for the Hilbert series that counts gauge invariant chiral operators in a large class of 3d ${\cal N} \ge 2$  Yang-Mills-Chern-Simons theories. The formula counts 't Hooft monopole operators dressed by gauge invariants of a residual gauge theory of massless fields in the monopole background. 
We  provide a general formula for the case of abelian theories, where nonperturbative corrections are absent, and consider a few examples of nonabelian theories where nonperturbative corrections are well understood. We also analyze in detail nonabelian ABJ(M) theories as well as worldvolume theories of M2-branes probing Calabi-Yau fourfold and hyperK\"ahler twofold singularities with ${\cal N} = 2$ and ${\cal N} = 3$ supersymmetry. }

\begin{document}
\setcounter{tocdepth}{2}
\maketitle

\section{Introduction}\label{sec:intro}
Three-dimensional gauge theories have several features that are absent in their four-dimensional counterparts, such as the non-trivial dynamics of abelian gauge groups, the presence of Chern-Simons couplings and interesting effects of real masses and Fayet-Iliopoulos parameters.  The moduli spaces of supersymmetric theories have rich structures and provide insights into the strongly coupled dynamics \cite{BoerHoriOz1997,AharonyHananyIntriligatorEtAl1997,Tong2000, Intriligator:2013lca}.  The interesting structure is largely due to the role of 't Hooft monopole operators \cite{'tHooft:1977hy}, which can be realised as follows in the weakly coupled regions of the moduli space, where the gauge group is spontaneously broken to its Cartan subgroup.  For each Cartan element of the gauge group, one can dualise the abelian gauge field into a periodic scalar, whose exponentiation is a well-defined local field.  The insertion of the latter as a local operator at a spacetime point modifies the boundary conditions of fields in the path integral and introduces a magnetic flux on any two-sphere surrounding that point.  Moreover, as was pointed out in \cite{Borokhov:2002ib, Borokhov:2002cg, Borokhov:2003yu}, the definition of a 
monopole operator by a singular boundary condition also holds at the origin of the moduli space, where the coupling becomes infinite.  This approach bypasses the dualisation of nonabelian gauge fields and hence allows for several exact calculations, including the enhancement of the global symmetry at infinite coupling \cite{Gaiotto:2008ak, Bashkirov:2010kz, Bashkirov:2010hj} and the quantum moduli space of three-dimensional supersymmetric gauge theories \cite{Cremonesi:2013lqa, Cremonesi:2014kwa, Cremonesi:2014vla, Cremonesi:2014xha, Cremonesi:2014uva, Cremonesi2015a, Hanany:2015via, Mekareeya:2015bla}.  

A systematic method to study the moduli space and the chiral ring is to compute the Hilbert series, a generating function that counts bosonic gauge invariant chiral operators that are annihilated by two supercharges $\bar{\mathcal{Q}}_\alpha$ of a 4-supercharges superalgebra.  For three dimensional gauge theories with $\CN=4$ supersymmetry, the Hilbert series can be used to study the Higgs and Coulomb branches as well as the corresponding chiral rings \cite{Hanany:2011db, Cremonesi:2013lqa, Cremonesi:2014kwa, Cremonesi:2014vla, Cremonesi:2014xha, Cremonesi:2014uva, Mekareeya:2015bla, Razamat:2014pta}. For the Higgs branch, which is protected against quantum corrections, the Hilbert series can be evaluated using the Molien formula involving an integral of a rational function (see \eg~ \cite{Hanany:2011db}).  For the Coulomb branch, which receives quantum corrections, the Hilbert series can be computed exactly using the ``monopole formula'' \cite{Cremonesi:2013lqa}.  The latter counts $\CN = 2$ chiral
monopole operators dressed by the adjoint chiral multiplet that arises in the decomposition of the $\CN = 4$ vector multiplet into $\CN = 2$ multiplets.  It is worth emphasising that the Coulomb branch chiral ring relations that involve monopole operators are purely quantum relations that do not follow from a superpotential.

There has also been a recent progress in the computation of Hilbert series for CP-invariant three-dimensional $\CN=2$ gauge theories, which have vectorlike matter and no Chern-Simons interactions \cite{Cremonesi2015a, Hanany:2015via}.  In this case, the Hilbert series counts monopole operators dressed by gauge invariants of a residual gauge theory of massless fields in the monopole background.  The Hilbert series provides information about the quantum moduli space, without relying on the effective superpotential as in the traditional semi-classical analysis.

The main goal of this paper is to generalise the previous results to general three-dimensional $\CN\geq 2$ supersymmetric gauge theories with generic matter and Chern-Simons interactions.  Several insights on the low energy dynamics and dualities of such theories can be gained using the Hilbert series.  Worldvolume theories of M2-branes probing Calabi-Yau fourfold singularities are also of our interest.  The Hilbert series of the geometric branch of their moduli space leads to a deeper understanding of the connection between the field theory and corresponding Calabi-Yau singularity.

The paper is organised as follows.  In section \ref{sec:abelian}, several aspects of the dynamics of $\CN=2$ abelian theories are discussed along with the Hilbert series.  We provide a general prescription for computing the Hilbert series for $\CN=2$ gauge theories in the presence of background charges and Chern-Simons couplings.  Several examples, including the Dorey-Tong theories \cite{DoreyTong2000} and $\CN=2$ mirror symmetry, are presented in detail.  Subsequently, we move on to discuss the Hilbert series for $\CN=2$ nonabelian gauge theories in section \ref{sec:nonabelian}.  We then apply this to the ABJM theory \cite{Aharony:2008ug} and its variants in section \ref{ABJM}.  In section \ref{sec:N3CS}, we explore $\CN=3$ gauge theories that are obtained from $\CN=4$ theories by turning on Chern-Simons couplings.  The discussion includes $\CN=3$ necklace Chern-Simons quivers that can be realised as the worldvolume theories of M2-branes probing a product of two ALE singularities. For certain special values of the Chern-Simons level, such theories are dual to $\CN=4$ Kronheimer-Nakajima quivers \cite{kronheimer1990yang} via an $SL(2,\BZ)$ transformation \cite{Gaiotto:2008ak, Assel:2014awa}.  In section \ref{sec:geoabM2}, we focus on the geometric branch of the worldvolume of a single M2-brane probing a Calabi-Yau fourfold singularity.  The discussion encompasses large classes of theories with quantum corrected chiral ring, including flavoured toric quiver gauge theories \cite{Benini2010, Cremonesi2011, Jafferis2013} as well as the worldvolume theories of M2-branes probing the cones over $Y^{p,q}(\mathbb{CP}^2)$ \cite{Benini2011,Closset2012} and $V^{5,2}$ \cite{Martelli:2009ga, Jafferis:2009th}.

\section{Hilbert series of abelian 3d $\cN=2$ gauge theories}\label{sec:abelian}

To introduce our formalism, let us first consider the class of 3d $\cN=2$ abelian gauge theories. Since the gauge group is abelian, there are no nonperturbative effects that may correct the semiclassical analysis. We will often assume that the superpotential vanishes to keep the presentation simpler. We will comment on the inclusion of a nontrivial superpotential in subsection \ref{subsec:W}. 

Abelian 3d $\cN=2$ gauge theories are defined by the following data: a matter content, specifying the chiral multiplets $X^a$ ($a=1,\dots,N$) in the theory; abelian gauge, flavor and topological  symmetry groups $G=U(1)^r$, $F$ and $G_J=U(1)^r$, to which we associate dynamical, background and background abelian vector multiplets respectively; a $U(1)_R$ $R$-symmetry, with no associated background multiplet since we work in flat space; mixed Chern-Simons (CS) interactions involving the gauge and global symmetries; and a superpotential, that we will take to vanish in most of this section. 
Real scalars in the background abelian vector multiplets for the flavor symmetry group define real mass parameters for the matter fields, whereas real scalars in the background abelian vector multiplets for the topological symmetry group define Fayet-Iliopoulos parameters.

We wish to compute the Hilbert series, a generating function that counts gauge invariant chiral operators of the theory: 
\be\label{HS_definition}
H(t,z,\h x) = \Tr\nolimits_{\cH} \bigg(t^R \prod_i z_i^{J_i} \prod_{\h i} \h x_{\h i}^{\h Q_{\h i}}  \bigg)~,
\ee
where $\cH$ is the vector space of gauge invariant chiral operators. $R$, $J_i$ and $\h Q_{\h i}$ are the $R$-charge, topological charges for the (abelian) topological symmetry group $G_J$ and flavor charges for the maximal torus of the continuous flavor symmetry group $U(1)^{N-r}$. 
 $t$, $z_i$ and $\h x_{\h i}$ are the corresponding fugacities. In the next subsection we will generalize the standard notion of Hilbert series \eref{HS_definition} to include certain background charges associated to the global symmetry group.

The gauge invariant chiral operators of the theory are 't Hooft monopole operators dressed by matter fields. The insertion of a bare chiral monopole operator $V_m$ is defined by imposing a Dirac singularity for gauge field configurations in the path integral, so that $\frac{1}{2\pi}\int F=m$ over a 2-sphere surrounding the insertion point, along with a singularity $\sigma\sim \frac{m}{2r}$ for the real scalar $\sigma$ in the vector multiplet to ensure that the Bogomol'nyi equation required by supersymmetry is obeyed \cite{Borokhov:2002cg,Borokhov:2003yu}. $m=(m_1,\dots,m_r)$ is the magnetic charge of the monopole operator, which by Dirac quantization belongs to the integer lattice $\bZ^r$ if the gauge group is $U(1)^r$.

Monopole operators are charged under the topological symmetry group  $G_J=U(1)^r$. For each $U(1)$ factor in the gauge group with gauge connection $A_i$, there is a topological symmetry $U(1)_{J_i}$ with conserved current $J_i = \frac{1}{2\pi}\ast dA_i$. 
The topological charges of monopole operators in the abelian gauge theory therefore coincide with the magnetic charges 
\be\label{topological_abel}
J_i[V_m] \equiv J(m) = m_i~.
\ee

Chern-Simons (CS) couplings $\frac{k_{ij}}{4\pi} \int A_i \wedge d A_j + \dots$, where the ellipses denote supersymmetric completion, induce classical electric charges for monopole operators
\be\label{elec_charge_from_CS_abel}
Q^{\rm class}_i[V_m] \equiv Q^{\rm class}_i(m) = -\sum_j k_{ij} m_j~, 
\ee
where $Q_i$ is the electric charge under the $i^{\rm th}$ $U(1)$ gauge factor.
Similarly, mixed global-gauge Chern-Simons couplings induce classical global charges
\be\label{classical_global_charges}
\begin{split}
	\h Q^{\rm class}_{\h i}[V_m] \equiv \h Q^{\rm class}_{\h i}(m) &= -\sum_j k_{\h i j} m_j~, \\
	R^{\rm class}[V_m] \equiv R^{\rm class}(m) &= -\sum_j k_{R j} m_j~.
\end{split}
\ee
Note that the topological charges \eref{topological_abel} may be interpreted as coming from mixed topological-gauge CS couplings with levels $k_{J_i i}=-1$.

Due to the spectral asymmetry of the Dirac operator in the monopole background, the global and gauge charges of bare monopole operators $V_m$ acquire quantum corrections from all the fermions in the theory that are charged under the monopole background. In an abelian gauge theory the quantum corrections are only due to fermions in matter chiral multiplets $X^a$, $a=1,\dots,N$. The quantum correction to the gauge or global charges $Q_A[V_m]\equiv Q_A(m)$ of the monopole operator is \cite{Borokhov:2002cg,ImamuraYokoyama2011,Benini2011}
\be\label{quantum_correct_charge}
Q_A^{\rm quant}[V_m] \equiv Q_A^{\rm quant}(m) = -\frac{1}{2}\sum_{a=1}^N Q_A[\psi^a]\bigg|\sum_i Q_i^a m_i \bigg|~, 
\ee
where the sum runs over all fermion matter fields $\psi^a$ in the theory, $Q^a_i = Q_i[\psi^a] = Q_i[X^a]$ are the \emph{gauge} electric charges of the fermions, whereas $Q_A$ stand for any gauge or global  charge, including the $R$-charge, for which $R[\psi^a]=R[X^a]-1\equiv r_a-1$ and 
\be\label{quantum_correct_R}
R^{\rm quant}[V_m] \equiv R^{\rm quant}(m) = -\frac{1}{2}\sum_{a=1}^N (r_a-1)\bigg|\sum_i Q_i^a m_i \bigg|~.
\ee
Topological charges do not receive quantum corrections since matter fields are not charged under the topological symmetry.

Adding the classical and quantum contributions, the total charges of monopole operators take the general form
\be\label{total_charge}
Q_A[V_m] \equiv Q_A(m) = Q_A^{\rm class}(m) + Q_A^{\rm quant}(m) = - \sum_j k_{A j}^{\rm eff}(m) m_j~,
\ee
in terms of the quantum corrected \emph{effective Chern-Simons couplings}
\be\label{eff_CS_1}
k_{Aj}^{\rm eff}(m) = k_{Aj}+\frac{1}{2}\sum_a Q^a_A Q^a_j \sign(\sum_i Q_i^a m_i)~.
\ee  
Even though $k_{Aj}^{\rm eff}(m)$ is ill-defined when $\sum_i Q_i^a m_i=0$ for some $a$, the charge $Q_A(m)$ is well-defined. 

To compute the Hilbert series \eref{HS_definition}, we decompose the vector space of chiral operators $\cH =\oplus_m \cH_m$ in vector spaces of chiral  operators of fixed magnetic charge. There is a unique bare chiral monopole operator $V_m$ defined in terms of the vector multiplet for each magnetic charge $m$ \cite{Borokhov:2002cg}, but it can be dressed by nonnegative powers of matter fields which are massless in the monopole background to form gauge invariants. These are all the matter fields $X^\alpha$ such that $\sum_i Q_i^\alpha m_i=0$: we call them the \emph{residual matter fields} in the magnetic sector of charge $m$. 
(See section 2 of \cite{Cremonesi2015a} for a detailed introduction to the formalism.)
Assuming for simplicity that there is no superpotential, powers of the residual matter fields are counted by the generating function%
\footnote{The plethystic exponential ($\PE$) of a multi-variate function $f(x_1, \ldots, x_n)$ is defined as
	\bea
	\PE \left[f(x_1, \ldots, x_n) \right] = \exp \left(\sum_{p=1}^\infty \frac{1}{p} f(x_1^p, \ldots, x_n^p) \right)~. \nn
	\eea}  
\be
\PE\bigg[\sum_{\alpha} t^{r_\alpha}\prod_i x_i^{Q^\alpha_i} \prod_{\h i} {\h x}_{\h i}^{\h Q^\alpha_{\h i}} \bigg] \equiv \frac{1}{\prod_\alpha \left(1-t^{r_\alpha}\prod_i x_i^{Q^\alpha_i} \prod_{\h i} {\h x}_{\h i}^{\h Q^\alpha_{\h i}}\right)},
\ee 
where we introduced fugacities $x_i$ for the $U(1)^r$ gauge group in addition to $t$ for the $U(1)_R$ symmetry and $\h x_{\h i}$ for the flavor symmetry.

Taking all these facts into account, we conclude that in the absence of a superpotential, the Hilbert series that counts gauge invariant dressed chiral monopole operators of a $U(1)^r$ gauge theory takes the general form:
\be\label{HS_standard_abelian}
\begin{split}
H(t,z,\h x) &=\sum_{m \in \bZ^r}  t^{R(m)} \prod_{i=1}^r z_i^{m_i}  \prod_{\h i=1}^{N-r} \h x_{\h i}^{\h Q_{\h i}(m)} \cdot \\
& \cdot \prod_{i=1}^r \left(\oint \frac{d x_i}{2\pi i x_i} x_i^{Q_i(m)} \right) \PE\bigg[\sum_{a=1}^N \delta_{\sum_i Q^a_i m_i,0} ~t^{r_a} \prod_i x_i^{Q^a_i} \prod_{\h i} {\h x}_{\h i}^{\h Q^a_{\h i}} \bigg]~.
\end{split}
\ee
Let us explain the ingredients. $x_i$ and $m_i$ are fugacities and magnetic charges for the $U(1)^r$ gauge group. The sum over magnetic charges $m$ and the integral over $x_i$ imposes $U(1)^r$ gauge invariance. Bare monopole operators $V_m$ are weighted by their global charges $R(m)$, $J_i(m)=m_i$ and $\h Q_{\h i}(m)$, and by their gauge charges $Q_i(m)$. They are dressed by nonnegative powers of the residual matter fields, counted by the plethystic exponential in the second line (with the Kronecker delta functions enforcing the masslessness condition). Finally, the dressed monopole operators are made gauge invariant by averaging over the gauge group.

It is important to note that the supersymmetry condition $\sigma\sim \frac{m}{2r}$, where $r$ is the distance from the insertion point, relates the \emph{real} scalar $\sigma$ that enters in the semiclassical analysis of the moduli space to the \emph{integer} magnetic charge $m$ that defines the monopole operator $V_m$.% 
\footnote{On the moduli space where $\sigma$ takes expectation value giving mass to the matter fields, the monopole operator is obtained by dualizing the abelian vector multiplet to a chiral multiplet with periodic imaginary part, and exponentiating the latter  \cite{BoerHoriOz1997}: $V_m= \exp\left[-m\left(\int^\sigma \frac{dx}{2g^2(x)}+i\tau\right)\right]$, where $g(\sigma)$ is the effective Yang-Mills coupling that includes one-loop corrections from integrated out massive matter fields, and $\tau$ is the periodic dual photon defined by $\ast F = \frac{g^2(\sigma)}{2\pi} d \tau$.}
As a result, our Hilbert series formalism is closely related to the old semiclassical analysis of the moduli space \cite{BoerHoriOz1997,AharonyHananyIntriligatorEtAl1997,DoreyTong2000} (see also the more recent \cite{Intriligator:2013lca}), but with the added benefit of providing a general formula to count gauge invariant chiral operators.

Geometrically, the Hilbert series \eqref{HS_standard_abelian}   counts holomorphic functions on the moduli space of supersymmetric vacua of the gauge theory in the absence of Fayet-Iliopoulos and real mass parameters. This is the moduli space of the infrared SCFT at the endpoint of an RG flow starting from an ultraviolet Maxwell-Chern-Simons theory, and has the structure of a cone by dilatation invariance. For generic abelian theories, the moduli space of the CFT is rather poor and often it only consists of the origin. Correspondingly, the only chiral gauge invariant counted by the Hilbert series is the identity operator. We will consider non-generic theories, such as M2-brane theories, that flow to CFT's with interesting conical moduli spaces parametrized by dressed monopole operators in sections \ref{ABJM}, \ref{sec:N3CS} and \ref{sec:geoabM2}.

\subsection{The Hilbert series with background magnetic charges}\label{subsec:HS_bg_charges}

Abelian $\cN=2$ gauge theories can also have interesting moduli spaces, including compact branches, when real masses and Fayet-Iliopoulos (FI) parameters are turned on \cite{BoerHoriOz1997,AharonyHananyIntriligatorEtAl1997,DoreyTong2000}. Real masses and FI parameters can be regarded as real scalars in background vector multiplets for the flavor and topological symmetry groups. 
In light of the correspondence between real scalars in vector multiplets and magnetic charges of monopole operators, in order to study the moduli spaces of vacua of gauge theories perturbed by real masses and FI parameters it is natural to consider a generalization of the Hilbert series \eref{HS_definition} and \eref{HS_standard_abelian} where we include background monopole operators for the global non-$R$ symmetry group. These are defined by inserting supersymmetric Dirac monopole singularities for background vector multiplets associated to the flavor or topological symmetry. 
We will refer to their magnetic charges as \emph{background magnetic charges} in the following.

Denoting by $-B_i$ the background magnetic charges for the topological symmetries and by $\h m_{\h i}$ the background magnetic charges for the flavor symmetries, the Hilbert series with background magnetic charges is 
\be\label{HS_bg_definition}
H(t,z,\h x; B, \h m) = \Tr\nolimits_{\cH_{B,\h m}} \bigg(t^R \prod_i z_i^{J_i} \prod_{\h i} \h x_{\h i}^{\h Q_{\h i}}  \bigg)~,
\ee
where $\cH_{B,\h m}$ denotes the vector space of gauge invariant chiral dressed monopole operators with fixed background magnetic charges $B_i$ and $\h m_{\h i}$. Decomposing $\cH_{B,\h m}= \oplus_m \cH_{m; B,\h m}$ in terms of magnetic sectors of the dynamical gauge group, the Hilbert series with background magnetic charges takes a similar form to \eref{HS_standard_abelian}, namely
\be\label{HS_standard_abelian2}
\begin{split}
		H(t,z,\h x; B, \h m) &=\sum_{m \in \bZ^r}  t^{R(m,\h m, B)}  \prod_{\h i=1}^{N-r} \h x_{\h i}^{\h Q_{\h i}(m,\h m, B)} \prod_{i=1}^r \left( z_i^{m_i} \oint \frac{d x_i}{2\pi i x_i} x_i^{Q_i(m,\h m, B)} \right) \\
		& \qquad \PE\bigg[\sum_{a=1}^N \delta_{\sum_i Q^a_i m_i+\sum_{\h i} \h Q^a_{\h i} \h m_{\h i},0} ~t^{r_a} \prod_i x_i^{Q^a_i} \prod_{\h i} {\h x}_{\h i}^{\h Q^a_{\h i}} \bigg]~.
\end{split}
\ee
The differences with \eref{HS_standard_abelian} are in the delta function inside the $\PE$, that determines the residual matter fields in the monopole background, and in the charges of monopole operators, that are affected by the background charges as follows:
\be\label{charges_general}
\begin{split}
\hspace{-3pt}	Q_A(m,\h m, B) &= -\sum_C k^{\rm eff}_{A C}(M) M_C = -\sum_C k_{A C} M_C - \frac{1}{2} \sum_{a} Q_A[\psi^a]|m^a_{\rm eff}(M)|~.
\end{split}
\ee
Here indices $A, C$ label global or gauge $U(1)$, as in \eref{total_charge}\eref{eff_CS_1}, $Q_A$ are electric charges and $M_A$ magnetic charges, namely $M_i=m_i$ for gauge $U(1)$ groups, $M_{\h i}=\h m_{\h i}$ for flavor $U(1)$ groups, $M_{J_i}=- B_i$ for topological $U(1)$ groups and $M_R=0$ for $U(1)_R$. Note the last minus sign and recall that $k_{i J_i}=-1$. This implies that $-B_i$ contribute bare background electric charges for the gauge groups. Finally we introduced the \emph{effective mass} 
\be\label{eff_mass_0}
m^a_{\rm eff}(m,\h m)= \sum_i Q^a_i m_i+\sum_{\h i} \h Q^a_{\h i} \h m_{\h i}~
\ee
of the $a^{\rm th}$ matter chiral multiplet, which is a function of the magnetic charges, in analogy with the effective real mass in the semiclassical analysis of the moduli space. We also 
recall that For $U(1)$ symmetries with integer charges, bare Chern-Simons levels obey the quantization law
\be\label{CS_quantiz}
k_{AB}+ \frac{1}{2}\sum_a Q^a_A Q^a_B \in \bZ
\ee
that ensures that the effective CS levels 
\be\label{CS_eff_1}
k_{AB}^{\rm eff}(m, \h m) = k_{AB}+ \frac{1}{2}\sum_a Q^a_A Q^a_B \sign(m^a_{\rm eff}(m,\h m))
\ee
are integer when they are well-defined.

To be precise, the discussion in this subsection needs to be corrected to account for the possibility of torsion in the magnetic charges of the flavor symmetry. To understand this subtle issue, in the next subsections we take a detour towards a more systematic definition of the Hilbert series with background magnetic charges. We will start from the ungauged theory in the presence of general background charges for its flavor symmetry, and then explain how to gauge an abelian subgroup of its flavor symmetry. A more careful analysis of Dirac quantization will show how torsion magnetic charges $m_\Gamma$ arise. Readers not interested in these technical details might skip to the examples of subsection \ref{subsec:examples} in a first reading, and also neglect torsion magnetic charges in section \ref{subsec:DoreyTong}.

\subsection{The ungauged theory}

The ungauged theory consists of $N$ chiral multiplets $X^a$, $a=1,\dots, N$, with charges $F_b[X^a]=\delta^a_b$ ($b=1,\dots,N$) under a $U(1)^N$ flavor symmetry and $R[X^a]=r_a$ under the $R$-symmetry. 
We couple the flavor symmetry to $N$ abelian background vector multiplets $U^b$. We call $\mu_b\in \bZ$ the associated magnetic charges and $u_b$ the associated $U(1)$-valued fugacities. We also introduce a fugacity $t$ for the $U(1)_R$ symmetry, but no background magnetic charge. The theory has flavor-flavor CS levels $k_{ab}$, flavor-$R$ CS levels $k_{aR}$, and $R$-$R$ CS levels $k_{RR}$, which satisfy the quantization law \eref{CS_quantiz} provided charges are integer.% 
\footnote{The quantization law for a general $R$-symmetry can be obtained by mixing a fiducial $U(1)$ $R$-symmetry with integer charges with (possibly broken) abelian non-$R$ symmetries.} 
The bare Chern-Simons couplings induce electric charges for monopole operators, given by $Q_A^{\rm class}[V_m]=-\sum_B k_{AB}m_B$. Thus the weight in the Hilbert series from the classical charges of a monopole operator is 
\be\label{HS_CS}
\prod_A w_A^{-\sum_B {k_{AB} m_B}} \equiv t^{-\sum_b k_{Rb} \mu_b} \prod_a u_a^{-\sum_b k_{ab}\mu_b} ~,
\ee
where we set $w_R=t$, $m_R=0$ for the $R$-symmetry, and $w_a=u_a$, $m_a=\mu_a$ for the flavor symmetries.

In the monopole background for the flavor symmetry, the chiral multiplet $X^a$ has ``mass'' $\mu_a$.%
\footnote{We slightly abuse terminology: as we explained, the effective real mass in the background of a chiral monopole operator is $\mu^a/(2r)$.}
If $\mu_a=0$, $X^a$ is a \emph{residual matter field}, a modulus of the monopole configuration that can be used to dress the background BPS monopole operator. Powers of $X^a$ are counted in the Hilbert series by $\PE[t^{r_a} u_a]$. If instead $\mu_a\neq 0$, $X^a$ is massive and induces a one-loop correction to the charges of the monopole operator, or equivalently to the effective Chern-Simons levels, leading to a weight $(t^{r_a-1} u_a)^{-\frac{1}{2}|\mu_a|}$. In summary, matter chiral multiplets contribute to the Hilbert series a factor
\be\label{HS_chiral}
\prod_a (t^{r_a-1} u_a)^{-\frac{1}{2}|\mu_a|} \PE[\delta_{\mu_a,0}~t^{r_a} u_a]~.
\ee

Altogether, the Hilbert series of the ungauged theory, in the presence of background magnetic charges $\mu_a$, reads 
\be\label{HS_ungauged}
\begin{split}
H_{\rm ug}(t, u_a;\mu_a) &=  t^{-\sum_b k_{Rb} \mu_b} \prod_a u_a^{-\sum_b k_{ab}\mu_b} \prod_a (t^{r_a-1} u_a)^{-\frac{1}{2}|\mu_a|} \PE[\delta_{\mu_a,0}~t^{r_a} u_a]=\\
&\equiv t^{-\sum_b k^{\rm eff}_{Rb}(\mu) \mu_b} \prod_a u_a^{-\sum_b k^{\rm eff}_{ab}(\mu)\mu_b}\PE[\delta_{\mu_a,0}~t^{r_a} u_a]~,
\end{split}
\ee
where the effective Chern-Simons levels are
\be\label{eff_CS}
k_{AB}^{\rm eff}(\mu) = k_{AB}+\frac{1}{2}\sum_a Q^a_A Q^a_B \sign(\mu_a)~.
\ee

\subsection{The superpotential}\label{subsec:W}

So far we have assumed for simplicity that the theory has no superpotential. Then the contribution to the Hilbert series of the massless matter fields in \eref{HS_chiral} reads $\PE[\sum_a \delta_{\mu_a,0}t^{r_a} u_a]=\PE[\sum_\alpha t^{r_\alpha} u_\alpha]$, which counts elements of the graded ring of polynomials $\bC[X^\alpha]$ in the massless fields $X^\alpha$, which have $\mu_\alpha=0$.

If the superpotential $W(X)$ does not vanish, there are two changes. First of all, the global symmetry $F\times U(1)^R$ is broken to a subgroup. This enforces constraints on the fugacities and magnetic charges: the weight associated to a superpotential term is $(t')^2$, where $t'$ is a fugacity for a preserved $U(1)_R$ symmetry, and the magnetic charge (or ``mass'') associated to the superpotential term is $0$. 

Secondly, we must impose the $F$-term relations induced by the superpotential. The $F$-term of a field of mass $\mu$ has mass $-\mu$ and vanishes when all massive fields are set to zero. We are thus left with a \emph{residual theory} $T_{\mu}$ of massless matter fields $X^\alpha$ (such that $\mu_\alpha=0$), with a \emph{residual superpotential} $W_{\mu}(X^\alpha)=W(X)|_{X^a=0\, {\rm if} \, \mu_a\neq 0}$ that is obtained by setting all massive fields to zero in the original superpotential. The contribution of the residual matter fields to the whole Hilbert series is the Hilbert series of the ring $\bC[X^\alpha]/\langle \partial_\alpha W_{\mu}\rangle$ and  takes the form $\PE[\sum_\alpha t^{r_\alpha} u_\alpha] N(t,u)$, where $N(t,u)$ is a polynomial that enforces $F$-term equations. If the $F$-term equations are independent --- that is, if there are no higher syzygies ---, then $N$ takes the simple factorized form $N(t,u)=\prod_{\alpha}(1-t^{2-r_\alpha}u_\alpha^{-1})= \PE[-\sum_{\alpha}t^{2-r_\alpha} u_\alpha^{-1}]$, otherwise it can be computed for instance  using software such as \texttt{Macaulay2} \cite{M2}.

\subsection{Gauging}\label{subsec:gauging}

In the rest of this section we will consider abelian gauge theories without superpotential. We gauge a $U(1)^r$ subgroup of the flavor symmetry $U(1)^N$ of the ungauged theory, introducing $r$ dynamical abelian vector multiplets $V^i$, $i=1,\dots,r$.%
\footnote{The gauge group may include a finite group, but we postpone its discussion to section \ref{subsec:DoreyTong}.} 

The matter fields carry integer charges $Q_i[X^a]=Q^a_i$ under the gauge group. If ${\rm span}_\bZ \{Q_i\} \neq {\rm span}_\bR \{Q_i\}\cap \bZ^r$, there is an ambiguity in the definition of the gauge group. 
Following \cite{SeibergTaylor2011}, we define the gauge group by the lattice of its allowed electric charges, rather than the electric charges of matter fields that are actually present in the theory. In our case, by $G=U(1)^r$ we mean that the allowed electric charge lattice is $\bZ^r$. By Dirac quantization, the magnetic charge lattice is the dual $\bZ^r$. We denote by $x_i$ and $m_i$ the fugacities and magnetic charges for the $U(1)^r$ gauge group. 

The flavor group is $F=U(1)^N/U(1)^r$, where $U(1)^r$ acts on the fields with charges $Q$. The integer kernel $\h Q = (\h Q^a_{\h i})$ of the charge matrix $Q$ defines the charge matrix for a $U(1)^{N-r}$ flavor symmetry. (Note that a common subgroup of $U(1)^r\times U(1)^{N-r}$ might not act on the matter fields.) 
We denote by $\h x_{\h i}$ and $\h m_{\h i}$ the fugacities and magnetic charges for the $U(1)^{N-r}$ flavor symmetry. If $\Gamma=\bZ^N/{\rm span}_\bZ (Q_i, \h Q_{\h i})$ is nontrivial, the flavor magnetic charges of the matter fields include a further torsion term $m_\Gamma \in \Gamma$.

The gauging of $U(1)^r$ in $U(1)^N$ is then achieved in the Hilbert series by replacing
\be\label{gauging_1}
\begin{split}
	u_a &\mapsto  x_a^{\rm eff} = \prod_{i=1}^r x_i^{Q^a_i} \prod_{\h i=1}^{N-r} \h x_{\h i}^{\h Q^a_{\h i}}~, \\
	\mu^a &\mapsto m^a_{\rm eff} =  \sum_i Q^a_i m_i + \sum_{\h i} \h Q^a_{\h i} \h m_{\h i} + m^a_\Gamma \equiv \sum_i Q^a_i m_i + \h m^a
\end{split}
\ee
in \eref{HS_ungauged}, and Fourier transforming over the $U(1)^r$ gauge group associated to $x_i$, $m_i$: 
\be\label{gauging_2}
H(t,\h x,z;\h m, m_\Gamma, B) = \sum_{m \in \bZ^r} \prod_{i=1}^r \left(z_i^{m_i} \oint \frac{dx_i}{2\pi i x_i}x_i^{-B_i} \right) H_{\rm ug}(t,x_a^{\rm eff};m^a_{\rm eff})~,
\ee
where the integral is over the unit torus. The integral over $x$ restricts the counting of chiral operators to gauge invariants (in the presence of a background monopole operator for the global symmetry); the sum over $m$ takes into account the dynamical monopole operators for the gauge group.

$z_i$ and $-B_i$ in \eref{gauging_2} are fugacities and magnetic charges for the $G_J=U(1)^r$ topological symmetry group, whose conserved currents are the Hodge duals of the gauge field strengths.  We will follow a common abuse of terminology and refer to $B_i$ as \emph{``baryonic charges''} \cite{Forcella:2007wk}. They are discrete counterparts of the FI parameters $\xi_i$, which insert background electric charges $-B_i$ for the $U(1)$ gauge factors. 

The equations of motion for the dynamical vector multiplets in the Chern-Simons theory give Gauss constraints $k_{ij} F_j + k_{i \h j} \h F_{\h j} + F^{(J)}_{i} = 2\pi \ast J_i$ (along with supersymmetric partners), where $F_j$, $\h F_{\h j}$ and $F^{(J)}_{i}$ are field strengths for the gauge, flavor and topological symmetries, whereas $J_i$ are conserved currents for the gauge symmetries, that involve the matter fields. These equations of motion impose linear relations among the conserved currents for the topological symmetries associated to the gauge and global symmetry groups and the conserved currents for the gauge symmetry. In our formalism, we introduce independent fugacities and magnetic charges for the gauge and global symmetries, without enforcing these constraints. The constraints are implemented by the Fourier transform over the gauge group.  

The gauging of a $U(1)$ factor as in \eref{gauging_2}, which involves a mixed Chern-Simons coupling at level $-1$ between the gauge $U(1)$ and its topological $U(1)$, plays a prominent role in understanding abelian mirror symmetry as a functional Fourier transform \cite{Kapustin1999} and defines the action of the $S$ element of $SL(2,\bZ)$ on the space of field theories with a $U(1)$ global non-R symmetry \cite{Witten2003}. We will elaborate on this $SL(2,\bZ)$ action at the level of the Hilbert series in appendix \ref{SL2Z}.

Note that the final formula \eqref{gauging_2} for the Hilbert series of the abelian gauge theory without superpotential may be written as 
\be\label{gauging_3}
H = \sum_{m \in \bZ^r} \prod_{i=1}^r \left(z_i^{m_i} \oint \frac{dx_i}{2\pi i x_i}x_i^{-B_i^{\rm eff}} \right) \prod_a \h x_a^{-\h B_a^{\rm eff}} \cdot t^{-B_R^{\rm eff}} \PE[\sum_a \delta_{m^a_{\rm eff},0}t^{r_a} \h \prod_i x_i^{Q^a_i} \prod_{\h i} \h x_{\h i}^{\h Q^a_{\h i}} ]
\ee
in terms of the \emph{effective masses} of the chiral fields $m^a_{\rm eff}$ introduced in \eref{gauging_1} and of \emph{effective baryonic charges}
\be\label{B_eff}
\begin{split}
	B_i^{\rm eff} &= B_i + \sum_j k_{ij}^{\rm eff} m_j + \sum_{\h j} k_{i \h j}^{\rm eff} \h m_{\h j}+ \sum_b k_{ib}^{\rm eff}  m^b_\Gamma \equiv - Q_i(m,\h m, m_\Gamma,B) \\  
	\h B_{\h i}^{\rm eff} &=  \sum_j k_{\h ij}^{\rm eff} m_j + \sum_{\h j} k_{\h i \h j}^{\rm eff} \h m_{\h j}+ \sum_b k_{\h i b}^{\rm eff}  m^b_\Gamma  \equiv - \h Q_{\h i}(m,\h m, m_\Gamma,B)\\  
	B_R^{\rm eff} &=  \sum_j k_{Rj}^{\rm eff} m_j + \sum_{\h j} k_{R \h j}^{\rm eff} \h m_{\h j}+ \sum_b k_{Rb}^{\rm eff} m^b_\Gamma \equiv -R(m,\h m, m_\Gamma,B)~, %\\  
   % 	B_R^{\rm eff} &= \sum_j k_{Rj}^{\rm eff} m^j + \sum_b k_{Rb}^{\rm eff} \h m^b ~.
\end{split}
\ee
which are equal and opposite to the quantum corrected charges of the monopole operator of magnetic charges $m$, $\h m$, $m_\Gamma$ and $B$.

The \emph{effective Chern-Simons levels} $k^{\rm eff}_{AB}$ take the form \eref{eff_CS_1} with the ``effective mass'' \eref{gauging_1}.
The bare CS levels involving the $U(1)^r$ gauge group read 
\be\label{CS_gauge}
k_{ij}=\sum_{a,b} k_{ab}Q^a_i Q^b_j~, \qquad k_{i \h j}=\sum_{a,b} k_{ab}Q^a_i \h Q^b_{\h j}~, \qquad k_{Rj}=\sum_{b} k_{Rb}Q^b_j~.
\ee
Similar formulas hold for mixed CS levels involving global symmetries only.

As we anticipated, the formula \eref{gauging_2} or \eref{gauging_3} for  the Hilbert series that counts chiral operators is closely related to the semiclassical analysis of the vacuum moduli space based on the 1-loop corrected scalar potential (see  section 2 of \cite{DoreyTong2000}), through the correspondence between integer magnetic charges $m$ and real scalars $\sigma$ in vector multiplets that is required for the supersymmetry of monopole operators. In particular, the dynamical magnetic charges $m_i$ corresponds to the dynamical real scalars $\sigma_i$ for the gauge symmetry; the background magnetic charges $\h m_{\h i}$ for the flavor symmetry corresponds to background scalars (or real mass parameters) $\h \sigma_{\h i}$; the ``effective mass'' $m^a_{\rm eff}=\sum_i Q^a_i m^i+\sum_{\h i} \h Q^a_{\h i} \h m_{\h i}+m^a_\Gamma$ correspond to the effective real mass $\mu^a_{\rm eff}=\sum_i Q^a_i \sigma_i+\sum_{\h i} \h Q^a_{\h i} \h \sigma_{\h i}$ of the matter field $X^a$; the background magnetic charges $B_i$ for the topological symmetries correspond to the bare FI parameters $\xi_i$; the effective baryonic charges $B_A^{\rm eff}$ in \eref{B_eff} correspond to effective FI parameters $\xi_A^{\rm eff}$. Note that the torsion element $m_\Gamma$ has no continuous counterpart.

The insertion of $\delta_{m^a_{\rm eff},0}$ inside the plethystic exponentials in \eref{gauging_3} corresponds to setting to zero the mass terms $\sum_a |\mu^a_{\rm eff} X^a|^2$ in the scalar potential: $X^a$ can take expectation value only if $\mu^a_{\rm eff}=0$. The integral over the gauge group in the presence of effective baryonic charges $B_i^{\rm eff}$ corresponds to imposing the $D$-term constraint with the effective FI parameters $\sum_a Q^a_i |X^a|^2 = \xi_i^{\rm eff}$ and modding out by the gauge group. If there is a superpotential, the insertion of the numerator $N(t,x,\h x)$ discussed in section \ref{subsec:W} corresponds to imposing the $F$-term constraints of the residual theory of massless fields.

\subsection{Examples}\label{subsec:examples}

In this section we provide a few examples of Hilbert series of 3d $\cN=2$ abelian gauge theories. We begin with theories without Chern-Simons interactions, partially discussed in \cite{Cremonesi2015a,Cremonesi:2014kwa}, emphasizing here the role of background magnetic charges. Then we move on to discuss Chern-Simons theories. In the next subsection we will discuss in detail a large class of abelian Chern-Simons theories studied in \cite{DoreyTong2000,Tong2000}.

\subsubsection{3d $\cN=2$ SQED with vectorlike flavors}

Our first example is 3d $\cN=2$ SQED theory with $N$ flavors of charge $1$ matter fields $Q^a$ and charge $-1$ fields $\tilde{Q}_a$. 
The standard Hilbert series without background magnetic charges was computed in \cite{Cremonesi2015a}. Here we turn on background magnetic charges for the vectorlike part of the flavor group, but not for the axial part of the flavor symmetry (nor the topological symmetry), so that no Chern-Simons terms are generated. We denote by $n_1, \dots, n_N$ the background flavor magnetic charges, so that $m^a_{\rm eff}=\pm(m-n_a)$, with $+$ sign for $Q^a$ and $-$ sign for $\t Q_a$. (A common shift of the $n_a$ can be undone by a shift of $m$.)  
The Hilbert series reads
\be\label{HS_SQED}
\hspace{-5pt} H(t,z,y;n)=\sum_{m \in \bZ} z^m \prod_{a=1}^N(t^{1-r}y_a^{-1})^{|m-n_a|} \oint \frac{dx}{2\pi i x}\PE\bigg[\sum_{a=1}^N \delta_{m,n_a}t^r y_a\big(\frac{x}{u_a}+\frac{u_a}{x}\big)\bigg]
\ee
where $u_a$ and $y_a$ are fugacities for the vector and axial part of the flavor group. 

In the following we consider for simplicity the case $n_1>n_2>\dots > n_N$, corresponding to $N$ different real masses.%
\footnote{ Cases where some of the background charges are equal can be discussed by a mixture of the analysis of this section and that of \cite{Cremonesi2015a}.}
The sum over $m$ in \eref{HS_SQED} separates in $N+1$ regions according to the signs of $m-n_a$. The Hilbert series with background magnetic charges is easily computed to be
\be\label{SQED_diff_masses}
\begin{split}
\hspace{-5pt}	H(t,z,y;n)&= z^N \prod_{a=1}^N (t^{1-r} y_a^{-1})^{n_a-n_N}\PE\left[z^{-1}t^{(1-r)N} \frac{1}{\prod_a y_a}\right]+\\
	&+ \sum_{h=1}^{N-1} z^{n_{h+1}}\prod_a (t^{1-r}y_a^{-1})^{|n_{h+1}-n_a|} \sum_{l=0}^{n_{h+1}-n_h}\left(z ~t^{(1-r)(N-2h)} \frac{\prod\limits_{a\leq h}y_a}{\prod\limits_{a>h}y_a} \right)^l \\
	&+ z^{n_1}\prod_a (t^{1-r}y_a^{-1})^{n_1-n_a}\PE\left[z~ t^{(1-r)N} \frac{1}{\prod_a y_a}\right]+\\
	&+ \sum_{b=1}^N z^{n_b} \prod_a (t^{1-r}y_a^{-1})^{|n_b-n_a|}\left(\PE[(t^r y_b)^2]-2\right)~.
\end{split}
\ee
This result reproduces the structure of the moduli space found in \cite{AharonyHananyIntriligatorEtAl1997,BoerHoriOz1997}, namely that of a one-dimensional Coulomb branch which is split into $N+1$ components by the intersection with $N$ one-dimensional Higgs branches, as we now explain. 

The Coulomb branch is parametrized by monopole operators $V_{m;n}$, where $m$ and $n$ denote the dynamical and background magnetic charges. The contributions in the first and third line of \eref{SQED_diff_masses} correspond to the two noncompact components of the Coulomb branch, which algebraically are two copies of the complex plane $\bC$. The operators parametrizing these components are $V_{m=n_N-p;n}= V_{m=n_N;n} Y_N^p$ and $V_{m= n_1+p;n}= V_{m=n_1;n} X_1^p$ with $p\geq 0$. $Y_N$ and $X_1$ generate the two $\bC$ factors. The second line corresponds to the $N-1$ compact components of the Coulomb branch, each of which is algebraically a $\bP^1$. Each term in the sum corresponds to a $\bP^1$ component, and the chiral operators are $V_{m= n_{h+1}+l;n}= V_{m=n_{h+1};n} X_{h+1}^l=V_{m=n_h;n} Y_h^{n_{h+1}-n_h-l}$ with $0\leq l\leq n_{h+1}-n_h$. The count of these operators in the Hilbert series gives the character of an $SU(2)$ representation $[n_{h+1}-n_h]$ of dimension $n_{h+1}-n_h+1$, up to an overall weight. $X_{h+1}$ and $Y_h$, subject to $X_{h+1} Y_h=1$, can be viewed as coordinates for the two patches of $\bP^1$, and the monopole operators with $n_{h+1}\leq m \leq n_h$ are holomorphic sections of the line bundle $\cO_{\bP^1}(n_{h+1}-n_h)$.

Finally, the terms involving plethystic exponentials in the last line of \eref{SQED_diff_masses} count the chiral operators taking expectation values in the $N$ components of the Higgs branch: $V_{m=n_a;n} (\t Q^a Q_a)^p= V_{m=n_a;n} (M^a_a)^p$, $p\geq 0$. The monopole operator $V_{m=n_a;n}$ determines the origin of the Higgs branch component on the Coulomb branch, while the mesons $M^a_a= \t Q^a Q_a$ generate the Higgs branch component, which algebraically is $\bC$. The subtraction of $2$ in the last line of \eref{SQED_diff_masses} ensures that the operators $V_{m=n_a;n}$, corresponding to the points at the intersections of two Coulomb and one Higgs branch component, are counted once. This structure  implies that $M^a_a X_a = M^a_a Y_a = X_a Y_a=0$  for all $a$, in addition to $ X_{a+1} Y_a=1$,  reproducing the findings of \cite{AharonyHananyIntriligatorEtAl1997}.

It is also possible to include complex masses, treating them as spurions. A superpotential term $W_h = m_h \tQ^h Q_h$ has the effect of lifting the $h$-th component of the Higgs branch, parametrized by $M^h_h$, in the last line of \eref{SQED_diff_masses}. It then follows that $X_h Y_h = m_h$, where the complex mass $m_h$, viewed as a spurion, carries the fugacity weight $t^{2(1-r)}y_h^{-2}$. This is interpreted as the merger of two $\bP^1$'s into a single one.

Finaly, if a baryonic charge $B$ is introduced (\emph{i.e} a background magnetic charge $-B$ for the topological symmetry, corresponding to an FI parameter $\xi$), leading to the insertion of $x^{-B}$ in the integrand of \eref{HS_SQED}, the Coulomb branch is lifted and one is left with $N$ one-dimensional Higgs branch components, with Hilbert series
\be\label{HS_SQED_B}
H(t,z,y,u;n,B)= \sum_{b=1}^N z^{n_b} \prod_a (t^{1-r}y_a^{-1})^{|n_b-n_a|}\cdot (t^r y_b)^{|B|} u_a^{-B}\PE[(t^r y_b)^2]~.
\ee

\subsubsection{Coulomb branch of 3d $\cN=4$ SQED}

A similar analysis can be performed for the Coulomb branch of 3d $\cN=4$ SQED with $N$ flavors of charge $1$ hypermultiplets. The $\cN=4$ Coulomb branch is parametrized by monopole operators and the neutral chiral multiplet $\Phi$ belongs to the $\cN=4$ vector multiplet. When the hypermultiplets are massless, the Coulomb branch of 3d $\cN=4$ SQED is $\bC^2/\bZ_N$ \cite{Intriligator:1996ex}. The Hilbert series was computed in \cite{Cremonesi:2013lqa} to be
\be\label{Neq4_SQED}
H(t=\tau^2,z)=\frac{1}{1-\tau^2} \sum_{m \in \bZ} z^m \tau^{N|m|} = \PE[\tau^2 + (z+z^{-1})\tau^N-\tau^{2N}]~.
\ee
\ref{Neq4_SQED} is indeed the Hilbert series of $\bC^2/\bZ_N$: the generators are $\Phi$, $V_+\equiv V_{+1}$ and $V_-\equiv V_{-1}$, which are subject to the relation $V_+ V_- = \Phi^N$.

When the hypermultiplets have $N$ distinct real masses, the Coulomb branch is the resolution of the $\bC^2/\bZ_N$ singularity. The Hilbert series of the Coulomb branch with background magnetic charges is \cite{Cremonesi:2014kwa}
\be\label{Neq4_SQED_res}
\begin{split}
\hspace{-5pt}	&H(\tau^2,z;n)=\frac{1}{1-\tau^2} \sum_{m \in \bZ} z^m \tau^{\sum_{a=1}^N|m-n_a|} =\\
	&\qquad=\PE[\tau^2]\bigg( z^{n_N} \tau^{\sum_a(n_a-n_N)}\PE\left[z^{-1}\tau^N\right]+ z^{n_1} \tau^{\sum_a(n_1-n_a)}\PE\left[z\tau^N\right] +\\
	&\qquad+ \sum_{h=1}^{N-1} z^{n_{h+1}} \tau^{\sum_a|n_{h+1}-n_a|} \sum_{l=0}^{n_{h+1}-n_h}\left(z \tau^{N-2h}\right)^l - \sum_{b=1}^N z^{n_b} \tau^{\sum_a|n_b-n_a|} \bigg) ~. 
\end{split}
\ee 
The neutral field $\Phi$ parametrizes a complex plane, whereas monopole operators parametrize a cylinder pinched at $N$ points. As in the previous subsection, we can define $X_a$ and $Y_a$ as in the previous subsection: $V_{m=n_a+p;n}=V_{m=n_a;n} X_a^p$ for $0\leq p \leq n_{a-1}-n_a$, and $V_{m=n_a-q;n}=V_{m=n_a;n} Y_a^q$ for $0\leq q \leq n_{a}-n_{a+1}$. Their weights in the Hilbert series are $z \tau^{N-2a+2}$ for $X_a$ and $z^{-1} \tau^{N-2a}$ for $Y_a$. They are now subject to the relations $X_a Y_a = \Phi$ for all $a$, in agreement with the fact that the effective theory near the locus where a single flavor is massless is SQED with 1 flavor, as well as $X_{a+1}Y_a=1$. 

Altogether, we have recovered the description of the resolution of $\bC^2/\bZ_N$ as a smooth variety covered by $N$ patches parametrized by $(X_a,Y_a)$, $a=1,\dots,N$, with covering maps 
\be\label{covering_maps}
(X_a, Y_a) \mapsto \begin{cases}
	V_+ = X_a^a Y_a^{a-1}\\
	V_- = X_a^{N-a} Y_a^{N-a+1}\\
	\Phi = X_a Y_a
\end{cases}
\ee
and transitions given by 
\be\label{transitions}
X_a Y_a = X_{a+1} Y_{a+1}~,\qquad\qquad X_{a+1}Y_a=1~.
\ee
(See for instance appendix B of \cite{Benini2011}.)

\subsubsection{$U(1)_k$ pure Chern-Simons theory}

We now move on to Chern-Simons theories, starting with the pure abelian theory with no charged matter. 
The Hilbert series of the pure $\cN=2$ $U(1)$ Chern-Simons theory at level $k$ is
\be\label{U1k}
\hspace{-3pt} H(t,z;B)=\sum_{m\in\bZ} z^m \oint\frac{dx}{2\pi i x}x^{-B-km} = \sum_{m\in\bZ} z^m \delta_{B+km,0}=  
\begin{cases}
	z^{-B/k}~, & B \in k \bZ\\
	0~, & B \notin k \bZ 
\end{cases}~.
\ee
The Hilbert series counts gauge invariant monopole operators of dynamical magnetic charge $m$ and background magnetic charge $B$ for the topological symmetry. Gauge invariance requires $B \in k\bZ$ and determines $m=-B/k$. This is a consequence of the equation of motion for the dynamical gauge field $k \ast F + \ast F_J=0$, where $F$ and $F_J$ are respectively dynamical and background field strengths for the gauge and the topological symmetry. 

The fact that the Hilbert series is non-vanishing only when $B \in k \bZ$ is related to the fact that the dynamical monopole operators for the gauge group have electric charges which are multiples of $k$ and break the $U(1)$ gauge group down to a residual $\bZ_k$ when taking expectation value. 
In the semiclassical analysis of the moduli space, one has the $D$-term equation $k\sigma+\xi=0$, which is related by supersymmetry to the aforementioned equation of motion for the gauge field. The $U(1)$ gauge transformation $e^{i\alpha}$ shifts the dual photon $\tau$ as $\tau \to \tau + k \alpha$, so that the monopole operator $V_m\propto e^{-i m \tau}$ has electric charge $-km$.

\subsubsection{$U(1)_{-1/2}$ with a charge $1$ chiral and the free chiral}\label{subsec:basic_duality}

Let us consider a free chiral of flavor charge $1$ and $R$-charge $1$. We take the global Chern-Simons levels to be $k_{FF}=\frac{1}{2}$ and $k_{RF}=1$. The Hilbert series reads
\be\label{H_chiral}
H_{\rm chiral}(t,u;\mu)= (t^2 u)^{-\frac{1}{2}\mu} u^{-\frac{1}{2}|\mu|}\PE[\delta_{\mu,0}~tu] = \begin{cases}
	(t u)^{-\mu}~, & \mu>0 \\
	\PE[t u]~, & \mu=0\\
	t^{-\mu}~, & \mu<0
\end{cases} ~,
\ee 
where $u$ and $\mu$ are the fugacity and background magnetic charge for the flavor symmetry. When the background flavor magnetic charge vanishes ($\mu=0$) the chiral operators counted by the Hilbert series are powers of the free chiral. When $\mu\neq 0$, the matter field is massive and the Hilbert series counts the background monopole operator for the flavor symmetry.

It is easy to show by direct evaluation that 
\be\label{basic_duality}
(t^2 u)^{-\frac{1}{2}\mu} u^{-\frac{1}{2}|\mu|}\PE[\delta_{\mu,0}~tu]=\sum_{\nu \in\bZ} u^{\nu} \oint \frac{dv}{2\pi i v} v^{\mu}         (t^2 v)^{\frac{1}{2}\nu} v^{-\frac{1}{2}|\nu|}\PE[\delta_{\nu,0}~tv]~.
\ee
The left-hand side is the Hilbert series \eref{H_chiral} of the free chiral of $R$-charge 1 and flavor charge $1$ with $k_{FF}=\frac{1}{2}$ and $k_{RR}=1$. The right-hand side is the Hilbert series of a $U(1)_{-1/2}$ Chern-Simons theory with a charge $1$ chiral of $R$-charge $1$, and Chern-Simons levels $k_{gg}=-\frac{1}{2}$ and $k_{Rg}=-1$, where $g$ stands for the gauge $U(1)$. The magnetic charge $\mu$ for the flavor symmetry of the free chiral maps to the baryonic charge of the dual theory (\emph{i.e.} minus the magnetic charge for the topological symmetry); the fugacity $u$ for the flavor symmetry of the chiral maps to the fugacity for the topological symmetry of the dual $U(1)$ theory. The free chiral maps to the monopole operator $V_{+}\equiv V_{+1}$ of the gauge theory.

We will see in section \ref{subsec:DoreyTong} that the identity \eref{basic_duality}, which encodes the duality between a free chiral and a $U(1)_{-1/2}$ theory with a charge $1$ chiral and maps a flavor symmetry to a topological symmetry, lies at the basis of mirror symmetry for 3d $\cN=2$ abelian Chern-Simons theories.

\subsubsection{$U(1)_0$ gauge theory with two charge $1$ chirals }
As a final example, let us consider a $U(1)$ gauge theory with two matter fields $X_1$, $X_2$ of charge $1$ and $R$-charge $1$, and vanishing bare Chern-Simons couplings. The flavor symmetry is $PSU(2)=SO(3)$, to which we associate a fugacity $y$.
For the sake of presentation in this section, we avoid the notation with the torsion element and instead introduce background magnetic charges $n^1=n/2$ and $n^2=-n/2$, with $n \in\bZ$, so that the effective masses of the matter fields are $m^1_{\rm eff}=m+ n/2$, $m^1_{\rm eff}=m- n/2$. Dirac quantization requires that $m \in \bZ + n/2$. Odd $n$ corresponds to having a nontrivial torsion element. It follows that the baryonic charge $B \in \bZ+n/2$ too, therefore we will set $B=\h B-n/2$ in the following. The Hilbert series reads
\be\label{HS_2flavors}
\begin{split}
	H(t,y,z;n,\h B) &= \sum_{m \in \bZ+\frac{n}{2}} z^m \oint \frac{dx}{2\pi i x} x^{-\h B + \frac{n}{2}} \left(x/y\right)^{-\frac{1}{2}|m-\frac{n}{2}|}
	\left(x y\right)^{-\frac{1}{2}|m+\frac{n}{2}|}\cdot \\
	&\qquad\qquad\qquad \cdot \PE[ \delta_{m,\frac{n}{2}} t x/y + \delta_{m,-\frac{n}{2}} t xy]~.
\end{split}
\ee

We evaluate \eref{HS_2flavors} for $n\geq 0$ by adding up the cases $m> \frac{n}{2}$, $m=\frac{n}{2}>0$, $|m|<\frac{n}{2}$, $m=-\frac{n}{2}<0$ and $m< -\frac{n}{2}$, and finally $m=0$ if $n=0$. The $n\leq 0$ case can be obtained noting that $H(t,y,z;n,\h B)=H(t,1/y,z;-n,\h B -n)$, that is inverting $y$ and changing sign to $n$ keeping $z$ and $B=\h B-n/2$ fixed. This corresponds to permuting the two flavors. The result is 
\be\label{HS_2flavors_result}
H(t,y,z;n,\h B) = \begin{cases}
	\left(z/y\right)^{\frac{n}{2}-\h B} y^{-\h B} + {\rm inverse}  & \qquad \h B<0 \wedge n > \h B \\
	\chi_{[n]}((z/y)^{1/2})   & \qquad \h B=0 \wedge n\geq 0 \\
	t^{\h B}\left(\left(z/y\right)^{\frac{n}{2}} y^{-\h B} + {\rm inverse} \right) & \qquad \h B>0 \wedge n> 0 \\
	t^{\h B} \chi_{[\h B]}(y)   & \qquad \h B\geq 0 \wedge n= 0  \\
	t^{\h B-n}\left((zy)^{-\frac{n}{2}} y^{\h B-n} + {\rm inverse} \right) & \qquad \h B> n  \wedge n< 0 	\\
	\chi_{[-n]}((zy)^{1/2})   & \qquad \h B=n\leq 0 ~,
\end{cases}
\ee
where $\chi_{[m]}(x)= \sum_{h=0}^m x^{-m+ 2h}$ is the character of the $(m+1)$-dimensional (spin $m/2$) representation of $SU(2)$. 

The right-hand side of \eref{HS_2flavors_result} has the following interpretation. We denote a monopole operator $V_{m;n,B}$ by its dynamical magnetic charge $m$ for the gauge $U(1)$ and background magnetic charges $n$ and $B$ for the flavor and topological symmetry. Then the first line of \eref{HS_2flavors_result} counts two gauge invariant monopole operators $V_{m=\pm B; n, B}$, corresponding to two isolated Coulomb vacua. 
The second line counts $n+1$ gauge invariant monopole operators $V_{m=-n/2+j; n, B=-n/2}$, $j=0,1,\dots,n$, that reconstruct an $SU(2)$ character and parametrize a $\bP^1$ Coulomb branch. The third line counts the two gauge invariants $V_{m=- n/2; n, B} X_1^{B+n/2}$ and $V_{m=+ n/2;n, B} X_2^{B+n/2}$, where $X_{1,2}$ are the matter fields, corresponding to two isolated Higgs vacua. The fourth line counts the $B+1$ gauge invariants $V_{m=0; n=0, B} X_1^{j} X_2^{B-j}$, $j=0,1,\dots,B$, that reconstruct an $SU(2)$ character and parametrize a $\bP^1$ Higgs branch. The fifth line counts two gauge invariants $V_{m=n/2;n, B} X_1^{B-n/2}$ and $V_{m=-n/2;n, B} X_2^{B-n/2}$,  corresponding to two isolated Higgs vacua. The sixth line counts the gauge invariant monopole operators $V_{m=n/2+j; n,B=n/2}$, $j=0,1,\dots,-n$, corresponding to a $\bP^1$ Coulomb branch. The common case of the second, fourth and sixth line, that is $\h B=n=0$, simply counts the identity operator, corresponding to a moduli space consisting of the origin only.

Our results are consistent with the semiclassical analysis of \cite{DoreyTong2000}. The phase diagram of the theory, encoded in the different lines of \eref{HS_2flavors_result}, has a natural interpretation in terms of a type IIB realization of the field theory in terms of a D3-brane interval suspended between two webs of five-branes \cite{Aharony:1997ju}. One such realization was discussed in Appendix A of \cite{DoreyTong2000}, and involves a D3-brane suspended between the following five-brane webs along a direction orthogonal to the two planes:
\bea
\begin{tikzpicture}[baseline, baseline,font=\footnotesize, scale=0.9, transform shape]
\draw[red] (-1,0.25)--(0,0.25); 
\draw[red] (-1,-0.25)--(0,-0.25); 
\draw[red] (0,0.25)--(0,-0.25);
\draw[red] (0,0.25)--(0.5,0.75);
\draw[red] (0,-0.25)--(0.5,-0.75);
\node at (-2,0.6) {\red{D5 = (1,0) 5-brane}};
\node at (0.7,-1) {\red{(1,1) 5-brane}};
\node at (1.8,0) {\red{NS5 = (0,1) 5-brane}};
\end{tikzpicture}
\qquad  %\qquad 
\text{and} \qquad %\qquad 
\begin{tikzpicture}[baseline, baseline,font=\footnotesize, scale=0.9, transform shape]
\draw[blue] (0,1)--(0,-1);
\node at (0,-1.5) {\blue{NS$5'$ = (0,1) $5'$-brane}};
\end{tikzpicture}
\eea  
Unprimed and primed five-brane systems are rotated with respect to one another, so that they intersect when the D3-brane interval collapses to a point.

As stressed in \cite{Cremonesi2011}, the same field theory can be also realized using
\bea
\begin{tikzpicture}[baseline, baseline,font=\footnotesize, scale=0.9, transform shape]
\draw[red] (-1,0)--(0,0); 
\draw[red] (0,0)--(0,-1); 
\draw[red] (0,0)--(1,1); 
\node at (-2,0.3) {\red{D5 = (1,0) 5-brane}};
\node at (1.5,1.5) {\red{(1,1) 5-brane}};
\node at (0,-1.5) {\red{NS5 = (0,1) 5-brane}};
\end{tikzpicture}
\qquad  %\qquad 
\text{and} \qquad %\qquad 
\begin{tikzpicture}[baseline, baseline,font=\footnotesize, scale=0.9, transform shape]
\draw[blue] (0,0)--(0,1); 
\draw[blue] (0,0)--(1,0); 
\draw[blue] (0,0)--(-1,-1); 
\node at (2,0.3) {\blue{D5 = (1,0) 5-brane}};
\node at (-1.5,-1.5) {\blue{(1,1) 5-brane}};
\node at (0,1.5) {\blue{NS5 = (0,1) 5-brane}};
\end{tikzpicture}
\eea
The advantage of this second brane configuration is to manifest %the toric structure of the continuous Higgs branch and 
the self-triality of the theory, that is realized by the subgroup of the $SL(2,\bZ)$ S-duality of type IIB string theory that interchanges $(1,0)$, $(0,1)$ and $(1,1)$ five-branes. 
The magnetic charges $n$ and $B$ (or rather the difference of the bare real masses of the matter fields and the FI parameter) are geometrized as the displacement of the two five-brane junctions in the vertical and horizontal directions.

The brane configurations corresponding to each line of \eref{HS_2flavors_result} are as follows.  Here the black dot and line denote the (toric base of the) moduli space of vacua, which coresponds to the allowed positions of the D3-brane.
\begin{table}[H]
	%\caption{default}
	\begin{center}
		\begin{tabular}{|c|c|c|c|}
			\hline
			Line of \eref{HS_2flavors_result} & The 1st config. & The 2nd config. & Comment \\
			\hline
			%%%%%%%%%%
			1st line &
			\begin{tikzpicture}[baseline, baseline,font=\footnotesize, scale=0.9, transform shape]
			\draw[red] (-1,0.25)--(0,0.25); 
			\draw[red] (-1,-0.25)--(0,-0.25); 
			\draw[red] (0,0.25)--(0,-0.25);
			\draw[red] (0,0.25)--(0.75,1);
			\draw[red] (0,-0.25)--(0.75,-1);
			\draw[blue] (0.3,1)--(0.3,-1);
			\node at (0.3,0.55) {$\bullet$};
			\node at (0.3,-0.55) {$\bullet$};
			\end{tikzpicture}  
			&
			\begin{tikzpicture}[baseline, baseline,font=\footnotesize, scale=0.9, transform shape]
			\draw[red] (-1,0.25)--(0,0.25); 
			\draw[red] (0,0.25)--(0,-1); 
			\draw[red] (0,0.25)--(0.75,1); 
			\draw[blue] (0.3,-0.25)--(0.3,1); 
			\draw[blue] (0.3,-0.25)--(1,-0.25); 
			\draw[blue] (0.3,-0.25)--(-0.45,-1);  
			\node at (0.3,0.55) {$\bullet$};
			\node at (0,-0.55) {$\bullet$};
			\end{tikzpicture}  
			&
			\shortstack{Two isolated \smallskip \\ Coulomb vacua} \\
			\hline
			%%%%%%%%%%
			%%%%%%%%%%
			2nd line &
			\begin{tikzpicture}[baseline, baseline,font=\footnotesize, scale=0.9, transform shape]
			\draw[red] (-1,0.25)--(0,0.25); 
			\draw[red] (-1,-0.25)--(0,-0.25); 
			\draw[red] (0,0.25)--(0,-0.25);
			\draw[red] (0,0.25)--(0.75,1);
			\draw[red] (0,-0.25)--(0.75,-1);
			\draw[blue] (0,1)--(0,-1);
			\draw[black,ultra thick] (0,0.25)--(0,-0.25);
			\end{tikzpicture}  
			&
			\begin{tikzpicture}[baseline, baseline,font=\footnotesize, scale=0.9, transform shape]
			\draw[red] (-1,0.25)--(0,0.25); 
			\draw[red] (0,0.25)--(0,-1); 
			\draw[red] (0,0.25)--(0.75,1); 
			\draw[blue] (0,-0.25)--(0,1); 
			\draw[blue] (0,-0.25)--(1,-0.25); 
			\draw[blue] (0,-0.25)--(-0.75,-1);  
			\draw[black,ultra thick] (0,0.25)--(0,-0.25);
			\end{tikzpicture}  
			&
			\shortstack{$\mathbb{P}^1$ Coulomb branch} \\
			\hline
			%%%%%%%%%%
			%%%%%%%%%%
			3rd line &
			\begin{tikzpicture}[baseline, baseline,font=\footnotesize, scale=0.9, transform shape]
			\draw[red] (-1,0.25)--(0,0.25); 
			\draw[red] (-1,-0.25)--(0,-0.25); 
			\draw[red] (0,0.25)--(0,-0.25);
			\draw[red] (0,0.25)--(0.75,1);
			\draw[red] (0,-0.25)--(0.75,-1);
			\draw[blue] (-0.3,1)--(-0.3,-1);
			\node at (-0.3,0.25) {$\bullet$};
			\node at (-0.3,-0.25) {$\bullet$};
			\end{tikzpicture}  
			&
			\begin{tikzpicture}[baseline, baseline,font=\footnotesize, scale=0.9, transform shape]
			\draw[red] (-1,0.25)--(0,0.25); 
			\draw[red] (0,0.25)--(0,-1); 
			\draw[red] (0,0.25)--(0.75,1); 
			\draw[blue] (-0.25,-0.25)--(-0.25,1); 
			\draw[blue] (-0.25,-0.25)--(1,-0.25); 
			\draw[blue] (-0.25,-0.25)--(-1,-0.95);  
			\node at (-0.25,0.25) {$\bullet$};
			\node at (0,-0.25) {$\bullet$};
			\end{tikzpicture}  
			&
			\shortstack{Two isolated \smallskip \\ Higgs vacua} \\
			\hline
			%%%%%%%%%%
			%%%%%%%%%%
			4th line &
			\begin{tikzpicture}[baseline, baseline,font=\footnotesize, scale=0.9, transform shape]
			\draw[red,ultra thick] (-1,0)--(0,0); 
			\draw[red] (0,0)--(1,1);
			\draw[red] (0,0)--(1,-1);
			\draw[blue] (-0.3,1)--(-0.3,-1);
			\node at (-0.3,0.0) {$\bullet$};
			\end{tikzpicture}  
			&
			\begin{tikzpicture}[baseline, baseline,font=\footnotesize, scale=0.9, transform shape]
			\draw[red] (-1.3,0)--(0,0); 
			\draw[red] (0,0)--(0,-1); 
			\draw[red] (0,0)--(1,1); 
			\draw[blue] (-0.3,0)--(-0.3,1); 
			\draw[blue] (-0.3,0)--(1,0); 
			\draw[blue] (-0.3,0)--(-1.3,-1);  
			\draw[black,ultra thick] (-0.3,0)--(0,0);
			\end{tikzpicture}  
			&
			\shortstack{$\mathbb{P}^1$ Higgs branch \\ (manifest in the 2nd config.)} \\ % \\ but not in the 1st one)} \\
			\hline
			%%%%%%%%%%
			%%%%%%%%%%
			5th line &
			\begin{tikzpicture}[baseline, baseline,font=\footnotesize, scale=0.9, transform shape]
			\draw[red] (-1,0.25)--(0,0.25); 
			\draw[red] (-1,-0.25)--(0,-0.25); 
			\draw[red] (0,0.25)--(0,-0.25);
			\draw[red] (0,0.25)--(0.75,1);
			\draw[red] (0,-0.25)--(0.75,-1);
			\draw[blue] (-0.3,1)--(-0.3,-1);
			\node at (-0.3,0.25) {$\bullet$};
			\node at (-0.3,-0.25) {$\bullet$};
			\end{tikzpicture}  
			&
			\begin{tikzpicture}[baseline, baseline,font=\footnotesize, scale=0.9, transform shape]
			\draw[red] (-1.55,-0.25)--(-0.5,-0.25); 
			\draw[red] (-0.5,-0.25)--(-0.5,-1); 
			\draw[red] (-0.5,-0.25)--(0.75,1); 
			\draw[blue] (-0.3,0.25)--(-0.3,1); 
			\draw[blue] (-0.3,0.25)--(0.75,0.25); 
			\draw[blue] (-0.3,0.25)--(-1.55,-1);  
			\node at (0,0.25) {$\bullet$};
			\node at (-0.8,-0.25) {$\bullet$};
			\end{tikzpicture}  
			&
			\shortstack{Two isolated \smallskip \\ Higgs vacua} \\ \hline 
			%%%%%%%%%%
			%%%%%%%%%%
			6th line &
			\begin{tikzpicture}[baseline, baseline,font=\footnotesize, scale=0.9, transform shape]
			\draw[red] (-1,0.25)--(0,0.25); 
			\draw[red] (-1,-0.25)--(0,-0.25); 
			\draw[red] (0,0.25)--(0,-0.25);
			\draw[red] (0,0.25)--(0.75,1);
			\draw[red] (0,-0.25)--(0.75,-1);
			\draw[blue] (0,1)--(0,-1);
			\draw[black,ultra thick] (0,0.25)--(0,-0.25);
			\end{tikzpicture}  
			&
			\begin{tikzpicture}[baseline, baseline,font=\footnotesize, scale=0.9, transform shape]
			\draw[red] (-1,-0.25)--(-0.25,-0.25); 
			\draw[red] (-0.25,-0.25)--(-0.25,-1); 
			\draw[red] (-0.25,-0.25)--(1,1); 
			\draw[blue] (0.25,0.25)--(0.25,1); 
			\draw[blue] (0.25,0.25)--(1,0.25); 
			\draw[blue] (0.25,0.25)--(-1,-1);  
			\draw[black,ultra thick] (-0.25,-0.25)--(0.25,0.25);
			\end{tikzpicture}  
			&
			\shortstack{$\mathbb{P}^1$ Coulomb branch} \\
			\hline
		\end{tabular}
	\end{center}
	%\label{default}
\end{table}%
In the first brane configuration, the 5th and 6th line are simply obtained from the 3rd and 2nd line by permuting the two half-D5 branes.

The common degenerate case of the second, fourth and sixth line, namely
\bea
\begin{tikzpicture}[baseline, baseline,font=\footnotesize, scale=0.9, transform shape]
\draw[red,ultra thick] (-1,0)--(0,0); 
\draw[red] (0,0)--(0.5,0.5);
\draw[red] (0,0)--(0.5,-0.5);
\draw[blue] (0,1)--(0,-1);
\node at (0,0.0) {$\bullet$};
\end{tikzpicture}  
\qquad \qquad {\rm or} \qquad \qquad 
\begin{tikzpicture}[baseline, baseline,font=\footnotesize, scale=0.9, transform shape]
\draw[red] (-1,0)--(0,0); 
\draw[red] (0,0)--(0,-1); 
\draw[red] (0,0)--(1,1); 
\draw[blue] (0,0)--(0,1); 
\draw[blue] (0,0)--(1,0); 
\draw[blue] (0,0)--(-1,-1); 
\node at (0,0.0) {$\bullet$};
\end{tikzpicture} \qquad,
\eea
corresponds to the single vacuum at the origin for the CFT.

\subsection{Dorey-Tong theories and mirror symmetry}\label{subsec:DoreyTong}

A class of 3d $\cN=2$ abelian Chern-Simons theories which have interesting toric moduli spaces and enjoy mirror symmetries that swap Coulomb and Higgs branches of dual pairs was studied by Dorey and Tong in \cite{DoreyTong2000} (see also \cite{Tong2000,AganagicHoriKarchEtAl2001}). These abelian $\cN=2$ theories can be obtained from abelian $\cN=4$ theories by gauging a $U(1)$ subgroup of the $SU(2)\times SU(2)$ $R$-symmetry of the latter \cite{Tong2000}. $\cN=2$ mirror symmetry then follows from $\cN=4$ mirror symmetry. 

Due to the $R$-gauging, the Chern-Simons levels of the $\cN=2$ theories are such that nontrivial Coulomb branches exist for vanishing FI parameters.
In this section we discuss the maximal dimensional Coulomb and Higgs branches of such theories and the equality of the moduli spaces of vacua of mirror pairs from the point of view of their Hilbert series.

The matter content consists of $N$ chiral multiplets $X^a$, which we all take to have $R$-charge $1$ (other $R$-charge assignment may be obtained by mixing with other symmetries). The bare flavor-flavor Chern-Simons levels of the ungauged theory are taken to be $k_{ab}=\frac{1}{2}\delta_{ab}$, so that the effective levels of the ungauged theory $k_{aa}^{\rm eff}=\frac{1}{2}(1+\sign(\mu^a))$ vanish for $\mu^a <0$. We also take $k_{Ra}=1$ for all $a$. We then gauge a $U(1)^r$ subgroup of the flavor symmetry, under which the matter fields have charges $Q^a_i$, $i=1,\dots,r$, as in section \ref{subsec:gauging}. The Chern-Simons levels involving the gauge group \eref{CS_gauge} then read
\be\label{CS_gauge_DT}
k_{ij}=\sum_{a} \frac{1}{2}Q^a_i Q^a_j~, \qquad k_{i\h j}=\sum_{a} \frac{1}{2}Q^a_i \h Q^a_{\h j}~, \qquad k_{i b}=\frac{1}{2}Q^b_i~. \qquad k_{iR}=\sum_{b} Q^b_i~,
\ee
Similar formulae hold for the flavor $U(1)^{N-r}$ group with charge matrix $\h Q^a_{\h i}$.

Following the prescription of section \ref{subsec:gauging}, the Hilbert series 
can be written as
\be\label{HS_DT}
\begin{split}
	&H(t,\h x,z;\h m, m_\Gamma, B) =\\
	& \quad = \sum_{(m_i) \in \bZ^r} \prod_{i=1}^r \left(z_i^{m_i} \oint \frac{dx_i}{2\pi i x_i}x_i^{-B_i} \right) \prod_{a=1}^N \left(t^2 x_a^{\rm eff}(x,\h x)\right)^{-\frac{1}{2}m^a_{\rm eff}(m,\h m, m_\Gamma)} \\ 
	& \quad\quad\cdot \prod_{a=1}^N x_a^{\rm eff}(x,\h x)^{-\frac{1}{2}|m^a_{\rm eff}(m,\h m, m_\Gamma)|} 
	\PE\big[ \sum_a \delta_{m^a_{\rm eff}(m,\h m, m_\Gamma),0} ~t x_a^{\rm eff}(x,\h x)\big]~, 
\end{split}
\ee
with $x_a^{\rm eff}$ and $m^a_{\rm eff}$ as defined in \eref{gauging_1}.

Note that since $x_a^{\rm eff}$ depends on $x_i$ only through $\prod_i x_i^{Q^a_i}$ and $m^a_{\rm eff}$ are integer, the electric charges of both monopole operators for the gauge symmetry and matter fields are in ${\rm span}_\bZ (Q^a)$, that is they take the form $\sum_a Q^a_i l_a$ where $l_a$ are some integers. Since the background monopole operator for the topological symmetry carries electric charge $-B$, the Hilbert series, that counts gauge invariant dressed monopole operators, vanishes unless $B\in {\rm span}_\bZ (Q^a)$, that is $B_i=-\sum_a Q^a_i \h n^a$, with $(\h n^a) \in \bZ^N$.%
\footnote{The minus sign is for later convenience. The $\h n^a$ can be taken of the form $\h n^a = \sum_a Q^a_i \h n_i + n^a_\Gamma$, with $(n_i)\in\bZ^r$ and $n_\Gamma \in \Gamma$, since terms proportional to $\h Q^a_{\h i}$ do not change $B_i$. \label{footnote_hn}}  
 We will only consider baryonic charges of this form in the following.  

Rather than discussing the evaluation of the Hilbert series as a function of the background magnetic charges, which reflects the dependence of the moduli space of vacua on the real masses and FI parameters, we focus in the following on the highest dimensional Coulomb and Higgs branch operators that arise for special choices.

\subsubsection{Coulomb branch}\label{subsec:DT_Coulomb}

In the interior of the maximal ({\emph{i.e.} $r$-) dimensional Coulomb branch, all matter fields vanish whereas the real scalars in the $r$ dynamical vector multiplets are moduli. The maximal dimensional Coulomb branch exists when all the effective FI parameters for the gauge symmetries vanish, so that all the vector multiplets are massless. 
The chiral operators that parametrize this Coulomb branch are undressed monopole operators. In our operator language, the effective baryonic charges for the gauge symmetry have to vanish to ensure that these undressed monopole operators are gauge invariant. The vanishing of the effective Chern-Simons couplings $k_{ij}^{\rm eff}$ requires $ m^a_{\rm eff}=\h m^a+\sum_i Q^a_i m_i\leq 0$ for all $a=1,\dots,N$, where $\h m^a \equiv \sum_{\h i}\h Q^a_{\h i} \h m_{\h i}+ m^a_\Gamma$. The mixed gauge-flavor effective Chern-Simons also vanish, therefore requiring $B_i=0$ for all $i=1,\dots,r$ ensures that the corresponding monopole operators are gauge invariant. Therefore, defining 
\be\label{nabla_Coulomb}
\nabla_\bZ(-\h m^a)= \{ (m_i) \in \bZ^r ~|~ \sum_i Q^a_i m_i \leq -\h m^a ~~~\forall a=1,\dots,N\}~,
\ee
the Hilbert series of the Coulomb branch is
\be\label{HS_Coulomb_DT}
\begin{split}
H_{C}(t,z_i;\h m^a) &= \sum_{(m_i) \in \nabla_\bZ(-\h m)} t^{-\sum_a(\h m^a + \sum_i Q^a_i m_i)} \prod_i z_i^{m_i}  ~. %\\
%&= t^{-\sum_a\h m^a} \sum_{m \in \nabla_\bZ(-\h m)}  \prod_i \left(t^{-\sum_a Q^a_i } z_i \right)^{m^i} ~.
\end{split}
\ee

Note that in the semiclassical analysis of the moduli space \cite{DoreyTong2000}, calling the bare real masses of the matter fields $\h \sigma^a \equiv \sum_{\h i}\h Q^a_{\h i} \h m_{\h i}$, the Coulomb branch is a toric $\mathbb{T}^r$ fibration over the base
\be\label{nabla_Coulomb_real}
\nabla_\bR(-\h \sigma)= \{ \sigma \in \bR^r ~|~ \sum_i Q^a_i \sigma^i \leq -\h \sigma^a ~~ \forall a=1,\dots,N\}~.
\ee
Its discretization \eqref{nabla_Coulomb} determines the gauge invariant monopole operators of the theory, in the presence of background magnetic charges $\h m^a$ for the flavor symmetry.

As an example, let us consider a $U(1)^2$ gauge theory with $3$ chirals. The charge matrix for the gauge and flavor symmetries are 
\be\label{example_charges}
Q=\begin{pmatrix}
	1 & -1 & 0 \\ 0 & 1 & -1
\end{pmatrix}~,\qquad\qquad 
\h Q=\begin{pmatrix}
	1 & 1 & 1
	\end{pmatrix}~,
\ee
and the torsion in the flavor magnetic charges is $\Gamma\cong\bZ_3$ generated by $m_{\Gamma,1}=(1,0,0)$,
so that $(\h m^a)=(\h m_{\h 1}+ \alpha, \h m_{\h 1}, \h m_{\h 1})$, with $\alpha=0,1,2$. Then, introducing a fugacity map and multiplying the Coulomb branch Hilbert series by an appropriate prefactor to bring the result to a more suggestive form, we find
\be
\hspace{-3pt} \prod_{i=1}^2 y_i^{-(\h m^i-\h m^{i+1})} H_C\Big(t,z_1=\frac{y_2}{y_1^2},z_2=\frac{y_1}{y_2^2};\h m^a\Big) = t^{-\sum_a \h m^a} [-\sum_a \h m^a,0]_{y_1,y_2}~,
\ee
where $[n,0]_y$ is a shorthand for the character of the representation $[n,0]$ of $SU(3)$, with $[1,0]_y=y_1 + \frac{y_2}{y_1} + \frac{1}{y_2}$. This result suggests that the $U(1)^2$ topological symmetry enhances to $SU(3)$ and reflects the fact that the Coulomb branch is a $\bP^2$ of K\"ahler class proportional to $-\sum_a \h \sigma^a$ \cite{Tong2000}.

\subsubsection{Higgs branch}\label{subsec:DT_Higgs}

On the maximal ($N-r$) dimensional Higgs branch, the matter fields take expectation value whereas scalars in the gauge vector multiplets do not. In terms of magnetic charges, we set $\h m^a=0$ for the flavor symmetries, and will focus on $m_i=0$. Then all the matter fields are massless and the Higgs branch is given by the quotient $\bC^N//_{\xi} \,U(1)^r$, where the $U(1)^r$ gauge group acts with charges $Q^a_i$ and the quotient is done at levels $\xi_i$, which translate into baryonic charges $B_i$ in our discussion of operators. Imposing the constraint on the baryonic charges that is necessary for the Hilbert series not to vanish, the Hilbert series of the Higgs branch reads
\bea\label{HS_Higgs_DT_1}
	&H_{H}(t,\h x_{\h i}; B_i=-\sum_a Q^a_i \h n^a) = \prod_{i=1}^r \left( \oint \frac{dx_i}{2\pi i x_i} x_i^{\sum_a Q^a_i \h n_a}\right) \PE[\sum_a t \prod_i x_i^{Q^a_i}\prod_{\h i} \h x_{\h i}^{\h Q^a_{\h i}}]=\nonumber \\
	&\qquad\qquad\qquad = \sum_{(l^a)\in \bZ_{\geq 0}^N} \left(\prod_i \delta_{-\sum_a Q^a_i \h n^a, \sum_a Q^a_i l^a} \right) 
	\prod_a  \bigg(t \prod_{\h i} \h x_{\h i}^{\h Q^a_{\h i}}\bigg)^{l^a}~,
\eea
where we expanded the $\PE$ in geometric series and integrated over the gauge group to reach the second line. The delta functions require that $l^a = -\h n^a - \sum_{\h i=1}^{N-r} \h Q^a_{\h i} n_{\h i}$, where $n_{\h i}$ are integers. The $l^a$ are nonnegative provided $(n_{\h i})$ belong to 
\be\label{delta_Higgs}
\Delta_\bZ(-\h n^a)= \{ (n_{\h i}) \in \bZ^{N-r} ~|~ \sum_{\h i} \h Q^a_{\h i} n_{\h i} \leq -\h n^a ~~~ \forall a=1,\dots,N\}~. 
\ee

The Higgs branch Hilbert series is therefore
\be\label{HS_Higgs_DT_2}
\begin{split}
	H_{H}(t,\h x_{\h i}; B_i=-\sum_a Q^a_i \h n^a) &= \sum_{(n_{\h i}) \in \Delta_\bZ(-\h n^a)}  \prod_a \bigg(t \prod_{\h i} \h x_{\h i}^{\h Q^a_{\h i}}\bigg)^{- (\h n^a+\sum_{\h i} \h Q^a_{\h i} n_{\h i})}~,
\end{split}
\ee
counting gauge invariant operators of the form $V_{m_i=0; B_i=\sum_a Q^a_i l^a} \prod_a X_a^{l_a}$, where $V_{m_i=0;B_i}$ is a background monopole operator for the topological symmetry and the matter fields appear with nonnegative powers $l^a$ as required by holomorphy. 

In the standard analysis of the moduli space, the $D$-term equations $\sum_a Q^a_i |X^a|^2 = \xi_i = -\sum_a Q^a_i \h \nu^a$ are solved by $|X^a|^2 = - (\h\nu^a+\sum_{\h i} \h Q^a_{\h i} \nu_{\h i})$ provided $(\nu_{\h i})_{\h i=1}^{N-r}$ belongs to  
\be\label{delta_Higgs_real}
\Delta_\bR(-\h \nu^a)= \{ (\nu_{\h i}) \in \bR^{N-r} ~|~ \sum_{\h i} \h Q^a_{\h i} \nu_{\h i} \leq -\h \nu^a ~~~ \forall a=1,\dots,N\}~.
\ee
This is the base of a toric fibration, the fibers of which are parametrized by the phases of $X^a$ modulo gauge equivalence.  Once again, the discretization \eqref{delta_Higgs} of the base \eqref{delta_Higgs_real} controls the gauge invariant operators that parametrize the Higgs branch. Note that if $B\neq 0$, as is required for instance to have a compact Higgs branch, the gauge invariant operators involve a background monopole operator for the topological symmetry, that cancels the electric charges of the matter fields. 
  
As an example, let us consider again the $U(1)^2$ gauge theory with $3$ chirals of charge matrices \eqref{example_charges}. Computing \eref{HS_Higgs_DT_2}, we obtain the Higgs branch Hilbert series
\be\label{Higgs_example}
H_{H}(t,\h x_{\h 1}; B_1=\h n^2 - \h n^1, B_2=\h n^3 - \h n^2) = (t \h x_{\h 1})^{\sum_a (\max(\h n^b)-\h n^a)} \PE[(t \h x_{\h 1})^3]~,
\ee
that counts gauge invariant operators of the form 
\be
V_{m_i=0;B_i=\h n^{i+1} - \h n^i} \prod_{a=1}^3 (X^a)^{\max(\h n^b)-\h n^a} (X^1 X^2 X^3)^{p}~, \qquad p \in \bZ_{\geq 0}~,
\ee
which parametrize a $\bC$ Higgs branch.

\subsubsection{Mirror symmetry}\label{subsec:DT_mirror}

The abelian $\cN=2$ Chern-Simons theories of \cite{DoreyTong2000} enjoy a mirror symmetry that relates a gauge theory with charge matrix $Q$ to  a gauge theory with charge matrix $\h Q$, where $\h Q$ is given by the integer kernel of $Q$ \cite{DoreyTong2000,Tong2000,AganagicHoriKarchEtAl2001}.%
\footnote{As we will see, the dual gauge group might include a discrete factor.} 
In this section we show how to relate the Hilbert series of mirror theories by applying to all chiral multiplets the basic duality discussed in section \ref{subsec:basic_duality} between a free chiral with global Chern-Simons couplings and a $U(1)$ Chern-Simons theory with one chiral and appropriate gauge Chern-Simons couplings.

We start with the ungauged theory of $N$ chirals with background flavor and $R$ Chern-Simons couplings, and apply to each chiral the basic duality \eref{basic_duality}:
\be\label{basic_duality_N_chirals}
\begin{split}
	&\prod_a(t^2 u_a)^{-\frac{1}{2}\mu^a} u_a^{-\frac{1}{2}|\mu^a|}\PE[\delta_{\mu^a,0}~tu_a]=\\
	& \qquad \qquad\qquad  =\prod_a \sum_{\nu^a \in\bZ} u_a^{\nu^a} \oint \frac{dv_a}{2\pi i v_a} v_a^{\mu^a}         (t^2 v_a)^{\frac{1}{2}\nu^a} v_a^{-\frac{1}{2}|\nu^a|}\PE[\delta_{\nu^a,0}~tv_a]~.
\end{split}
\ee
The mirror of theory $A$ consisting of $N$ free chirals is then theory $B$, a $U(1)^N$ Chern-Simons theory with $N$ chirals. 

To obtain more general mirror pairs, we gauge a $U(1)^r$ subgroup with charge matrix $Q$ of the $U(1)^N$ flavor symmetry of theory $A$, which corresponds to the topological symmetry of theory $B$. Gauging is achieved by substituting \eref{gauging_1} and Fourier transforming the $U(1)^r$ gauge group \eref{gauging_2} in both sides of  \eref{basic_duality_N_chirals}. On the left-hand side, we obtain the Hilbert series of theory $A$, a $U(1)^r$ Chern-Simons theory with $N$ chiral multiplets and charge matrix $Q$. To identify theory $B$ on the right-hand side, we perform the integration over $x_i$ and the summation over $m_i$, which lead to the delta functions
\bea
\sum_{m_i \in \bZ} \bigg( z_i \prod_a v_a^{Q^a_i} \bigg)^{m_i} &= 2\pi i \delta \bigg( z_i \prod_a v_a^{Q^a_i} - 1\bigg) \label{delta_fugacities} \\
\oint \frac{d x_i}{2\pi i x_i} x_i^{-B_i} x_i^{\sum_a Q^a_i \nu^a} &= \delta_{B_i, \sum_a Q^a_i \nu^a} ~.\label{delta_charges}
\eea

The delta functions determine the dual gauge group as follows. Defining the map
\be\label{mapU1}
\begin{split}
	\cQ:\quad  U(1)^N &\to U(1)^r \\
	 v_a~~& \mapsto \prod_a v_a^{Q^a_i}~,
\end{split}
\ee
the dual gauge group is the abelian group $\h G = \ker \cQ = U(1)^{N-r} \times \Gamma_Q$, where the $U(1)^{N-r}$ continuous factor has charge matrix $\h Q$, and there might also be a finite abelian multiplicative group $\Gamma_Q$. 
Indeed, the delta function \eref{delta_fugacities} imposes 
\be\label{v_a_solution}
v_a = V_a(z) \epsilon_a \prod_{\h i} y_{\h i}^{\h Q^a_{\h i}} ~,
\ee
where $(y_{\h i})$ are $U(1)^{N-r}$ fugacities, $V_a(z)$ is a particular solution of the inhomogeneous equation, \emph{i.e.}  $\prod_a V_a(z)^{Q^a_i}= z_i^{-1}$,  and the discrete $(\epsilon_a) \in \Gamma_Q$ satisfy $\prod_a \epsilon_a^{Q^a_i}= 1$ for all $i$. 

Note that, at this stage, $\Gamma_Q$ is distinct from the torsion group $\Gamma$ that was introduced in section \ref{subsec:gauging}}. In particular $\Gamma_Q$ depends on $Q$ only, whereas $\Gamma$ depends on both $Q$ and $\h Q$. However, using a discrete subgroup of the $U(1)^{N-r}$ freedom, one can further impose $\prod_a V_a(z)^{\h Q^a_{\h i}}= 1$ and similarly $\prod_a \epsilon_a^{\h Q^a_{\h i}}= 1$ for all $\h i$, so that $(\epsilon_a)$ subject to both $Q$ and $\h Q$ constraints now belong to the multiplicative finite abelian group associated to the torsion $\Gamma$ that we defined in section \ref{subsec:gauging}. 
Then we might say that the gauge group of theory $B$ is $U(1)^{N-r}\times \Gamma$, even though a common subgroup of $U(1)^{N-r}$ and $\Gamma$ does not act on the matter fields.  At any rate, once the delta function is imposed, the gauge group average reduces to an integration over $y$ and a summation over the allowed values of $\epsilon$. 

As a simple example, if $N=r=1$ with $Q=(q)$, then the dual gauge group is $\ker \cQ = \{e^{2\pi i n/q}, ~~n=0,1,\dots,q-1 \} = \bZ_q = \Gamma_Q = \Gamma$. An example where $\Gamma_Q$ and $ \Gamma$ differ is $N=2$, $r=1$, with $Q=(1,1)$. Then $\h Q = (1,-1)$ and $\Gamma=\bZ_2$. The dual group is $\ker \cQ= U(1)$ with charge matrix $\h Q=(1,-1)$, so $v_1= y z^{1/2}$ and $v_2=y^{-1} z^{1/2}$ where $y$ is the fugacity of the dual $U(1)$ group. Here $\Gamma_Q=\{1\}$, whereas $\Gamma=\bZ_2=\{\pm 1\}$ acts in the same way as a $\bZ_2$  subgroup of the $U(1)$ group with charge matrix $\h Q$. We are free to average over $ U(1)^{N-r} \times \Gamma$: this has the same effect as averaging over $\h G= U(1)^{N-r}\times \Gamma_Q$.

As for the magnetic charges, recalling that $B_i = -\sum_a Q^a_i \h n^a$ with $\h n^a\in \bZ$ is required in order for the Hilbert series not to vanish (see also footnote \ref{footnote_hn}), the delta function \eref{delta_charges} imposes 
\be\label{nu_a_solution}
\nu^a = - (\h n^a + \sum_{\h i} \h Q^a_{\h i} n_{\h i}) = - ( \sum_{\h i} \h Q^a_{\h i} n_{\h i} + \sum_i Q^a_i \h n_i + n^a_\Gamma)\equiv - n^a_{\rm eff}~,
\ee
and one is left with a sum over magnetic charges $(n_{\h i})\in \bZ^{N-r}$.

In conclusion, slightly overparametrizing the gauge group by a discrete factor that does not act on the dual matter fields, the Hilbert series of theory $B$ reads 
\be\label{HS_theoryB}
\begin{split}
	H^B&= \prod_{\h j} \h x_{\h j}^{-\sum_a \h Q^a_{\h j} n^a_\Gamma} \prod_a V_a(z)^{m^a_\Gamma} \cdot 	
		 \\
	&\quad \cdot \sum_{(n_{\h i})\in\bZ^{N-r}}
	\prod_{\h i} \h z_{\h i}^{n_{\h i}} \frac{1}{|\Gamma|} \sum_{\epsilon \in \Gamma} \prod_a \epsilon_a^{\h m^a} \prod_{\h i}\left(\oint\frac{d y_{\h i}}{2\pi i y_{\h i}} y_{\h i}^{-\h B_{\h i}}\right)  \\
	& \quad \cdot \prod_a \big(t^2 y_a^{\rm eff}\big)^{-\frac{1}{2}n^a_{\rm eff}}
	\prod_a \big(y_a^{\rm eff}\big)^{-\frac{1}{2}|n^a_{\rm eff}|} \cdot  \PE[\sum_a \delta_{n^a_{\rm eff},0}~ t y_a^{\rm eff}]~,
\end{split}
\ee
in terms of dual fugacities and charges \be\label{dual_data}
\begin{split}
	\h z_{\h i} &= \prod_{\h j} \h x_{\h j}^{-\sum_a \h Q^a_{\h j} Q^a_{\h i}}~, \qquad \h B_{\h i}= - \sum_a \h Q^a_{\h i} (\sum_{\h j}\h Q^a_{\h j} \h m_{\h j}+m^a_\Gamma)\equiv - \sum_a \h Q^a_{\h i} \h m^a\\
	y_a^{\rm eff} &= V_a(z)\epsilon_a \prod_{\h i} y_{\h i}^{\h Q^a_{\h i}}~, \qquad\qquad n^a_{\rm eff}= \sum_{\h i} \h Q^a_{\h i} n_{\h i} + \sum_i Q^a_i \h n_i + n^a_\Gamma~. 
\end{split}
\ee
The Hilbert series \eqref{HS_theoryB} of theory $B$ takes a similar form to the Hilbert series \eref{HS_DT} of theory $A$ that it is equal to. The only differences in form are an extra average over a finite gauge group and the prefactors in the first line which are due to torsion. 

It is straightforward to check by direct computation as in sections \ref{subsec:DT_Coulomb} and \ref{subsec:DT_Higgs} that the Hilbert series of the Coulomb (Higgs) branch of theory $B$ equals the Hilbert series of the Higgs (Coulomb) branch of theory $A$.

\section{Hilbert series of nonabelian 3d $\cN=2$ gauge theories}\label{sec:nonabelian}

The Hilbert series formalism introduced in the previous section can be also applied to nonabelian $\cN=2$ gauge theories, with some modifications. 

We first discuss the modifications due to classical and perturbative effects. For each simple factor $G$ of rank $r$ in the gauge group, one can introduce a Chern-Simons interaction $\frac{k}{4\pi} \int \Tr (A dA + \frac{2}{3}A^3 + \dots)$, where the ellipses denote superpartners. Fixing the normalization, this bare Chern-Simons term leads to the following factor in the integrand of the Hilbert series:
\be\label{CS_nonab}
\prod_{\alpha} (x^\alpha)^{-\frac{k}{2 h^\vee}\alpha(m)}= \prod_{\alpha} \bigg(\prod_{i=1}^r x_i^{\langle \alpha, \alpha_i^\vee \rangle}\bigg)^{-\frac{k}{2 h^\vee}\alpha(m)}= \prod_{\alpha>0} \bigg(\prod_{i=1}^r x_i^{\langle \alpha, \alpha_i^\vee \rangle}\bigg)^{-\frac{k}{h^\vee}\alpha(m)}~.
\ee
Here the product is over roots $\alpha$, $x$ are gauge fugacities, $h^\vee$ is the dual Coxeter number of the simple gauge group factor $G$  and $m$ is the magnetic charge, which is a weight of the Langlands or GNO dual $G^\vee$ of the gauge group \cite{Goddard:1976qe}, modulo Weyl reflections. 
In the second expression we have written $x^\alpha=\prod_i x_i^{\langle \alpha, \alpha_i^\vee \rangle}$ in the basis of fundamental weights, which are dual to the coroots $\alpha_i^\vee$. The third expression is in terms of positive roots only. For example, for an $SU(2)$ gauge group we have $(x^2)^{-\frac{k}{2}2m}=x^{-2km}$, where the magnetic charge $m$ is integer and nonnegative. (For $SO(3)$ gauge group the same expression holds but $m$ is half-integer.) For $SU(N)$ gauge group, we have $\prod_{i=1}^{N-1} x_i^{-k C_{ij}m_j}$ with $m_i\in \bZ_{\geq 0}$.

In the presence of dynamical magnetic charges $m$ for the gauge symmetry, the gauge group $G$ is broken to a \emph{residual gauge group} $H_m$ of rank $r$ like $G$, given by the commutant of the magnetic charge $m$ in $G$. The $W$-bosons supermultiplets with $\alpha(m)\neq 0$ are integrated out, and their gauginos lead to a shift of the $R$-charge of monopole operators
\be\label{R_monopole_nonab}
R^{\rm quant}_{gauge}(m) = - \sum_{\alpha>0}|\alpha(m)| ~.
\ee
Grouping this with the Haar measure, the $W$-boson supermultiplets contribute to the Hilbert series the factor
\be\label{Wbosons_nonab}
\prod_{\alpha>0} \bigg( t^{-|\alpha(m)|} (1-x^\alpha)^{\delta_{\alpha(m),0}}\bigg)~.
\ee
The Kronecker delta function reduces the Haar measure of $G$ to that of  $H_m$.

Matter fields transforming in a representation $\cR$ of the gauge group and $\h \cR$ of the flavor symmetry group acquire an \emph{effective mass} 
\be\label{eff_mass_nonab_0}
m_{\rm eff}^{\rho, \h \rho} = \rho(m)+\h \rho(\h m)~,
\ee
where $\rho$ and $\h\rho$ are weights of the gauge and flavor symmetry representations, in the presence of dynamical and background magnetic charges $m$ and $\h m$. Note that we are making the by now common abuse of terminology of referring to the magnetic charges as masses, in view of the BPS condition $\sigma\sim m/(2r)$. To be precise, the effective real mass of the component $X_{\rho,\h\rho}$ of the matter field $X$ with weights $\rho$, $\h\rho$ are due to the potential term
\be
\sum_{\substack{\rho\in \cR\\ \h\rho\in \h\cR}} | (\rho(\sigma)+\h\rho(\h\sigma))X_{\rho,\h\rho}|^2~.
\ee

Matter fields with $m_{\rm eff}^{\rho,\h\rho}=0$ in a monopole background are \emph {residual matter fields} and define a \emph{residual theory} $T_{m,\hat m}$. They  can take VEV and contribute a plethystic exponential to the Hilbert series. Matter fields with $m_{\rm eff}^{\rho,\h\rho}\neq 0$ in a monopole background are massive and are integrated out, leading to a quantum correction of the charges of monopole operators. Summarizing, matter fields of $R$-charge $r$ transforming in the representation $(\cR,\h\cR)$ contribute to the Hilbert series 
\be\label{matter_nonab}
\prod_{\substack{\rho \in \cR\\ 		\h\rho \in \h \cR}}\bigg( t^{r-1}x^{\rho} \h x^{\h \rho}  \bigg)^{-\frac{1}{2}|\rho(m)+\h \rho(\h m)|} \PE[\delta_{\rho(m)+\h \rho(\h m),0}t^r x^{\rho} \h x^{\h\rho}]~.
\ee
The product is over weights $\rho$ of the representation $\cR$ of the gauge symmetry, and $x^\rho = \prod_i x_i^{\rho_i} =  \prod_i x_i^{\langle \rho,\alpha_i^\vee\rangle}$ using the basis of fundamental weights (similarly for the flavor symmetry). The first factor in \eqref{matter_nonab} can be interpreted as a shift of the Chern-Simons levels, as in the abelian case. 

In particular, when matter fields in a representation $\cR$ of the gauge group $G$ acquire a large real mass $\h m_\cR$ associated to the flavor symmetry and are integrated out, they lead to a shift of the bare Chern-Simons level $k$ for the gauge group $G$. In the Hilbert series, this is due to the factor 
\be\label{1loop_shift_nonab}
\prod_{\rho \in \cR} (x^\rho)^{-\frac{1}{2}|\rho(m)+\h m_{\cR}|}\xrightarrow[\h m_{\cR} \to \pm \infty]{}  \prod_{\rho \in \cR} (x^\rho)^{-\frac{1}{2}\sign(\h m_\cR)(\rho(m)+\h m_{\cR})}~.
\ee
The bare  Chern-Simons level $k$ in \eref{CS_nonab} then receives the 1-loop shift
\be\label{1loop_shift_2}
k \to k + \frac{1}{2}\sum_{\cR}\sign(\h m_\cR) T_\cR~,
\ee
where $T_\cR$ is twice the Dynkin index of the matter field representation $\cR$, that is the quadratic Casimir normalized so that $T_{ad}=2h^\vee$ for the adjoint representation. (For instance, $T_{ad}=2N$ and $T_{fund}=1$ for  $SU(N)$.)

Finally, as in \cite{Cremonesi2015a} we need to take into account nonperturbative corrections that may lift perturbative vacua.  Instantons can induce a superpotential that lifts partially or totally a semiclassical Coulomb branch that exists if the effective Chern-Simons couplings vanish for simple gauge group factors. In addition, isolated semiclassical vacua of supersymmetric Chern-Simons theories can also be lifted: for instance, pure $\cN=2$ $SU(N)_k$ Chern-Simons theories break supersymmetry for $k<N$.
The net effect of these nonperturbative effects on the Hilbert series is to reduce the magnetic charge lattice from a Weyl chamber of the weight lattice of the dual group, $\Gamma_{G^\vee}/\cW_G$, to a sublattice $\Gamma_q$, that we dub the \emph{quantum sublattice}. In the next subsections we will analyze a few examples, incorporating nonperturbative corrections in our Hilbert series formalism. The general analysis of these nonperturbative corrections is an interesting open problem that we leave for future work.

\subsection{$SU(2)$ theories with doublets}\label{subsec:SU2}

In this subsection we compute the Hilbert series of certain $SU(2)$ gauge theories whose moduli spaces were first discussed in \cite{Tong2000}. Interesting phenomena such as nonperturbative superpotentials and quantum deformed moduli spaces will appear.

We begin by discussing the $SU(2)$ Yang-Mills theory with $N_f=1$ fundamental flavor, that is $n_f=2N_f=2$ doublets $Q_1$ and $Q_2$. The theory has a $U(1)_A\times SU(2)_F$ flavor symmetry. If the flavors are massless, instantons cannot generate a superpotential due to an excess of fermionic zero modes \cite{AharonyHananyIntriligatorEtAl1997,BoerHoriOz1997}: the semiclassical Coulomb branch is not lifted, and we sum over all magnetic charges $m\in \bZ_{\geq 0}$ for monopole operators $V_m$. The nontrivial charges of monopole operators are 
\be\label{monop_charges_SU2_1fl}
R[V_m]=-2rm~, \qquad\qquad A[V_m]=-2m~,
\ee
where $r$ is the $R$-charge of the flavors. The effective masses of matter fields are $\pm m$ for the two components of each doublet, therefore the residual theory is $SU(2)$ with two doublets if $m=0$, and a pure $U(1)$ theory if $m>0$. Adding up the two contributions, we obtain the Hilbert series 
\be\label{HS_SU2_1fl}
\begin{split}
	H(t,y,u)&=\oint \frac{dx}{2\pi i x} (1-x^2)\PE[t^r y(x+x^{-1})(u+u^{-1})]+ \sum_{m=1}^\infty (t^r y)^{-2m}=\\
	&=\PE[(t^r y)^2]+\PE[(t^r y)^{-2}]-1= \sum_{n \in \bZ}(t^r y)^n= 2\pi i \delta((t^r y)^2-1)~,
\end{split}
\ee
which does not depend on the $SU(2)_F$ fugacity $u$. The first $\PE$ in \eref{HS_SU2_1fl} counts nonnegative powers of the meson $M=M_{12}=\epsilon_{ij}Q^i_1 Q^j_2$, where $i$ and $j$ are $SU(2)$ gauge indices. The second $\PE$ counts monopole operators $V_m = (V_1)^m\equiv Y^m$, with $m\geq 0$. $1$ is subtracted in order not to overcount the identity operator. The two geometric series have different regions of convergence, but we can add them formally to obtain the final expression in terms of a delta function. The structure of the Hilbert series \eref{HS_SU2_1fl} shows that the moduli space is generated by the meson $M$ and the monopole operator $Y$, subject to the relation $YM=c$, with $c$ a constant. By the standard lore that what is not forbidden is compulsory, we assume that $c$ does not vanish, therefore it can be rescaled to $1$, leading to $YM=1$. The moduli space is thus algebraically a cylinder $\bC^*$, obtained by merging the Coulomb and mesonic branch quantum-mechanically \cite{AharonyHananyIntriligatorEtAl1997}.%
\footnote{This is the same structure as the moduli space of the pure $U(1)_0$ theory, where the Hilbert series is $H(z)=\sum_{m \in \bZ} z^m=2\pi i \delta(z-1)$. The moduli space consists of a Coulomb branch which is a cylinder, algebraically generated by the monopole operators $V_\pm\equiv V_{\pm 1}$ subject to $V_+ V_-=1$.}

Next, we turn on real masses, or in our language background magnetic charges, for the $U(2)_F \cong U(1)_A\times SU(2)_F$ flavor symmetry. We are interested in the case where the real masses are $-\mu$ and $+M$, with $M\geq \mu\geq 0$. We denote the background magnetic charges as $-n$ and $+N$, with $N\geq n \geq 0$. 
The charge of monopole operators under the Cartan of the $SU(2)$ gauge symmetry is
\be\label{el_SU2_1fl}
\begin{split}
Q[V_{m;-n,N}]&=-\frac{1}{2}(|m-n|-|-m-n|+|m+N|-|-m+N|)=\\
&=\begin{cases}
	0 & \quad 0\leq m \leq n\\
	-(m-n) & \quad n\leq m \leq N\\
	-(N-n) & \quad N\leq m 	
\end{cases}~.
\end{split}
\ee
The dependence of the electric charge on $m$  for $n\leq m\leq N$ signals the presence of an effective Chern-Simons coupling at level $1$.

The electric charges \eref{el_SU2_1fl} indicate the existence of a semiclassical Coulomb branch, parametrized by monopole operators of dynamical magnetic charges $0\leq m\leq n$ if $n \neq N$, and $m\geq 0$ if $n=N$. However, a superpotential is dynamically generated in the one instanton sector, lifting the semiclassical Coulomb branch $0\leq \sigma \leq \mu$ \cite{Tong2000}. This is because in the one instanton background the gauginos provide exactly two fermionic zero modes as required to generate a superpotential, but each quark $Q_a$ of real mass $\mu_a$ provides an extra fermionic zero mode for $\sigma\geq |\mu_a|$ that does not allow the superpotential to be generated. 

Restricting correspondingly the sum over magnetic charges to $m \in\Gamma_q=\bZ_{\geq n}$, the Hilbert series of the $SU(2)_0$ theory with $2$ doublets and background magnetic charges $n_1=-n$, $n_2=N$ for the $U(2)$ flavor symmetry is
\be\label{HS_SU2_bg}
\begin{split}
\hspace{-4pt}	H(t,y_1,y_2;-n,N)&= \sum_{m=n}^\infty \oint \frac{dx}{2\pi i x} (1-x^2)^{\delta_{2m,0}} t^{-2m}\cdot\\
	&\cdot \prod_{a=1}^{2}\prod_{s_a=\pm 1} (t^{r-1}x^{s_a}y_a)^{-\frac{1}{2}|s_am+n_a|}\PE[\delta_{s_am+n_a,0}t^r x^{s_a}y_a]\Big|_{\substack{n_1=-n\\ n_2=N}}
\end{split}
\ee
where $y_a$ are $U(2)$ flavor fugacities. For $0\leq n <N$ the Hilbert series \eref{HS_SU2_bg} 
\be
H(t,y_1,y_2;-n,N) = t^{-2n} (t^{r-1}y_1)^{-n} (t^{r-1}y_2)^{-N}
\ee
counts the bare monopole operator $V_{m=n;-n,N}$, which is made gauge invariant by averaging over the Weyl group, and corresponds to an isolated Coulomb vacuum. For $0<n=N$ the Hilbert series is
\be\label{HS_SU2_mass_deformed}
\begin{split}
&H(t,y_1,y_2;-n,N) = \\
&\qquad =(t^{2r}y_1 y_2)^{-n} \oint \frac{dx}{2\pi i x} \PE[t^r (x y_1+x^{-1} y_2)] +\sum_{m=n+1}^\infty (t^{2r}y_1 y_2)^{-m} = \\
&\qquad = (t^{2r}y_1 y_2)^{-n} \sum_{h\in \bZ}(t^{2r}y_1 y_2)^h =(t^{2r}y_1 y_2)^{-n} \cdot 2\pi i \delta(t^{2r}y_1 y_2-1)~.
\end{split}
\ee
This result has the following interpretation. The prefactor $(t^{2r}y_1 y_2)^{-n}$ corresponding to $h=0$ counts the bare monopole operator $V_{m=n;-n,n}$, made gauge invariant by averaging over the Weyl group. A term with $h>0$ counts the bare monopole operator dressed by the $h$-th power of the abelian meson $Q^1_1 Q^2_2$, and then made gauge invariant. A term with  $h<0$ counts the bare monopole operator $V_{m=n-h;-n;n}$, made gauge invariant. The form of the Hilbert series \eref{HS_SU2_mass_deformed} implies that, up to coefficients, these gauge invariant operators take the form $V_{m=n;-n,n} \cM^h$ for $h\geq 0$ and $V_{m=n;-n,n} \cY^{-h}$ for $h\leq 0$, with $\cM \cY=1$.  The moduli space is a cylinder arising from a quantum merger of a mesonic branch (generated by $\cM)$ and a Coulomb branch (generated by $\cY$), like for $n=0$ (in that case $\cM=M$ and $\cY=Y$). The result deduced from the computation of the Hilbert series \eref{HS_SU2_mass_deformed} is in exact agreement with the analysis of \cite{Tong2000}.

Incidentally, we note that the decoupling of matter fields with large real mass can be easily understood. Consider for instance the limit $N\to +\infty$ in \eref{HS_SU2_bg}. The massive doublet contributes the factor $x^{-m} (t^{r-1}y_2)^N$, due to radiative Chern-Simons terms. In particular, a Chern-Simons interaction at level $k=1/2$ is generated for the $SU(2)$ gauge group, according to \eref{1loop_shift_nonab}. The Hilbert series of the $SU(2)_{1/2}$ Chern-Simons theory with a single doublet is then obtained as 
\be
H'(t,y_2;-n)=\lim_{N\to + \infty} (t^{r-1}y_2)^N H(t,y_1,y_2;-n,N) = (t^{r+1}y_1)^{-n}~.
\ee

Finally, let us briefly discuss as in \cite{Tong2000} the addition of a vectorlike pair of doublets with equal and opposite real masses $\t n_1 =\t n$ and $\t n_2 =-\t n$, with $0\leq \tilde n \leq n$. The effective masses of the components of the two doublets are therefore $\pm(\pm m +\tilde n)$. The addition of this vectorlike pair of doublet matter fields has a few interesting consequences: the $R$-charges of monopole operators are changed, a new mesonic branch opens up on top of $m=\tilde n$, and most importantly the quantum sublattice is now $\Gamma_q=\bZ_{\geq \tilde n}$. Therefore the Coulomb branch is parametrized by monopole operators of magnetic charge $\tilde n\leq m\leq n$. Adding the vectorlike pair of doublets to the $SU(2)_{1/2}$ theory with $1$ doublet discussed above, the Hilbert series is
\be\label{HS_SU2_mass_deformed_2}
\begin{split}
	&H(t,y_1,\t y_1,\t y_2;-n,\t n, -\t n) = \\
	&\qquad =t^{-2\t n}(t^{r-1}y_1)^{-n}(t^{2(r-1)}\t y_1 \t y_2)^{-\t n} \oint \frac{dx}{2\pi i x} \PE[t^r (x^{-1} \t y_1+x \t y_2)] +\\
	&\qquad +\sum_{m=\t n}^n t^{-2m}(t^{r-1}y_1)^{-n}(t^{2(r-1)}\t y_1 \t y_2)^{-m} - t^{-2\t n}(t^{r-1}y_1)^{-n}(t^{2(r-1)}\t y_1 \t y_2)^{-\t n}=\\
	&\qquad =(t^{3r-1}y_1\t y_1 \t y_2)^{-n} \PE[t^{2r}\t y_1 \t y_2]~.
\end{split}
\ee
The result shows how the Higgs branch $\bC$ (corresponding to the second line of \eref{HS_SU2_mass_deformed_2}), and the Coulomb branch $\bP^1$ (corresponding to the first entry in the third line), which semiclassically intersect at a point (hence the negative term in the third line), merge into a single branch $\bC$ (the fourth line) due to a quantum deformation. Once again, the Hilbert series is consistent with the analysis of \cite{Tong2000}.

\subsection{The duality appetizer}\label{subsec:appetizer}

We now analyze from the point of view of the Hilbert series the duality appetizer of \cite{JafferisYin2011}: an $SU(2)_1$ Chern-Simons theory with an adjoint $\Phi$, dual to a free chiral $X=\Tr(\Phi^2)$ plus a topological sector. While the Hilbert series is not sensitive to topological vacua, which arise when the theory is defined on a Riemann surface \cite{Intriligator:2013lca}, it captures related information through discrete torsion. For that purpose, it will be useful to distinguish the theories with gauge group $SU(2)$ and $SO(3)$. Crucial in our analysis will be the input that pure $\cN=2$ $SU(2)_{1}$ Chern-Simons theory breaks supersymmetry, whereas $\SU(2)_3$ does not (similarly for $SO(3)$).

Let us first consider $SO(3)_1$ Chern-Simons theory with an adjoint $\Phi$ of $R$-charge $1$ and flavor charge $1$, and background Chern-Simons levels $k_{FF}=1/2$, $k_{RF}=0$. Since the dual group of $SO(3)$ is $SU(2)$, the magnetic charge is $m\in \frac{1}{2}\bZ_{\geq 0}$, corresponding to half-integer spin $m$. (This is to be contrasted with $G=SU(2)$, in which case $m \in \bZ_{\geq 0}$.) 
Monopole operators of half-odd $m$ are charged under the topological symmetry $\bZ_2=Z(SU(2))$. We introduce a fugacity $\epsilon$ such that $\epsilon^2=1$ for the topological symmetry, and weigh the sum over the magnetic charge lattice by $\epsilon^{2m}$. We also introduce a fugacity $y$ and an integer background magnetic charge $n$ for the $U(1)$ flavor symmetry that rotates $\Phi$. The Hilbert series of the $SO(3)_1$ theory is given by 
\be\label{SO3_appetizer}
\begin{split}
	H_{SO(3)}(t,y,\epsilon;n) &= \sum_{m \in \Gamma_q(n)} \epsilon^{2m} y^{-\frac{1}{2} n} \oint\frac{dx}{2\pi i x} x^{-2m} \left(1-x^2\right)^{\delta _{2 m,0}} t^{-2 m}\cdot\\
	&\cdot \left(x^2 y\right)^{-\frac{1}{2} \left| 2 m+n\right| } y^{-\frac{1}{2}|n| } \left(x^{-2} y\right)^{-\frac{1}{2} \left|-2 m+n\right| } \cdot \\
	& \cdot \PE[\delta _{2 m+n,0} t x^2 y+\delta_{n,0} t y + \delta_{-2 m+n,0} t x^{-2} y ] ~,
\end{split}
\ee
where the sum is over the quantum sublattice $\Gamma_q(n)$ of $\frac{1}{2}\bZ_{\geq 0}$, which we now determine. 

Since the effective theory obtained by integrating out massive fields has always a nonvanishing effective Chern-Simons level, there is no semiclassical Coulomb branch therefore we need not worry about instanton corrections. (The power of $x$ in the integrand of \eqref{SO3_appetizer} is a piecewise linear function which is nowhere locally constant.) We need to take into account supersymmetry breaking, however. 
If $m\neq 0$, the residual gauge group is $U(1)$, and supersymmetry is preserved regardless of the value of the effective CS level. If $m=0$ instead, the residual gauge group is $SU(2)$, and supersymmetry is broken if $|k_{\rm eff}(m=0,n)|<2$ when there is no residual matter \cite{Bergman1999}. Since $k_{\rm eff}(m=0,n)=1+2\sign(n)$ in our theory, there is no supersymmetric vacuum corresponding to $m=0$ and $n<0$, whereas a supersymmetric vacuum at the origin exists for $n\geq 0$. Thus the quantum sublattice is 
\be\label{Gamma_q_app}
\Gamma_q(n)= \begin{cases}
	\frac{1}{2}\bZ_{>0}~, & n<0\\
	\frac{1}{2}\bZ_{\geq 0}~, & n\geq 0
\end{cases}~.
\ee

Evaluating  \eref{SO3_appetizer} using \eref{Gamma_q_app}, we find the Hilbert series
\be\label{HS_SO3_app_res}
H_{SO(3)}(t,y,\epsilon;n)=\begin{cases}
	(t^2 y^2)^n~, & n<0\\
	\PE[t^2 y^2] ~,& n=0\\
	y^{-2n}~, & n>0
\end{cases}\quad 
= t^n (t y^2)^{-\frac{1}{2}|2n|}\PE[\delta_{2n,0} t^2 y^2]
\ee
which equals the Hilbert series of a free chiral field $X$ of $R$-charge and flavor charge $2$, to be identified as $X=\Tr(\Phi^2)$, with global Chern-Simons couplings at levels $k_{RF}=-1$, $k_{FF}=0$. 
We note that the result \eref{HS_SO3_app_res} does not depend on the $\bZ_2$ fugacity $\epsilon$, provided the magnetic charge $n$ for the flavor symmetry is an integer, as we assumed to have integer flavor charges for gauge variant operators.%
\footnote{If the integrality of flavor charges is only required for gauge invariants, then $n\in \frac{1}{2}\bZ$ is allowed and the result gets multiplied by the sign $\epsilon^{n-|n|}$. 	The result of \eref{SU(2)_appetizer} would then be modified too. }

Next, we move to the $SU(2)_1$ theory with an adjoint. The magnetic charge $m$ is now an integer rather than a half-integer, but we can turn on a torsion magnetic charge $m_\Gamma\in\{0,1\}$, by replacing $m\to m+\frac{1}{2}m_\Gamma$ in all formulae. Hence trivial (nontrivial) torsion in the $SU(2)$ case corresponds to $m$ integer (half-odd) in the $SO(3)$ case. The Hilbert series of the $SU(2)$ theory is obtained by projecting the Hilbert series of the $SO(3)$ theory \eref{SO3_appetizer} to even or odd $\bZ_2$ sectors:
\be\label{SU(2)_appetizer}
\begin{split}
	H_{SU(2)}(t,y;n;m_\Gamma) &= \frac{1}{2}\sum_{\epsilon=\pm 1} \epsilon^{m_\Gamma}H_{SO(3)}(t,y,\epsilon;n) = \\
	&=\begin{cases}
		H_{SO(3)}(t,y,1;n)~, & m_{\Gamma}=0 \\
		0~, & m_{\Gamma}=1
	\end{cases}~.
\end{split}
\ee
This result can be explained on the dual free chiral side by a topological sector given by a pure $U(1)_2/\bZ_2$ CS theory, whose Hilbert series is
\be
H_{top}(\sigma;m_\Gamma)=\sum_{m \in \bZ_{\geq 0}} \sigma^m \oint \frac{du}{2\pi i u} u^{-2m-m_\Gamma}=\begin{cases}
	1~, & m_\Gamma=0\\	0~, & m_\Gamma=1
\end{cases}~,
\ee
where $m_{\Gamma}$ is a $\bZ_2$ baryonic charge, or background magnetic charge for a $\bZ_2$ topological symmetry with fugacity $\sigma$.

\subsection{$U(N)$ theories with fundamentals and antifundamentals}
\label{subsec:UN}

We close this section by computing the Hilbert series of $U(N)_k$ Chern-Simons theories with $N_f$ fundamental and $N_a$ antifundamental flavors, setting all background magnetic charges to zero  for simplicity.
The gauge Chern-Simons level $k$ satisfies  $k+\frac{1}{2}(N_f-N_a) \in \bZ$ due to a parity anomaly. Here we only consider theories whose Seiberg dual group has nonnegative rank, so that supersymmetry is unbroken. We refer to \cite{Benini2011a} for details of the duality and a review of this class of theories.

We first compute the mesonic Hilbert series of $U(N)$ with $N_f$ fundamentals and $N_a$ antifundamentals, setting the $R$-charges of squarks to $1$ for simplicity:
\be\label{mes}
\begin{split}
&H^{\rm mes}_{N,N_f,N_a}(t,y,u,v)=\prod_{i=1}^N \oint \frac{dx_i}{2\pi i x_i} \prod_{i<j}(1-\frac{x_i}{x_j}) \PE\bigg[t y\big(\sum_{a=1}^{N_f} \frac{x_i}{u_a}+\sum_{\t a=1}^{N_a} \frac{v_{\t a}}{x_i}\big)\bigg]=\\
& \qquad\qquad= \sum_{n_1,\dots,n_N=0}^\infty  [0^{N_f-N},n_N,\dots,n_1; n_1,\dots,n_N,0^{N_a-N}]_{u,v} (ty)^{\sum_j {j n_j}}~,
\end{split}
\ee
where $y$ is a $U(1)_A$ fugacity, $u$ and $v$ are $SU(N_f)$ and $SU(N_a)$ flavor fugacities, subject to $\prod_a u_a=\prod_{\t a} v_{\t a}=1$, and we denoted characters of the flavor symmetry group by the Dynkin labels of the representation. 
The mesonic chiral ring is generated by the $N_a\times N_f$ meson matrix $M^{\t a}_a = \t Q^{\t a}_i Q^i_a$ of rank at most $N$. We note that the mesonic moduli space is a Calabi-Yau cone only when $N_f=N_a$. 

In addition to the mesons counted in \eref{mes}, there are gauge invariant chiral dressed monopole operators if a Coulomb branch exists, that is if  $k^+=0$ or $k^-=0$, where  
\be\label{kpm}
k^\pm = k \pm \frac{1}{2}(N_f-N_a)
\ee
are the effective Chern-Simons levels. The case where both $k^+$ and $k^-$ vanish (that is $k=0$ and $N_f=N_a$) was studied in \cite{Cremonesi2015a}, therefore we focus here on the remaining cases, where at most one of $k^+$ and $k^-$ vanish. The complex dimension of this Coulomb branch is $1$ rather than $N$, due to an instanton generated superpotential that lifts the remaining $N-1$ directions. (See \cite{Cremonesi2015a} for a detailed discussion of these effects in the case $k=0$ and $N_f=N_a$.)
Taking into account the perturbative effects due to Chern-Simons couplings and the nonperturbative effects due to instantons, the quantum sublattice $\Gamma_q$ corresponding to the unlifted Coulomb branch is 
\be
\Gamma_q=\{(m_1,0,\dots,0,m_N)\in \bZ^N| ~k^+ m_1=0 \land k^- m_N=0 ~ \}~.
\ee

Adding up the mesonic contribution and the contribution of dressed monopole operators, the full Hilbert series of the $U(N)_k$ theory with $N_f$ fundamentals and $N_a$ antifundamentals (with $k\neq 0$ or $N_f\neq N_a$) is then
\be\label{HS_SQCD_chiral}
\begin{split}
&H_{N,N_f,N_a}(t,y,u,v,z)= H^{\rm mes}_{N,N_f,N_a}(t,y,u,v) +\\
&\qquad\qquad + (\delta_{k^+,0}\sum_{m=1}^\infty a_+^m + \delta_{k^-,0} \sum_{-m=1}^\infty a_-^{-m} ) H^{\rm mes}_{N-1,N_f,N_a}(t,y,u,v)~,
\end{split}
\ee
where the sums in the second line count monopole operators $V_+^m\equiv V_{(m,0^{N-1})}$ and $V_-^{-m}\equiv V_{(0^{N-1},m)}$, dressed by mesons of a residual $U(N-1)$ theory with $N_f$ fundamentals and $N_a$ antifundamentals. We denoted by 
\be\label{apm_SQCD}
a_\pm=z^{\pm 1} t^{-(N-1)} y^{\mp k_{gA}-\frac{1}{2}(N_f+N_a)}
\ee
the fugacity weights of $V_\pm$. $k_{gA}$ is the mixed Chern-Simons level between the central gauge $U(1)$ and the axial $U(1)_A$ symmetry, which is quantized to ensure the exponents of $y$ in \eref{apm_SQCD} are integer.

We conclude from this analysis that the chiral ring of the theory is generated by the $N_a\times N_f$ meson matrix $M$, of rank at most $N$, and possibly by bare monopole operators $V_+$ or $V_-$ (if $k^+$ or $k^-$ vanish), subject to the extra relations that the rank of $V_\pm M$ is at most $N-1$. Note that, as in \cite{Cremonesi2015a}, we reached this conclusion with no need of  postulating a singular dynamical generated superpotential.

\section{ABJM and its variants}\label{ABJM}

In this section we consider the variants of the ABJM theory \cite{Aharony:2008ug,Aharony:2008gk,GaiottoTomasiello2010} with gauge group $U(N_1)_{k_1} \times U(N_2)_{k_2}$, bifundamental fields described by the quiver 
\bea\label{abjmquiver}
\begin{tikzpicture}[baseline, baseline,font=\footnotesize, scale=0.9]
\begin{scope}[auto,%
every node/.style={draw, minimum size=1.5cm}, node distance=2cm];
% the vertices
\node[circle] (USp2k) at (-0.1, 0) {$U(N_1)_{k_1}$};
\node[circle, right=of USp2k] (BN)  {$U(N_2)_{k_2}$};
\end{scope}
\draw[draw=black,solid,line width=0.5mm,->>]  (USp2k) to[bend right=30] node[midway,above] {$A_{1,2}$}node[midway,above] {}  (BN) ; 
\draw[draw=black,solid,line width=0.5mm,<<-]  (USp2k) to[bend left=30] node[midway,above] {$B_{1,2}$} node[midway,above] {} (BN) ;    
\end{tikzpicture}
\eea
\\
and superpotential
\bea \label{supA1}
W= \tr( A_1 B_1 A_2 B_2 - A_1 B_2 A_2 B_1)  ~.
\eea
There is a global symmetry $SU(2)\times SU(2)\times U(1)_R$ under which the fields $A_i$ and $B_i$ transform as $({\bf 2}, {\bf 1})_{1/2}$ and $({\bf 1}, {\bf 2})_{1/2}$, respectively.
There is also a topological symmetry $U(1)_M$ corresponding to the diagonal abelian factor in the gauge group under which matter fields are neutral. The topological symmetry for the other abelian factor acts trivially.
When $k_1=-k_2$ the theory has $\cN=6$ supersymmetry but it useful to discuss it in $\CN =2$ notation.  

We can write the Hilbert series for the ABJM theory using the formalism discussed in Section \ref{sec:nonabelian}, with  some simplifications.
First, in ABJM there are no nonperturbative corrections and there is no corresponding lifting of the Coulomb branch. Secondly,  we set to zero all the background charges. With these simplifications, the Hilbert series  takes the form of a sum over monopole operators $V_{{\vec m},\tilde {\vec m}}$, where ${\vec m}$ and $\tilde {\vec m}$ are the magnetic charges of  $U(N_1)$ and $U(N_2)$, respectively, dressed by fields in the residual matter theory $T_{{\vec m},\tilde {\vec m}}$. The gauge charge of the monopole $V_{{\vec m},\tilde {\vec m}}$ is
\bea - {\vec B}({\vec m},  \tilde {\vec m}) = - ( k_1 {\vec m}, k_2 \tilde {\vec m}) =  - (B^{\rm eff}_{i,1}({\vec m}),B^{\rm eff}_{i,2}(\tilde {\vec m}) ) \, ,\eea
and therefore it should be dressed by residual matter field with electric charge  ${\vec B}({\vec m} ,  \tilde {\vec m})$. Notice that ${\vec B}({\vec m} ,  \tilde {\vec m})$ gives an homomorphism from   the magnetic lattice to the center of the residual group $H_{{\vec m},\tilde {\vec m}}$. Therefore, the electric charges of the matter fields live in the abelian part of $H_{{\vec m},\tilde {\vec m}}$
and can be interpreted as {\it baryonic charges} for the residual  theory  $T_{{\vec m},\tilde {\vec m}}$, as discussed in details  in the examples below. Compared with Section \ref{sec:nonabelian}, we have the complication that the residual theory $T_{{\vec m},\tilde {\vec m}}$ has a non-trivial superpotential. This implies that, sometimes, we will need to use  Macaulay2 \cite{M2} to evaluate the   Hilbert series of $T_{{\vec m},\tilde {\vec m}}$. 

In what follows,  it is convenient  to rescale the fugacity associated with the superconformal $R$-charge (or scaling dimension) $t\rightarrow t^2$. In this way the fields $A_i$ and $B_i$, with dimension $1/2$, are weighted by a factor of $t$. The monopole  $V_{{\vec m},\tilde {\vec m}}$ is then weighted by a factor $t^{2 R({\vec m} ; \tilde {\vec m})}$, where 
$R({\vec m} ; \tilde {\vec m})$ is its $R$-charge, which can be computed by combining the contributions \eref{quantum_correct_R} and (\ref{R_monopole_nonab}) and reads
\bea
\label{dimformulamonopole}
R({\vec m}; \tilde {\vec  m}) =   \sum_{a=1}^{N_1}\sum_{b=1}^{N_2} |m_a -   \tilde m_b| -  \sum_{1\le a<b\le N_1} |m_a-m_b|  -   \sum_{1\le a<b\le N_2} |\tilde m_a-\tilde m_b |~.
\eea 

\subsection{The ABJM theory: $k_1=-k_2=k$ and $N_1=N_2=N$}

It is well known from its brane realization that the moduli space of the ABJM theory is $\Sym^N(\BC^4/\BZ_k)$. We now recover this result in our formalism. The case $N=1$ offers no surprise and
it is  well known. The case $N>1$ is strongly connected with a conjecture made in \cite{Forcella:2007wk}. 

\subsubsection{The Hilbert series for $N=1$}

For $N=1$ the gauge theory is $U(1)_k\times U(1)_{-k}$. The $R$-charge of a bare monopole operator with  magnetic fluxes $(m,\tilde m)$ is
\bea
R(m,\tilde m) =  |m-\tilde m|
\eea
and its $U(1)\times U(1)$ gauge charge is 
\bea
(-k m, k \tilde m)\, .
\eea
The gauge group is unbroken by the magnetic fluxes, being Abelian.  We can dress the monopole with residual bifundamental fields which remain massless:
\bea m A_{1,2} =A_{1,2} \tilde m \, ,\quad\qquad B_{1,2}\tilde m = m B_{1,2} \, .\eea
For $\tilde m \ne m$ these equations cannot be satisfied and we cannot construct any gauge invariant, since %we have 
no fields %that 
can compensate the gauge charge of the monopole.
For $\tilde m = m$, on the other hand, the bare monopole operator can be dressed by bifundamental fields. The residual theory is the conifold quiver with gauge group $U(1)\times U(1)$. Notice that the $R$-charge of the bare monopole operator vanishes for $\tilde m=m$.

Let us introduce some notations that will be useful in the following. Given the conifold quiver \cite{Klebanov:1998hh}
\bea
\label{conifoldequalrank}
\begin{tikzpicture}[baseline, font=\footnotesize, scale=1]
\begin{scope}[auto,%
every node/.style={draw, minimum size=1cm}, node distance=2cm];
% the vertices
\node[circle] (USp2k) at (-0.1, 0) {$U(r)$};
\node[circle, right=of USp2k] (BN)  {$U(r)$};
\end{scope}
% the edges
\draw[draw=black,solid,line width=0.5mm,->>]  (USp2k) to[bend right=30] node[midway,above] {$A_{1,2}$}node[midway,above] {}  (BN) ; 
\draw[draw=black,solid,line width=0.5mm,<<-]  (USp2k) to[bend left=30] node[midway,above] {$B_{1,2}$} node[midway,above] {} (BN) ;    
\end{tikzpicture}
\eea
with gauge group $U( r)\times U( r)$, no Chern-Simons level, and superpotential (\ref{supA1}), we denote with
\bea
g_r (t,x,y; B)
\eea
the generating function for  operators with charge $(B,-B)$ under the $U(1)\times U(1)$ center of the gauge group. $g_r (t,x,y; B)$ can
be interpreted as the generating function for the $SU( r)\times SU( r)$ conifold theory in  a sector with baryonic charges $(B,-B)$ and, for this reason, we call it a {\it baryonic generating function}.%
\footnote{$A_i$ and $B_i$ are normalized to carry baryonic charges $(1,-1)$ and $(-1,1)$ respectively.} As discussed in Section \ref{subsec:HS_bg_charges}, $g_r (t,x,y; B)$ can be also interpreted as the Hilbert series for the $U( r)\times U( r)$ conifold theory with background magnetic charges $(-B,B)$ for the topological symmetry $U(1)\times U(1)$. We call  $x$ and $y$ the fugacities for the $SU(2)\times SU(2)$ global symmetry.

The Hilbert series for $N=1$ is then given by 
\bea\label{ABJM_abelian}
H_{N=1,k}(t, x,y,z ) = \sum_{m=-\infty}^\infty g_{1}(t, x,y; km) z^{m}
\eea
where $z$ is a fugacity for the topological symmetry $U(1)_M$.%

Since at rank one there is no superpotential, the baryonic generating function  $g_{1}(t; x,y; B) $ can be
evaluated using the Molien formula
\bea
g_1(t, x,y; B) &= \oint_{|b| =1} \frac{d b}{2 \pi i b^{B+1}} \PE \Big[ (x+x^{-1}) bt + (y+y^{-1})b^{-1} t \Big]~. \label{g1tBABJMref}
\eea

Note that the unrefined expression for $g_1$ is given by
\bea
g_1(t, x=1, y=1; B) &= \frac{t^{|B|}(1+|B|+t^2-|B|t^2)}{(1-t^2)^3}~. \label{g1tBABJM}
\eea
We can easily re-interpret this formula in terms of operators. Consider $B>0$. The generic operator in the $U(1)\times U(1)$ conifold theory with baryonic charge
$(B,-B)$ is 
\bea
A_{i_1}\cdots A_{i_{n+B}} \, B_{j_1}\cdots B_{j_n}
\eea 
for arbitrary $n\ge 0$. There are $(n+B+1)(n+1)$ such inequivalent operators and the corresponding Hilbert series is
\bea
\sum_{n=0}^\infty (n+B+1) (n+1) t^{2 n+B}
\eea
which resums to (\ref{g1tBABJM}). Restoring $x$ and $y$ replaces the dimensions $(n+B+1) (n+1)$ by characters of the representations $[n+B;n]$ of $SU(2)_x\times SU(2)_y$ \cite{Forcella:2007wk}. The case $B<0$ is obtained by exchanging the role of $A_i$ and $B_i$. 

It is easy to show that $H_{N=1,k}$ is the  Hilbert series of $\BC^4/\BZ_k$. Firstly, from \eref{ABJM_abelian} and the expression of $g_1$ in terms of $SU(2)$ characters one finds that the Hilbert series of the moduli space of the abelian ABJM theory at level $1$ is given by
\bea\label{C4}
H_{N=1,k=1}(t, x,y,z ) =\frac{1}{(1- t z x)(1- t z/x)(1- t z^{-1} y)(1- t z^{-1} /y)} \,, 
\eea
that is the Hilbert series of $\bC^4$. Then one uses the identity 
\be
\frac{1}{k}\sum_{i=0}^{k-1} \omega_k^{im} =\begin{cases}
	1~, & m\in k\bZ\\
	0~, & m\notin k\bZ
\end{cases}
\ee 
to show that 
\be\label{C4/Zk_2}
\begin{split}
	\hspace{-7pt}H_{N=1,k}(t,x,y,z)&= \sum_{m' \in\bZ} g_{1}(t, x,y; km')z^{m'} =\sum_{m\in\bZ} g_{1}(t, x,y; m) z^{m/k} \frac{1}{k} \sum_{i=0}^{k-1} \omega_k^{im} \\
	&= \frac{1}{k}\sum_{i=0}^{k-1}H_{N=1,k=1}(t, x,y,z^{1/k} \omega_k^i )~.
\end{split}
\ee
Since the average in the last expression realizes the $\bZ_k$ quotient, we see that $H_{N=1,k}$ is the Hilbert series of $\bC^4/\bZ_k$, as expected. 

The chiral ring is generated by the mesons $x_{ij}=A_i B_j$ and the dressed monopole operators $u_{i_1 \dots i_k} = T A_{i_1}\dots A_{i_k}$, $v_{j_1 \dots j_k} = \t T B_{j_1}\dots B_{j_k}$. The ideal of chiral ring relations is generated by the conifold relation $x_{11}x_{22}=x_{12}x_{21}$ involving the mesons, and the relations $u_{i_1 \dots i_k} v_{j_1 \dots j_k} = \prod_{a=1}^k x_{i_a j_a}$ involving the monopole operators, which follow from $T \t T=1$.

\subsubsection{The Hilbert series for $N=2$}

Let ${\vec m}= (m_1, m_2)$ and $\tilde {\vec  m} =(\tilde m_1, \tilde m_2)$ be the magnetic charges for $U(2)_k$ and $U(2)_{-k}$ respectively.  The $R$-charge of the monopole operator is
\bea
\label{dimformula}
R({\vec m}; \tilde {\vec  m}) =  \sum_{a,b=1}^2 |m_a - \tilde m_b| -   |m_1-m_2|  -   |\tilde m_1-\tilde m_2|~.
\eea
We can dress the monopole with matter fields according to our general prescription
\bea
\label{ABJMHilbert2}
H_{N=2,k}(t,x,y,z) =\sum_{m_1 \geq m_2}\,\,  \sum_{\tilde m_1\geq \tilde m_2} t^{2 R({\vec m}; \tilde {\vec m}) }  z ^{\frac{1}{2}\sum_{i=1}^2 (m_i+\tilde m_j)} g_{T_{({\vec m},\tilde {\vec m})}}(t,x,y; B( {\vec m},\tilde {\vec m}))
\eea
where the dressing factor is the baryonic generating function for the residual theory $T_{({\vec m}, \tilde {\vec m})}$ which has  gauge group $H_{({\vec m},\tilde {\vec m})}$ equal to the commutant of $({\vec m},\tilde {\vec m})$ inside $U(2)\times U(2)$ and fields consisting of the subset of the original matter fields $X=\{A_I,B_i\}$  that satisfy $\rho_X({\vec m},\tilde {\vec m})\circ X=0$.  The baryonic charges for the abelian factors are given   by the embedding ${\vec B}( {\vec m},\tilde {\vec m}) = ( k\, {\vec m} \, ,\,  -k\, \tilde {\vec m})$ into $H_{({\vec m},\tilde {\vec m})}$.
We  introduced fugacities $x$ and $y$ for the $SU(2)\times SU(2)$ global symmetry and %a fugacity 
$z$ for the topological symmetry.

The commuting conditions 
\bea
&m_a (A_{1,2})_{ab} = (A_{1,2})_{ab} \tilde m_b \\
&\tilde m_a (B_{1,2})_{ab} = (B_{1,2})_{ab}  m_b 
\eea
imply that the matter fields can take VEV only if some eigenvalues of ${\vec m}$ and $\tilde {\vec m}$ are paired. We will show in the next section that because a diagonal gauge $U(1)$ does not act on the matter fields, gauge invariants can only be constructed if $\tilde {\vec m} ={\vec m}$. 

There are then two possible cases to be considered:
\ben
\item $({\vec m}; \widetilde{\vec m})=(m, m; m, m)$, with $m \in \BZ$. The residual gauge symmetry is $U(2) \times U(2)$ and
\bea
R({\vec m}; \widetilde{\vec m}) = 0~.
\eea
The gauge charge of the monopole operator under the abelian factors is specified by the embedding
\bea
(-km, -km; km, km)~\, ,
\eea
and selects the sector of baryonic charge $(2 k m, -2 k m)$ in the $U(2)\times U(2)$ theory. 
The Hilbert series for this case is thus equal to
\bea
H^{(1)}_{N=2,k}(t, x,y ) = \sum_{m=-\infty}^\infty g_{2}(t,x,y; 2km) z^{2 m} \, .
\eea
The expression for $g_2(t; x,y; B)$ has been computed in \cite{Forcella:2007wk}.  An explicit expression is (for simplicity, we set the fugacities $x$ and $y$ to unity)
\be
\begin{split}
	&g_2(t, x=1, y=1; B) \\
	&= \oint_{|b|=1} \frac{d b}{2\pi i b^{B+1}} \frac{b^2 t^8+\frac{t^8}{b^2}-3 b^2 t^6-\frac{3 t^6}{b^2}+4 t^8-3 t^6+t^4+t^2+1}{\left(1-t^2\right)^3 \left(1-\frac{t^2}{b^2}\right)^3 \left(1-b^2 t^2\right)^3}
\end{split}
\ee
where the integrand is given by (3.47) of \cite{Forcella:2007wk} with $t_1 =t b$ and $t_2 = t b^{-1}$. The integrand is 
the Hilbert series of the $SU(2)\times SU(2)$ Klebanov-Witten theory \cite{Klebanov:1998hh}. It is obtained by first finding the Hilbert series for the space of solutions of the $F$-term equations for the superpotential (\ref{supA1}) using the software {\tt Macaulay2} \cite{M2}, and  averaging it over the nonabelian gauge group $SU(2)\times SU(2)$. Since the Hilbert series of the $SU(2)\times SU(2)$ theory only depends on $b^2$, the baryonic generating function $g_2$ vanishes unless $B\in 2\bZ$.

As noticed in \cite{Forcella:2007wk},  $g_2(t,x,y; 2B)$ is the 2nd symmetric power of the baryonic generating function for the abelian gauge groups with a fixed $B$:
\be
\begin{split}
	g_2(t, x,y; 2B) &= {\rm Sym}^2 \left ( g_1(t,x,y; B) \right )  \\
	&= \frac{1}{2} \Big[ g_1(t, x, y; B)^2 + g_1(t^2, x^2, y^2; B) \Big]~.
\end{split}
\ee 
%%%%%%%%%%%%
\item $({\vec m}; \widetilde{\vec m})=(m_1, m_2; m_1, m_2)$ with $m_1 \neq m_2$. The residual gauge symmetry is $U(1)^2 \times U(1)^2$ and
\bea
R({\vec m}; \widetilde{\vec m}) = 0~.
\eea
The gauge charge of the monopole operator under the abelian factors is
\bea
(-km_1, -km_2; km_1, km_2)~.
\eea
The Hilbert series for this case is therefore 
\bea
H^{(2)}_{N=2,k}(t, x,y,z) = \sum_{m_1 > m_2} g_1(t, x, y ;km_1) g_1(t;x ,y ;km_2)  z^{m_1+m_2}
\eea
where $g_1(t, x,y; B)$ is given by \eref{g1tBABJMref} and the summand is the baryonic generating function when the gauge group is broken to $U(1)^2 \times U(1)^2$.
\een
The Hilbert series of the moduli space of the $U(2)_k\times U(2)_{-k}$ ABJM theory is the sum of the two contributions above:
\bea
H_{N=2,k}(t, x,y, z) = H^{(1)}_{N=2,k}(t, x,y, z)+ H^{(2)}_{N=2,k}(t, x,y, z)~.
\eea

We find that $H_{N=2,k}(t, x,y, z)$ is in fact the Hilbert series of the 2nd symmetric product of $\BC^4/\BZ_k$. The $N=2$ case has an $SU(4)$ symmetry \cite{Forcella:2008bb} which 
can be made manifest in the expression of the Hilbert series by using the fugacity map
\bea
y_1 = xz,\qquad y_2 = z^2, \qquad y_3 = y z~,
\eea
where $y_1$, $y_2$ and $y_3$ are $SU(4)$ fugacities. We provide examples for low $k$ below.

\paragraph{Examples.}
For $k=1$, we have
\be
\begin{split}
	H_{k=1}(t,\vec y)  &= H[\Sym^2 (\BC^4)](t;,\vec y) \\
	&= \frac{1}{2} \Big( \PE[ (y_1+y_2y_1^{-1} + y_3 y_2^{-1} + y_3^{-1})t]^2  \\
	& \qquad +\PE[ (y_1^2+y_2^2y_1^{-2} + y_3^2 y_2^{-2} + y_3^{-2}))t^2] \Big) \\
	&= \PE[ [1,0,0] t + [2,0,0]t^2 -  [0,2,0] t^4 + \ldots]~, \\
	H_{k=1}(t, \vec y=1) &= 1 + 4 t + 20 t^2 + 60 t^3 + 170 t^4 + 396 t^5 + 868 t^6 + \ldots~.
\end{split}
\ee
For $k=2$, we have
\be
\begin{split}
	H_{k=2}(t, \vec y)  &= H[\Sym^2 (\BC^4/\BZ_2)](t, \vec y) \\
	&= \frac{1}{2} \Big( H[\BC^4/\BZ_2](t, \vec y) ^2 + H[\BC^4/\BZ_2](t^2, \vec y^2)\Big)~,
\end{split}
\ee
where
\bea
H[\BC^4/\BZ_2](t, \vec y) = \frac{1}{2} \Big( \PE[ (y_1+y_2y_1^{-1} + y_3 y_2^{-1} + y_3^{-1})t] +( t \rightarrow -t) \Big)~.
\eea
Explicitly, the first few terms are given by
\be
\begin{split}
	H_{k=2}(t, \vec y)  &= \PE[ [2,0,0] t^2 + [4,0,0]t^4  - ([2,2,0]+[0,0,2]) t^6 +  \\
	& \qquad+ ([1,2,1]+[1,0,1]-[0,4,0]-[4,2,0])t^8 + \ldots]~, \\
	H_{k=2}(t, \vec y=1) &= 1 + 10 t^2 + 90 t^4 + 434 t^6 + 1635 t^8 + 4876 t^{10} + \ldots~.
\end{split}
\ee

\subsubsection{The Hilbert series for $N=2$ -- half ABJM} 

The previous computation relied on the use  of a computer software in order to take into accounts the $F$-term constraints and becomes more and more
involved if not impossible to perform for larger values of $N$. We can make an analytic computation in the case where we set $A_2=B_2=0$ and we
count only operators involving the fields $A_1$ and $B_1$. Following \cite{Forcella:2007wk}, we call this sub-branch  of the theory {\it half-ABJM}.

Since the global symmetry $SU(2)\times SU(2)$ is broken by our choice $A_2=B_2=0$, we set $x=y=1$. 
The $F$-term equations following from the superpotential 
(\ref{supA1}) are trivial for $A_2=B_2=0$ and the fields $A_1$ and $B_1$ are unconstrained. The baryonic generating functions $g_r(t,B)$ can then be explicitly  computed using a Molien integral from the Hilbert series freely generated by $A_1$ and $B_1$. We have
\bea
g^{{\rm ABJM}/2}_1(t; B) =  \oint_{|b|=1} \frac{d b}{2\pi i b^{B+1}} \PE \left[(b + b^{-1}) t\right] = \frac{t^{|B|}}{1-t^2}~, \label{abelianhalfABJM}
\eea
which can be easily interpreted as counting the operators $A_1^{n+B} B_1^n$ for $B>0$ and $A_1^n B_1^{n-B}$ for $B<0$.
Moreover
\be
\begin{split}
	g^{{\rm ABJM}/2}_2(t; 2 B) &= \oint_{|b|=1} \frac{d b}{2\pi i b^{2 B+1}} \oint_{|z_1|=1} \frac{d z_1}{2\pi i z_1} (1-z_1^2) \oint_{|z_2|=1} \frac{d z_2}{2\pi i z_2} (1-z_2^2) \\
	& \qquad \PE \left[ (b+ b^{-1}) (z_1+z_1^{-1})(z_2+z_2^{-1})t \right] = \frac{ t^{2 |B|}}{(1-t^2)(1-t^4)}~, \label{ghalfABJM2}
\end{split}
\ee
where $z_1, z_2$ are fugacities for the gauge group $SU(2) \times SU(2)$ and $b$ is the fugacity for the (non-decoupled) $U(1)$ gauge symmetry.
The result can be understood for $B>0$ as counting the operator $(\det A_1)^B$ dressed by powers of $\tr A_1B_1$ and  $\tr A_1B_1A_1B_1$ without
constraints. The result for $B<0$ is obtained by exchanging $A_1$ and $B_1$. 

Notice that indeed %the baryonic Hilbert series 
$g^{{\rm ABJM}/2}_2(t; 2 B)$ % for $N=2$ half-conifold 
is the 2nd symmetric power of %the baryonic Hilbert series  
$g^{{\rm ABJM}/2}_1(t; B)$: % of $N=1$ half-conifold,
\bea\label{2ndsym_g_half}
g^{{\rm ABJM}/2}_2(t; 2B)  = \frac{1}{2} \left[ g^{{\rm ABJM}/2}_1(t; B)^2 + g^{{\rm ABJM}/2}_1(t^2; B)  \right]~. 
\eea

The $N=1$ Hilbert series for half-ABJM is 
\bea
H_{N=1,k}(t,z) = \sum_{m=-\infty}^\infty g_{1}(t; km) z^{m} = \frac{1-t^{2 k}}{(1-t^2)(1-t^k z)(1-t^k/z)}\, ,
\eea
which is indeed the Hilbert series for $\BC^2/\BZ_k$. The generators are $u=TA_1^k$, $v=\t T B_1^k$ and $w=A_1 B_1$, subject to the relation $uv=w^k$.

The $N=2$ Hilbert series for half-ABJM is given by two contributions
\be
\begin{split}
	H_{N=2,k}(t,z) &= H^{(1)}_{N=2,k}(t,z)+ H^{(2)}_{N=2,k}(t,z) = \\
	&= \sum_{m=-\infty}^\infty g_{2}(t; 2km) z^{2 m} + \sum_{m_1 > m_2} g_1(t ;km_1) g_1(t;km_2)  z^{m_1+m_2}\, .
\end{split}
\ee
An explicit computation shows that it is the 2nd symmetric power of $H_{N=1,k}(t,z)$:
\bea
H_{N=2,k}(t,z) =  \frac{1}{2} \left[ H_{N=1,k}(t,z)^2 + H_{N=1,k}(t^2,z^2) \right]~.
\eea

\subsubsection{The Hilbert series for arbitrary $N$}\label{sec:arbitrary_N}

Let ${\vec m}= (m_1, m_2, \cdots)$ and $\tilde {\vec  m} =(\tilde m_1, \tilde m_2, \cdots)$ be the magnetic charges for $U(N)_k$ and $U(N)_{-k}$ respectively.  The $R$-charge of a bare monopole operator is
\bea
\label{dimformula}
R({\vec m}; \tilde {\vec  m}) =  \sum_{a,b=1}^N |m_a - \tilde m_b| -  \sum_{1\le a<b\le N} |m_a-m_b|  -  \sum_{1\le a<b\le N} |\tilde m_a-\tilde m_b|~.
\eea

The monopole operators can be dressed with matter fields according to our general prescription
\bea
\label{ABJMHilbert}
H_{N,k}(t,x,y,z) =\sum_{m_1 \ge m_2\ge \cdots}\,\,  \sum_{\tilde m_1\ge \tilde m_2\ge \cdots} t^{2 R({\vec m}; \tilde {\vec m}) }  z ^{\frac{1}{2}\sum_{i=1}^N (m_i+\tilde m_j)} g_{T_{({\vec m},\tilde {\vec m})}}(t,x,y; B( {\vec m},\tilde {\vec m}))
\eea
where the dressing factor is the baryonic generating function for the residual theory $T_{({\vec m}, \tilde {\vec m})}$ which has  gauge group $H_{({\vec m},\tilde {\vec m})}$ equal to the commutant of $({\vec m},\tilde {\vec m})$ inside $U(N)\times U(N)$ and fields consisting of the subset of the original matter fields that satisfy $\rho({\vec m},\tilde {\vec m})=0$.  The baryonic charges for the abelian factors are given   by the embedding ${\vec B}( {\vec m},\tilde {\vec m}) = ( k\, {\vec m} \, ,\,  -k\, \tilde {\vec m})$ into $H_{({\vec m},\tilde {\vec m})}$.
We  introduced fugacities $x$ and $y$ for the $SU(2)\times SU(2)$ global symmetry and %a fugacity 
$z$ for the topological symmetry.

The condition $\rho_X({\vec m},\tilde {\vec m})\circ X=0$ for the fields $X\in \{A_i,B_i\}$ becomes 
\bea
&m_a (A_{1,2})_{ab} = (A_{1,2})_{ab} \tilde m_b \\
&\tilde m_a (B_{1,2})_{ab} = (B_{1,2})_{ab}  m_b 
\eea
which implies $(A_{1,2})_{ab}=(B_{1,2})_{ab} =0$ if $m_a \ne \tilde m_b$. Since we need to turn on the bifundamentals $A$ and $B$ to form gauge invariants, we see that the fluxes in ${\vec m}$ and $\tilde {\vec m}$ must be paired: each integer flux $m$ in ${\vec m}$ should correspond to an equal flux in $\tilde{\vec m}$ . If the number $m$ appears in ${\vec m}$ and $\tilde {\vec m}$ with multiplicity $r$ and $\tilde r$, respectively, 
the reduced theory $T_{({\vec m}, \tilde {\vec m})}$ contains a subquiver isomorphic to the conifold quiver
\bea
\label{conifold}
\begin{tikzpicture}[baseline, font=\footnotesize, scale=1]
\begin{scope}[auto,%
every node/.style={draw, minimum size=1cm}, node distance=2cm];
% the vertices
\node[circle] (USp2k) at (-0.1, 0) {$U(r)$};
\node[circle, right=of USp2k] (BN)  {$U(\tilde r)$};
\end{scope}
% the edges
\draw[draw=black,solid,line width=0.5mm,->>]  (USp2k) to[bend right=30] node[midway,above] {$A_{1,2}$}node[midway,above] {}  (BN) ;  
\draw[draw=black,solid,line width=0.5mm,<<-]  (USp2k) to[bend left=30] node[midway,above] {$B_{1,2}$} node[midway,above] {} (BN) ;    
\end{tikzpicture}
\eea
of which we need to compute the  generating function in the sector of baryonic charges $(k r m, -k \tilde r m)$. Since the overall $U(1)$ in this sub-quiver is decoupled and no field is charged under it, 
the baryonic generating function is nonvanishing only for $\tilde r= r$. Since the fluxes in ${\vec m}$ and $\tilde {\vec m}$ are paired and have the same multiplicity we conclude that the  sum in (\ref{ABJMHilbert}) is restricted to ${\vec m}=\tilde {\vec m}$.

The Hilbert series  (\ref{ABJMHilbert}) drastically simplifies for ${\vec m}=\tilde {\vec m}$ since the $R$-charge (\ref{dimformula}) of a monopole operator vanishes, $R({\vec m};  {\vec  m})=0$. 
By denoting $r_i$  with $\sum_i r_i=N$ the multiplicities of equal fluxes in ${\vec m}=\tilde {\vec m}$,  the residual theory $T_{({\vec m},  {\vec m})}$ has gauge group $\prod_i U(r_i)^2$ and it is a collection of conifold quivers (\ref{conifold}) with equal ranks.

More explicitly, a flux of the form 
\bea
{\vec m} = \tilde {\vec m} = (\underbrace{m_{r_1}=m_{r_1}=\cdots =m_{r_1}}_\text{$r_1$},\underbrace{m_{r_2}=m_{r_2}=\cdots =m_{r_2}}_\text{$r_2$}, \cdots ) 
\eea
contributes to the Hilbert series  (\ref{ABJMHilbert}) the factor
\bea
\label{Hilbertnew}
z^{\sum_\alpha r_\alpha m_{r_\alpha}}  \prod_\alpha g_{r_\alpha} (t,x,y; k r_\alpha m_{r_\alpha})  
\eea
where, following the previous notations,  we denoted with
\bea
g_r (t,x,y; B)
\eea
the  generating function of the conifold quiver (\ref{conifoldequalrank})  in the sector with baryonic charge $(B,-B)$. 

The crucial ingredient now is the observation made in \cite{Forcella:2007wk} that, for the conifold theory,  the baryonic generating function $g_{r} (t,x,y; r B)$ is the $r$-fold symmetric product of
$g_{1} (t,x,y; B)$:
\bea\label{baryonic_sym}
g_{r} (t,x,y; r B) = {\rm Sym}^r (g_{1} (t,x,y; B)) \, ,
\eea 
where ${\rm Sym}^r f(t_1, t_2, \ldots, t_n)$ is the coefficient of $\nu^r$ in a power series expansion of
\bea
\PE \left[\nu f(t_1, t_2, \ldots, t_n) \right] = \exp \left( \sum_{p=1}^\infty \frac{1}{p}\nu^p f(t_1^p, t_2^p, \ldots, t_n^p) \right)~.
\eea
This observation follows from a brane construction and has been tested  for small values of $r$. It stands as a  conjecture for large $r$.% 
\footnote{For half-ABJM it is easy to show that $g_r^{\rm ABJM/2}(t;rB)=\frac{t^{r|B|}}{\prod_{i=1}^r (1-t^{2i})}={\rm Sym}^r\left( \frac{t^{|B|}}{1-t^2} \right)$. }

The fact that the baryonic functions are symmetric products of the abelian ones implies that the Hilbert series for ABJM with ranks $N$ is the $N$-fold  symmetric product of the ABJM theory with $N=1$. Indeed, the Hilbert series is obtained by summing  the contributions \eref{Hilbertnew} for $m_1\ge m_2\ge \cdots \ge m_N$. 
When all the fluxes $m_i$ are different, we are just counting the symmetric products of the states encoded in the abelian baryonic functions $g_1(t;k m_i)$. When some of the fluxes are equal, say for example $m_1=m_2=m$, we insert the contribution $g_2(t;2km)$ which again counts the symmetric product of the states in two copies of the abelian 
function $g_1(t;k m)$. We thus have
\bea
\hspace{-4pt} H_{N,k}(t,x,y,z)  = {\rm Sym}^N \left ( \sum_{m=-\infty}^\infty z^{m} g_1(t,x,y; k m) \right ) = {\rm Sym}^N \left ( H_{1,k}(t,x,y,z)\right ) \, .
\eea
A more detailed proof is provided in Appendix \ref{multiple_branes}.

\subsection{Fractional branes and arbitrary Chern-Simons levels}

The theory \eref{abjmquiver} with arbitrary ranks and Chern-Simons couplings has $\CN=2$ supersymmetry. The Hilbert series is given by 
\be\label{ABJMHilbertuneq}
\begin{split}
H_{N_1,N_2,k_1,k_2}(t,x,y,z) &=\sum_{m_1 \ge m_2\ge \cdots}\,\,  \sum_{\tilde m_1\ge \tilde m_2\ge \cdots} t^{2 R({\vec m}; \tilde {\vec m}) }  z ^{\frac{1}{2}(\sum_{i=1}^{N_1} m_i + \sum_{i=1}^{N_2} \tilde m_j)} \cdot\\
 &\qquad \qquad \cdot g_{T_{({\vec m},\tilde {\vec m})}}(t,x,y; B( {\vec m},\tilde {\vec m}))
\end{split}
\ee
where  ${\vec B}( {\vec m},\tilde {\vec m}) = ( k_1\, {\vec m} \, ,\,  k_2\, \tilde {\vec m})$ and the $R$-charge of a bare monopole operator is
\bea
\label{dimformulauneq}
R({\vec m}; \tilde {\vec  m}) =  \sum_{a=1}^{N_1}\sum_{b=1}^{N_2} |m_a - \tilde m_b| -  \sum_{1\le a<b\le N_1} |m_a-m_b|  -  \sum_{1\le a<b\le N_2} |\tilde m_a-\tilde m_b|~.
\eea
The condition $\rho({\vec m},\tilde {\vec m})=0$ can be again solved only if the fluxes in ${\vec m}$ and $\tilde{\vec m}$ are paired. If the number $m$ appears in ${\vec m}$ and $\tilde {\vec m}$ with multiplicity $r$ and $\tilde r$, respectively, 
the residual theory $T_{({\vec m}, \tilde {\vec m})}$ contains a subquiver isomorphic to (\ref{conifold}), of which we need to compute the  generating function is the sector of baryonic charges $(k_1 r m, k_2 \tilde r m)$. Since the overall $U(1)$ in this sub-quiver is decoupled and no field is charged under it, the baryonic generating function is non-vanishing only for 
\bea
k_1 r + k_2 \tilde r = 0 \,. 
\eea
This conclusion applies for $m\neq 0$, so that the baryonic charge is non-vanishing; if $m=0$ we compute instead the mesonic Hilbert series and no condition on the ranks arises. The previous discussion implies that
\bea\label{Dterm}
k_1 \sum_i m_i +  k_2 \sum_i \tilde m_i =0 \, .
\eea
We present two simple examples.

\subsubsection{The ABJ theory: $k_1=-k_2=k$ and $N_1\ne N_2$}

The ABJ theory obtained by adding fractional branes and thus considering theories with quiver \eref{abjmquiver}, $k_1=-k_2=k$ and different ranks. It  is well known that the theory is still $\CN=6$ and the moduli space is the same of ABJM. We consider first the simple
case of the $U(1)_{-k} \times U(2)_{+k}$ theory and unrefine the Hilbert series to simplify formulae, and discuss the general case at the end of the subsection. 

Let $m$ and $(n_1,n_2)$ be the monopole fluxes for $U(1)_{-k}$ and $U(2)_{+k}$ respectively. 
The $R$-charge of the monopole operator is
\bea
R(m; n_1, n_2) =   |m - n_1|+  |m - n_2| - |n_1-n_2|~.
\eea
Given (\ref{Dterm}) and the fact that fluxes must be paired, we must have
\bea
-m+n_1+n_2 = 0~, \qquad n_1=m ~\text{or}~ n_2=m~.
\eea
There are two cases to consider:
\ben
\item   The gauge group $U(2)_k$ is unbroken.  In this case, $n_1=n_2$.  Thus, we have
\bea
m= n_1=n_2=0~.
\eea
Hence, the $R$-charge of the monopole operator is $R(m; n_1, n_2) =0$.
The gauge charges of the monopole operator is $(km; -kn_1, -kn_2) = (0; 0,0)$.  
The baryonic generating function can be computed using {\tt Macaulay2}, which yields
\bea
g^{\rm (1)}(t;B)  &= \oint_{|z_1|=1} \frac{d z_1}{2 \pi i z_1} \oint_{|z_2|=1} \frac{d z_2}{2 \pi i z_2} (1-z_2^2) \oint_{|b|=1} \frac{d b}{2\pi i b^{B+1}} \times \nn \\
& \quad  \frac{1}{\left(1-\frac{t}{b z_1 z_2}\right)^2 \left(1-\frac{b t z_1}{z_2}\right)^2 \left(1-\frac{t z_2}{b z_1}\right)^2 \left(1-b t z_1 z_2\right)^2} \times \nn\\
& \quad \Big[ 1-t^3 \left(2 b z_2 z_1+\frac{2 b z_1}{z_2}+\frac{2 z_2}{b z_1}+\frac{2}{b z_2 z_1}\right)+t^4 \left(2 z_2^2+\frac{2}{z_2^2}+9\right) \nn \\
&\quad  -t^5 \left(2 b z_2 z_1+\frac{2 b z_1}{z_2}+\frac{2 z_2}{b z_1}+\frac{2}{b z_2 z_1}\right)+t^6 \left(b^2 z_1^2+\frac{1}{b^2 z_1^2}\right) \Big] \\
&= \frac{1+t^2}{(1-t^2)^3}~,
\eea
independent of $B$.  Thus, the contribution of this case to the Hilbert series is 
\bea
H^{(1)}_{N=1,k}(t)  = g^{\rm (1)}(t;0) = \frac{1+t^2}{(1-t^2)^3}~.
\eea
\item   The gauge group $U(2)_k$ is broken to $U(1)^2$.  In this case, $n_1 \neq n_2$.  We take
\bea
(m; n_1, n_2) = (m;m, 0)~, \qquad m \neq 0~.
\eea
The $R$-charge of the monopole operator is $R(m; m,0) =0$
and the gauge charges are $(km; -km,0)$.  The baryonic generating function in this case is
\bea
g^{\rm (2)} (t; B_1, B_2)& =  \oint_{|z_1|=1} \frac{d z_1}{2 \pi i z_1^{B_1+1}} \oint_{|z_2|=1} \frac{d z_2}{2 \pi i z_2^{B_2+1}} \PE \left[2 (z_1 z_2^{-1} + z_2 z_1^{-1}) t \right] \nn \\
& = \frac{t^{\left| B_2\right| } \left(1+ t^2- t^2 \left|
	B_2\right| +\left| B_2\right| \right)
}{\left(1-t^2\right)^3}~.
\eea
Thus, the contribution of this case to the Hilbert series is 
\bea
H^{(2)}_k(t)  = \sum_{m \in \BZ-\{0\}}  g^{\rm (2)}(t;km, -km)~.
\eea
\een
The Hilbert series is the sum of the two contributions:
\bea
H^{\rm ABJ}_k(t) &= H^{(1)}_k(t) + H^{(2)}_k(t) \nn \\
&= H [ \BC^4/\BZ_k] = \frac{1}{k} \sum_{m=0}^{k-1} \frac{1}{(1- \omega_k^{m} t)^2(1- \omega_k^{-m} t)^2} ~, \quad \omega_k= e^{2\pi i/k}~,
\eea
and indeed reproduces the Hilbert series for the ABJM theory $U(1)_k\times U(1)_{-k}$. 

 The general case of the $U(N_1)_k\times U(N_2)_{-k}$ theory can be discussed along similar lines. As we recalled in section \ref{sec:arbitrary_N} and saw explicitly in the previous example, the baryonic generating function of the conifold theory vanishes unless the ranks of the two gauge groups are equal. If $N_1\neq N_2$, the  magnetic charges for the unbalanced rank $|N_1-N_2|$ must vanish. The Hilbert series is then the same as for ABJM  with ranks $N=\min(N_1,N_2)$, corresponding to $N$ regular M2-branes. In addition, $|N_2-N_1|\leq |k|$ is needed to preserve SUSY for the $U(|N_1-N_2|)_{\pm k}$ subquiver of vanishing magnetic charges, corresponding to fractional M2-branes.

\subsubsection{The case $k_1\ne -k_2$}
The case $k_1+k_2\ne 0$ is truly $\CN=2$ and the moduli space is not hyperk\"ahler anymore. The case with equal ranks 
is not so interesting. Consider for example $U(1)_{k_1}\times U(1)_{k_2}$ with $k_1\ne -k_2$. Condition (\ref{Dterm}) implies
$m=\tilde m=0$ and the Hilbert series is the same as the Hilbert series of the conifold, of complex dimension three.%
\footnote{In general, for all the models
	which can be obtained using  tilings \cite{Hanany:2008cd,Hanany:2008fj},  the Hilbert series for $N$ branes and $k_1\ne -k_2$ is the $N$-fold symmetric product of the Calabi-Yau threefold associated with the tiling.}

The first interesting case is $U(1)_2\times U(2)_{-1}$. Let $m$ and $(n_1,n_2)$ be the monopole fluxes for $U(1)_{2}$ and $U(2)_{-1}$ respectively. 
The $R$-charge of the monopole operator is
\bea\label{dimun}
R(m; n_1, n_2) =   |m - n_1|+  |m - n_2| - |n_1-n_2|~.
\eea
The condition (\ref{Dterm}) and the fact that fluxes must be paired leaves us with the only possibility $(m;m,m)$ for the flux. 

We can understand what to expect about the moduli space by analyzing the solutions of the $F$-terms.
By writing the fields as two by two matrices
\bea
{\cal A}_{i a}=(A_i)_{1,a}\, , \qquad   {\cal B}_{a i} = (B_i)_{a,1}\, \qquad\qquad a=1,2\,  \,, \,\, i=1,2~,
\eea
the $F$-term conditions can be derived from the superpotential $W=\det {\cal A} \det {\cal B}$.  There are three branches:
\ben
\item $\det {\cal A}=\det {\cal B}=0$.  Only the flux $(0;0,0)$ contributes. We can form the four meson gauge invariants encoded in the matrix ${\cal M}={\cal A}{\cal B}$.
They satisfy $\det {\cal M}=0$ and we have the Hilbert series  
\bea
H^{\rm (I)}(t) = \frac{1-t^2}{(1-t)^4}~.
\eea
\item ${\cal B}=0$ and ${\cal A}\ne 0$. The flux $(m;m,m)$ can be dressed with fields in the sector of  baryonic charge $(2 m, -2 m)$ of the residual  $U(1)\times U(2)$ theory.
Such fields exist only for $m\ge 0$ and are given by \bea \det {\cal A}^m = ((A_1)_{11} (A_2)_{12} - (A_2)_{11} (A_1)_{12})^m\, , \eea which are gauge invariant under $SU(2)$ and have charge $(2 m, -2m)$ under
the abelian factors.   Since the $R$-charge of the monopole (\ref{dimun}) is $0$ we find
\bea
H^{\rm (II)}(t) = \sum_{m=0}^\infty t^{2 |m|}  = \frac{1}{1-t^2} \, .
\eea
The existence of these operators is related to the fact that  the BPS equations in a monopole background
with unbroken group $U(n_1)\times U(n_2)$ 
\bea
A_i A_i^\dagger = k_1 \sigma {\bf 1}_{n_1\times n_1} \, , \qquad A_i^\dagger  A_i = k_2 \sigma {\bf 1}_{n_2\times n_2}
\eea
can be solved by rectangular matrices when $n_2=n_1+1$ and $k_1/k_2= (n_1+1)/n_1$. The corresponding gauge invariant operators
can be  interpreted as baryons in the conifold theory with unequal ranks \cite{Aharony:2000pp, Dymarsky:2005xt} and have a simple
geometric characterization in terms of harmonic oscillators (see for example \cite{Furuuchi:2010gu}).  

\item ${\cal A}=0$ and ${\cal B}\ne 0$. This case is obtained from the previous one by interchanging $A_i$ and $B_i$: 
\bea
H^{\rm (III)}(t) = \sum_{m=0}^\infty t^{2 |m|} = \frac{1}{1-t^2} \, .
\eea
\een

\section{ $\CN =3$  Chern-Simons theories} \label{sec:N3CS}

$\CN=3$ theories can be obtained by adding Chern-Simons couplings to  $\CN=4$ theories. They have the same multiplets as $\CN=4$ theories,
in particular the vector multiplet contains a triplet of real scalars ${\vec \sigma}$. In a BPS monopole configuration, the matter fields $X$ must satisfy the BPS equations
\be\label{BPSN=3}
\begin{split}
	& {\vec D} (X)= - \frac{k}{2\pi} {\vec \sigma} \\
	& {\vec \sigma}\circ X =0 \, ,
\end{split}
\ee
where in the first equation ${\vec D}$ is the triplet of $D$-terms, in the second equation ${\vec \sigma}$ acts on the matter fields $X$ in the appropriate representation, and we wrote for simplicity the equations for a single gauge group $G$ with Chern-Simons coupling $k$. 

We work, as usual,  by selecting an $\CN=2$ subalgebra under which ${\vec \sigma}$
splits into a real scalar $\sigma$ and a complex scalar $\Phi$, and ${\vec D}$ into a real $D$-term $D$ and
a complex $F$-term $F$. In a BPS monopole configuration, $\sigma$ is identified as before with the magnetic flux $m$.
The real equations in (\ref{BPSN=3}) tell us as before that we can dress the monopole with matter fields satisfiying $m\circ X=0$
and with baryonic charge specified by $k m$. The complex equations in (\ref{BPSN=3}) can be derived from the superpotential
\bea
W= \tr \tilde Q \Phi Q +\frac{1}{2} k \Phi^2
\eea
where $(Q,\tilde Q)$ is the chiral multiplet matter content of a hypermultiplet and the traces in the previous expression are taken in the appropriate representation. 

For particular values of the Chern-simons couplings, the $F$-term equations can be solved with nonvanishing adjoint fields  and we can have the analogue of $\CN=4$ Coulomb branches. There are also mixed Coulomb-Higgs branches, as we will see.

\subsection{General results on the affine $A_{n-1}$ quiver with CS levels}
An interesting example is the $\CN=3$ $A_{n-1}$ quiver with a Chern-Simons level associated with each node.  This quiver is depicted in \eref{genquiv}.
\bea 
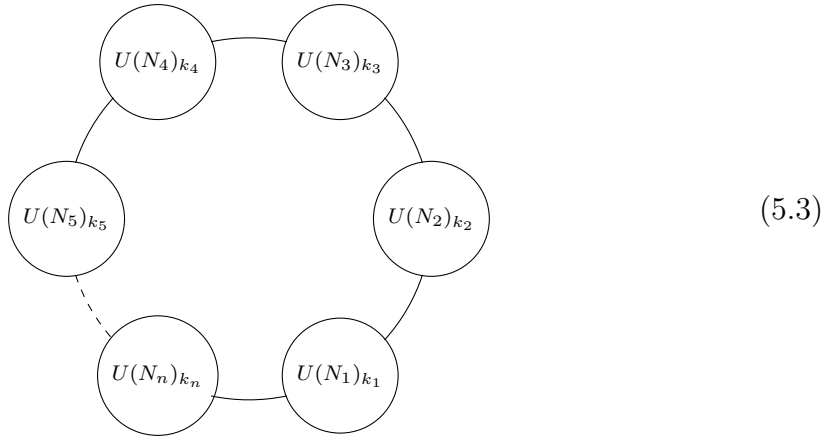
\label{genquiv}
\begin{tikzpicture}[baseline, scale=0.8,font=\scriptsize]
\def \n {6}
\def \radius {3cm}
\def \margin {18} % margin in angles, depends on the radius
\foreach \s in {1,...,5}
{
	\node[draw, circle] at ({360/\n * (\s - 2)}:\radius) {{$U(N_{\s})_{k_{\s}}$}};
	\draw[-, >=latex] ({360/\n * (\s - 3)+\margin}:\radius) 
	arc ({360/\n * (\s - 3)+\margin}:{360/\n * (\s-2)-\margin}:\radius);
}
\node[draw, circle] at ({360/3 * (3 - 1)}:\radius) {{$U(N_{n})_{k_{n}}$}};
\draw[dashed, >=latex] ({360/6 * (5 -2)+\margin}:\radius) 
arc ({360/6 * (5 -2)+\margin}:{360/6 * (5-1)-\margin}:\radius);
\end{tikzpicture}
\eea

These theories represent M2 branes probing hypertoric hyperK\"ahler cones over 3-Sasakian manifolds \cite{Jafferis:2008qz}
and we expect a branch isomorphic to the symmetric product of the hyperK\"ahler cone. It was argued in \cite{Hosomichi:2008jd,Imamura:2008nn} that for some specific values of the Chern-Simons couplings the theories are actually $\CN=4$ in the IR. In some specific cases we can explicitly map them to an $\CN=4$ theory via mirror symmetry.

For some specific values  of the Chern-Simons couplings, the moduli space has  an interesting structure of branches. We list here a series of general results  that can be obtained using the monopole formula. Some explicit examples of computations are given in Appendix \ref{Aquiver} for the $A_2$ quiver.

\paragraph{Example 1: Two non-zero CS levels with opposite signs.} \label{ex1}
Let us first take each gauge group to be $U(N)$ and the CS levels associated with a pair of adjacent nodes to be $k$ and $-k$, while those associated with other nodes are zero.  For definiteness, we take
\be \label{paramtheory1}
\begin{split}
	&N_1 = N_2 = \ldots = N_{n} =N~, \\
	&k_1 = -k_2 = k \neq 0 ~, \qquad k_i = 0 \quad \text{for $i \neq 1, 2$}~.
\end{split}
\ee
The theory corresponds to M2 branes probing (quotients of) $\BC^2 \times \BC^2/\BZ_{n-1}$.
It has 2 interesting branches of the moduli space:
\ben
\item When the vacuum expectation values (VEVs) of all bifundamental hypermultiplets are non-zero,  the moduli space is $\Sym^N [(\BC^2 \times \BC^2/\BZ_{n-1})/\BZ_k]$, where the action of the $\BZ_k$ orbifold is described below. 
This is a direct generalization of  the result for ABJM, which is recovered for $n=2$, and describes M2 branes which can be separated in a BPS way on $(\BC^2 \times \BC^2/\BZ_{n-1})/\BZ_k$.

Let the holomorphic coordinates of the first $\BC^2$ factor be $(z_1, z_2)$ and the second $\BC^2$ be $(w_1, w_2)$.  The $\BZ_{n-1}$ orbifold acts on $(w_1, w_2)$ as
\bea
\BZ_{n-1}: \quad (w_1, w_2) \rightarrow (\omega_{n-1} w_1, \omega_{n-1}^{-1} w_2)~, \qquad \omega_{n-1} = \exp(2 \pi i/ (n-1))~.
\eea
Hence, the invariant quantities under this $\BZ_{n-1}$ action is
\bea
w_1^{n-1}~, \qquad  w_1 w_2~, \qquad w_2^{n-1}~.
\eea
The $\BZ_k$ orbifold acts on $(z_1, z_2)$ and $(w_1^{n-1}, w_1 w_2, w_2^{n-1})$ as
\be
\begin{split}
	\BZ_k: \quad &(z_1, z_2) \rightarrow (\omega_k z_1, \omega_k^{-1} z_2)~,  \qquad \omega_{k} = \exp(2 \pi i/k) \\
	& (w_1^{n-1}, w_1 w_2, w_2^{n-1}) \rightarrow ( \omega_k w_1^{n-1}, w_1 w_2, \omega_k^{-1} w_2^{n-1})~.
\end{split}
\ee
%%%%%%%%
\item When the VEVs of the hypermultiplet between nodes $1$ and $2$ are non-zero and those of the others are zero, the moduli space is identified to that of $N$ $SU(n-1)$ instantons on $\BC^2/\BZ_k$ with framing $(n-1,0, \ldots,0)$.   This branch of the moduli space is isomorphic to the {\it Higgs branch} of the $\CN=4$ Kronheimer-Nakajima (KN) quiver \cite{kronheimer1990yang} depicted in \eref{KNn1flv}; see \cite{Mekareeya:2015bla} for a review.
\bea \label{KNn1flv}
\begin{tikzpicture}[baseline]
\def \n {6}
\def \radius {1.5cm}
\def \margin {16} % margin in angles, depends on the radius
\foreach \s in {1,...,5}
{
	\node[draw, circle] (\s) at ({360/\n * (\s - 2)}:\radius) {{\footnotesize $N$}};
	\draw[-, >=latex] ({360/\n * (\s - 3)+\margin}:\radius) 
	arc ({360/\n * (\s - 3)+\margin}:{360/\n * (\s-2)-\margin}:\radius);
}
\node[draw, circle] at ({360/3 * (3 - 1)}:\radius) {{\footnotesize $N$}};
\draw[dashed, >=latex] ({360/6 * (5 -2)+\margin}:\radius) 
arc ({360/6 * (5 -2)+\margin}:{360/6 * (5-1)-\margin}:\radius);
\node[draw, rectangle] (flv) at (3,0) {{\footnotesize $n-1$}};
\draw[-, >=latex] (2) to (flv);
\node[draw=none] at (4,-1) {{\footnotesize ($k$ circular nodes)}};
\end{tikzpicture}
\eea

Moreover, this branch of the moduli space is isomorphic to the {\it Coulomb branch} of the  mirror theory which is given by the $\CN=4$ affine $A_{n-2}$ quiver with all gauge groups being $U(N)$ and with $k$ flavours of fundamental hypermultiplet under one of the gauge groups, as depicted in \eref{Coulquiv1}.  The identification of the Coulomb branch of \eref{Coulquiv1} with the moduli space of $SU(n-1)$ instantons on $\BC^2/\BZ_k$ is discussed in (2.7) of \cite{Mekareeya:2015bla} (see also \cite{deBoer:1996mp,Porrati:1996xi}) and will be reviewed in Appendix \ref{app:inst}.
\bea 
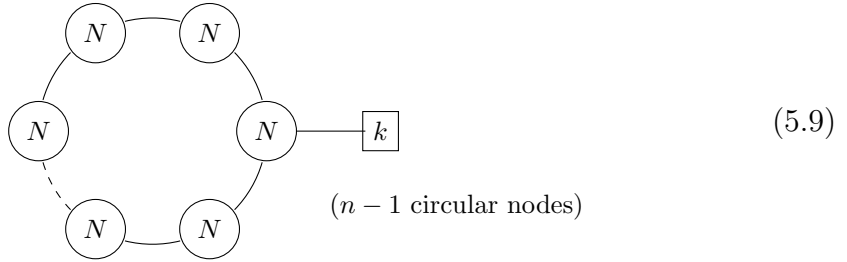
\label{Coulquiv1}
\begin{tikzpicture}[baseline]
\def \n {6}
\def \radius {1.5cm}
\def \margin {16} % margin in angles, depends on the radius
\foreach \s in {1,...,5}
{
	\node[draw, circle] at ({360/\n * (\s - 2)}:\radius) {{\footnotesize $N$}};
	\draw[-, >=latex] ({360/\n * (\s - 3)+\margin}:\radius) 
	arc ({360/\n * (\s - 3)+\margin}:{360/\n * (\s-2)-\margin}:\radius);
}
\node[draw, circle] at ({360/3 * (3 - 1)}:\radius) {{\footnotesize $N$}};
\draw[dashed, >=latex] ({360/6 * (5 -2)+\margin}:\radius) 
arc ({360/6 * (5 -2)+\margin}:{360/6 * (5-1)-\margin}:\radius);
\node[draw, rectangle] at (3,0) {{\footnotesize $k$}};
\draw[-, >=latex] (1+0.9,0) to (3-0.2,0);
\node[draw=none] at (4,-1) {{\footnotesize ($n-1$ circular nodes)}};
\end{tikzpicture}
\eea
\een

For $k=1$ we can actually relate the  quiver \eref{genquiv} with the parameters \eref{paramtheory1}  to the $\CN =4$ quiver  \eref{Coulquiv1} using mirror symmetry. We use a brane construction.  We start with the brane configuration corresponding to \eref{Coulquiv1}. 
 Such a system consists of 
\bi
\item $N$ coincident D3-branes wrapping $\BR^{1,2}_{0,1,2} \times S^1_6$ (where the subscripts indicate the direction in $\BR^{10}$), 
\item $n-1$ NS5-branes, wrapping $\BR^{1,2}_{0,1,2} \times \BR^3_{7,8,9}$, located at different positions along the circular $x^6$ direction, and 
\item $k=1$ D5-branes, wrapping $\BR^{1,2}_{0,1,2} \times \BR^3_{3,4,5}$, located along the circular $x^6$ direction within one of the NS5-brane intervals.
\ei
We then perform $SL(2,\BZ)$ action $T^T=-TST$ on this brane system, where $T$ and $S$ are the generators of $SL(2,\BZ)$ such that $S^2=-1$ and $(ST)^3=1$.  Under this action, the D5-brane transform into the $(1,1)$ five-brane, whereas NS5-branes and D3-branes remain invariant (see \eg~ \cite{Gaiotto:2008ak,Assel:2014awa}).  The $(1,1)$ five-branes induce the CS levels $1$ and $-1$ to a pair of the adjacent nodes in \eref{genquiv}, while the other nodes have zero CS levels. The branches 1 and 2 that we have found can be then re-interpreted
as the Higgs and Coulomb branches of the $\CN =4$ quiver  \eref{Coulquiv1}. The quiver  \eref{Coulquiv1}  can be in fact re-interpreted as describing $N$ D2-branes probing
a  $\BC^2 \times \BC^2/\BZ_{n-1}$ singularity. We know that its Higgs branch is the $ \BC^2/\BZ_{n-1}$ singularity and its Coulomb branch the moduli space of $N$ $SU(n-1)$ instantons on $\BC^2$ \cite{Intriligator:1996ex, deBoer:1996mp, Porrati:1996xi, Dey:2013nf, Cremonesi:2014xha}.

Note that for $k>1$ it is not possible to to obtain quiver \eref{genquiv} from \eref{Coulquiv1} via an $SL(2,\BZ)$ action.

\paragraph{Example 2: Non-zero CS levels with alternating signs.} \label{ex2}
We take the number of circular nodes
\bea
n=2m
\eea
to be an even number and
\bea \label{exam2}
&N_1 = N_2 = \ldots = N_{n} =N~, \\
&k_i = k ~ \text{for $i$ odd}~, \qquad k_j = -k ~ \text{for $j$ even}~.
\eea
The theory corresponds to M2 branes probing (quotients of ) $(\BC^2/\BZ_{m})^2/\BZ_k$.
It has 2 interesting branches of the moduli space:
\ben
\item When the VEVs of all bifundamental hypermultiplets are non-zero,  the moduli space is $\Sym^N [(\BC^2/\BZ_{m})^2/\BZ_k]$, where $\BZ_k$ acts on the complex coordinates as
\be
\begin{split}
\BZ_k: \quad & (z_1^{m-1}, z_1 z_2, z_2^{m-1}) \rightarrow ( \omega_k z_1^{m-1}, z_1 z_2, \omega_k^{-1} z_2^{m-1}) \\
& (w_1^{m-1}, w_1 w_2, w_2^{m-1}) \rightarrow ( \omega_k w_1^{m-1}, w_1 w_2, \omega_k^{-1} w_2^{m-1})~.
\end{split}
\ee
This branch describes $N$ M2 branes which can be separated in a BPS way  on $(\BC^2/\BZ_m)^2/\BZ_k$.
\item When the VEVs of the bifundamental hypermultiplets vanish alternately,  the moduli space is  isomorphic to the moduli space of $N$ $SU(m)$ instantons on $\BC^2/\BZ_{mk}$ with framing $(0^{k-1},1,0^{k-1},1,\ldots,0^{k-1},1)$, where $1$'s are in the $k$-th, $2k$-th, \ldots, $mk$-th positions.

Note that this branch is identical to the {\it Coulomb branch} of the $3d$ $\CN=4$ gauge theory whose quiver is given by the affine $A_{m-1}$ quiver with all gauge groups being $U(N)$ and with $k$ flavours of fundamental hypermultiplets under each gauge group, as depicted in \eref{flvallgauge}.  The identification of the Coulomb branch of this quiver with the moduli space of $SU(m)$ instantons on $\BC^2/\BZ_{mk}$ is discussed in \cite{Mekareeya:2015bla} (see a brief review in Appendix \ref{app:inst}).
\bea 
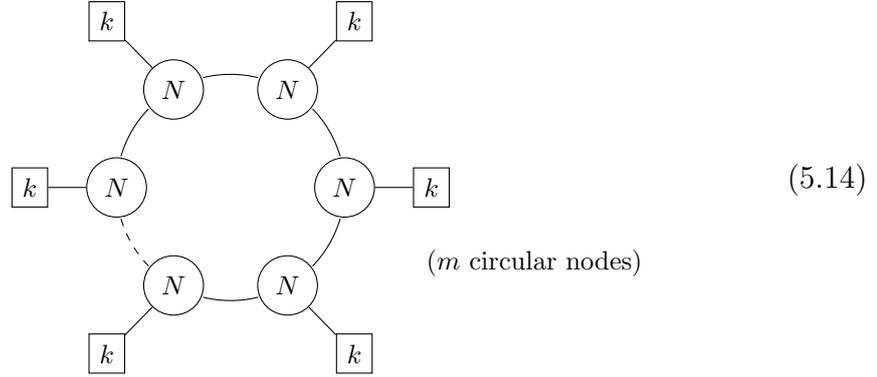
\label{flvallgauge}
\begin{tikzpicture}[baseline, align=center,node distance=0.5cm]
\def \n {6}
\def \radius {1.5cm}
\def \margin {16} % margin in angles, depends on the radius
\foreach \s in {1,...,5}
{
	\node[draw, circle] (\s) at ({360/\n * (\s - 2)}:\radius) {{\footnotesize $N$}};
	\draw[-, >=latex] ({360/\n * (\s - 3)+\margin}:\radius) 
	arc ({360/\n * (\s - 3)+\margin}:{360/\n * (\s-2)-\margin}:\radius);
}
\node[draw, circle] (last) at ({360/3 * (3 - 1)}:\radius) {{\footnotesize $N$}};
\draw[dashed, >=latex] ({360/6 * (5 -2)+\margin}:\radius) 
arc ({360/6 * (5 -2)+\margin}:{360/6 * (5-1)-\margin}:\radius);
\node[draw, rectangle,  below right= of 1] (f1) {{\footnotesize $k$}};
\node[draw, rectangle, right= of 2] (f2) {{\footnotesize $k$}};
\node[draw, rectangle, above right= of 3] (f3) {{\footnotesize $k$}};
\node[draw, rectangle, above left= of 4] (f4) {{\footnotesize $k$}};
\node[draw, rectangle,  left= of 5] (f5) {{\footnotesize $k$}};
\node[draw, rectangle,  below left= of last] (f6) {{\footnotesize $k$}};
\draw[-, >=latex] (1) to (f1);
\draw[-, >=latex] (2) to (f2);
\draw[-, >=latex] (3) to (f3);
\draw[-, >=latex] (4) to (f4);
\draw[-, >=latex] (5) to (f5);
\draw[-, >=latex] (last) to (f6);
\node[draw=none] at (4,-1) {{\footnotesize ($m$ circular nodes)}};
\end{tikzpicture}
\eea
For $k=1$, theory \eref{exam2} can be obtained by performing $SL(2,\BZ)$ action $T^T$ to the brane configuration of \eref{flvallgauge}, as in \cite{Gaiotto:2008ak,Assel:2014awa}.  For $k>1$, one cannot obtained the former from the latter in this way.
\een

\paragraph{Example 3: A more general configuration.} 
Let us now consider a more general configuration by taking quiver \eref{genquiv} with
\bea
& N_1 = N_2 = \ldots = N_n = N~,
\eea
and $p$ pairs of adjacent nodes with CS levels $k$ and $-k$ and $q$ nodes with zero CS level.
The number of circular nodes $n$ is then
\bea
n = 2p+q~.
\eea
The previous two examples are special cases of this theory: Example 1 corresponds to $p=1$ and Example 2 corresponds to $p=m$ and $n=2m$.

An example of quivers of this type is depicted below.  In this example,  we have $n=6$ and $p=2$.  
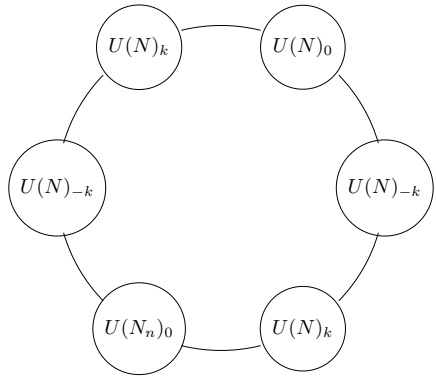
\begin{figure}[htb]
	\centering
	\resizebox{0.37\textwidth}{!}{%
		\begin{tikzpicture}[baseline]
		\def \n {6}
		\def \radius {2.8cm}
		\def \margin {16} % margin in angles, depends on the radius
		%%%%%%%%%%
		\node[draw, circle] at ({360/\n * (1 - 2)}:\radius) {{\footnotesize $U(N)_{k}$}};
		\draw[-, >=latex] ({360/\n * (1 - 3)+\margin}:\radius) 
		arc ({360/\n * (1 - 3)+\margin}:{360/\n * (1 -2)-\margin}:\radius);
		\node[draw, circle] at ({360/\n * (2 - 2)}:\radius) {{\footnotesize $U(N)_{-k}$}};
		\draw[-, >=latex] ({360/\n * (2 - 3)+\margin}:\radius) 
		arc ({360/\n * (2 - 3)+\margin}:{360/\n * (2 -2)-\margin}:\radius);
		\node[draw, circle] at ({360/\n * (3 - 2)}:\radius) {{\footnotesize $U(N)_{0}$}};
		\draw[-, >=latex] ({360/\n * (3 - 3)+\margin}:\radius) 
		arc ({360/\n * (3 - 3)+\margin}:{360/\n * (3 -2)-\margin}:\radius);  
		\node[draw, circle] at ({360/\n * (4 - 2)}:\radius) {{\footnotesize $U(N)_{k}$}};
		\draw[-, >=latex] ({360/\n * (4 - 3)+\margin}:\radius) 
		arc ({360/\n * (4 - 3)+\margin}:{360/\n * (4 -2)-\margin}:\radius);    
		\node[draw, circle] at ({360/\n * (5 - 2)}:\radius) {{\footnotesize $U(N)_{-k}$}};
		\draw[-, >=latex] ({360/\n * (5 - 3)+\margin}:\radius) 
		arc ({360/\n * (5 - 3)+\margin}:{360/\n * (5 -2)-\margin}:\radius);        
		%%%%%%%%%    
		\node[draw, circle] at ({360/3 * (3 - 1)}:\radius) {{\footnotesize $U(N_{n})_{0}$}};
		\draw[-, >=latex] ({360/6 * (5 -2)+\margin}:\radius) 
		arc ({360/6 * (5 -2)+\margin}:{360/6 * (5-1)-\margin}:\radius);
		\end{tikzpicture}}
\end{figure}

This theory has 2 interesting branches of the moduli space:
\ben
\item When the VEVs of all bifundamental hypermultiplets are non-zero, the moduli space is 
\bea
& \Sym^N [ (\BC^2/ \BZ_p \times \BC^2/\BZ_{m})/\BZ_k] =\Sym^N [ (\BC^2/ \BZ_p \times \BC^2/\BZ_{n-p})/\BZ_k]~,
\eea
where the second expression follows from the fact that $m=p+q$ and $n=2p+q$.   This branch of the moduli space in the case of $N=1$ was studied in \cite{Imamura:2008nn}.%
\footnote{Here $p$ is the number of $(1,k)$ five-branes and $n-p$ is the number of NS5-branes.}
\item When the VEVs of all bifundamental hypermultiplets between the nodes with CS levels $k$ and $-k$ are non-zero and the others are zero, the moduli space is isomorphic to the {\it Coulomb branch} of quiver \eref{gencase3dN4}, where the ranks $\ell_i$ (with $i=1,\ldots,m$) of the flavour symmetries can be obtained from the original theory as follows. 
\bi
\item Replace every node with zero CS levels in the original theory by a circular node with zero flavour $\ell_i=0$.
\item Replace every pair of adjacent nodes with non-zero CS levels $(k,-k)$ by a circular node attached to the flavour node with $\ell_i=k$.
\ei
\bea 
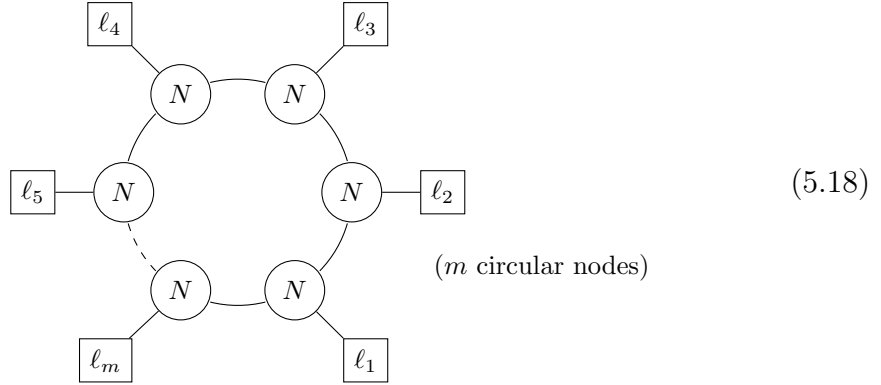
\label{gencase3dN4}
\begin{tikzpicture}[baseline, align=center,node distance=0.5cm]
\def \n {6}
\def \radius {1.5cm}
\def \margin {16} % margin in angles, depends on the radius
\foreach \s in {1,...,5}
{
	\node[draw, circle] (\s) at ({360/\n * (\s - 2)}:\radius) {{\footnotesize $N$}};
	\draw[-, >=latex] ({360/\n * (\s - 3)+\margin}:\radius) 
	arc ({360/\n * (\s - 3)+\margin}:{360/\n * (\s-2)-\margin}:\radius);
}
\node[draw, circle] (last) at ({360/3 * (3 - 1)}:\radius) {{\footnotesize $N$}};
\draw[dashed, >=latex] ({360/6 * (5 -2)+\margin}:\radius) 
arc ({360/6 * (5 -2)+\margin}:{360/6 * (5-1)-\margin}:\radius);
\node[draw, rectangle,  below right= of 1] (f1) {{\footnotesize $\ell_1$}};
\node[draw, rectangle, right= of 2] (f2) {{\footnotesize $\ell_2$}};
\node[draw, rectangle, above right= of 3] (f3) {{\footnotesize $\ell_3$}};
\node[draw, rectangle, above left= of 4] (f4) {{\footnotesize $\ell_4$}};
\node[draw, rectangle,  left= of 5] (f5) {{\footnotesize $\ell_5$}};
\node[draw, rectangle,  below left= of last] (f6) {{\footnotesize $\ell_m$}};
\draw[-, >=latex] (1) to (f1);
\draw[-, >=latex] (2) to (f2);
\draw[-, >=latex] (3) to (f3);
\draw[-, >=latex] (4) to (f4);
\draw[-, >=latex] (5) to (f5);
\draw[-, >=latex] (last) to (f6);
\node[draw=none] at (4,-1) {{\footnotesize ($m$ circular nodes)}};
\end{tikzpicture}
\eea
According to (2.7) of \cite{Mekareeya:2015bla} (see also Appendix \ref{app:inst}), this branch of the moduli space can be identified with
\bea
&\text{the moduli space of $N$ $SU(m)$ instantons on $\BC^2/\BZ_{kp}$} \nn\\
&\text{with framing $(0^{\ell_1-1},1,0^{\ell_2-1},1,\ldots,0^{\ell_m-1},1)$ and $\sum_{i=1}^m \ell_i = kp$}
\eea 
As before, for $k=1$, the original theory can be obtained by applying $T^T$ action to the brane configuration of \eref{gencase3dN4}, as in \cite{Gaiotto:2008ak,Assel:2014awa}.  For $k>1$, one cannot obtained the former from the latter in this way.
\een

\section{Geometric moduli spaces of abelian M2-brane theories} \label{sec:geoabM2}

The formalism for computing Hilbert series of the moduli space of $\cN\geq 2$ Chern-Simons theories that we have introduced applies both to abelian and to nonabelian theories. In this section however we focus our attention on the worldvolume theories of a single mobile M2-brane probing a CY$_4$ cone. We show how previously known results on the \emph{geometric} branch of the moduli space, that is the CY$_4$ cone transverse to the M2-brane, can be neatly reproduced in our formalism, and provide a counting of chiral operators in these theories.

In our formalism, the \emph{geometric} branch of the moduli space is obtained when monopole operators are only turned on for the diagonal $U(1)$ factor in the $U(1)^G$ quiver and %, that is $m_i=m$ for all $i=1,\dots, G$, and 
fundamental flavors, if present, do not take expectation value.
We will see that the Hilbert series of the geometric branch of the moduli space takes the general form
\be\label{geometric_general}
H= \sum_{m\in\bZ_{\leq 0}} a_-^{-m} g_1(k^- m) + \sum_{m\in\bZ_{\geq 0}} a_+^{m} g_1(k^+m) - g_1(0)   ~,
\ee
where the weights $a_\pm$ keep track of the global charges of the monopole operators $T$, $\t T$ of magnetic charge $\pm 1$ for the diagonal gauge $U(1)$, and  $g_1$ is the baryonic Hilbert series for the abelian quiver. The sum is restricted to baryonic charges lying along the rays $\theta^\pm \equiv \pm k^\pm$ defined by the effective gauge Chern-Simons levels
\be\label{kpm}
k^\pm_i = k_i \pm \frac{1}{2}(F_i - A_i)~,
\ee
where $k_i$ is the diagonal bare Chern-Simons level of the $i$-th gauge group, and $F_i$ and $A_i$ are the number of fundamentals and antifundamentals for the $i$-th gauge group. 

This result is consistent with the analysis of section 4 of \cite{Closset2012} and translates it into an explicit counting formula, with no need of knowing the quantum $F$-term relation determining $T\t T$ in terms of the bifundamental fields of the quiver.

We also note that the argument of section \ref{sec:arbitrary_N} and appendix \ref{multiple_branes}, assuming the conjecture \eqref{baryonic_sym} of \cite{Forcella:2007wk}, shows that the geometric moduli space of the theory on a stack of $N$ mobile M2-branes is the $N$-th symmetric product of the geometric moduli space of the theory on a single mobile M2-brane.

\subsection{M2-brane theories without quantum corrections}

Let us first consider an abelian $\prod_{i=1}^G U(1)_{k_i}$ quiver Chern-Simons theory without fundamental flavors and with $\sum_i k_i=0$. Neglecting Chern-Simons interactions, the quiver gauge theory with superpotential is the worldvolume theory on a D2-brane transverse to a CY$_3$ cone, which is the dimensional reduction of the worldvolume theory on a D3-brane probing the CY$_3$ cone. Due to the higher dimensional origin, each node of the quiver has as many incoming as outgoing arrows. As a result, the charges of monopole operators do not receive quantum corrections. 

We focus on the geometric branch of the moduli space, which in our formalism corresponds to having equal magnetic charges all gauge groups, $m_i=m$ for all $i=1,\dots,G$. If $z$ is the fugacity of the topological symmetry associated to the overall $U(1)$ and $q_i$, $i=1,\dots,G$ are fugacities for the gauge $U(1)$ groups in the quiver, the Hilbert series depends on $z$ through
\be\label{id_geometric}
\sum_{m\in\bZ} z^m \prod_{i=1}^G q_i^{-k_i m} =2\pi i z \cdot \delta\left(\prod_{i=1}^G q_i^{k_i}-z \right)~.
\ee
The equality follows from $\sum_{m\in\bZ} e^{i m\alpha} =2\pi\sum_{n\in\bZ}\delta(\alpha-2\pi n)$, passing from the angle $\alpha$ to the $U(1)$ valued fugacity $x=e^{i\alpha}$ to get $\sum_{m\in\bZ} x^m = 2\pi i \delta(x-1)$, and setting $x=\prod_i q_i^{k_i} /z$.

The delta function in the RHS of \eqref{id_geometric} allows us to integrate over one of the $G-1$ nontrivially acting $U(1)$ gauge groups. (The overall $U(1)$ does not act on the matter fields and its Molien integral gives $1$.) The effectively acting gauge group consists of linear combinations of the $G$ $U(1)$ groups in the integer kernel of $(1,1,\dots,1)$ and $(k_1,k_2,\dots,k_G)$. Explicitly, defining $k=\gcd\{k_i\}$ and $q_M=\prod_i q_i^{k_i/k}$, \eqref{id_geometric} becomes
\be
2\pi i z \delta\left(q_M^k-z \right)= 2\pi i q_M \frac{1}{k}\sum_{n=0}^{k-1} \delta (q_M-\omega_k^n z^{1/k})~.
\ee
The average in the RHS shows that the gauge $U(1)_M$ associated to the fugacity $q_M$, namely the linear combination of $U(1)$ gauge groups parallel to the vector of Chern-Simons couplings $(k_1,\dots,k_G)$, is Higgsed to a residual $\bZ_k$ gauge symmetry on the geometric moduli space where the charged monopole operator $T$ or $\t T$ acquire vev. This reproduces the results of \cite{Martelli2008d,Hanany:2008cd}. The CY$_3$ cone moduli space of the $D3$-brane theory is the symplectic quotient CY$_4//U(1)_M$ of the CY$_4$ cone moduli space of the M2-brane theory. Conversely, using the LHS of \eref{id_geometric}, the Hilbert series of the geometric moduli space of the 3d quiver Chern-Simons theory (the CY$_4$ cone) can be expressed as a sum of baryonic Hilbert series with baryonic charges $B_i=k_i m$ of the associated 4d quiver theory (corresponding to partial resolutions of the CY$_3$ cone).

\subsubsection{Example: $\cN=3$ circular abelian Chern-Simons quivers}
\label{sec:circular}

As an application, let us consider $\cN=3$ circular $\prod_{i=1}^G U(1)_{k_i}$ Chern-Simons quivers as depicted in \eref{genquiv1}.  These can be engineered in type IIB brane by a D3-brane wrapping a circle and intersecting $G$ $(1,p_i)$5-branes, with $k_i=p_i-p_{i-1}$ \cite{Bergman1999}. 
\bea 
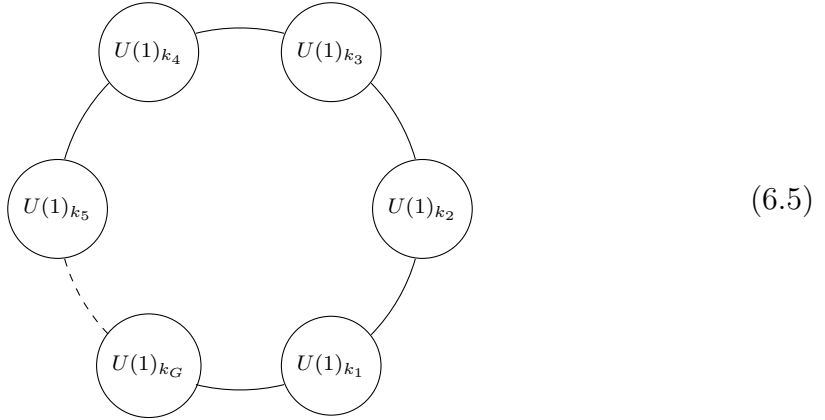
\label{genquiv1}
\begin{tikzpicture}[baseline, scale=0.8,font=\scriptsize]
\def \n {6}
\def \radius {3cm}
\def \margin {16} % margin in angles, depends on the radius
\foreach \s in {1,...,5}
{
	\node[draw, circle] at ({360/\n * (\s - 2)}:\radius) {{$U(1)_{k_{\s}}$}};
	\draw[-, >=latex] ({360/\n * (\s - 3)+\margin}:\radius) 
	arc ({360/\n * (\s - 3)+\margin}:{360/\n * (\s-2)-\margin}:\radius);
}
\node[draw, circle] at ({360/3 * (3 - 1)}:\radius) {{$U(1)_{k_{G}}$}};
\draw[dashed, >=latex] ({360/6 * (5 -2)+\margin}:\radius) 
arc ({360/6 * (5 -2)+\margin}:{360/6 * (5-1)-\margin}:\radius);
\end{tikzpicture}
\eea
The geometric moduli spaces of these theories are hypertoric hyperK\"ahler cones of quaternionic dimension $2$, given by hyperK\"ahler quotients of $\bH^{G}$ by an abelian group $\mathbf{N}=\ker(\beta)$, where \cite{Jafferis:2008qz}
\be
 \beta:~ U(1)^G \to U(1)^2 
 \qquad \beta=\begin{pmatrix}
	1 & 1 & \dots & 1 \\
	p_1 & p_2 & \dots & p_G
\end{pmatrix}~.
\ee

We now show this from the point of view of the Hilbert series, starting from the Hilbert series of the hyperK\"ahler quotient
\be\label{HS_HK_quot_JT}
\begin{split}
	H(t,X,Y)&=\left(\prod_{i=1}^G \oint \frac{dx_i}{2\pi i x_i}\right) \PE[-G t^2+t \sum_{i=1}^G (x_i+x_i^{-1})]\cdot\\
	&\cdot 2\pi i X \delta(\prod_{i=1}^G x_i-X) \cdot 2\pi i Y \delta(\prod_{i=1}^G x_i^{p_i}-Y)\cdot \PE[2t^2] ~.
\end{split}
\ee
The first line would be the Hilbert series for the hyperK\"ahler quotient by the whole $U(1)^G$ group. The delta functions and the $\PE$s in the second line reduce the group in the quotient from $U(1)^G$ to $\mathbf{N}=\ker(\beta)$. $X$ and $Y$ are fugacities for the two triholomorphic symmetries of the hyperK\"ahler cone. 

Let us change integration variables from $(x_1,\dots,x_G)$ to $(q_1,\dots,q_{G-1},u)$, with $x_i=u^{1/G} q_i q_{i+1}^{-1}$. Being the Hilbert series independent of $q_G$, we can average over it to obtain 
\be\label{final_JT}
\begin{split}
	H(t,X,Y)&=\PE[-(G-2)t^2] \left(\prod_{i=1}^G \oint \frac{dq_i}{2\pi i q_i}\right) \oint\frac{du}{2\pi i u} 2\pi i X\delta(u-X)\cdot\\
	&\cdot 2\pi i Y\delta\left(u^{\frac{1}{G}\sum_i p_i} \prod_{i=1}^G q_i^{k_i} -Y\right) \cdot \PE\left[t \sum_{i=1}^G \left(u^\frac{1}{G} \frac{q_i}{q_{i+1}}+u^{-\frac{1}{G}} \frac{q_{i+1}}{q_i}\right)\right]\\
	&= \PE[-(G-2)t^2]  \cdot\\
	&\cdot \sum_{m\in\bZ} z^m \left(\prod_{i=1}^G \oint \frac{dq_i}{2\pi i q_i}q_i^{-k_i m}\right) \PE\left[t \sum_{i=1}^G \left(X^{\frac{1}{G}} \frac{q_i}{q_{i+1}}+X^{-\frac{1}{G}} \frac{q_{i+1}}{q_i}\right)\right],
\end{split}
\ee
where we defined $Y=X^{\sum_i p_i/G} z$. The result is nothing but the Hilbert series of the geometric moduli space of the Abelian quiver Chern-Simons theory. This can be seen as follows. The superpotential of the theory is
$W=\sum_i \phi_i(A_i B_i-B_{i-1}A_{i-1})+ \sum_i \frac{k_i}{2}\phi_i^2$. The $F$-term equations for $A$ and $B$ are solved on the geometric branch by $\phi_i=\phi ~\forall i=1,\dots,G$. The $F$-term equations for $\phi_i$ then read $k_i \phi=B_{i-1}A_{i-1}-A_i B_i$. The $F$-term for the $U(1)$ parallel to the vector of Chern-Simons level determines $\phi=\frac{1}{\|k\|^2}\sum_i(k_{i+1}-k_i)A_i B_i$, leaving us with $A_i$ and $B_i$ subject to $G-2$ independent $F$-term (or complex moment map) equations. Together with the Chern-Simons interactions at levels $k_i$, this precisely reproduces the structure of the final expression in \eqref{final_JT}.

\subsection{M2-brane theories with quantum corrected chiral ring}

Let us now add flavors to the theories considered above, along the lines of \cite{Hohenegger2009,Gaiotto2012} for $\cN=3$ theories and \cite{Benini2010,Jafferis2013} for $\cN=2$ theories. (We will consider M2-brane theories with quantum corrections but no flavors in section \ref{sec:wrapped_D6}.) 
We add pairs of fundamental and antifundamental flavors $(p_a,q_a)$, attached to possibly different nodes of the quiver, with superpotential interactions $\Delta W_a = p_a M_a(X) q_a$. The effective complex masses $M_a(X)$ are polynomials of the bifundamental fields $X$, given by linear combinations of open paths in the quiver. Due to the extra matter fields, the charges of monopole operators $T$ and $\t T$ acquire one-loop corrections. The quantum correction of the charge of a monopole operator $V_m$ of magnetic charge $(m;m;\dots;m)$ (so that $T=V_1$ and $\t T=V_{-1}$) under a $U(1)$ global or gauge symmetry is \cite{Benini2010,Jafferis2013}
\be
Q^{\rm quant}[V_m]= -\frac{|m|}{2} \sum_a (Q[\psi_{p_a}]+Q[\psi_{q_a}])~,
\ee
where the sum runs over fermions of the chiral multiplets of fundamental and antifundamental flavors. In particular, the one-loop correction to the $R$-charge is 
\be\label{R_monopole}
R^{\rm quant}[V_m]= -\frac{|m|}{2} \sum_a (R[p_a]+R[q_a]-2)=\frac{|m|}{2} \sum_a R[M_a(X)]~
\ee
and the one-loop correction to the charge under the $i$-th gauge $U(1)$ in the quiver is 
\be\label{Qi_monopole}
Q_i^{\rm quant}[V_m]= -\frac{|m|}{2} \sum_a (Q_i[p_a]+Q_i[q_a])=-\frac{|m|}{2} (F_i-A_i)=\frac{|m|}{2} \sum_a Q_i[M_a(X)]~,
\ee
where $F_i$ ($A_i$) is the number of (anti)fundamental flavors of the $i$-th gauge group. Similar formulas hold for all other global symmetries, that we suppress in the following discussion.

Let $u=z t^{2R[T]} \prod_i q_i^{-k^+_i}\equiv a_+ \prod_i q_i^{-k^+_i}$ and $v=z^{-1} t^{2R[\t T]} \prod_i q_i^{k^-_i}\equiv a_- \prod_i q_i^{k^-_i}$ be the fugacity weights associated to $T$ and $\t T$ respectively. Here $R[T]$, $R[\t T]$ are the $R$-charges of the monopole operators, that we can take to equal the 1-loop correction $R^{\rm quant}[V_{\pm 1}]$ in \eqref{R_monopole}, choosing vanishing classical $R$-charges for the monopole operators.%
\footnote{The classical $R$-charges of monopole operators are due to bare mixed $R$-gauge Chern-Simons terms and are opposite for $T$ and $\t T$. We set them to zero here by appropriately mixing the $R$-symmetry with the topological symmetry.} $k^\pm_i=k_i\pm \frac{1}{2}(F_i-A_i)$  are the effective Chern-Simons levels for positive/negative values of the real scalar $\sigma$ in the diagonal $U(1)$ gauge group.  Brought inside the Molien integrals, the sum over magnetic fluxes leads to%
\footnote{Note that $\frac{1-uv}{(1-u)(1-v)}\to 2\pi i \delta(u-1)$ as $v\to u^{-1}$, reproducing $\sum_m u^m = 2\pi i \delta(u-1)$. This can be seen setting $u=ax$, $v=a/x$ with $|a|<1$, integrating against a test function $f(x)$ and taking $a\to 1$.}
\be\label{id_geometric_2}
\sum_{m\leq 0} v^{-m} +\sum_{m\geq 0} u^m -1 = \frac{1}{1-v} + \frac{1}{1-u} -1 = \frac{1-uv}{(1-u)(1-v)}~.
\ee
The first expression is the sum over magnetic charges $m \in \bZ$, split into $m\leq 0$ and $m \geq 0$, with $1$ subtracted not to overcount $m=0$. The second expression resums the two geometric series, both of which converge for small $|t|$ provided $R[T],R[\t T]>0$. The final expression shows that summing over all monopole operators for the diagonal $U(1)$ gauge group is equivalent to adding to the classical analysis of the $F$-flat moduli space two new fields $T$, $\t T$ subject to a quantum $F$-term relation that determines their product. The existence of such a quantum $F$-term relation, along with $V_m=T^m$ and $V_{-m}=\t T^{m}$ for $m>0$, is tied to the existence of a single bare BPS monopole operator for each magnetic charge $m$, which is a crucial input in the formula for the Hilbert series.  Taking into account the charges of $T$ and $\t T$ and the vanishing of the  circle parametrized by the dual photon when flavors are massless, one can conclude, consistently with the proposals of \cite{Gaiotto2012,Benini2010,Jafferis2013}, 
that the quantum $F$-term relation is 
\be \label{quantum_$F$-term}
T\t T = \prod_a M_a(X) ~.
\ee

\subsubsection{$\cN=3$ circular abelian Chern-Simons quivers with flavors}
\label{sec:circular_flavors}

Let us endow the $\cN=3$ circular $\prod_{i=1}^G U(1)_{k_i}$ Chern-Simons quiver \eref{genquiv1} with $F$ flavors of fundamental hypermultiplets, as depicted below. 
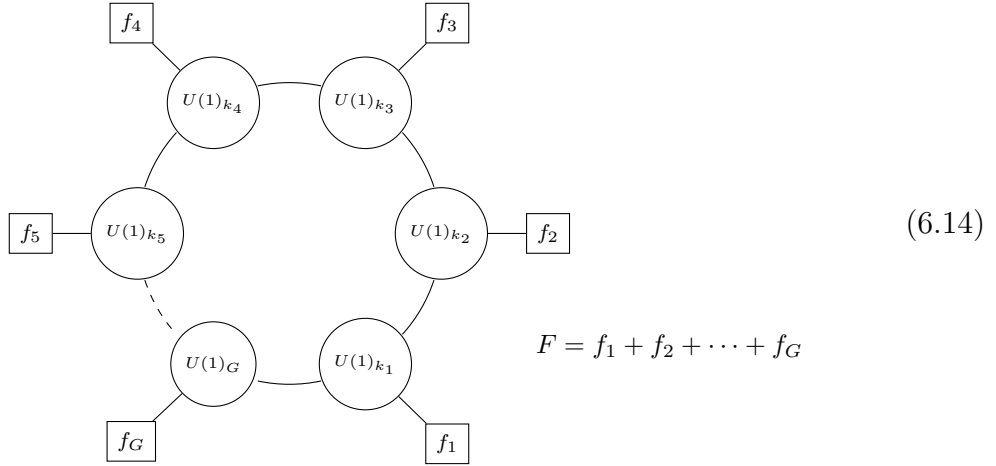
\bea \label{flvallgauge1}
\begin{tikzpicture}[baseline, align=center,node distance=0.5cm, font=\tiny]
\def \n {6}
\def \radius {2cm}
\def \margin {18} % margin in angles, depends on the radius
\foreach \s in {1,...,5}
{
	\node[draw, circle] (\s) at ({360/\n * (\s - 2)}:\radius) {{$U(1)_{k_{\s}}$}};
	\draw[-, >=latex] ({360/\n * (\s - 3)+\margin}:\radius) 
	arc ({360/\n * (\s - 3)+\margin}:{360/\n * (\s-2)-\margin}:\radius);
}
\node[draw, circle] (last) at ({360/3 * (3 - 1)}:\radius) {{$U(1)_G$}};
\draw[dashed, >=latex] ({360/6 * (5 -2)+\margin}:\radius) 
arc ({360/6 * (5 -2)+\margin}:{360/6 * (5-1)-\margin}:\radius);
\node[draw, rectangle,  below right= of 1] (f1) {{\scriptsize$f_1$}};
\node[draw, rectangle, right= of 2] (f2) {{\scriptsize$f_2$}};
\node[draw, rectangle, above right= of 3] (f3) {{\scriptsize$f_3$}};
\node[draw, rectangle, above left= of 4] (f4) {{\scriptsize $f_4$}};
\node[draw, rectangle,  left= of 5] (f5) {{\scriptsize $f_5$}};
\node[draw, rectangle,  below left= of last] (f6) {{\scriptsize $f_G$}};
\draw[-, >=latex] (1) to (f1);
\draw[-, >=latex] (2) to (f2);
\draw[-, >=latex] (3) to (f3);
\draw[-, >=latex] (4) to (f4);
\draw[-, >=latex] (5) to (f5);
\draw[-, >=latex] (last) to (f6);
\node[draw=none] at (5,-1.5) {{\footnotesize $F= f_1+f_2+ \cdots+f_G$}};
\end{tikzpicture}
\eea
In the type IIB engineering, we are adding $F$ D5-branes to the original brane configuration. The precise partition of the $F$ flavors among the $G$ gauge groups does not affect the geometric moduli space, which is a hyperK\"ahler quotient of $\bH^{G+1}$ by an abelian group $\mathbf{N}=\ker(\beta)$, where now \cite{Gaiotto2012}
\be
\beta: U(1)^{G+1}\to U(1)^2\qquad \beta=\begin{pmatrix}
	1 & 1 & \dots & 1 & 0 \\
	p_1 & p_2 & \dots & p_G & F
\end{pmatrix}~.
\ee

We can recover this result with the Hilbert series, starting from the hyperK\"ahler quotient description
\be\label{HS_HK_quot_JT_flav}
\begin{split}
	H(t, X,Y)&=\left(\prod_{i=1}^G \oint \frac{dx_i}{2\pi i x_i}\right) \PE[t \sum_{i=1}^G (x_i+x_i^{-1})-G t^2]\cdot 2\pi i X \delta(\prod_{i=1}^G x_i-X) \\
	& \oint \frac{dy}{2\pi i y}\PE[t(y+y^{-1})-t^2] \cdot 2\pi i Y \delta(y^F\prod_{i=1}^G x_i^{p_i}-Y)\cdot\PE[2t^2] ~.
\end{split}
\ee
The $y$ integral can be computed using the delta function,
\bea
&\oint \frac{dy}{2\pi i y}\PE[t(y+y^{-1})-t^2] \cdot 2\pi i Y \delta(y^F\prod_{i=1}^G x_i^{p_i}-Y)=\\
%&=\PE[-t^2]\oint dy\PE[t(y+y^{-1})] \frac{1}{F}\sum_{n=0}^{F-1}\delta(y-\omega_F^n b^\frac{1}{F})=\\
&= \frac{1}{F}\sum_{n=0}^{F-1} \PE[t(\omega_F^n b^\frac{1}{F}+\omega_F^{-n}b^{-\frac{1}{F}})-t^2]=\PE[t^F(b+b^{-1})-t^{2F}]=
\sum_{m\in\bZ}t^{F|m|}b^m~, \nonumber
\eea
where we set $b=Y/\prod_{i=1}^G x_i^{p_i}$ and used the expression for the Hilbert series of $\bC^2/\bZ_F$ as a Coulomb branch \cite{Cremonesi:2013lqa} in the last equality. Setting $Y=X^{\sum_i p_i/G}z$ and changing integration variables as in section \ref{sec:circular}, we obtain the Hilbert series of the geometric moduli space of the flavored circular quiver:
\be\label{HS_JT_flav_result}
\begin{split}
	H(t,X,z)&=\PE[-(G-2)t^2] \sum_{m\in\bZ} z^m t^{F|m|} \cdot\\
	&\cdot \left(\prod_{i=1}^G \oint \frac{dq_i}{2\pi i q_i}q_i^{-k_i m}\right) \PE\left[t \sum_{i=1}^G \left(X^{\frac{1}{G}} \frac{q_i}{q_{i+1}}+X^{-\frac{1}{G}} \frac{q_{i+1}}{q_i}\right)\right].
\end{split}
\ee

As an example, the unrefined Hilbert series of the geometric moduli space of the flavored ABJM theory of \cite{Hohenegger2009,Gaiotto2012}, engineered by a D3-brane intersecting an NS5-brane, a $(1,k)5$-brane and $F$ D5-branes along a circle, is 
\be\label{Ga-Ja}
H(t)=\frac{1+t^2+2 k t^{F+k}-2 k t^{F+k+2}-t^{2 F+2 k}-t^{2 F+2 k+2}}{(1-t)^3 (1+t)^3 \left(1-t^{F+k}\right)^2}~
\ee
and the volume of the triSasakian base $M_7$ of the cone, in agreement with \cite{Lee2007}, is
\be
\Vol(M_7)/\Vol(S^7)=\lim_{t\to 1} (1-t)^4 H(t) = \frac{F+2k}{2(F+k)^2}~.
\ee

Let us further specialize to the case $k=F=1$ discussed in detail in \cite{Gaiotto2012}: the geometric moduli space is the cone over $N^{0,1,0}$, which is nothing but the reduced moduli space of one $SU(3)$ instanton. We can compute the Hilbert series, with $w=X^{1/2}$ an $SU(2)$ fugacity and $z$ the fugacity of the topological $U(1)$. Setting $w=x_1 x_2^{-1/2}$ and $z =x_2^{3/2}$, the refined Hilbert series is 
\be\label{N010}
\begin{split}
	H(t,x_1,x_2)&=\PE[([1,1]-2)t^2]\cdot(1 + 2 t^2 + (2-[1,1]) t^4 + 2 t^6 + t^8)=\\
	&=\sum_{n=0}^\infty [n,n]t^{2n}~,
\end{split}
\ee
where $[n,n]$ is a shorthand for the character of the representation $[n,n]$ of $SU(3)$, expressed in terms of $SU(3)$ fugacities $x_1$, $x_2$ \cite{Benvenuti:2010pq}. The Hilbert series \eref{N010} manifests the enhancement of the $SU(2)\times U(1)$ global symmetry to $SU(3)$, not only at the level of scalar partners of conserved currents that correspond to the term $[1,1]t^2$, but for the entire spectrum of chiral operators. The generators are the $4$ mesons $M_{ij}=A_i B_j$ and the $4$ dressed monopole operators $u_i=T A_i$, $v_j=\t TB_j$, which altogether form the adjoint representation of $SU(3)$.

\subsubsection{Toric flavored ABJM theories}\label{sec:flavABJM}

An interesting and rich class of $\cN=2$ theories with quantum corrected chiral rings is provided by the toric flavored abelian ABJM models of \cite{Benini2010,Cremonesi2011}.
\bea
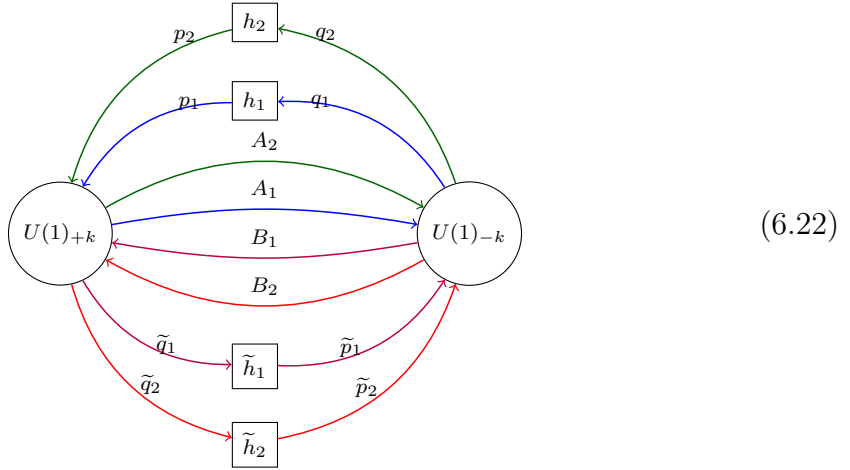
\begin{tikzpicture}[baseline, font=\scriptsize, scale=0.8]
\begin{scope}[auto,%
every node/.style={draw, minimum size=0.5cm}, node distance=4cm];
% the vertices
\node[circle] (UN1) at (0, 0) {$U(1)_{+k}$};
\node[circle, right=of UN1] (UN2)  {$U(1)_{-k}$};
\node[rectangle] at (3.2,2.2) (UNa1)  {$h_1$};
\node[rectangle] at (3.2,3.5) (UNa2)  {$h_2$};
\node[rectangle] at (3.2,-2.2) (UNb1)  {$\tilde{h}_1$};
\node[rectangle] at (3.2,-3.5) (UNb2)  {$\tilde{h}_2$};
\end{scope}
% the edges
\draw[draw=red,solid,line width=0.2mm,<-]  (UN1) to[bend right=30] node[midway,above] {$B_2 $}node[midway,above] {}  (UN2) ;
\draw[draw=blue,solid,line width=0.2mm,->]  (UN1) to[bend right=-10] node[midway,above] {$A_1$}node[midway,above] {}  (UN2) ; 
\draw[draw=purple,solid,line width=0.2mm,<-]  (UN1) to[bend left=-10] node[midway,above] {$B_1$} node[midway,above] {} (UN2) ;  
\draw[draw=black!60!green,solid,line width=0.2mm,->]  (UN1) to[bend left=30] node[midway,above] {$A_2$} node[midway,above] {} (UN2) ;    
\draw[draw=purple,solid,line width=0.2mm,->]  (UN1)  to[bend right=30] node[midway,right] {$\tilde{q}_1$}   (UNb1);
\draw[draw=purple,solid,line width=0.2mm,->]  (UNb1) to[bend right=30] node[midway,left] {$\tilde{p}_1$} (UN2) ; 
\draw[draw=red,solid,line width=0.2mm,->]  (UN1)  to[bend right=30]  node[midway,right] {$\tilde{q}_2$} (UNb2);
\draw[draw=red,solid,line width=0.2mm,->]  (UNb2) to[bend right=30] node[midway,left] {$\tilde{p}_2$} (UN2); 
\draw[draw=blue,solid,line width=0.2mm,->]  (UN2)  to[bend right=30] node[pos=0.9,right] {$q_1$}   (UNa1);
\draw[draw=blue,solid,line width=0.2mm,->]  (UNa1) to[bend right=30] node[pos=0.1,left] {$p_1$} (UN1) ; 
\draw[draw=black!60!green,solid,line width=0.2mm,->]  (UN2)  to[bend right=30]  node[pos=0.9,right] {$q_2$} (UNa2);
\draw[draw=black!60!green,solid,line width=0.2mm,->]  (UNa2) to[bend right=30] node[pos=0.1,left] {$p_2$} (UN1); 
\end{tikzpicture}
\eea
These are $U(1)_k\times U(1)_{-k}$ Chern-Simons quiver gauge theories, specified by the quiver diagram depicted above and the superpotential 
\be
\begin{split}
	W&=A_1B_1A_2B_2-A_1B_2A_2B_1+\\
	&+\sum_{i=1}^{h_1} p^i_{1} A_1 q_{1,i}+\sum_{i=1}^{h_2} p^i_{2} A_2 q_{2i} +\sum_{i=1}^{\t h_1} \t p^i_{1} B_1 \t q_{1,i}+\sum_{i=1}^{\t h_2} \t p^i_{2} B_2 \t q_{2i}~.
\end{split}
\ee
The coupling to flavors generically breaks the mesonic $SU(2)\times SU(2)$ to its maximal torus $U(1)\times U(1)$, to which we associate fugacities $x$ and $y$. 

The quantization condition for gauge Chern-Simons levels is
\be
k^\pm \equiv k \mp \frac{1}{2}(h_1+h_2-\t h_1-\t h_2) \in \bZ~.
\ee
Similarly, mixed Chern-Simons couplings $k_x$, $k_y$ between the mesonic $U(1)$ symmetries and the diagonal gauge $U(1)$ are subject to the quantization conditions
\be
\begin{split}
	k_x^\pm &\equiv k_x \mp \frac{1}{2}(h_1-h_2) \in \bZ~\\
	k_y^\pm &\equiv k_y \mp \frac{1}{2}(\t h_1- \t h_2) \in \bZ~.
\end{split}
\ee
We assign equal $R$-charges $1/2$ to all bifundamentals and set to zero the classical $R$-charges of monopole operators.

The Hilbert series of the geometric moduli space is given by 
\be\label{HS_flavored_ABJM}
H(t,x,y,z)=\sum_{m\in\bZ_{\geq 0}} a_+^m g_1(k^+ m) + \sum_{m\in\bZ_{\leq 0}} a_-^{-m} g_1(k^-m) - g_1(0)   ~,
\ee
where $a_\pm=z^{\pm 1}x^{\mp k_x^\pm}y^{\mp k_y^\pm}t^{\frac{1}{2}(h_1+h_2+\t h_1+\t h_2)}$ keep track of the charges of the monopole operators $T$, $\t T$, and $g_1$ is the baryonic Hilbert series for the abelian conifold quiver,
\be\label{g1_conifold}
g_1(B) = \PE[[1;1]t^2] t^{|B|}\cdot \begin{cases}
	[B;0]-t^2[B-1;1]+t^4[B-2;0]~, & B\geq 0\\
	[0;|B|]-t^2[1;|B|-1]+t^4[0;|B|-2]~, & B\leq 0
\end{cases}~.
\ee
Here $[m;n]\equiv[m]_x [n]_y$, where $[m]_x= \frac{x^{m+1}-x^{-(m+1)}}{x-x^{-1}}$ is the character of the $(n+1)$-dimensional representation of $SU(2)$. 

The terms in \eqref{HS_flavored_ABJM} can be computed using 
\be
\begin{split}
	\sum_{m\geq 0} a^m g_1(Km)&= \PE[[1;1]t^2 +a t^{K}(x^{K}+x^{-K})]\cdot\\
	&\cdot (1-t^4+a t^{K}([K-2;0]-t^2[K-1;1]+t^4[K;0]))
\end{split}
\ee
for $K\geq 0$. The result for $K\leq 0$ is obtained replacing $(K,x,y) \leftrightarrow (-K,y,x)$. Finally, $g_1(0)=\PE[[1;1]t^2-t^4]$.

Let us briefly consider a couple of examples discussed in \cite{Benini2010,Cremonesi2011}. For $k=h_1=h_2=0$, $\t h_1=\t h_2=1$, the geometric moduli space is the Calabi-Yau cone over $Q^{1,1,1}$. 
Choosing $k_x=k_y=0$ and setting $x=\alpha$, $y=\gamma\beta$ and $z=\gamma/\beta$, the Hilbert series is 
\be
H=\sum_{n=0}^\infty [n;n;n]_{\alpha,\beta,\gamma} t^{2n} \xrightarrow[\alpha,\beta,\gamma\to 1]{} \frac{1+4t^2+t^4}{(1-t^2)^4}~, 
\ee
which manifests an $SU(2)^3$ symmetry and reproduces the Hilbert series of the cone over $Q^{1,1,1}$, in agreement with (A.7) of \cite{Hanany:2008fj}.  We will discuss the field theory counterparts of resolutions of the cone in appendix \ref{sec:Q111}.

For $k=\frac{3}{2}$, $h_1=1$, $h_2=\t h_1=\t h_2=0$, the geometric moduli space is the Calabi-Yau cone over $Y^{1,2}(\bC \bP^2)$. Setting $k_x=\frac{1}{2}$ and $k_y=0$, the Hilbert series is 
\be\label{HS_Y12}
\begin{split}
	H=&\PE[t^2 x^{-1} (y+y^{-1})+t^{5/2} x (y^2+y^{-2}) z^{-1}+ t^{3/2} z (x+x^{-1})]
	\cdot\\
	&\cdot (1+ t^2 x (y + y^{-1})+t^{5/2} x z^{-1} - t^{7/2} (y + y^{-1}) z - 
	t^4 (y + y^{-1})^2 +\\
	&- t^{9/2} (y + y^{-1}) z^{-1} + t^{11/2} x^{-1} z  
	+ t^6 x^{-1} (y + y^{-1}) + t^8 )~.
\end{split}
\ee
The plethystic logarithm%
\footnote{The plethystic logarithm of a multi-variate function $f(x_1, \ldots, x_n)$ such that $f(0, \ldots, 0)=1$ is 
	\bea
	\PL[f(x_1, \ldots, x_n)] =  \sum_{k=1}^\infty \frac{1}{k} \mu(k) \log f(x_1^k, \ldots, x_n^k) ~. \nn
	\eea}
of the Hilbert series 
\be
\begin{split}
	\PL[H] &= t^2 (x +x^{-1})[1]_y +t^{3/2} z (x +x^{-1})+ t^{5/2} z^{-1} x [2]_y +\\
	&-t^{7/2} z [1]_y -t^4 (1+(x^2+1) [2]_y) - t^{9/2}z^{-1}(x^2+1)[1]_y- t^5 z^{-2}x^2+ \dots
\end{split}
\ee
suggests that there are $8$ generators $M_{ij}=A_iB_j$, $u_i=TA_i$, $v_{j_1 j_2}=v_{j_2 j_1}=\t T B_{j_1} B_{j_2}$ subject to $14$ independent relations, as proposed in \cite{Benini2010}, that can be summarized as 
\be
\epsilon^{i_1 i_2} M_{i_1 j}u_{i_2}=0, ~~ \det M=0, ~~ u_i v_{j_1 j_2}=M_{i j_1} M_{1 j_2}, ~~ \epsilon^{j j_1} M_{ij} v_{j_1 j_2}=0, ~~ \det v=0.
\ee
We checked using {\tt Macaulay2} that \eqref{HS_Y12} is indeed the Hilbert series of this quotient ring. 
The superconformal $R$-symmetry can also be determined from the Hilbert series by volume minimization \cite{Martelli2006}: it is obtained by the fugacity map $x=t^{a}$, $y=1$ and $z=t^{1/2}$, with 
\be 
a=\frac{1}{3}(5-23 c^{-1/3}+c^{1/3})~, \qquad c=24 \sqrt{78}-181~.
\ee

\subsubsection*{The Hilbert series from geometry}

We will now show how the field theory monopole formula \eref{HS_flavored_ABJM} for the abelian toric flavored ABJM theories can be obtained geometrically by appropriately rewriting the Hilbert series of the toric CY$_4$ cones. We use the K\"ahler quotient description of the toric variety, viewed as the vacuum moduli space of an abelian gauged linear sigma model (GLSM). (We provide an alternative computation that starts directly from the toric data in appendix \ref{app:HS_from_toric}.) 
The trick is to gauge and ungauge the $U(1)_M$ symmetry of the CY$_4$ cone
that was used in \cite{Benini2010} to reduce M-theory on the CY$_4$ to type IIA on the conifold fibered over $\bR$, with D6-branes and RR 2-form flux. 

The CY$_4$ cones in this class have toric diagrams consisting of four columns of points,%
\footnote{The columns of toric points can be replaced by their bottom and top points if the $\bC^2/\bZ_h$ fibers over the toric divisors of the conifold are not resolved. Even though we do not consider these resolutions in this section, we keep all the toric points since the associated GLSM fields are useful to describe monopole operators.}
with coordinates $a_{n_1+i_1}=(0,0,n_1+i_1)$, $b_{\t n_1+\t i_1}=(1,0,\t n_1+\t i_1)$, $c_{n_2+i_2}=(1,1,n_2+i_2)$, $d_{\t n_2+\t i_2}=(0,1,\t n_2+\t i_2)$, where $i_1=0,\dots,h_1$ and similarly for the other columns.  The CY$_4$ is the vacuum moduli space of a GLSM with $h_1+h_2+\t h_1+\t h_2$ $U(1)$ gauge factors. To obtain the field theory formula \eref{HS_flavored_ABJM}, it is convenient to gauge and ungauge the $U(1)_M$ symmetry that corresponds to the vertical direction in the toric diagram. Projecting the 3d toric diagram of the CY$_4$ vertically (forgetting the last coordinate), we obtain the 2d toric diagram of the conifold, made of 4 points with coordinates $a=(0,0)$, $b=(1,0)$, $c=(1,1)$, $d=(0,1)$.

Up to mixing with the genuine gauge symmetries of the GLSM, we can take the charges of this `vertical' $U(1)_M$ to be $\pm 1$ for the GLSM fields corresponding to two consecutive toric points in a column, \emph{e.g.} $-1$ for $a_{n_1}$, $+1$ for $a_{n_1+1}$, and $0$ for all other GLSM fields. Denoting by $\zeta$ the fugacity of this extra $U(1)$, its gauging and ungauging is achieved by writing $f(\zeta)=\sum_{m\in\bZ}\zeta^m \oint \frac{du}{2\pi u} u^{-m} f(u)$. If we focus on GLSM fields belonging to the $a$ column (other columns are treated similarly) and drop subscripts, the resulting GLSM has gauge fugacities, charge matrix and baryonic charges 
\be
\begin{tabular}{c|ccccccc|c}
	& $a_{n}$ & $a_{n+1}$ & $a_{n+2}$ & $a_{n+3}$ & $\dots$ & $a_{n+h-1}$ &  $a_{n+h}$ & \\ \hline
	$q$ & $n+1$ & $-n$ & $0$ & $0$ & $\dots$ & $0$ & $0$ & $0$ \\ 
	$u_1$ & $-1$ & $1$ & $0$ & $0$ & $\dots$ & $0$ & $0$ & $m$ \\ 
	$u_2$ & $0$ & $-1$ & $1$ & $0$ & $\dots$ & $0$ & $0$ & $m$ \\ 
	$u_3$ & $0$ & $0$ & $-1$ & $1$ & $\dots$ & $0$ & $0$ & $m$ \\ 
	$\vdots$ & $\vdots$ & $\vdots$ & $\vdots$ & $\vdots$ & $\ddots$ & $\vdots$ & $\vdots$ &$\vdots$ \\ 
	$u_h$ & $0$ & $0$ & $0$ & $0$ & $\dots$ & $-1$ & $1$ & $m$ 
\end{tabular}
\ee
The common baryonic charge $m$ will eventually be summed over. The integrals over the fugacities $u_1, \dots, u_h$ at fixed $m$ can be evaluated, giving
\be\label{integral_I}
\begin{split}
	&I_a(m) \equiv \prod_{j=1}^h \left(\oint \frac{du_j}{2\pi i u_j^{1+m}}\right) \PE\left[\alpha_n \frac{1}{u_1}+\alpha_{n+1} \frac{u_1}{u_2}+\alpha_{n+2} \frac{u_2}{u_3}+\dots+\alpha_{n+h} \frac{1}{u_h}\right]\\
	&~~= \PE\bigg[\prod_{j=0}^h \alpha_{n+j}\bigg] \bigg(\prod_{j=0}^h \alpha_{n+j}^j\bigg)^{\frac{|m|+m}{2}} \bigg(\prod_{j=0}^h \alpha_{n+j}^{h-j}\bigg)^{\frac{|m|-m}{2}}\equiv \PE[\alpha] \alpha_+^{\frac{|m|+m}{2}}\alpha_-^{\frac{|m|-m}{2}}~.
\end{split}
\ee
Here $\alpha_{n+j}$ are monomials in the fugacities $q$, $t$, $x$ and $y$ keeping track of the charges of $a_{n+j}$. We take $\alpha_{n}=(tx)^\frac{1}{h+1} q^{n+1}$, $\alpha_{n+1}=(tx)^\frac{1}{h+1} q^{-n}$ and $\alpha_{n+j}=(tx)^\frac{1}{h+1}$ for $j=2,\dots,h$,%
\footnote{Different assignments of global charges are obtained by redefining $\zeta$.} so that
\be
\alpha 
=txq~,\qquad
\alpha_+ 
= (tx)^\frac{h}{2}q^{-n}~,\qquad
\alpha_- 
= (tx)^\frac{h}{2}q^{n+h}~.
\ee
Restoring subscripts, the integral \eref{integral_I} is then
\be
I_a(m)=\PE[txq] ~ q^{-\left(n_1+\frac{h_1}{2}\right)m}(txq)^{\frac{h_1}{2}|m|}~.
\ee

Taking into account all the four columns, integrating over the remaining gauge fugacity $q$ corresponding to the conifold GLSM of charges $(1,-1,1,-1)$ for $a,b,c,d$, and summing over $m$, the Hilbert series of the CY$_4$ cone can be written as
\be\label{H_flavABJM_geom}
\begin{split}
	&H_{\rm geom}(t,x,y,\zeta)=\sum_{m\in\bZ}  \oint \frac{dq}{2\pi i q} \PE\left[tq(x+x^{-1})+tq^{-1}(y+y^{-1})\right]  \cdot\\
	& \qquad \cdot\left(\zeta  q^{-(n_1+n_2-\t n_1-\t n_2)}\right)^m \left(q^{h_1+h_2-\t h_1-\t h_2} t^{h_1+h_2+\t h_1+\t h_2} x^{h_1-h_2} y^{\t h_1-\t h_2}\right)^\frac{|m|}{2} ~.
\end{split}
\ee
Using the dictionary between toric data and field theory data \cite{Benini2010}
\be\label{dict_flavABJM}
k=n_1+n_2-\t n_1-\t n_2 +\frac{1}{2}(h_1+h_2-\t h_1-\t h_2)
\ee
and making the identification $\zeta=z x^{-k_x} y^{-k_y}$, we see that the geometric formula \eref{H_flavABJM_geom} for the Hilbert series reproduces precisely the field theory formula \eref{HS_flavored_ABJM}. The first factor in the second line of \eref{H_flavABJM_geom} corresponds to the classical contribution of Chern-Simons interactions, the second factor corresponds to the quantum corrections. 

Note that, as proposed in \cite{Benini2010}, the bare monopole operators $T=V_1$ and $\t T=V_{-1}$ can be expressed in terms of the GLSM fields of the CY$_4$ cone as
\be\label{monop_flavABJM}
\begin{split}
	T &=\prod_{j_1=0}^{h_1} a_{n_1+j_1}^{j_1} \prod_{\t j_1=0}^{\t h_1} b_{\t n_1+\t j_1}^{\t j_1} \prod_{j_2=0}^{h_2} c_{n_2+j_2}^{j_2} \prod_{\t j_2=0}^{\t h_2} d_{\t n_2+\t j_2}^{\t j_2}\\
	\t T &=\prod_{j_1=0}^{h_1} a_{n_1+j_1}^{h_1-j_1} \prod_{\t j_1=0}^{\t h_1} b_{\t n_1+\t j_1}^{\t h_1-\t j_1} \prod_{j_2=0}^{h_2} c_{n_2+j_2}^{h_2-j_2} \prod_{\t j_2=0}^{\t h_2} d_{\t n_2+\t j_2}^{\t h_2-\t j_2}
\end{split}
\ee
and are therefore counted in the Hilbert series \eref{H_flavABJM_geom} with the weights 
\be
\begin{split}
	w[T]&= \zeta \alpha_+ \beta_+\gamma_+\delta_+ = z t^{\frac{1}{2}(h_1+h_2+\t h_1+\t h_2)} x^{-k_x^+}  y^{-k_y^+}q^{-k^+}\\
	w[\t T]&= \zeta^{-1} \alpha_- \beta_-\gamma_-\delta_- = z^{-1} t^{\frac{1}{2}(h_1+h_2+\t h_1+\t h_2)} x^{k_x^-}  y^{k_y^-}q^{k^-}~.
\end{split}
\ee
Similarly, for the bifundamentals $A_{1,2}$, $B_{1,2}$ we have
\be\label{bifund_flavABJM}
\begin{split}
	&A_1 = a= \prod_{j_1=0}^{h_1} a_{n_1+j_1} \qquad\qquad 	A_2 = c= \prod_{j_2=0}^{h_2} c_{n_2+j_2} \\
	&B_1 = b= \prod_{\t j_1=0}^{\t h_1} b_{\t n_1+\t j_1} \qquad\quad\, B_2 = d= \prod_{\t j_2=0}^{\t h_2} d_{\t n_2+\t j_2}~.
\end{split}
\ee
and the weights
\be\label{bifund_flavABJM_w}
\begin{split}
	&w[A_1]=\alpha=txq \qquad\qquad 	w[A_2]=\gamma=tx^{-1}q\\
	&w[B_1]=\beta=t y q^{-1} \qquad\quad\, w[B_2] = \gamma=t y^{-1} q^{-1}~.
\end{split}
\ee
The quantum $F$-term relation $T \t T=A_1^{h_1}A_2^{h_2}B_1^{\t h_1}B_2^{\t h_2} $  of \cite{Benini2010}, which is consistent with these identifications, arises in the Hilbert series from the sum over $m$, as in \eref{id_geometric_2}.

\subsubsection{M2-brane theories from wrapped D6-branes: $Y^{p,q}(\bC\bP^2)$}
\label{sec:wrapped_D6}

Another large class of M2-brane theories with quantum corrected chiral rings arises when the reduction to type IIA leads to D6-branes wrapping exceptional divisors in a CY$_3$ cone \cite{Benini2011,Closset2012}, rather then noncompact 4-cycles as was the case for the flavored quiver gauge theories discussed in the previous section. To show how our methods can be applied to these theories as well, we consider here the worldvolume theory of a single mobile regular M2-brane probing the cone over $Y^{p,q}(\bC\bP^2)$ \cite{Gauntlett2004b,Gauntlett2005,Martelli2008c}, initially in the absence of fractional M2-branes.

The theory was obtained in \cite{Benini2011} by reducing M-theory on the cone over $Y^{p,q}(\bC\bP^2)$ to type IIA string theory on a resolved $\bC^3/\bZ_3$ foliated over $\bR$, with Ramond-Ramond 2-form flux and $p$ anti-D6-branes wrapping the exceptional $\bC\bP^2$ in the singular $\bC^3/\bZ_3$ leaf over the origin of $\bR$. The quiver is the one for regular branes probing $\bC^3/\bZ_3$, but the presence of the anti-D6-branes and the RR flux changes the gauge groups and Chern-Simons levels to $U(1)_0\times U(1+p)_{\frac{3}{2}p-q} \times U(1+p)_{-\frac{3}{2}p+q}$, see the figure below.\footnote{This is actually the quiver for $p+1$ regular D2-branes and $p$ wrapped anti-D6-branes: $p$ 2-branes are stuck at the singularity and only one of them is free to explore the transverse geometry \cite{Benini2011}.}
\bea 
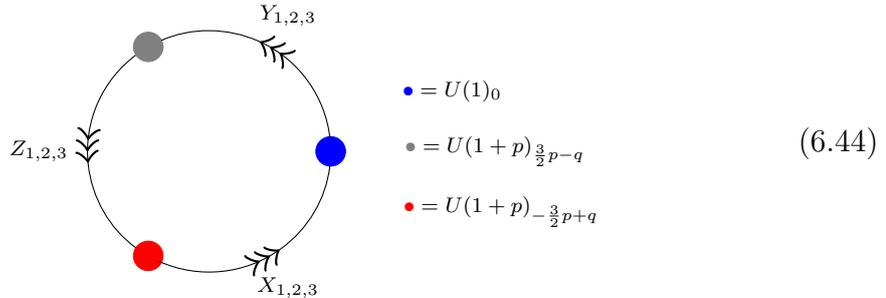
\begin{tikzpicture}[baseline, scale=0.8, font=\scriptsize]
\def \n {3}
\def \radius {2cm}
\def \margin {6} % margin in angles, depends on the radius
\foreach \s in {1,...,1}
{
	\node[draw=red, circle, fill=red] at ({360/\n * (\s - 2)}:\radius) {};
	\draw[decoration={markings, mark=at position 0.45 with {\arrow[scale=2.5]{>}}, mark=at position 0.5 with {\arrow[scale=2.5]{>}}, mark=at position 0.55 with {\arrow[scale=2.5]{>}}}, postaction={decorate}, shorten >=0.7pt] ({360/\n * (\s - 3)+\margin}:\radius) 
	arc ({360/\n * (\s - 3)+\margin}:{360/\n * (\s-2)-\margin}:\radius) node at (60:1.3*\radius) {$Y_{1,2,3}$};
}
\foreach \s in {2,...,2}
{
	\node[draw=blue, circle, fill=blue] at ({360/\n * (\s - 2)}:\radius) {};
	\draw[decoration={markings, mark=at position 0.45 with {\arrow[scale=2.5]{>}}, mark=at position 0.5 with {\arrow[scale=2.5]{>}}, mark=at position 0.55 with {\arrow[scale=2.5]{>}}}, postaction={decorate}, shorten >=0.7pt] ({360/\n * (\s - 3)+\margin}:\radius) 
	arc ({360/\n * (\s - 3)+\margin}:{360/\n * (\s-2)-\margin}:\radius) node at (180:1.4*\radius) {$Z_{1,2,3}$};
}
\foreach \s in {3,...,3}
{
	\node[draw=gray, circle, fill=gray] at ({360/\n * (\s - 2)}:\radius) {};
	\draw[decoration={markings, mark=at position 0.45 with {\arrow[scale=2.5]{>}}, mark=at position 0.5 with {\arrow[scale=2.5]{>}}, mark=at position 0.55 with {\arrow[scale=2.5]{>}}}, postaction={decorate}, shorten >=0.7pt] ({360/\n * (\s - 3)+\margin}:\radius) 
	arc  ({360/\n * (\s - 3)+\margin}:{360/\n * (\s-2)-\margin}:\radius) node at (300:1.3*\radius) {$X_{1,2,3}$};
}
\node[draw=none] at (4,1) {${\blue \bullet} = U(1)_0$};
\node[draw=none] at (4.7,0) {${\gray \bullet} = U(1+p)_{\frac{3}{2}p-q}$};
\node[draw=none] at (4.8,-1) {${\red\bullet} = U(1+p)_{-\frac{3}{2}p+q}$};
\end{tikzpicture}
\eea
The superpotential $W=\epsilon^{abc}\tr(X_aY_bZ_c)$ preserves an $SU(3)$ mesonic symmetry.

The cancellation of parity anomalies requires mixed Chern-Simons couplings between the abelian factors of the gauge group, which escaped the stringy derivation of \cite{Benini2011}. Like in section 6 of \cite{Benini2011}, we will choose the mixed Chern-Simons couplings in such a way that they do not affect the charges of diagonal monopole operators, and therefore the geometric branch of the moduli space.

We will compute the Hilbert series of the geometric branch of the moduli space, on which only monopole operators of magnetic charge $(m;m,0^p;m,0^p)$, $m \in \bZ$, the first component of $X_a$, $Y_b$ and the $11$ entry of $Z_c$ acquire expectation values \cite{Benini2011}. We choose to assign non-democratic $R$-charges $R[X]=R[Y]=\frac{3}{4}$ and $R[Z]=\frac{1}{2}$, to ensure the $R$-charges of unit charge monopole operators are positive:%
\footnote{Assigning $R$-charge $2/3$ to all bifundamentals leads to zero $R$-charge for monopole operators.}
\be
R[T]=R[\t T]=p-\frac{3}{2}p R[Z]=\frac{1}{4}p~.
\ee
The charges of monopole operators under the active $U(1)^3$ gauge group of the geometric branch are
\bea
Q[T]&=-(k_1^+,k_2^+,k_3^+)=(0,-3p+q,3p-q) \\
Q[\t T]&=(k_1^-,k_2^-,k_3^-)=(0,-q,q)~.
\eea
The constraint $0\le q \le 3p$ on this class of geometries implies that $\pm k_2^\pm =\mp k_3^\pm\geq 0$. (In the limiting cases $q=0,3p$, the CY$_4$ cones are orbifolds of flat space \cite{Martelli2008c} .)

The Hilbert series of the $F$-flat moduli space of the abelian quiver for $\bC^3/\bZ_3$ with a democratic $R$-charge assignment is \cite{Forcella:2008bb}
\be\label{F-flat_C3Z3}
g(t;x,y)=\sum_{n=0}^\infty [0,n;n,0]_{x,y} t^{\frac{4}{3}n}
\ee
where $x_1$, $x_2$ are fugacities for an $SU(3)$ under which $(X,Y,Z)$ transform in the $[0,1]$ representation, whereas $y_1$ and $y_2$ are fugacities for the $SU(3)$ respected by the superpotential, under which each of $X_a$, $Y_a$ and $Z_a$ is a triplet. Our non-democratic $R$-charge assignment is obtained by replacing $x_1\to t^{-1/6} x_1$, $x_2\to t^{-1/3} x_2$ in \eqref{F-flat_C3Z3}. The fugacities counting the bare monopole operators $T$ and $\t T$ are 
\be
u=z t^\frac{p}{2}\left(\frac{q_2}{q_3}\right)^{-3p+q}=z t^\frac{p}{2}x_2^{-3p+q}~, \qquad  v=z^{-1} t^\frac{p}{2}\left(\frac{q_2}{q_3}\right)^{-q}=z^{-1} t^\frac{p}{2} x_2^{-q}~,
\ee
where $q_1$, $q_2$, $q_3$ are fugacities for the $U(1)^3$ gauge group of the residual abelian quiver gauge theory on the geometric branch. The overall $U(1)$ is decoupled and the relative $U(1)$'s are identified with the Cartan of $SU(3)_x$ according to $x_1=q_1/q_3$, $x_2=q_2/q_3$. 

Performing the sum over magnetic charges as in \eqref{id_geometric_2}, the Hilbert series of the geometric moduli space is given by the formula
\be
\begin{split}
	H(t,y_1,y_2,z)&=\oint \frac{dx_1}{2\pi i x_1} \oint \frac{dx_2}{2\pi i x_2} \sum_{n=0}^\infty [0,n;n,0]_{x,y} t^{\frac{4}{3}n}\bigg|_{\substack{x_1\to t^{-1/6} x_1\\ x_2\to t^{-1/3} x_2}}\cdot\\
	&\qquad \qquad \cdot \PE[z t^\frac{p}{2}x_2^{-3p+q}+z^{-1} t^\frac{p}{2} x_2^{-q}- t^p x_2^{-3p}]~.
\end{split}
\ee
We first perform the $x_1$ integral, picking the residues at $x_1=t^{3/2}y_1$, $x_1=t^{3/2}y_2/y_1$ and $x_1=t^{3/2}/y_2$, then the $x_2$ integral, picking minus the residues at $x_2=t^{-1} /y_1$, $x_2=t^{-1} y_1/y_2$ and $x_2=t^{-1} y_2$. The result is 
\be 
\begin{split}
	H&=\frac{y_1^3 y_2}{\left(y_1^2-y_2\right) (y_1 y_2-1)}\PE[t^4 y_1^3+ z t^{\frac{p}{2}+3p-q} y_1^{3 p-q} + z^{-1} t^{\frac{p}{2}+q} y_1^{q} - t^{4p} y_1^{3p}] \\
	&+ \left(~~~(y_1,y_2)\to(1/y_2,1/y_1)~~~ \right) + \left(~~~(y_1,y_2)\to(y_2/y_1,1/y_1)~~~ \right)~,
\end{split}
\ee
where the sum over the three terms leads to symmetrization. Since 
\be
\begin{split}
	&\left(y_1^n \frac{y_1^3 y_2}{\left(y_1^2-y_2\right) (y_1 y_2-1)}\right)+ \left(~~~(y_1,y_2)\to(1/y_2,1/y_1)~~~ \right) +\\
	&\qquad+ \left(~~~(y_1,y_2)\to(y_2/y_1,1/y_1)~~~ \right)= [n,0]_{y_1,y_2}~,
\end{split}
\ee
we find that the $SU(3)$ highest weight generating function \cite{Hanany:2014dia} for the Hilbert series of the geometric moduli space takes the very simple form 
\be\label{HWG}
{\rm{HWG}}(t;z;\mu_1,\mu_2)= \PE[t^4 \mu_1^3+ z t^{\frac{p}{2}+3p-q} \mu_1^{3 p-q} + z^{-1} t^{\frac{p}{2}+q} \mu_1^{q} - t^{4p} \mu_1^{3p}] ~,
\ee
where $\mu_1$, $\mu_2$ are highest weight fugacities for $SU(3)$. The Hilbert series is obtained by Taylor expanding \eqref{HWG} in $t$ and replacing $\mu_1^{n_1} \mu_2^{n_2} \mapsto [n_1,n_2]_{y_1,y_2}$. Note that only symmetric powers $[n,0]$ of the fundamental representation appear.  
The democratic $R$-charge assignment can be restored by the fugacity map $z\to z t^{\frac{p}{2}-\frac{q}{3}}$, giving 
\be\label{HWG_dem}
{\rm{HWG}}_{dem}(t;z;\mu_1,\mu_2)= \PE[(t^\frac{4}{3} \mu_1)^3+ z (t^\frac{4}{3}\mu_1)^{3 p-q} + z^{-1} (t^\frac{4}{3}\mu_1)^{q} - (t^\frac{4}{3}\mu_1)^{3p}] ~.
\ee

This result agrees with the Hilbert series of the cone over $Y^{p,q}(\bC\bP^2)$ computed from the toric description. \eqref{HWG}--\eqref{HWG_dem} hold even when $p$, $q$ are not coprime. 

An interesting example is for $p=2$, $q=3$, in which case the Sasaki-Einstein 7-fold is known as $M^{1,1,1}$ or $M^{3,2}$. Setting $z=w^2$, we see that the topological symmetry enhances to $SU(2)$ and the Hilbert series of the geometric moduli space  
\be\label{HS_M111}
H(t;y_1,y_2,w)=\sum_{n=0}^\infty [3n,0;2n]_{y_1,y_2;w} t^{4 n}
\ee
reproduces the Hilbert series of $C(M^{1,1,1})$ computed in \cite{Hanany:2008fj} using a theory with different Chern-Simons levels (that was argued in \cite{Benini2011} to correspond to regular and fractional M2-branes), as well as the Kaluza-Klein spectrum obtained in  \cite{Fabbri:1999hw}.

\subsubsection*{A general argument}

Following the logic of \cite{Benini2011,Closset2012}, it is possible to reformulate the computation of the Hilbert series of the geometric moduli space in a general way that also holds in the presence of fractional M2-branes. We still focus on a single mobile M2-brane. (The generalization to multiple mobile M2-branes goes along the lines of appendix \ref{multiple_branes}.)

Let us first recall that, in the language of \cite{Closset2012}, the quiver for D-branes transverse to $\bC^3/\bZ_3$ has three open string K\"ahler chambers in Fayet-Iliopoulos parameter space: chamber $\bf X$ corresponding to $-\xi_3\equiv\xi_1+\xi_2\geq 0$ and $\xi_1\ge 0$; chamber $\bf Y$ corresponding to $-\xi_1\geq 0$ and $\xi_2\ge 0$; chamber $\bf Z$ corresponding to $-\xi_2\geq 0$ and $\xi_3=-\xi_1-\xi_2\ge 0$. If there are no fractional M2-branes as considered so far, the effective FI parameters for positive and negative $\sigma$ of the theory on the mobile M2-brane lie on the wall between chambers $\bf X$ and $\bf Y$. 

The three chambers can be seen using the baryonic Hilbert series as follows. Replacing irreducible representations of the $SU(3)$ symmetry that acts on indices $a,b,c$ by their highest weight states, the Hilbert series of the master space becomes 
\be
\sum_{n=0}^\infty [0,n]_{x_1,x_2} \tau^n = \PE\left[ \tau[0,1]_{x_1,x_2}\right]=  \PE\left[ \tau \left(x_2 + \frac{x_1}{x_2}+\frac{1}{x_1}\right)\right]
\ee
where $\tau=t^{4/3} \mu_1$, using democratic $R$-charges. Letting $B=(B_1,B_2,B_3)$ with $B_1+B_2+B_3=0$ be the vector of baryonic charges, the baryonic Hilbert series is
\be\label{HWG_Ypq}
\begin{split}
	{\rm HWG}[g_1&(B)] =\oint \frac{dx_1}{2\pi i x_1^{1+B_1}} \oint \frac{dx_2}{2\pi i x_2^{1+B_2}} \PE\left[ \tau[0,1]_{x_1,x_2}\right]= \PE[\tau^3] \tau^{\chi(B)}~,
\end{split}
\ee
where 
\be
\chi(B) =\begin{cases}
	B_1-B_3 ~, & -B_3\ge 0 \wedge B_1\ge 0 \Leftrightarrow B\in {\bf X}\\
	B_2-B_1 ~, & -B_1\ge 0 \wedge B_2\ge 0 \Leftrightarrow B\in {\bf Y}\\
	B_3-B_2 ~, & -B_2\ge 0 \wedge B_3\ge 0 \Leftrightarrow B\in {\bf Z}
\end{cases}
\ee
is the Chern class of the line bundle $\cO(\chi(B))$ on $\bC\bP^2$ of which $g_1(B)$ counts holomorphic sections. Translating quiver data to geometric data, the baryonic charge $B$ is mapped to $\chi(B)$; the map depends on which chamber ($\bf X$, $\bf Y$ or $\bf Z$) $B$ belongs to.

Given a theory for regular and fractional M2-branes at the cone over $Y^{p,q}(\bC\bP^2)$ (see section 5 of \cite{Benini2011}), we compute the 3-vectors $\theta^\pm\equiv \pm k^\pm=\pm (k^\pm_1,k^\pm_2,k^\pm_3)$, where $k^\pm_i$ are the effective Chern-Simons levels of the $i$-th gauge group of the worldvolume theory on a mobile M2-brane. The sum over magnetic charges then becomes a sum over $\theta^- \bZ_{\geq 0}\cup \theta^+ \bZ_{\geq 0}$ in the lattice of baryonic charges, giving the Hilbert series
\be
H=\sum_{n=0}^\infty z^n g_1(\theta^+ n) + \sum_{n=0}^\infty z^{-n} g_1(\theta^- n) - g_1(0)~.
\ee
Here we used the fact that monopole operators of charge $(m,0^{N_1-1};m,0^{N_2-1};m,0^{N_3-1})$ have vanishing $R$-charge if all bifundamentals have $R$-charge $2/3$, for all $N_1$, $N_2$, $N_3$. 

Using \eref{HWG_Ypq}, the highest weight generating function for the Hilbert series is then 
\be\label{HWG_Ypq_2}
\begin{split}
	{\rm HWG}&(t,z;\mu_1,\mu_2)=\PE[\tau^3]\left(\PE[z \tau^{\chi(\theta^+)}] +\PE[z^{-1} \tau^{\chi(\theta^-)}]-1 \right)\\
	&=\PE[\tau^3+z \tau^{\chi(\theta^+)}+z^{-1} \tau^{\chi(\theta^-)}-\tau^{\chi(\theta^+)+\chi(\theta^-)} ]~, \qquad \tau =t^{4/3}\mu_1~.
\end{split}
\ee
For all the quiver gauge theories on a regular M2-brane in the presence of any number of fractional M2-branes in \cite{Benini2011}, it is straightforward to derive that $\chi(\theta^+)=3p-q$ and $\chi(\theta^-)=q$. Thus \eref{HWG_Ypq_2} becomes \eref{HWG_dem}. Note that while $\theta^\pm$ depend on the torsion flux sourced by the fractional M2-branes,%
\footnote{The pair of chambers to which $\theta^\pm$ belong determine \emph{windows} in the lattice that defines the torsion cohomology of $Y^{p,q}(\bC\bP^2)$ \cite{Benini2011}. In particular, $(\theta^+,\theta^-)\in ({\bf X}, {\bf X})$ in window $[0,0]$ of \cite{Benini2011}; $(\theta^+,\theta^-)\in ({\bf X}, {\bf Z})$ in window $[1,0]$; and $(\theta^+,\theta^-)\in ({\bf X}, {\bf Y})$ in window $[-1,0]$.} 
$\chi(\theta^\pm)$ only depend on the purely geometric data $p$ and $q$ of the CY$_4$ cone.

\subsubsection{The cone over $V^{5,2}$}
In this subsection we study the field theories on M2-branes probing the cone over $V^{5,2}$, a homogenous Sasaki-Einstein 7-manifold that can be described as a coset $V^{5,2} = SO(5)/SO(3)$.  The supergravity solution thus possesses an $SO(5) \times U(1)_R$ isometry \cite{Fabbri:1999hw}. There are two known gauge theories corresponding to this geometry: one with classical chiral ring, proposed by Martelli and Sparks (MS model henceforth) \cite{Martelli:2009ga}, the other with quantum corrected chiral ring, proposed by Jafferis (J model) \cite{Jafferis:2009th}.  We will see that the total moduli spaces of the two abelian theories agree assuming that the superpotential of the J model is appropriately modified.

\subsection*{The MS model} 
The quiver diagram of the MS model is
\bea
\begin{tikzpicture}[baseline, font=\scriptsize, scale=0.8]
\begin{scope}[auto,%
every node/.style={draw, minimum size=0.5cm}, node distance=2cm];
% the vertices
\node[circle] (USp2k) at (-0.1, 0) {$U(N)_{+k}$};
\node[circle, right=of USp2k] (BN)  {$U(N)_{-k}$};
\end{scope}
% the edges
\draw[draw=black,solid,line width=0.2mm,->>]  (USp2k) to[bend right=-50] node[midway,above] {$A_{1,2}$}node[midway,above] {}  (BN) ; 
\draw[draw=black,solid,line width=0.2mm,<<-]  (USp2k) to[bend left=-50] node[midway,below] {$B_{1,2}$} node[midway,below] {} (BN) ;  
\draw[black,-> ] (USp2k) edge [out={-150},in={150},loop,looseness=8] (USp2k) node at (-2,1.2) {$\phi_1$} ;
\draw[black,-> ] (BN) edge [out={30},in={-30},loop,looseness=8] (BN) node at (5.8,1.2) {$\phi_2$};
\end{tikzpicture}
\eea
and the superpotential is
\begin{equation}\label{W_{MS}}
W = \Tr\left[ s(\phi_1^3 + \phi_2^3) +\phi_1(A_1 B_1 + A_2 B_2) + \phi_2 (B_1 A_1+ B_2 A_2) \right] ~. 
\end{equation}
We set the coupling $s$ to unity in the following, by a rescaling of chiral superfields.

\paragraph{The case of $N=1$.} Let us focus on the moduli space of this theory for $N=1$.  Using the primary decomposition of {\tt Macaulay2}, we see that the $F$-flat moduli space has two branches:
\be \label{branchesV52MS}
\begin{split}
	{\rm (I)} & \qquad \{ 3 \phi_2^2 + A_1B_1+A_2 B_2=0,~ \phi_1+\phi_2 =0 \}~, \\
	{\rm (II)} & \qquad \{ A_1=A_2=B_1=B_2=\phi_1^2=\phi_2^2 =0 \}~.
\end{split}
\ee  
After introducing bare monopole operators and modding out by the complexified gauge group, branch I in \eref{branchesV52MS} corresponds to the cone over $V^{5,2}/\bZ_k$, in agreement with the discussion in \cite{Martelli:2009ga}. The Hilbert series of this branch is
\bea
H^{{\rm (I)}}_{\text{MS}}(t, x, y) = \sum_{m \in \bZ} y^m g^{(I)}_1(k m)~,
\eea
where $g^{(I)}_1$ is the baryonic Hilbert series for the abelian quiver
\bea
g^{(I)}_1(B) &= \oint
\frac{dz}{2\pi i z^{B+1}}
\PE \left[\{ 1+(x+x^{-1})(z+z^{-1}) \} t - t^2  \right]~,
\eea
$x$ is a fugacity for the $SU(2)$ flavour symmetry, $y$ is a fugacity for the topological symmetry $U(1)_M$, and $t$ keeps track of the $R$-charge in units of $2/3$.

For $k=1$, we have
\bea \label{MSHSk1}
H^{{\rm (I)}}_{\text{MS}; k=1}(t, x, y)  = \PE \left[ \chi^{SO(5)}_{[1,0]}(\vec a) t - t^2 \right]
\eea
where the character of the vector representation of $SO(5)$ is
\bea
\chi^{SO(5)}_{[1,0]}(\vec a) = 1+ a_1 + a_2 + a_1^{-1} + a_2^{-1} ~, \quad a_1= x y, ~ a_2 = x y^{-1}~.
\eea
Explicitly, the generators of the chiral ring that transform under the vector representation of $SO(5)$ are
\bea
\phi_2=-\phi_1~, \qquad  u_1= A_1 T~, \qquad u_2=A_2 T,  \qquad v_1 = B_1 \tilde{T}~, \qquad v_2 =B_2  \tilde{T}.  
\eea
where $T$ and $\tilde{T}$ are the monopole operators carrying magnetic fluxes $(1,1)$ and $(-1,-1)$ under the $U(1) \times U(1)$ gauge group.%
\footnote{$T$ and $\tilde{T}$ carry electric charges $(-1,1)$, $(1,-1)$ and $R$-charges $0$, $0$ respectively.}  They are subject to the relation
\bea
3 \phi_2^2+u_1 v_1 + u_2 v_2 =0~, \label{relV52MS}
\eea
which follows from $T\tilde{T}=1$ and the classical $F$-terms.

For $k>1$, the $\bZ_k$ quotient breaks $SO(5)$ to $SO(4) \cong SU(2) \times SU(2)$.  The Hilbert series can be written as
\bea
H^{{\rm (I)}}_{\text{MS}}(t, x, y)  &=\frac{1}{k} \sum_{p=0}^{k-1} \PE[\{1+ \omega^p_k (a_1+  a_1^{-1}) + \omega^{-p}_k (a_2 +a_2^{-1}) \}t - t^2] \nn \\
&=\frac{1}{k} \sum_{p=0}^{k-1} \PE[\{1+ \omega^p_k \chi^{SU(2)}_{[1]}(a_1)  + \omega^{-p}_k \chi^{SU(2)}_{[1]}(a_2) \}t - t^2]~,
\eea
where $\omega_k = \exp(2 \pi i/k)$.

\paragraph{The total moduli space.} It is also interesting to discuss the total moduli space, which includes the second branch. Looking at the $F$-flat moduli space in \eqref{branchesV52MS}, we see that the second branch provides two more gauge invariant operators: $\phi_1+\phi_2$ at order $t$ and $(\phi_1+\phi_2)^2=-(\phi_1-\phi_2)^2$ at order $t^2$. 

The baryonic generating function is
\be
\begin{split}
	g^{\text{tot}}_1(B) &= g^{(I)}_1(B) + \delta_{B,0}(t+t^2)
\end{split}
\ee
as can be checked using {\tt Macaulay2}. The Hilbert series of the total moduli space is
\bea\label{MS_total_k1}
H^{\text{tot}}_{\text{MS}}(t, x, y) = \sum_{m\in\bZ} y^m g^{\text{tot}}_1(k m) = H^{(I)}_{\text{MS}}(t, x, y)+t+t^2~.
\eea
The corresponding unrefined Hilbert series is
\bea
H^{\text{tot}}_{\text{MS}, k=1}(t; x=y=1) = 1+6 t+15 t^2+30 t^3+55 t^4+91 t^5+140 t^6+\ldots~. \label{MSunrefined}
\eea

\subsection*{The J model} 

This theory can be derived by reducing M-theory to type IIA along a different $U(1)$ isometry of $V^{5,2}$, this time with a fixed locus that leads to D6-branes \cite{Jafferis:2009th}. We introduce a $\bZ_h$ quotient along this circle direction. The theory is a flavored version of $\cN=8$ SYM, with quiver diagram 
\be
\begin{tikzpicture}[scale=3]
\begin{scope}[auto,%
every node/.style={draw, minimum size=0.6cm}, node distance=2cm];
\node[circle]  (UN)  at (-0.5,1.7) {$N$};
\node[rectangle, right=of UN] (Ur) {$h$};
\end{scope}
\path[decoration={ 
	markings, mark=at position .7  with {\arrow[blue,line width=1pt]{>}}}]  (UN) edge [out={-150},in={150},loop,looseness=10] (UN) node at (-1.2,1.7) {$\phi_{1,2,3}$} ;
\draw[draw=black,solid,line width=0.3mm,->]  (UN) to[bend right=30] node[midway,below] {$q$}node[midway,above] {}  (Ur) ; 
\draw[draw=black,solid,line width=0.3mm,<-]  (UN) to[bend left=30] node[midway,above] {$\t q$} node[midway,above] {} (Ur) ;    
\end{tikzpicture}
\ee
In \cite{Jafferis:2009th} the superpotential was argued to be 
\begin{equation}\label{superpotential fundamental}
W = \Tr \left( \phi_3 \left[ \phi_1, \phi_2 \right] + \sum_{j=1}^{h} q_j \tilde q^j \left( \phi_1^2 + \phi_2^2 + \phi_3^2 \right)  \right) \, .
\end{equation}
In these variables not all Cartan elements of the $SO(5)$ isometry of $V^{5,2}$ are manifest, as we shall see below. We amend this by introducing the following variables,%
\footnote{N.M. would like to thank Seyed Morteza Hosseini for a related discussion on this model in the context of the large $N$ limit.}
\bea \label{Xandphi}
X_1 =  \frac{1}{\sqrt{2}}(\phi_1 + i \phi_2)~, \qquad X_2 = \frac{1}{\sqrt{2}}(\phi_1 - i \phi_2)~, \qquad X_3 = i \phi_3~,
\eea
in terms of which the superpotential is
\bea \label{supV52JmodelX}
W = \Tr \left[ X_3  [X_1 , X_2] +   \sum_{j=1}^h q_j \tilde q^j ( X_1 X_2+ X_2 X_1 - X_3^2)  \right]~.
\eea

We will see below that an extra superpotential term consistent with the symmetry should be added to reproduce the total moduli space of the MS theory.

\paragraph{The case of $N=1$.} Let us focus on the moduli space of this theory for $N=1$.  Using the primary decomposition of {\tt Macaulay2}, we see that the $F$-flat moduli space has three branches:
\be \label{branchesV52J}
\begin{split}
	{\rm (I)} & \qquad \{ q = \tilde{q} = 0 \}~, \\
	{\rm (II)} & \qquad \{ q= 2X_1 X_2 - X_3^2=0\} \quad \text{or} \quad  \{ \tilde{q} = 2X_1 X_2 - X_3^2 =0 \} ~, \\
	{\rm (III)} & \qquad \{ X_1= X_2 =X_3 =0 \}~.
\end{split}
\ee  

Branch I leads to the geometric branch of the moduli space, once monopole operators are included. The Hilbert series of this branch is given by 
\bea
H^{{\rm (I)}}_{\text{J}}(t, x, y) = \sum_{m\in\bZ} y^m  t^{h |m|}  \PE \left[( 1+x+ x^{-1}) t   \right]~,
\eea
where $x^{1/2}$ is a fugacity for the $SU(2)$ flavour symmetry  that rotates $X_{1}$, $X_2$, $X_3$ as a triplet, $y$ is a fugacity for the topological symmetry, and $t$ keeps track of the $R$-charge in units of $2/3$.  

For $h=1$, we obtain
\bea \label{HSJmodelk1}
H^{{\rm (I)}}_{\text{J};h=1}(t, x, y) = \PE \left[ (1+ x+ x^{-1} + y+ y^{-1}) t - t^2 \right]~.
\eea
This is in agreement with \eref{MSHSk1}, identifying $a_1 = x$ and $a_2 = y$.  Explicitly, the generators of the chiral ring in the vector representation of $SO(5)$ are  
\bea\label{gen_J}
X_1 \; , \qquad X_2\; , \qquad X_3 \; , \qquad V_+ \; , \qquad V_-~,
\eea
where $V_\pm$ are the monopole operators of magnetic charge $\pm 1$.%
\footnote{$V_\pm$ are gauge invariant and carry $R$-charge $2/3$.}  
Note that had we used $\phi_{1,2,3}$ instead of the variables $X_{1,2,3}$ defined in \eref{Xandphi}, we would not have been able to make the Cartan variable $x$ in \eref{HSJmodelk1} manifest.  The generators are subject to the relation 
\bea
V_+ V_- =2X_1 X_2 - X_3^2~. \label{relV52Jaff}
\eea
The operator map between the MS model and the J model is as follows:
{\small
	\bea
	u_1 \; \leftrightarrow \; \sqrt{2} X_1~, \quad v_1 \; \leftrightarrow \; \sqrt{2} X_2~, \quad u_2 \; \leftrightarrow \; i V_+~, \quad v_2 \; \leftrightarrow \; i V_-~, \quad \phi_2 \; \leftrightarrow \;  i \sqrt{3} X_3~.
	\eea}
Under this map, \eref{relV52MS} and \eref{relV52Jaff} are transformed into each other as expected. 

The $\bZ_h$ quotient for $h>1$ breaks $SO(5)$ to $SO(3) \times SO(2)$.  The Hilbert series can be written as
\bea
H^{{\rm (I)}}_{\text{J}}(t; x, y)  = \PE \left[ t \chi^{SU(2)}_{[2]} (x^{1/2})  + t^h (y+y^{-1}) -t^{2h} \right]~,
\eea
and the chiral ring of the geometric branch is generated by \eqref{gen_J} subject to 
\bea
V_+ V_- = (2X_1 X_2 - X_3^2)^h~. \label{relV52Jaff_h}
\eea

\paragraph{The total moduli space.} Let us discuss the total moduli space without focusing on a particular branch.  In order to match this with that of the MS model at $k=1$, we need to modify the superpotential \eref{supV52JmodelX} by adding an extra term cubic in the flavor meson, that is allowed by the $U(1)_R \times SO(3) \times U(1)_M$ symmetry: 
\be
\begin{split}
	W &= \Tr \Big[ X_3  [X_1 , X_2] +  q \t q ( X_1 X_2+ X_2 X_1 - X_3^2) + (q \t q)^3 \Big] ~.
\end{split}
\ee

We focus again on the abelian $N=1$ theory. The new term does not affect the geometric branch, on which $q=\tilde{q}=0$. A branch where only one of $q$, $\t q$ vanishes does not lead to gauge invariant operators involving the quarks, and is therefore a sub-branch of the geometric branch. Finally, we have  the branch where $q$, $\t q$ and therefore $M=q\tilde q$ take vev. On this branch the monopole operators vanish and the $F$-term equations imply that $MX_i = 0$ and $M^3=0$. Therefore the only new operators are $M$ and $M^2$, which add $t+t^2$ to the Hilbert series of the geometric branch \eref{HSJmodelk1}. (We checked this conclusion by doing primary decomposition with {\tt Macaulay2}.) This result agrees with \eref{MS_total_k1} found in the dual MS model. 

Note that were the $(q\t q)^3$ superpotential term absent, all powers $M^n$ would be allowed and the extra flavor mesonic branch would be $\bC$, corresponding to branch (III) in \eref{branchesV52J}. This result would agree with a modification of the MS model where the superpotential of the abelian theory is taken to be 
\be\label{W_MS_modified}
W = (\phi_1+\phi_2)\left(\frac{3}{4}(\phi_1-\phi_2)^2 + A_1 B_1 + A_2 B_2 \right)~.
\ee
Such a modification was proposed in \cite{ChesterIliesiuPufuEtAl2015}, where it was argued that the further superpotential term $\frac{1}{4}(\phi_1+\phi_2)^3$, that is needed to obtain \eref{W_{MS}} with $s=1$, flows to zero at low energies. With this superpotential, the branch (II) of \eref{branchesV52MS} is modified to $A_1=A_2=B_1=B_2=\phi_1-\phi_2=0$ with $\phi_1+\phi_2$ unconstrained, in agreement with the result for the J model.

\section*{Acknowledgements}

We thank Nick Halmagyi, Seyed Morteza Hosseini, Dario Martelli for useful discussions, and the following institutes and workshops for hospitality and partial support: University of Milano-Bicocca, the Galileo Galilei Institute for Theoretical Physics and INFN in Florence, the LPTENS laboratory of the Ecole Normale Superieure, the LPTHE laboratory of the Universit\'e Pierre et Marie Curie, the Henri Poincar\'e Institute and the University of Bern. NM is grateful to Nick Halmagyi, Ruben Minasian and Claudius Klare for their kind hospitality during his academic visit in Paris.  NM also gratefully acknowledges support from the Simons Center for Geometry and Physics, Stony Brook University, as well as the 2016 Summer Workshop at which some of the research for this paper was performed.  AZ is partially supported by  INFN and the MIUR-FIRB grant RBFR10QS5J ``String Theory and Fundamental Interactions''.

\appendix

\section{Multiple membranes and symmetric products} \label{multiple_branes}

In this appendix, following a similar reasoning in \cite{Nakajima2015}, we show in detail that the Hilbert series of the moduli space of the ABJM theory of rank $N$ and level $k$ (the theory on $N$ M2-branes probing $\bC^4/\bZ_k$) is the $N$-th symmetric product of the Hilbert series of ABJM of rank $N=1$ and level $k$ (the theory on $1$ M2-brane probing $\bC^4/\bZ_k$). We shall prove that 
\be\label{aim_proof}
\sum_{N=0}^\infty H_{N,k}(t,x,y,z) \nu^N = {\rm PE}[\nu \sum_{m \in \bZ} z^m g_1(t,x,y;km)]~,
\ee
assuming the validity of the conjecture \eqref{baryonic_sym} of \cite{Forcella:2007wk} according to which the baryonic generating function $g_{r} (t,x,y; r B)$ is the $r$-fold symmetric product of $g_{1} (t,x,y; B)$. The proof immediately generalizes to the worldvolume theories on membranes probing the other singularities considered in this paper.

We start from $H_{N,k}$, which is given by a sum over integers $m_1\geq m_2\geq \dots\geq m_N$ in a Weyl chamber of $U(N)$, as shown in section \ref{sec:arbitrary_N}. (These are magnetic weights for the diagonal $U(N)$ factor of the gauge group.) It is convenient to change parametrization from $m_1\geq m_2\geq \dots\geq m_N$ in $\bZ^N/S_N$ to $(m;r_0,r_1,\dots)\in \bZ \times \bZ_{\geq 0}^\infty$, where $m=r_N$ and $r_\alpha=\#\{i| m_i-m_N=\alpha \}$. By definition $r_0\neq 0$ and $\sum_\alpha r_\alpha=N$, so there are finitely many non-vanishing $r_\alpha$ at fixed $N$. In the latter parametrization, the Hilbert series for the $N$ M2-brane theory reads 
\be\label{start_proof}
H_{N,k}(t,x,y,z) = \sum_{m\in\bZ} \sum_{\substack{r_0=1\\r_1,r_2,\dots=0\\\sum_\alpha r_\alpha=N}}^\infty z^{\sum_\alpha r_\alpha (m+\alpha)}
\prod_\alpha g_{r_\alpha}(t,x,y;k r_\alpha (m+\alpha))~,
\ee
therefore the LHS of \eqref{aim_proof} can be written as
\be\label{proof_1}
\begin{split}
	\hspace{-8pt}\sum_{N=0}^\infty H_{N,k}(t,x,y,z) \nu^N &= 1+\sum_{m \in\bZ} \sum_{\substack{r_0=1\\r_1,r_2,\dots=0}}^\infty \prod_{\alpha=0}^\infty g_{r_\alpha}(t,x,y;k r_\alpha (m+\alpha)) z^{r_\alpha (m+\alpha)} \nu^{r_\alpha}\\
	&= 1+\sum_{m \in\bZ} \prod_{\alpha=0}^\infty \sum_{r_\alpha=\delta_{\alpha,0}}^\infty  g_{r_\alpha}(t,x,y;k r_\alpha (m+\alpha)) z^{r_\alpha (m+\alpha)} \nu^{r_\alpha}~.
\end{split}
\ee
The conjecture \eqref{baryonic_sym} implies that  
\be\label{proof_2}
\sum_{r_\alpha=0}^\infty  g_{r_\alpha}(t,x,y;k r_\alpha (m+\alpha)) z^{r_\alpha (m+\alpha)} \nu^{r_\alpha}= {\rm PE}[\nu\, g_1(t,x,y;k(m+\alpha))z^{m+\alpha}]~,
\ee
from which we obtain
\be\label{proof_3}
\begin{split}
	\sum_{N=0}^\infty H_{N,k}(t,x,y,z) \nu^N = 1+\sum_{m \in\bZ}&\bigg(\prod_{\alpha=0}^\infty {\rm PE}[\nu\,g_1(t,x,y;k(m+\alpha))z^{m+\alpha}]\\
	&- \prod_{\alpha=1}^\infty {\rm PE}[\nu\, g_1(t,x,y;k(m+\alpha))z^{m+\alpha}] \bigg)~.
\end{split}
\ee
We split the sum over $m$ into $m\geq 0$ and $m<0$. The first sum is
\bea\label{proof_4}
&\sum_{m=0}^\infty \bigg(\prod_{\alpha=m}^\infty {\rm PE}[\nu\,g_1(t,x,y;k\alpha)z^{\alpha}] - \prod_{\alpha=m+1}^\infty {\rm PE}[\nu\,g_1(t,x,y;k\alpha)z^{\alpha}]\bigg) = \nonumber\\
=& \sum_{m=0}^\infty \bigg(\prod_{\alpha=m}^\infty {\rm PE}[\nu\,g_1(t,x,y;k\alpha)z^{\alpha}] -1\bigg) - \sum_{m=0}^\infty \bigg(\prod_{\alpha=m+1}^\infty {\rm PE}[\nu\,g_1(t,x,y;k\alpha)z^{\alpha}]-1\bigg) \nonumber\\
=& \prod_{\alpha=0}^\infty {\rm PE}[\nu\,g_1(t,x,y;k\alpha)z^{\alpha}] -1 = 
{\rm PE}[\nu \sum_{\alpha=0}^\infty g_1(t,x,y;k\alpha)z^{\alpha}] -1~,
\eea
where in the second line we subtracted and added $1$ so that the two sums converge. 

By similar manipulations, the sum over $m<0$ is
\be\label{proof_5}
\begin{split}
	&\sum_{m<0} \bigg(\prod_{\alpha=m}^\infty {\rm PE}[\nu\,g_1(t,x,y;k\alpha)z^{\alpha}] - \prod_{\alpha=m+1}^\infty {\rm PE}[\nu\,g_1(t,x,y;k\alpha)z^{\alpha}]\bigg) = \\
	&= \prod_{\alpha=0}^\infty {\rm PE}[\nu\,g_1(t,x,y;k\alpha)z^{\alpha}]\cdot \\
	&\cdot \sum_{m=1}^\infty \bigg(\prod_{\alpha=1}^m {\rm PE}[\nu\,g_1(t,x,y;-k\alpha)z^{-\alpha}] - \prod_{\alpha=1}^{m-1} {\rm PE}[\nu\,g_1(t,x,y;-k\alpha)z^{-\alpha}]\bigg) =\\
	&=  {\rm PE}[\nu \sum_{\alpha=0}^\infty g_1(t,x,y;k\alpha)z^{\alpha}]
	\bigg({\rm PE}[\nu\sum_{\alpha=1}^\infty g_1(t,x,y;-k\alpha)z^{-\alpha}]-1\bigg)~.
\end{split}
\ee
Adding up the various contributions to the RHS, \eqref{proof_3} becomes
\be\label{proof_end}
\begin{split}
	\sum_{N=0}^\infty H_{N,k}(t,x,y,z) \nu^N = {\rm PE}[\nu\,\sum_{\alpha\in\bZ} g_1(t,x,y;k\alpha)z^{\alpha}]={\rm PE}[\nu\, H_{1,k}(t,x,y,z)] ~,
\end{split}
\ee
which shows that 
\be
H_{N,k}(t,x,y,z)= {\rm Sym}^N \left(H_{1,k}(t,x,y,z)\right)~.
\ee

\section{$SL(2,\bZ)$ action on theories with an Abelian symmetry} \label{SL2Z}

In this appendix we discuss the $SL(2,\bZ)$ action on three-dimensional $\cN=2$ theories with a $U(1)$ symmetry \cite{Witten2003}. Let $x$ and $m$ be the fugacity and magnetic charge for the $U(1)$ symmetry, and $f(x;m)$ be the Hilbert series of the moduli space of the $\cN=2$ theory with the $U(1)$ symmetry. Fugacities and magnetic charges for other symmetries are spectators of this $SL(2,\bZ)$ action and will be suppressed in this appendix. 
The $SL(2,\bZ)$ group is generated by $S$ and $T$, subject to $S^2=-1$ and $(ST)^3=1$. The action of $-1$ on a theory with a $U(1)$ symmetry is meant to produce the same theory, but with the sign of the $U(1)$ charges reversed.

The action of $T$ is to introduce a Chern-Simons interaction at level $1$ for the $U(1)$ global symmetry, therefore
\be\label{T}
(T \circ f)(x;m)=f(x;m)x^{-m}~.
\ee

The action of $S$ is to couple the background gauge field for the original $U(1)$ symmetry to a new $U(1)$ background gauge field via a mixed Chern-Simons term at level $1$, and to gauge the original $U(1)$ symmetry \cite{Kapustin1999, Witten2003}. On the Hilbert series,
\be\label{S}
(S\circ f)(x';m')=\sum_{m\in \bZ} \oint \frac{dx}{2\pi i x} f(x;m)x^{-m'}x'^{-m}~.
\ee
The new $U(1)$ symmetry can be thought of as the topological symmetry of the gauged $U(1)$. In our standard notation, $x'=1/z$ and $m'=B$.

Let us check the relations among $SL(2,\bZ)$ generators. For $S^2$ we find
\be\label{S2}
\begin{split}
	(S^2 \circ f)(x'';m'')&=\sum_{m,m'\in \bZ} \oint \frac{dx}{2\pi i x} \oint \frac{dx'}{2\pi i x'}f(x;m)x^{-m'}x'^{-m-m''}x''^{-m'}=\\
	&=\sum_{m'\in \bZ} \oint \frac{dx}{2\pi i x} f(x;-m'')(x x'')^{-m'} = f(1/x'';-m'')~,
\end{split}
\ee
where the last equality can be derived using $f(x;m)= \sum_{n\in\bZ} f_n(m)x^n$. We see that indeed $S^2$ returns the Hilbert series of the same theory, but with the sign of $U(1)$ charges reversed, that is, $S^2=-1$. Similarly 
\bea\label{ST3}
&((ST)^3\circ f)(x''';m''') =\sum_{m,m',m''} \oint \frac{dx}{2\pi i x} \oint \frac{dx'}{2\pi i x'} \oint \frac{dx''}{2\pi i x''}
f(x;m)x^{-m-m'}x'^{-m-m'-m''}\cdot \nonumber\\
&\qquad \cdot x''^{-m'-m''-m'''}x'''^{-m''}=\sum_{m''} \oint \frac{dx}{2\pi i x} f(x;m''')(x/x''')^{m''} = f(x''';m''')~,
\eea
shows that $(ST)^3=1$.

\section{$\CN=3$ $A_k$ affine quivers} \label{Aquiver}

\subsection*{Affine $A_1$ quiver: $U(N)_k \times U(N)_{-k}$}

Let us consider the gauge theory given by the affine Dynkin diagram of $A_1$ with gauge group $U(N)_k \times U(N)_{-k}$. 
\bea
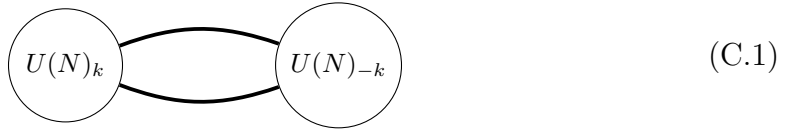
\begin{tikzpicture}[baseline, font=\footnotesize, scale=0.8]
\begin{scope}[auto,%
every node/.style={draw, minimum size=1.5cm}, node distance=2cm];
% the vertices
\node[circle] (USp2k) at (-0.1, 0) {$U(N)_k$};
\node[circle, right=of USp2k] (BN)  {$U(N)_{-k}$};
\end{scope}
% the edges
\draw[draw=black,solid,line width=0.5mm]  (USp2k) to[bend right=20] node[midway,below] {$$}node[midway,above] {}  (BN) ;
\draw[draw=black,solid,line width=0.5mm]  (USp2k) to[bend left=20] node[midway,above] {$$}
node[midway,below] {} (BN) ;  
\end{tikzpicture}
\eea
In $3d$ $\CN=2$ notation, this can be written as
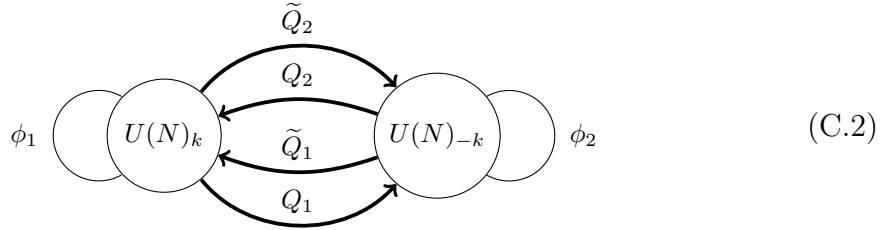
\bea
\begin{tikzpicture}[baseline, baseline, font=\footnotesize, scale=0.8]
\begin{scope}[auto,%
every node/.style={draw, minimum size=1.5cm}, node distance=2cm];
% the vertices
\node[circle] (USp2k) at (-0.1, 0) {$U(N)_k$};
\node[circle, right=of USp2k] (BN)  {$U(N)_{-k}$};
\end{scope}
% the edges
\draw[draw=black,solid,line width=0.5mm,<-]  (USp2k) to[bend right=20] node[midway,above] {$\tQ_1 $}node[midway,above] {}  (BN) ;
\draw[draw=black,solid,line width=0.5mm,->]  (USp2k) to[bend right=50] node[midway,above] {$Q_1$}node[midway,above] {}  (BN) ; 
\draw[draw=black,solid,line width=0.5mm,<-]  (USp2k) to[bend left=20] node[midway,above] {$Q_2$} node[midway,above] {} (BN) ;  
\draw[draw=black,solid,line width=0.5mm,->]  (USp2k) to[bend left=50] node[midway,above] {$\tQ_2$} node[midway,above] {} (BN) ;    
\draw(-0.8,0.65) arc (60:300:0.75cm) node at (-2.4,0) {$\phi_1$} ;
\draw(5.2,-0.65) arc (60:300:-0.75cm) node at (6.8,0) {$\phi_2$} ;
\end{tikzpicture}
\eea
with the superpotential
\bea \label{supA1bis}
W= \tr( Q_1 \phi_2 \tQ_1 -\tQ_1 \phi_1 Q_1 + Q_2 \phi_1 \tQ_2 -\tQ_2 \phi_2 Q_2) + \frac{1}{2} k\tr(\phi_1^2) - \frac{1}{2} k\tr(\phi_2^2)~.
\eea
Integrating out the adjoints, we obtain the ABJM theory with $A_i=(Q_1,\tilde Q_2)$ and $B_i =(Q_2,\tilde Q_1)$. We already discussed this case at length in section \ref{ABJM}.

\subsection*{Affine $A_2$ quiver: $U(N)_k \times U(N)_{-k} \times U(N)_0$}

In $3d$ $\CN=2$ notation, the quiver can be drawn as
% \\ 

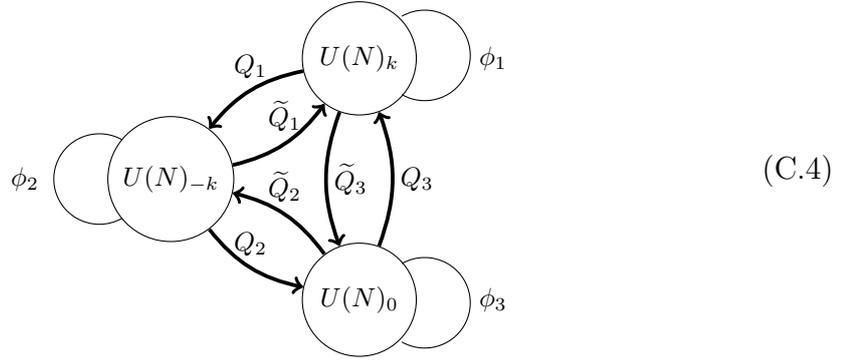
\bea
\begin{tikzpicture}[baseline, font=\footnotesize, scale=0.8]
\begin{scope}[auto,%
every node/.style={draw, minimum size=1.5cm}, node distance=2cm];
% the vertices
\node[circle] (USp2k) at (-0.1, 0) {$U(N)_{-k}$};
\node[circle] (BN)  at (3,2)  {$U(N)_{k}$};
\node[circle] (BN1)  at (3,-2)  {$U(N)_{0}$};
\end{scope}
% the edges
\draw[draw=black,solid,line width=0.5mm,<-]  (USp2k) to[bend left=20] node[midway,above] {$Q_1 $}node[midway,above] {}  (BN) ;
\draw[draw=black,solid,line width=0.5mm,->]  (USp2k) to[bend right=20] node[midway,above] {$\tQ_1$}node[midway,above] {}  (BN) ; 
\draw[draw=black,solid,line width=0.5mm,<-]  (USp2k) to[bend left=20] node[midway,above] {$\tQ_2 $}node[midway,above] {}  (BN1) ;
\draw[draw=black,solid,line width=0.5mm,->]  (USp2k) to[bend right=20] node[midway,above] {$Q_2$}node[midway,above] {}  (BN1) ; 
\draw[draw=black,solid,line width=0.5mm,<-]  (BN) to[bend left=20] node[midway] {\hspace{15pt} $Q_3 $}node[midway,above] {}  (BN1) ;
\draw[draw=black,solid,line width=0.5mm,->]  (BN) to[bend right=20] node[midway] {\hspace{15pt} $\tQ_3$}node[midway,above] {}  (BN1) ;  
\draw(3.7,1.4) arc (60:300:-0.75cm) node at (5.2,2) {$\phi_1$} ;
\draw(-0.9,0.65) arc (60:300:0.75cm) node at (-2.5,0) {$\phi_2$} ;
\draw(3.7,-2.7) arc (60:300:-0.75cm) node at (5.2,-2) {$\phi_3$} ;
\end{tikzpicture}
\eea
with the superpotential
\bea \label{supA2k0mk}
\hspace{-5pt} W= \sum_{i=1}^3 \tr( Q_i \phi_{i+1} \tQ_i -\tQ_{i-1} \phi_{i-1} Q_{i-1}) + \frac{1}{2}k  \tr(\phi_1^2) - \frac{1}{2}k \tr(\phi_2^2)~, \quad \phi_{i+3} \equiv \phi_i~.
\eea

Integrating out the massive adjoints, one is left with the quiver for branes probing the Suspended Pinch Point Calabi-Yau threefold singularity \cite{MorrisonPlesser1999}.

\subsubsection*{The case of $N=1$}\label{A2N=1}

The $R$-charge of monopole operator is 
\bea
R(m_1,m_2,m_3) = \frac12(|m_1-m_2| +|m_2-m_3|+|m_1-m_3|)~.
\eea
The sum of the triplet of $D$-terms in \eref{BPSN=3}, or equivalently the decoupling of the overall $U(1)$, imply that 
\bea
m_1 =m_2~,\qquad \phi_1 =\phi_2 ~. 
\eea
\paragraph{Branch I:  $Q_i, \tQ_i \neq 0~ (i=1,2,3)$ -- $(\BC^2 \times (\BC^2/\BZ_2))/\BZ_k$.}
On this branch,
\bea
\phi_1 = \phi_2= \phi_3 = \frac1k(Q_1 \tQ_1- Q_2 \tQ_2) ~, \quad \tQ_2 Q_2 = Q_3 \tQ_3~.
\eea
From  \eref{BPSN=3} we also have $m_1 =m_2=m_3 \equiv m$. 
Thus, the $R$-charge of the monopole operator is $R(m,m,m) = 0$,
and the gauge charge is $(-km,km,0)$.

Notice that the pattern of identifications of the fluxes $m$ corresponds to the pattern of identifications of the VEVs of the $\phi$. This is a general
fact that we will see again in the following examples and it  is a consequence of $\CN=3$ supersymmetry. 

Therefore, the Hilbert series is
\be
H^{\rm (I)}_{N=1,k}(t, x_1, x_2, x_3, z) = \sum_{m=-\infty}^\infty  g^{\rm (I)}_1 (t, x_1, x_2, x_3; km) z^{m}~, 
\ee
where $g^{\rm (I)}_1(t, x_1, x_2, x_3;B)$ is the baryonic generating function
\be \label{GN1brIA2}
\begin{split}
	g^{\rm (I)}_1(t; x_1, x_2, x_3; B) &= \oint \frac{d q_1}{2\pi i q_1^{B+1}} \oint \frac{d q_2}{2\pi i q_2^{-B+1}}\oint \frac{d q_3}{2\pi i q_3} \\
	&\quad \PE \Big[-t^2 + (q_1 q_2^{-1} x_1 + q_2 q_1^{-1} x^{-1}_1 +q_2 q_3^{-1} x_2 +  \\
	& \qquad \qquad+q_3 q_2^{-1} x^{-1}_2 +q_3 q_1^{-1} x_3 + q_1 q_3^{-1} x_3^{-1}) t \Big]~,
\end{split}
\ee
where $x_1$, $x_2$, $x_3$ are flavour fugacities corresponding to each edge of the quiver, using an overparametrization. Computing the integrals, we find that for $k=1$ 
\be \label{C2C2Z2}
\begin{split}
	H^{\rm (I)}_{N=1,k=1}(t; x,y, z) &=\PE \left[  (z x + z^{-1} x^{-1}) t\right] \PE\left[(1 + z y + z^{-1}y^{-1})t^2 - t^4 \right]\\
	&=  H[\BC^2](t;zx) H[\BC^2/\BZ_2](t;zy) ~,
\end{split}
\ee
where $x=x_1$ and $y=(x_2 x_3)^{-1}$. The arguments of the $\PE$ are interpreted as follows. First we see the $\bC^2$ generators: $z_1=TQ_1$ and $z_2=\tilde T\tilde Q_1$, where $T\equiv V_{(1;1;1)}$ and $\tilde T\equiv V_{(-1;-1;-1)}$. Then we see the $\bC^2/\bZ_2$ generators: $w=Q_2\tilde Q_2=Q_3\tilde Q_3$, $u=T\tilde Q_2 \tilde Q_3$ and $v=\tilde T Q_2 Q_3$, subject to the relation $uv=w^2$ corresponding to the negative term.

For higher $k$, the manipulation \eqref{C4/Zk_2} shows that the moduli space is a $\bZ_k$ quotient of the moduli space for $k=1$, 
with the $\bZ_k$ charge equal to the $U(1)_M$ charge: 
\be \label{C2C2Z2Zk}
H^{\rm (I)}_{N=1,k}(t; x,y,z) =\frac1k \sum_{n=0}^{k-1} H^{\rm (I)}_{N=1,k}(t; x,y,z^{1/k} \omega_k^n)=  H\left[ (\BC^2 \times (\BC^2/\BZ_2))/\BZ_k \right]~.
\ee

In this section we used the $\cN=3$ description for the quiver, which only manifests a $U(1)^2$ non-R symmetry (one mesonic and one topological symmetry), with associated fugacities $x$ and $y$. Integrating out the massive adjoints and making field redefinitions, it is possible to reach a toric $\cN=2$ description, which has a $U(1)^3$ non-R symmetry manifest (including the topological $U(1)_M$). One can then introduce an extra mesonic fugacity $w$, to find that $t\to w^{-1}t$ for the $\bC^2$ factor and $t\to wt$ for the $\bC^2/\bZ_2$ factor in the above formulae.

\paragraph {Branch II: $Q_1, \tQ_1 \neq 0$ and $Q_i, \tQ_i =0~ (i=2,3)$ --
	$(\BC^2/\BZ_k) \times (\BC^2/\BZ_2)$.} This branch corresponds to one $SU(2)$ instanton on $\BC^2/\BZ_k$ with framing $(2,0 \ldots, 0)$, or the Coulomb branch of the Kronheimer-Nakajima (KN) quiver $(1)=(1)-[k]$.
Explicitly, we have
\bea
\phi_1 = \phi_2 = \frac1k Q_1 \tQ_1 \neq 0,  \quad Q_i, \tQ_i =0~ (i=2,3)~.
\eea
We also have $m_1 =m_2 \equiv m$. The $R$-charge of the monopole operator is 
\bea
R(m,m,m_3) = |m-m_3|~,
\eea
and the gauge charge of the monopole operator is $(-km,km,0)$.

Therefore, the Hilbert series is
\bea
H^{\rm (II)}_{N=1,k}(t,x,v,w) = \sum_{m\in\bZ} 
\sum_{m_3\in \bZ} t^{2|m-m_3|} v^m w^{m_3} \frac{g^{\rm (II)}_1(t,x; km)}{1-t^2}~, 
\eea
where $(1-t^2)^{-1}$ is the contribution of $\phi_3^n$ operators, 
$x=x_1$ is the flavour fugacity introduced above, $v$ is the fugacity for the topological symmetry of the 12 subquiver, $w$ is the fugacity for the topological symmetry of gauge group 3,% 
\footnote{The fugacity $z$ for the topological $U(1)_M$ is related to $v$, $w$ by $z=vw$.}
and $g^{\rm (II)}_1(t,x; B)$ is the baryonic generating function
\be
\begin{split}
	g^{\rm (II)}_1(t, x; B) &= \oint \frac{d q_1}{2\pi i q_1^{B+1}} \oint \frac{d q_2}{2\pi i q_2^{-B+1}} \PE \left[(x q_1 q_2^{-1} +x^{-1}  q_2 q_1^{-1}) t \right] \\
	&=\frac{t^{|B|} x^{B}}{1-t^2}~.
\end{split}
\ee
The Hilbert series for $k=1$ is therefore
\be
\begin{split}
	H^{\rm (II)}_{N=1,k=1}(t, x,v,w) &=  \PE \left[\left(vwx +v^{-1}w^{-1}x^{-1}\right)t\right] \PE \Big[ \chi^{SU(2)}_{[2]} (w^{1/2})t^2 -t^4 \Big]  \\
	&=  H[\BC^2](t,vwx) ~H[\BC^2/\BZ_2](t, w) ~,
\end{split}
\ee
where $\chi^{SU(2)}_{[2]} (z)=z^{2}+1+z^{-2}$ is the character of the triplet of $SU(2)$. The generators of $\bC^2$ are $V_{(1;1;1)}Q_1$, $V_{-(1;1;1)}\tQ_1$; the generators of $\bC^2/\bZ_2$ are $\phi_3$ and $V_{\pm(0;0;1)}$, subject to the relation $V_{(0;0;1)}V_{-(0;0;1)}=\phi_3^2$. 
For $k>1$, 
\be
\hspace{-8pt}
\begin{split}
	H^{\rm (II)}_{N=1,k}&(t, x,v,w) = \frac{1}{k}\sum_{n=0}^{k-1}H^{\rm (II)}_{N=1,k=1}(t,x, v^{1/k} \omega_k^n,w)=\\
	&= \PE \left[t^2 + \left(v w^k x^k + v^{-1}w^{-k}x^{-k}\right)t^k - t^{2k}\right]\PE \Big[\chi^{SU(2)}_{[2]} (w^{1/2})t^2 -t^4\Big]  \\
	&= H[\BC^2/\bZ_k](t,v w^k x^k) \; H[\BC^2/\BZ_2](t, w)  ~,
\end{split}
\ee
and the generators of $\bC^2/\bZ_k$ are $c=Q_1\tQ_1$, $a=TQ_1^k$, $b=\t T\tQ_1^k$, subject to $ab=c^k$.

\subsubsection*{The case of $N=2$}
The $R$-charge of the monopole operator is
\bea
R({\vec m}^{(1)}; {\vec m}^{(2)}; {\vec m}^{(3)}) = \frac{1}{2}\sum_{1 \leq i<j\leq 3} \sum_{a,b=1,2} |m^{(i)}_a - m^{(j)}_b| - \sum_{i=1}^3  |m^{(i)}_1-m^{(i)}_2|~.
\eea

\paragraph{Branch I: $Q_i, \tQ_i \neq 0~ (i=1,2,3)$ -- $\Sym^2((\BC^2 \times \BC^2/\BZ_2)/\BZ_k)$.} 
There are two cases to consider.
\ben
\item $({\vec m}^{(1)}; {\vec m}^{(2)}; {\vec m}^{(3)})=(m, m;m,m; m, m)$, with $m \in \BZ$.
The gauge charge of the monopole operator is
\bea
(-km, -km; 0,0, km, km)~.
\eea
The residual gauge symmetry is $U(2) \times U(2) \times U(2)$ and
\bea
R({\vec m}^{(1)}; {\vec m}^{(2)}; {\vec m}^{(3)}) = 0~.
\eea
The Hilbert series for this case is therefore
\bea
H^{(1)}_{N=2,k}(t) = \sum_{m=-\infty}^\infty g^{(1)}_{2}(t; km)~.
\eea
Here $g^{(1)}_{2}(t; B)$ is the baryonic generating function when the gauge group $U(2) \times U(2) \times U(2)$ is preserved: 
\bea\label{gone2A2}
& g^{(1)}_2 (t; B) \nn \\
&=  \left(\prod_{i=1}^3 \oint \frac{d z_i}{2\pi iz_i} (1-z_i)^2 \right) \oint \frac{d q_1}{2 \pi iq_1^{B+1}} \oint \frac{d q_2}{2\pi iq_2} \oint \frac{d q_3}{2 \pi i q_3^{-B+1}} \times \nn \\
& \qquad \fflat(t; z_1,z_2,z_3, q_1, q_2,q_3)~,
\eea
where the function $\fflat(t; z_1,z_2,z_3, q_1, q_2, q_3)$, with $z_i$ fugacities for the $SU(2)$ gauge groups and $q_i$ fugacities for the $U(1)$ gauge factors, can be computed using {\tt Macaulay2}. Since the full result is too long to be reported here, we present only the first few terms:
\be
\begin{split}
	&\fflat(t, z_1,z_2,z_3, q_1, q_2 ,q_3) \\
	&= \PE \Big[ t\Big( \sum_{1 \leq i <j \leq3} (q_i q_j^{-1} +q_j q_i^{-1}) \chi^{SU(2)}_{[1]} (z_i)\chi^{SU(2)}_{[1]} (z_j)\Big)  \\
	& \qquad \quad -t^2 \Big( 1+ \chi^{SU(2)}_{[2]} (z_1)+\chi^{SU(2)}_{[2]} (z_2)+\chi^{SU(2)}_{[2]} (z_3)  \Big) + O(t^4) \Big]~.
\end{split}
\ee
Using \eref{gone2A2} and \eref{GN1brIA2}, one can indeed check that 
\bea
g^{(1)}_{2}(t; 2B) &= \frac{1}{2} \left[ g^{(1)}_{1}(t; B)^2 + g^{(1)}_{1}(t^2; B)\right]~,
\eea
where $g^{(1)}_{1}(t; B)$ is given by \eref{GN1brIA2}. 
%%%%%%%%
\item $({\vec m}^{(1)}; {\vec m}^{(2)}; {\vec m}^{(3)})=(m_1, m_2;m_1,m_2; m_1, m_2)$, with $m_1 \neq m_2$.
The gauge charge of the monopole operator is
\bea
(-km_1, -km_2; 0,0, km_1, km_2)~.
\eea
The residual gauge symmetry is $U(1)^2 \times U(1)^2 \times U(1)^2$ and
\bea
R({\vec m}^{(1)}; {\vec m}^{(2)}; {\vec m}^{(3)}) = 0~.
\eea
The Hilbert series for this case is therefore
\bea
H^{(2)}_{N=2,k}(t) = \sum_{m_1 > m_2 >-\infty}^\infty g^{(2)}_{2}(t; km_1, km_2),
\eea
where $g^{\rm (2)}_2(t; B_1, B_2)$ is the baryonic generating function when the gauge group is broken to $U(1)^2 \times U(1)^2 \times U(1)^2$.
\bea
g^{\rm (2)}_2(t; B_1, B_2) &= g^{\rm (I)}_1(t; B_1) g^{\rm (I)}_1(t; B_2)~,
\eea
where $g^{\rm (I)}_1(t; B)$ is given by \eref{GN1brIA2}.
\een
We can explicitly compute 
\bea 
H^{\rm (I)}_{N=2,k}(t) = H^{(1)}_{N=2,k}(t)  + H^{(2)}_{N=2,k}(t) = H[\Sym^2((\BC^2 \times \BC^2/\BZ_2)/\BZ_k)](t)~.
\eea

\paragraph{Branch II: $Q_3, \tQ_3 \neq 0$ and $Q_i, \tQ_i =0~ (i=1,2)$.}
This branch corresponds to two $SU(2)$ instanton on $\BC^2/\BZ_k$ with framing $(2,0 \ldots, 0)$, or, equivalently, the Coulomb branch of the KN quiver $(2)=(2)-[k]$. 
There are two cases to be considered:
\ben
\item $({\vec m}^{(1)}; {\vec m}^{(2)}; {\vec m}^{(3)}) = (m,m; n_1,n_2; m,m)$. The gauge charges for the monopole operators is $(-km,-km; 0,0; km ,km)$. The Hilbert series is therefore
\bea
H^{(1)}_{N=2,k}(t) &= \sum_{n_1 \geq n_2 > -\infty}~~ \sum_{m=-\infty}^\infty  t^{-2|n_1-n_2| +4(|n_1-m|+|n_2-m|)}  \times \nn \\
&\qquad \qquad  g^{{\rm ABJM}/2}_2(t; 2km) P_{U(2)} ({n_1,n_2})~,
\eea
where $g^{{\rm ABJM}/2}_2$ is given by \eref{ghalfABJM2} and $P_{U(2)}(n_1,n_2)$ is the generating functions for the Casimirs under the residual gauge symmetry of $U(2)$.
%%%%%%
\item $({\vec m}^{(1)}; {\vec m}^{(2)}; {\vec m}^{(3)}) = (m_1,m_2; n_1,n_2; m_1,m_2)$,  with $m_1 > m_2 $ and $n_1 \geq n_2 $.  The gauge charges for the monopole operators is $(-km_1,-km_2; 0,0; km_1 ,km_2)$. The Hilbert series is therefore
\bea
H^{(2)}_{N=2,k}(t) &= \sum_{n_1 \geq n_2 > -\infty}~~ \sum_{m_1 > m_2 > -\infty}  t^{2R}  g^{{\rm ABJM}/2}_1 (t; km_1)g^{{\rm ABJM}/2}_1 (t; km_2) P_{U(2)} ({n_1,n_2})~,
\eea
where the $R$-charge of the monopole operator is
\bea
R =  |m_1-m_2| +|m_1-n_1| +|m_2-n_1|+|m_1-n_2|+|m_2-n_2|-|n_1-n_2|~,
\eea
and $g^{{\rm ABJM}/2}_1$ is given by \eref{abelianhalfABJM}
\een
The Hilbert series is the sum of the two contributions:
\bea
H^{\rm (II)}_{N=2;k}(t) = H^{(1)}_{N=2,k}(t) +H^{(2)}_{N=2,k}(t)~.
\eea

\paragraph{Examples.} For $k=1$, we obtain
\bea
H^{(1)}_{N=2,k}(t) &= 1 + 6 t^2 + 24 t^4 + 73 t^6 +\ldots~, \nn \\
H^{(2)}_{N=2,k}(t) &= 2 t + 3 t^2 + 22 t^3 + 31 t^4 + 116 t^5 + 169 t^6+\ldots~, \nn \\
H^{\rm (II)}_{N=2;k}(t) &= 1 + 2 t + 9 t^2 + 22 t^3 + 55 t^4 + 116 t^5 + 242 t^6 +\ldots \nn \\
&=\frac{1 + t + 3 t^2 + 6 t^3 + 8 t^4 + 6 t^5 + 8 t^6 + 6 t^7 + 
	3 t^8 + t^9 + t^{10}}{(1 - t)^8 (1 + t)^4 (1 + t + t^2)^3} \nn \\
&= H \left[ \text{2 $SU(2)$ instantons on $\BC^2$} \right]~.
\eea
For $k=2$, we obtain
\bea
H^{(1)}_{N=2,k}(t) &= 1 + 4 t^2 + 16 t^4 + 43 t^6 + \ldots~, \nn \\
H^{(2)}_{N=2,k}(t) &= 2 t^2 + 19 t^4 + 88 t^6 + \ldots~, \nn \\
H^{\rm (II)}_{N=2;k}(t) &= 1+6 t^2+35 t^4+131 t^6+ \ldots \nn \\
&= \frac{1 + 2 t^2 + 13 t^4 + 15 t^6 + 28 t^8 + 15 t^{10} + 13 t^{12}+ 
	2 t^{14} + t^{16}}{(1 - t)^8 (1 + t)^8 (1 + t^2)^4} \nn \\
&= H \left[ \text{2 $SU(2)$ instantons on $\BC^2/\BZ_2$ with framing $(2,0)$} \right]~.
\eea

\subsection*{Affine $A_2$ quiver: $U(N)_{+1} \times U(N)_{+2} \times U(N)_{-3}$}

For generic $k_i$ the quiver is truly $\CN=3$. The $A$-type quiver with generic $k_i$  correspond to the theory of M2-branes at a CY$_3$  \cite{Jafferis:2008qz}
and, in the absence of quotient singularities that can give rise to other branches,   we expect the existence of a single branch. We verify it below in the case of $N=1$.

In $3d$ $\CN=2$ notation, the quiver can be written as
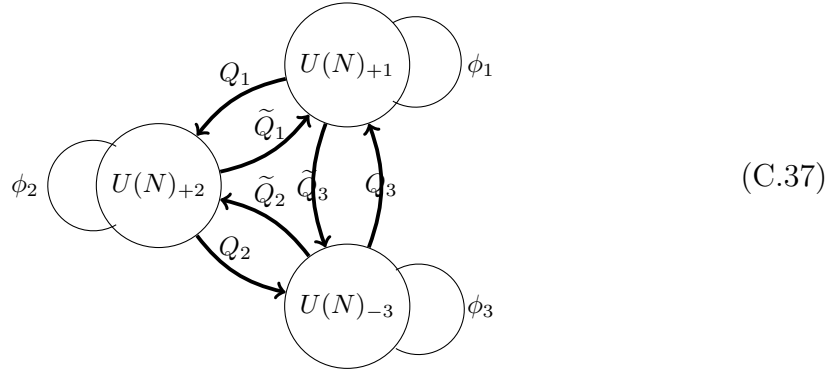
\bea
\begin{tikzpicture}[baseline, font=\footnotesize, scale=0.8]
\begin{scope}[auto,%
every node/.style={draw, minimum size=1.5cm}, node distance=2cm];
% the vertices
\node[circle] (USp2k) at (-0.1, 0) {$U(N)_{+2}$};
\node[circle] (BN)  at (3,2)  {$U(N)_{+1}$};
\node[circle] (BN1)  at (3,-2)  {$U(N)_{-3}$};
\end{scope}
% the edges
\draw[draw=black,solid,line width=0.5mm,<-]  (USp2k) to[bend left=20] node[midway,above] {$Q_1 $}node[midway,above] {}  (BN) ;
\draw[draw=black,solid,line width=0.5mm,->]  (USp2k) to[bend right=20] node[midway,above] {$\tQ_1$}node[midway,above] {}  (BN) ; 
\draw[draw=black,solid,line width=0.5mm,<-]  (USp2k) to[bend left=20] node[midway,above] {$\tQ_2 $}node[midway,above] {}  (BN1) ;
\draw[draw=black,solid,line width=0.5mm,->]  (USp2k) to[bend right=20] node[midway,above] {$Q_2$}node[midway,above] {}  (BN1) ; 
\draw[draw=black,solid,line width=0.5mm,<-]  (BN) to[bend left=20] node[midway] {$Q_3 $}node[midway,above] {}  (BN1) ;
\draw[draw=black,solid,line width=0.5mm,->]  (BN) to[bend right=20] node[midway] {$\tQ_3$}node[midway,above] {}  (BN1) ; 
\draw(3.75,1.4) arc (60:300:-0.75cm) node at (5.2,2) {$\phi_1$} ;
\draw(-0.8,0.65) arc (60:300:0.75cm) node at (-2.3,0) {$\phi_2$} ;
\draw(3.8,-2.7) arc (60:300:-0.75cm) node at (5.2,-2) {$\phi_3$} ;
\end{tikzpicture}
\eea
with the superpotential
\bea \label{supA2k12m3}
W= \sum_{i=1}^3 \tr( Q_i \phi_{i+1} \tQ_i -\tQ_i \phi_i Q_i) + \frac{1}{2}  \tr(\phi_1^2)+  \tr(\phi_1^2) - \frac{3}{2} \tr(\phi_3^2)~, \qquad \phi_4= \phi_1~.
\eea

The $R$-charge of a monopole operator is
\bea
R(m_1, m_2, m_3) = \frac{1}{2} \left( |m_1-m_2|+|m_2-m_3|+|m_3-m_1| \right) ~,
\eea
with the following condition from the $D$-terms:
\bea
m_1 + 2m_2 -3m_3=0~.
\eea
There is just one branch,  $Q_i, \tQ_i \neq 0~ (i=1,2,3)$  which corresponds to the  cone over $U(1) \symbol{92} U(3)/ U(1)$  \cite{Jafferis:2008qz}.
On this branch,
\bea
\phi_1 = \phi_2= \phi_3 =\frac{1}{3}( Q_2 \tQ_2- Q_3 \tQ_3) ~, \quad -3Q_1 \tQ_1 + Q_2 \tQ_2 + 2Q_3 \tQ_3=0~.
\eea
The second equality is in fact contained in (2.30) of \cite{Jafferis:2008qz}.

We also have $m_1 =m_2=m_3 \equiv m$. Thus, the $R$-charge of the monopole operator is $R(m,m,m) = 0$,
and the gauge charge of the monopole operator is $(-m,-2m,3m)$.
Therefore, the Hilbert series is
\bea
H^{\rm (I)}_{N=1,k}(t) = \sum_{m=-\infty}^\infty  g^{\rm (I)}_1 (t; km)~, 
\eea
where $g^{\rm (I)}_1(t,b)$ is the baryonic generating function
\be \label{GN1brIA2new}
\begin{split}
	g^{\rm (I)}_1(t; b) &= \oint \frac{d q_1}{2\pi i q_1^{b+1}} \oint \frac{d q_2}{2\pi i q_2^{2b+1}}\oint \frac{d q_3}{2\pi i q_3^{-3b+1}}  \\
	&\quad \PE \left[-t^2 + (q_1 q_2^{-1} + q_2 q_1^{-1}+q_2 q_3^{-1} + q_3 q_2^{-1}+q_1 q_3^{-1} + q_3 q_1^{-1}) t \right]~. \\
	&= \frac{t^{3 |b|} \left(1+t+t^2 -t^{|b|+1}-t^{2 |b|+1}\right)}{(1 - t)^3 (1 + t) (1 + t + t^2)}~.
\end{split}
\ee
For $k=1$, we have
\be
\begin{split}
	H^{\rm (I)}_{N=1,k=1}(t) &= \frac{1+t^2+3 t^3+4 t^4+4 t^5+4 t^6+3 t^7+t^8+t^{10}}{(1 - t)^4 (1 + t)^2 (1 + t^2) (1 + t + t^2) (1 + t + t^2 + t^3 + t^4)}  \\
	&= 1 + 2 t^2 + 4 t^3 + 7 t^4 + 10 t^5 + 16 t^6 +\ldots~.
\end{split}
\ee
For $k=2$, we have
{\small \bea
	H^{\rm (I)}_{N=1,k=2}(t) &= \frac{1}{(1- t)^4 (1 + t)^2 (1+3 t^2+6 t^4+9 t^6+11 t^8+11 t^{10}+9 t^{12}+6 t^{14}+3 t^{16}+t^{18})} \times \nn \\
	& \Big( 1 - 2 t + 4 t^2 - 4 t^3 + 5 t^4 - 4 t^5 + 7 t^6 - 4 t^7 + 8 t^8 - 
	4 t^9 + 8 t^{10} - 4 t^{11} + \text{palindrome} +t^{20}\Big) \nn \\
	&= 1 + 2 t^2 + 2 t^3 + 3 t^4 + 4 t^5 + 8 t^6 + \ldots~.
	\eea}
Observe that the numerators of the above Hilbert series are palindromic.  This implies that the moduli space is Calabi-Yau \cite{Forcella:2008bb, Gray:2008yu, Hanany:2008kn}, as expected (in fact it is even hyperK\"ahler by $\cN=3$ supersymmetry).

\subsection*{Affine $A_2$ quiver: $U(2)_k  \times U(1)_0 \times U(2)_{-k}$}
This is an $A_2$ quiver with fractional branes. We expect to recover the same moduli space as of the $A_2$ quiver with equal ranks for $N=1$ (see section \ref{A2N=1}) and we show here that this is indeed the case. 

In $3d$ $\CN=2$ notation, the quiver can be written as
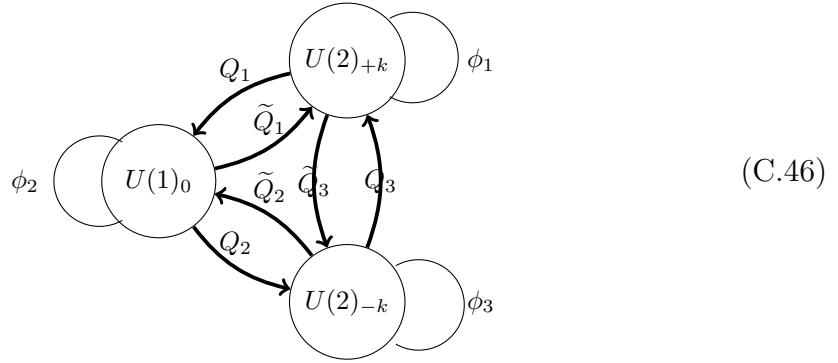
\bea
\begin{tikzpicture}[baseline, font=\footnotesize, scale=0.8]
\begin{scope}[auto,%
every node/.style={draw, minimum size=1.5cm}, node distance=2cm];
% the vertices
\node[circle] (USp2k) at (-0.1, 0) {$U(1)_{0}$};
\node[circle] (BN)  at (3,2)  {$U(2)_{+k}$};
\node[circle] (BN1)  at (3,-2)  {$U(2)_{-k}$};
\end{scope}
% the edges
\draw[draw=black,solid,line width=0.5mm,<-]  (USp2k) to[bend left=20] node[midway,above] {$Q_1 $}node[midway,above] {}  (BN) ;
\draw[draw=black,solid,line width=0.5mm,->]  (USp2k) to[bend right=20] node[midway,above] {$\tQ_1$}node[midway,above] {}  (BN) ; 
\draw[draw=black,solid,line width=0.5mm,<-]  (USp2k) to[bend left=20] node[midway,above] {$\tQ_2 $}node[midway,above] {}  (BN1) ;
\draw[draw=black,solid,line width=0.5mm,->]  (USp2k) to[bend right=20] node[midway,above] {$Q_2$}node[midway,above] {}  (BN1) ; 
\draw[draw=black,solid,line width=0.5mm,<-]  (BN) to[bend left=20] node[midway] {$Q_3 $}node[midway,above] {}  (BN1) ;
\draw[draw=black,solid,line width=0.5mm,->]  (BN) to[bend right=20] node[midway] {$\tQ_3$}node[midway,above] {}  (BN1) ; 
\draw(3.7,1.4) arc (60:300:-0.75cm) node at (5.2,2) {$\phi_1$} ;
\draw(-0.7,0.65) arc (60:300:0.75cm) node at (-2.3,0) {$\phi_2$} ;
\draw(3.8,-2.7) arc (60:300:-0.75cm) node at (5.2,-2) {$\phi_3$} ;
\end{tikzpicture}
\eea
with the superpotential
\bea \label{supA2U2U1U2}
W= \sum_{i=1}^3 \tr( Q_i \phi_{i+1} \tQ_i -\tQ_i \phi_i Q_i) + \frac{1}{2}k  \tr(\phi_1^2) - \frac{1}{2}k \tr(\phi_3^2)~, \qquad \phi_4= \phi_1~.
\eea
The $R$-charge of the monopole operator is
\bea
R({\vec m}^{(1)};  m^{(2)}; {\vec m}^{(3)}) &= \frac{1}{2} \left( \sum_{a,b=1,2} |m^{(1)}_a - m^{(3)}_b|+ \sum_{a=1}^2 |m^{(1)}_a - m^{(2)}| + \sum_{a=1}^2 |m^{(3)}_a - m^{(2)}| \right) \nn \\
& \qquad -  |m^{(1)}_1-m^{(1)}_2|-  |m^{(3)}_1-m^{(3)}_2|~.
\eea

\paragraph{Branch I: $Q_i, \tQ_i \neq 0~(i=1,2,3)$ -- $(\BC^2 \times \BC^2/\BZ_{2})/\BZ_k$}
The $D$-terms imply that
\bea
m^{(2)} &= m^{(i)}_1~\text{or}~ m^{(i)}_2~, \quad i=1,3~, \\
m^{(1)}_1 + m^{(1)}_2 &= m^{(3)}_1 + m^{(3)}_2~.
\eea
There are two cases to consider:
\ben
\item  {\bf Case 1: $(m^{(1)}_1, m^{(1)}_2; m^{(2)}; m^{(3)}_1, m^{(3)}_2) = (m,m; m; m,m).$}\\
The $R$-charge of the monopole operator is $R({\vec m}^{(1)};  m^{(2)}; {\vec m}^{(3)})=0$.  The baryonic generating function is given by
\be \label{GI1A2frac}
\begin{split}
	g^{(1)} (t; B) &=  \left(\prod_{i=1}^3 \oint \frac{d z_i}{z_i} (1-z_i)^2 \oint \frac{d b_i}{b_i^{B+1}} \right) \oint \frac{d z_2}{z_2} \times \\
	& \qquad \fflat(t, z_1,z_2,z_3, b_1, b_3)~,
\end{split}
\ee
where the function $\fflat(t, z_1,z_2,z_3, b_1, b_3)$ can be computed using {\tt Macaulay2}. Since the full result is too long to be reported here, we present only the first few terms:
\bea
&\fflat(t, z_1,z_2,z_3, b_1, b_3) \nn \\
&= \PE \Big[ t\Big( \sum_{i=1, 3}(b_i z_2^{-1}+b_i^{-1} z_2)\chi^{SU(2)}_{[1]} (z_i) +(b_1b_3^{-1} + b_1^{-1} b_3) \chi^{SU(2)}_{[1]} (z_1)\chi^{SU(2)}_{[1]} (z_3) \Big) \nn \\
& \quad -t^2 \Big( 1+ \chi^{SU(2)}_{[2]} (z_1)+\chi^{SU(2)}_{[2]} (z_3)  \Big)  + t^4 +t^5(2+ b_1^2 b_3^{-2} + b_1^{-2} b_3^2) \nn \\
& \quad -t^6\Big( (b_3^2 b_1^{-1} z_2^{-1} +  b_3^{-2} b_1 z_2)\chi^{SU(2)}_{[1]} (z_1)+ (b_1^2 b_3^{-1} z_2^{-1} +  b_1^{-2} b_3 z_2)\chi^{SU(2)}_{[1]} (z_3) \nn \\
& \qquad \qquad  + \text{the same terms that appear at order $t$}  \Big)+ O(t^7) \Big]
\eea
For reference, we present the unrefined Hilbert series of $\fflat$:
\bea
\fflat(t, \{ z_i =1 \}, \{b_j = 1\}) =\frac{(1 + t) (1 + 2 t - t^2) (1 + 3 t + 4 t^2)}{(1 - t)^{10}}~.
\eea
Upon the evaluation of the integral in \eref{GI1A2frac}, we find that
\bea
g^{(1)}  (t; B) &= \PE[ 2t^2 +2t^3 -t^6] \delta_{B,0}~.
\eea
The Hilbert series for this case is
\bea
H^{(1)}_k(t) =  \sum_{m \in \BZ} g^{(1)}  (t; km) = \PE[ 2t^2 +2t^3 -t^6]~,
\eea
independent of $k$.
\item  {\bf Case 2: $(m^{(1)}_1, m^{(1)}_2; m^{(2)}; m^{(3)}_1, m^{(3)}_2) = (m,0; m; m,0),~m\neq0$}\\
In this case, each $U(2)$ gauge group is broken to $U(1)^2$. The $R$-charge of the monopole operator is $R({\vec m}^{(1)};  m^{(2)}; {\vec m}^{(3)})=0$.  The baryonic generating function is given by
\be \label{GI2A2frac}
\begin{split}
	&g^{(2)} (t; B_1, B_3) \\
	&=   \oint  \frac{d z_1}{z_1^{B_1+1}} \oint \frac{d z_2}{z_2} \oint \frac{d z_3}{z_3^{B_3+1}} \PE \Bigg[ \Big( (z_1 + z_1^{-1})(z_3+z_3^{-1})  \\
	& \qquad + z_1 z_2^{-1}+z_2 z_1^{-1} + z_2 z_3^{-1}+z_3 z_2^{-1}  \Big) t - 2t^2 \Bigg]~,
\end{split}
\ee
The Hilbert series in this case is
\bea
H^{(2)}_k(t) =  \sum_{m \neq 0} g^{(2)}  (t; km, -km) ~.
\eea
For example, for $k=1$, we have
\bea
H^{(2)}_{k=1}(t) = 2 t + 4 t^2 + 8 t^3 + 16 t^4 + 24 t^5 + 38 t^6+\ldots~
\eea
and for $k=2$
\bea
H^{(2)}_{k=2}(t) = 2 t^2 + 2 t^3 + 8 t^4 + 8 t^5 + 18 t^6+\ldots~.
\eea
\een
The Hilbert series of this branch is the sum of the two contributions:
\bea
H^{\rm (I)}_k(t) &= H^{(1)}_k(t) + H^{(2)}_k(t) = H [ (\BC^2 \times \BC^2/\BZ_{2})/\BZ_k] ~,
\eea
equal to \eref{C2C2Z2Zk}.

\paragraph{Branch II: $Q_3, \tQ_3 \neq 0$ and $Q_i, \tQ_i =0~(i=1,2)$}. This branch is   the Coulomb branch of the KN quiver $(1)=(2)-[k]$.
The $D$-terms imply that
\bea
m^{(1)}_1 + m^{(1)}_2 &= m^{(3)}_1 + m^{(3)}_2~.
\eea
There are two cases to consider:
\ben
\item  {\bf Case 1: $(m^{(1)}_1, m^{(1)}_2; m^{(2)}; m^{(3)}_1, m^{(3)}_2) = (m,m; n; m,m).$}\\
The $R$-charge of the monopole operator is 
\bea
R({\vec m}^{(1)};  m^{(2)}; {\vec m}^{(3)})=2|m-n|~.
\eea
The gauge charges of the monopole operator is $(-km, -km ; 0; km, km)$.  
The baryonic generating function is the same as for the half-ABJM theory \eref{ghalfABJM2}:
\bea
g^{{\rm ABJM}/2}_2(t; B) &= \frac{(-1)^{-|B|} [1+(-1)^{|B|}] t^{|B|}}{2(1-t^2)^2(1+t^2)}~.
\eea
Thus, the Hilbert series for this case is
\bea
H^{(1)}_k(t) = \sum_{m, n \in \BZ} t^{4|m-n|} g^{{\rm ABJM}/2}_2(t; 2km) P_{U(1)}(n)
\eea
\item  {\bf Case 2: $(m^{(1)}_1, m^{(1)}_2; m^{(2)}; m^{(3)}_1, m^{(3)}_2) = (m_1,m_2; n; m_1,m_2),~ m_1 \neq m_2$} \\
The $R$-charge of the monopole operator is 
\bea
R({\vec m}^{(1)};  m^{(2)}; {\vec m}^{(3)})=|m_1-n|+|m_2-n|-|m_1-m_2|~.
\eea
The gauge charges of the monopole operator is $(-km_1, -km_2 ; 0; km_1; km_2)$.  
The baryonic generating function for this case is the same as for half-ABJM:
\bea
g^{{\rm ABJM}/2}_1(t; B_1) g^{{\rm ABJM}/2}_1(t; B_2) &= \frac{t^{|B_1|+|B_2|}}{(1-t^2)^2}~.
\eea
Thus, the Hilbert series for this case is
\bea
H^{(2)}_k(t) = \sum_{m_1 \neq m_2 \in \BZ} \sum_{n \in \BZ} t^{2\sum_{i=1}^2|m_i-n|-2|m_1-m_2|} \prod_{i=1,2}g^{{\rm ABJM}/2}_1(t; km_i)  P_{U(1)}(n)~.
\eea
\een
The Hilbert series is the sum of the two contributions:
\bea
H^{\rm (II)}_k(t) &= H^{(1)}_k(t) + H^{(2)}_k(t) ~.
\eea

\paragraph{Examples:} For $k=1$, we have
\be
\begin{split}
	H^{\rm (II)}_{k=1}(t) &= \PE[ 4 t+ 3t^2-t^4] \\
	&= H[ \BC^2 \times \BC^2/\BZ_2] \\
	&= H[\text{Coulomb branch of $(1)=(2)-[1]$}]~.
\end{split}
\ee
For $k=2$, we have
\be
\begin{split}
	H^{\rm (II)}_{k=2}(t) &= 1 + 6 t^2 + 29 t^4 + 89 t^6 + 236 t^8 + 521 t^{10}+\ldots \\
	&= H[\text{Coulomb branch of $(1)=(2)-[2]$}]~.
\end{split}
\ee

\section{The moduli space of instantons on $\BC^2/\BZ_n$} \label{app:inst}
In this appendix we give a brief summary of $3d$ $\CN=4$ gauge theories whose Higgs or Coulomb branch describes the moduli space of $SU(N)$ instantons on $\BC^2/\BZ_n$.  The reader can find more details in \cite{Mekareeya:2015bla}.

The instanton configuration is specified by the monodromies of the $SU(N)$ gauge field at infinity and at the origin of $\BC^2/\BZ_n$ \cite{Witten:2009xu}.  With these data specified, the moduli space of such instantons are described by the {\it Higgs branch} of a $3d$ $\CN=4$ gauge theory specified by the flavoured affine $A_{n-1}$ quiver diagram.  This is also known as the Kronheimer-Nakajima (KN) quiver \cite{kronheimer1990yang}:
\bea 
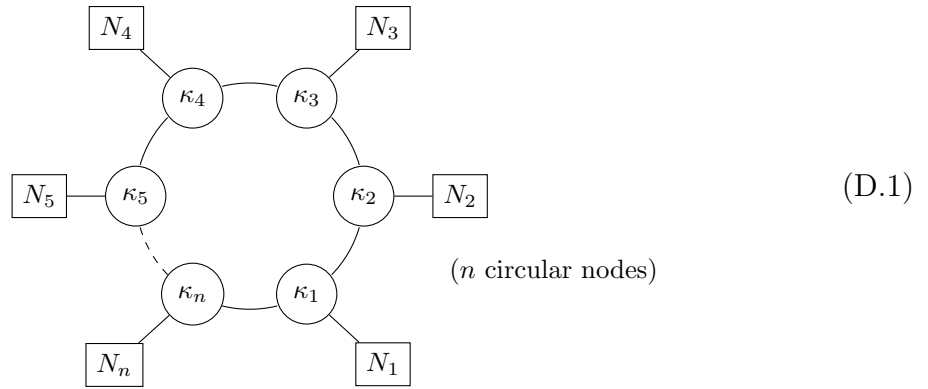
\label{KN}
\begin{tikzpicture}[baseline, align=center,node distance=0.5cm]
\def \n {6}
\def \radius {1.5cm}
\def \margin {16} % margin in angles, depends on the radius
\foreach \s in {1,...,5}
{
	\node[draw, circle] (\s) at ({360/\n * (\s - 2)}:\radius) {{\footnotesize $\kappa_{\s}$}};
	\draw[-, >=latex] ({360/\n * (\s - 3)+\margin}:\radius) 
	arc ({360/\n * (\s - 3)+\margin}:{360/\n * (\s-2)-\margin}:\radius);
}
\node[draw, circle] (last) at ({360/3 * (3 - 1)}:\radius) {{\footnotesize $\kappa_n$}};
\draw[dashed, >=latex] ({360/6 * (5 -2)+\margin}:\radius) 
arc ({360/6 * (5 -2)+\margin}:{360/6 * (5-1)-\margin}:\radius);
\node[draw, rectangle,  below right= of 1] (f1) {{\footnotesize $N_1$}};
\node[draw, rectangle, right= of 2] (f2) {{\footnotesize $N_2$}};
\node[draw, rectangle, above right= of 3] (f3) {{\footnotesize $N_3$}};
\node[draw, rectangle, above left= of 4] (f4) {{\footnotesize $N_4$}};
\node[draw, rectangle,  left= of 5] (f5) {{\footnotesize $N_5$}};
\node[draw, rectangle,  below left= of last] (f6) {{\footnotesize $N_n$}};
\draw[-, >=latex] (1) to (f1);
\draw[-, >=latex] (2) to (f2);
\draw[-, >=latex] (3) to (f3);
\draw[-, >=latex] (4) to (f4);
\draw[-, >=latex] (5) to (f5);
\draw[-, >=latex] (last) to (f6);
\node[draw=none] at (4,-1) {{\footnotesize ($n$ circular nodes)}};
\end{tikzpicture}
\eea
with
\bea
N= N_1 + N_2 + \ldots+ N_n~.
\eea
This theory can be realised as the low energy theory on the worldvolume of  D3-branes in the following configuration \cite{Hanany:1996ie}:
\bea 
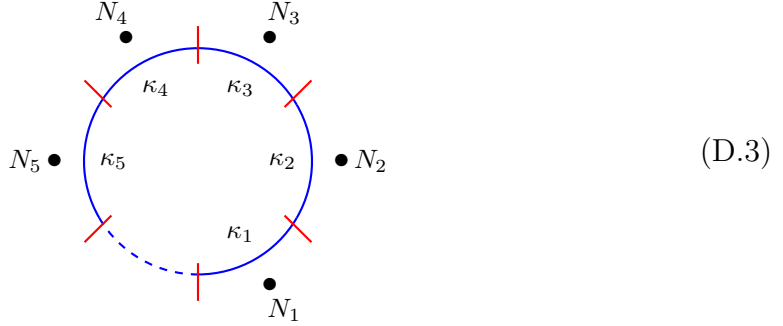
\label{braneKN}
\begin{tikzpicture}[baseline, align=center,node distance=0.5cm]
\def \n {6}
\def \radius {1.5cm}
\def \margin {0} % margin in angles, depends on the radius
\foreach \s in {1,...,5}
{
	\node[draw=none] (\s) at ({360/\n * (\s - 2)+30}:{\radius-10}) {};
	\node[draw=none]  at ({360/\n * (\s - 2)}:{\radius-11}) {{\footnotesize $\kappa_{\s}$}};
	\node[draw=none]  at ({360/\n * (\s - 2)}:{\radius+11}) {$\bullet$};
	\node[draw=none]  at ({360/\n * (\s - 2)}:{\radius+22}) {{\footnotesize $N_{\s}$}};
	\draw[-, >=latex,blue,thick] ({360/\n * (\s - 3)+\margin+30}:\radius) 
	arc ({360/\n * (\s - 3)+\margin+30}:{360/\n * (\s-2)-\margin+30}:\radius);
}
\node[draw=none, circle] (last) at ({360/3 * (3 - 1)+30}:{\radius-10}) {};
\draw[dashed, >=latex,thick,blue] ({360/6 * (5 -2)+\margin+30}:\radius) 
arc ({360/6 * (5 -2)+\margin+30}:{360/6 * (5-1)-\margin+30}:\radius);
\node[draw=none,  below right= of 1] (f1) {};
\node[draw=none, above right= of 2] (f2) {};
\node[draw=none, above = of 3] (f3) {};
\node[draw=none, above left= of 4] (f4) {};
\node[draw=none,  below left= of 5] (f5) {};
\node[draw=none,  below = of last] (f6) {};
\draw[-, >=latex,red, thick] (1) to (f1);
\draw[-, >=latex,red, thick] (2) to (f2);
\draw[-, >=latex,red, thick] (3) to (f3);
\draw[-, >=latex,red, thick] (4) to (f4);
\draw[-, >=latex,red, thick] (5) to (f5);
\draw[-, >=latex,red, thick] (last) to (f6);
\end{tikzpicture}
\eea
where each blue line denotes D3-branes along $\BR^{1,2}_{0,1,2} \times S^1_6$, with $\kappa_i$  ($i=1,\ldots,n$) denoting the number of D3-brane segments in the $i$-th interval; red lines denote NS5-brane along $\BR^{1,2}_{0,1,2} \times \BR^3_{7,8,9}$, located at different positions along the $S^1_6$ direction; and black dots with the label $N_i$ denote $N_i$ D5-branes along $\BR^{1,2}_{0,1,2} \times \BR^3_{3,4,5}$, located in the $i$-th interval of the $S^1_6$ direction. 

From the quiver \eref{KN}, the information about the gauge field at infinity $U_\infty$ and the gauge field at the origin $U_0$ can be obtained as follows \cite{Witten:2009xu}.  The number of eigenvalues of $U_\infty$ equal to $e^{2\pi i \ell/n}$ (for $\ell =1, \ldots, n$) is $N_\ell$.  Indeed, $N_\ell$ is also equal to the number of D5-branes with linking number $\ell$.  The number of eigenvalues of $U_0$ that equal $e^{2\pi i \ell/n}$ is 
\bea
\beta_\ell &= N_\ell + \kappa_{\ell+1} +\kappa_{\ell-1} -2\kappa_{\ell}~, \qquad \ell = 1,\ldots, n~,
\eea
which is the difference between the linking numbers of the $(\ell+1)$-th and the $\ell$-th NS5-branes.  From now on and in the main text, we refer to the partition $(N_1, \ldots, N_n)$ of $N$ as the {\it framing} of the $SU(N)$ instantons on $\BC^2/\BZ_n$.  Note the cyclicity of the framing.

For simplicity, let us take
\bea
\kappa_1 = \kappa_2 = \ldots = \kappa_n = \kappa~.
\eea
and use the terminology that $\kappa$ is the instanton number.  In this case, the monodromies $U_0$ and $U_\infty$ have the same eigenvalues.  Therefore, it is enough to specify the instanton configuration just by the framing, say $(N_1, \ldots, N_n)$.  In which case, the monodromy breaks $SU(N)$ into residual symmetry $S(U(N_1) \times U(N_2) \times \dots \times U(N_n))$.

Let us now focus on the {\it Coulomb branch} of the KN quiver \eref{KN}. It describes
\be
\begin{split}
	&\text{the moduli space of $\kappa$ $SU(n)$ instantons on $\BC^2/\BZ_N$}  \\
	& \text{with framing $(0^{N_1-1},1, 0^{N_2-1},1,\ldots, 0^{N_n-1},1)$.}
\end{split}
\ee
This can be seen by considering the Higgs branch of the quiver obtained by performing an S-duality on the configuration \eref{braneKN}, under which the NS5-branes and the D5-branes are exchanged \cite{Hanany:1996ie}. Indeed, one observes that the roles of the gauge group and the orbifold type get exchanged under mirror symmetry \cite{deBoer:1996mp, Porrati:1996xi, Dey:2013nf}.

If $N_i >0$ for all $i =1, 2, \ldots, n$, then $SU(n)$ is broken to its maximal abelian subgroup $U(1)^{n-1}$.  On the other hand, if some $N_i$ vanishes, certain nonabelian symmetries are restored. There are some interesting special cases to consider:
\bi
\item If one of the $N_i$'s is equal to one and the other $N_i$'s are zero, the Coulomb branch of \eref{KN} can be identified with the moduli space of $\kappa$ $SU(n)$ instantons on $\BC^2$, in agreement with \cite{Cremonesi:2014xha}.
\item If one of the $N_i$'s is equal to $N$ and the other $N_i$'s are zero, the symmetry of the Coulomb branch is $U(1) \times SU(n)$ for $N \geq 3$ and $SU(2) \times SU(n)$ for $N=1,2$.  

If in addition we set $\kappa=1$, the Coulomb branch of \eref{KN} is isomorphic to 
\bea \label{1pureSUN}\BC^2/\BZ_N \times \CN_{SU(n)}~, \eea 
where $\CN_{SU(n)}$ is the reduced moduli space of one $SU(n)$ instanton on $\BC^2$, which is the minimal nilpotent orbit of $SU(n)$ \cite{kronheimer1990, Br, KS1, Gaiotto:2008nz, Benvenuti:2010pq}.  The moduli space \eref{1pureSUN} was pointed out in (2.69) of \cite{Dey:2013fea}.
\ei

\section{Hilbert series of toric CY$_4$ cones from toric data}\label{app:HS_from_toric}

In this appendix we write a universal formula for the Hilbert series of toric CY$_4$ cones based on the toric data. Let $\{\vec{v}_s=(1,x_s,y_s,z_s)\in\bZ^4\}$ be the toric fan of the CY$_4$, $\{p_s\}$ a set of associated complex variables, 
and $T^2$, $X$, $Y$, $Z$ fugacities for the toric symmetries corresponding to the four axes of $\bZ^4$. Then the Hilbert series of the toric CY$_4$ cone may be computed as 
\be\label{HS_toric_fan}
\begin{split}
	&H_{\rm geom}(T,X,Y,Z)=\prod_s \left[\oint \frac{dp_s}{2\pi i p_s}\PE[p_s]\right]\cdot 2\pi i T^2 \delta(\prod_s p_s-T^2)\\
	& \qquad\qquad \cdot 2\pi i X \delta(\prod_s p_s^{x_s}-X)\cdot 2\pi i Y \delta(\prod_s p_s^{y_s}-Y)\cdot 2\pi i Z \delta(\prod_s p_s^{z_s}-Z)~.
\end{split}
\ee

Expanding the delta function involving $Z$ as in \ref{id_geometric}, the integral at fixed $m$ counts holomorphic sections of a line bundle on CY$_3=$ CY$_4//U(1)_M$, where $U(1)_M$ corresponds to the $z$ axis. 
To be more explicit, let us regroup the $p_s$ variables according to their $(x,y)$ coordinates: $t_{r,z_r}$, where $r$ labels the $(x,y)$ coordinates and $z_r$ is the $z$ coordinate of the column of points in the fan. Then we can write the delta function involving $Z$ as
\be
2\pi i Z \delta(\prod_s p_s^{z_s}-Z)=2\pi i Z \delta(\prod_r \prod_{z_r} t_{r,z_r}^{z_r}-Z)=\sum_m \Big(Z\prod_r \prod_{z_r} t_{r,z_r}^{-z_r} \Big)^m~,
\ee
while the arguments of the other delta functions only involve $t_r\equiv \prod_r t_{r,z_r}$. Then, if we focus on a column labelled by $r$, we have to compute the integral 
\be\label{identity_column}
\hspace{-5pt} \prod_{z_r} \left[\oint \frac{dt_{r,z_r}}{2\pi i t_{r,z_r}} t_{r,z_r}^{-z_r m}\right] \PE[\sum_{z_r} t_{r,z_r}]  f(\prod_{z_r} t_{r,z_r})=  \oint\frac{dt_{r}}{2\pi i t_{r}} t_{r}^{-\min(z_r m)} \PE[t_{r}]  f(t_{r}),
\ee
where the values of $z_r$ in the minimum correspond to the $z$ coordinates of the points in the column. The exponent of $t_r$ is
$-m \min(z_r)$ if $m\ge 0$, and $-m \max(z_r)$ if $m\le 0$.%
\footnote{An interesting generalization is to count holomorphic sections of line bundles over the CY$_4$. In that case, an extra factor of $t_{r,z_r}^{-n_{z_r}}$ is inserted in the LHS of \eref{identity_column}, and the exponent of $t_r$ in the RHS becomes $-\min(z_r m+n_{z_r})$.} 
The Hilbert series of the CY$_4$ cone can then be written as
\be\label{HS_toric_fan_final}
\begin{split}
	&H_{\rm geom}(T,X,Y,Z)=\sum_m Z^m \prod_r \left[\oint \frac{dt_r}{2\pi i t_r}t_r^{-\min(z_r m)}\PE[t_r]\right]\cdot \\
	& \qquad 2\pi i T^2 \delta(\prod_r t_r-T^2)\cdot 2\pi i X \delta(\prod_r t_r^{x_r}-X)\cdot 2\pi i Y \delta(\prod_r t_r^{y_r}-Y)~,
\end{split}
\ee
a sum of characters counting holomorphic sections of the line bundles (or rather sheaves) $\cO(\sum_r \min(z_r m) D_r)$ \cite{Martelli2006,Butti:2006au} over the CY$_3$ = CY$_4//U(1)_M$. Here $D_r$ is the toric divisor corresponding to the $r$-th point of the toric diagram of the CY$_3$, associated to the fugacity $t_r$. 

As a first example, let us consider the toric CY$_4$ cones that are geometric moduli spaces of the abelian flavored ABJM theories studied in section \ref{sec:flavABJM}. In this case, $\{t_r\}=\{a,b,c,d\}$ and $\{z_a\}=\{n_1,n_1+1,\dots,n_1+h_1\}$, $\{z_b\}=\{\t n_1,\dots,\t n_1+\t h_1\}$, $\{z_c\}=\{n_2,\dots,n_2+h_2\}$, $\{z_d\}=\{\t n_2,\dots,\t n_2+\t h_2\}$. 
Setting $T^{1/2}=t$, $X=t^2 x^{-1} y$ and $Y=t^2 x^{-1} y^{-1}$, the delta functions in \eref{HS_toric_fan_final} are solved by $a=txq$, $b=tyq^{-1}$, $c=tx^{-1}q$ and $d=ty^{-1}q^{-1}$. The fugacity $q$ corresponds to the $U(1)$ gauge group of the GLSM for the conifold (the CY$_3$ cone that is obtained as the K\"ahler quotient of the CY$_4$ by $U(1)_M$) and is to be integrated over. Setting also $\ell_i\equiv n_i+\frac{h_i}{2}$, $\t\ell_i\equiv \t n_i+\frac{\t h_i}{2}$, \eref{HS_toric_fan_final} reduces to
\be\label{HS_toric_flavABJM}
\begin{split}
	&H_{\rm geom}(t,x,y,Z)=\sum_m Z^m \oint \frac{dq}{2\pi i q} \PE[tq(x+x^{-1})+tq^{-1}(y+y^{-1})] \\
	& \quad \cdot (txq)^{-\ell_1 m+\frac{h_1}{2}|m|}
	(tyq^{-1})^{-\t \ell_1 m+\frac{\t h_1}{2}|m|} 
	(tx^{-1}q)^{-\ell_2 m+\frac{h_2}{2}|m|}
	(ty^{-1}q^{-1})^{-\t \ell_2 m+\frac{\t h_2}{2}|m|}
	~.
\end{split}
\ee
Identifying 
\be\label{Zzeta}
\begin{split}
	Z&\equiv\zeta (tx)^{\ell_1}(ty)^{\t \ell_1} (tx^{-1})^{\ell_2}(ty^{-1})^{\ell_2}= z t^{\ell_1+\ell_2+\t\ell_1+\t\ell_2} x^{-k_x+\ell_1-\ell_2} y^{-k_y+\t\ell_1-\t\ell_2} ~,
\end{split}
\ee
\eref{HS_toric_flavABJM} agrees with \eref{H_flavABJM_geom}, that we obtained directly from the GLSM description. In particular, choosing the mixed gauge-mesonic Chern-Simons levels to be $k_x=\ell_1-\ell_2$ and $k_y=\t\ell_1-\t\ell_2$, one has $Z=z t^{\ell_1+\ell_2+\t\ell_1+\t\ell_2}$. The power of $t$ can also be eliminated if appropriate mixed $R$-gauge Chern-Simons couplings are introduced.

As a second example, let us consider the cone over $Y^{p,q}(\bC\bP^2)$ studied in section \ref{sec:wrapped_D6}. In this the toric diagram of the CY$_3=\bC^3/\bZ_3$ consists of four points $\{t_r\}=\{t_0,t_1,t_2,t_3\}$ with coordinates $(0,0)$, $(1,0)$, $(0,1)$ and $(-1,-1)$. The $z$-coordinates of points of the toric diagram of $C(Y^{p,q}(\bC\bP^2))$ are $\{z_0\}=\{0,1,\dots,p\}$, $\{z_1\}=\{z_2\}=\{0\}$ and $\{z_3\}=\{q\}$ respectively. Then formula \eref{HS_toric_fan_final} reads
\be\label{HS_toric_Ypq}
\begin{split}
	&H_{\rm geom}(T,X,Y,Z)=\sum_m Z^m \oint \frac{dt_0}{2\pi i t_0}t_0^{-\min(p m)} \oint \frac{dt_1}{2\pi i t_1} \oint \frac{dt_2}{2\pi i t_2} \oint \frac{dt_3}{2\pi i t_3^{1+qm}} \\
	& \qquad \PE[\sum_r t_r] 2\pi i T^2 \delta(\prod_r t_r-T^2)\cdot 2\pi i X \delta(t_1/t_3-X)\cdot 2\pi i Y \delta(t_2/t_3-Y)~.
\end{split}
\ee
Setting $X=y_1^{-2}y_2$, $Y=y_1^{-1}y_2^{-1}$, $Z=z T^{\frac{2}{3}}x_1$ and $t_3=T^{\frac{2}{3}-\epsilon}uy_1$, we obtain
\be\label{HS_toric_Ypq_2}
H_{\rm geom}=\sum_{m\in\bZ}z^m\oint \frac{du}{2\pi i u}\PE\left[T^{\frac{2}{3}-\epsilon}u [1,0]_y+T^{3\epsilon}u^{-3}\right]  u^{-qm+3\min(pm,0)}~,
\ee
where $[1,0]_y=y_1+y_1^{-1}y_2+y_2^{-1}$ is the character of the triplet of $SU(3)$. We inserted the $\epsilon$ dependence, mixing the $R$-symmetry with the gauge symmetry of the GLSM, to have positive powers of $T$. We will send $\epsilon\to 0^+$ after expanding the $\PE$. Since
\be
\hspace{-3pt}
g(B)\equiv \lim_{\epsilon\to 0^+} \oint \frac{du}{2\pi i u^{1+B}}\PE\left[T^{\frac{2}{3}-\epsilon}u [1,0]_y+T^{3\epsilon}u^{-3}\right]=\sum_{l=0}^\infty T^{\frac{2}{3}(3l+B)}[3l+B,0]_y~
\ee
if $B\geq0$, we have that 
\be
{\rm HWG}[\sum_{m=0}^\infty v^m g(hm)]=\PE[\tau^3+v \tau^h]~, \qquad \tau=T^{\frac{2}{3}}\mu_1~,
\ee
where $\mu_1$, $\mu_2$ are highest weight fugacities for $SU(3)_y$. Then \eref{HS_toric_Ypq_2} leads to
\be\label{HS_toric_Ypq_final}
{\rm HWG}[H_{geom}]=\PE[\tau^3+z \tau^q +z^{-1} \tau^{3p-q}-\tau^{3p}]~,
\ee
which agrees with the field theory result \eref{HWG_Ypq_2}, up to relabelling $q\leftrightarrow 3p-q$.

\section{Flavored ABJM theory for $C(Q^{1,1,1})$ and resolutions} \label{sec:Q111}

In this appendix we further study the Hilbert series of the geometric moduli space of the flavored ABJM theory for a single M2-brane probing $C(Q^{1,1,1})$ \cite{Benini2010}. We turn on background magnetic charges for the topological symmetry and the flavor symmetry and match them to the ``baryonic charges'' corresponding to resolutions of the cone.%
\footnote{For simplicity we set to zero the background magnetic charges for the mesonic symmetries.}

We turn on magnetic charge $-B$ for the topological $U(1)$, and magnetic charges $\mu_1$, $\mu_2$ for the flavor symmetry. Since the latter is really $U(1)^2/U(1)$, we can set $\min(\mu_1,\mu_2)=0$ by shifting the magnetic charge $m$ of the diagonal gauge $U(1)$. Having turned on the flavor magnetic charges, we need to specify the values of some mixed Chern-Simons couplings required to cancel parity anomalies: we choose $k_{g_1 F_i}=-k_{g_2 F_i}=-\frac{1}{2}$ for the mixed gauge-flavor couplings, and $k_{y F_1}=-k_{y F_2}=\frac{1}{2}$ for the mixed coupling between the flavor symmetries and the mesonic symmetry that acts on $B_{1,2}$ and hence on the flavors. Then the Hilbert series of the geometric moduli space of the abelian theory is 
\be\label{Q111_FT}
\begin{split}
	\hspace{-5pt} 
	H(B,\mu_1,\mu_2) &=z^{-\frac{\mu_1+\mu_2}{2}}y^{\frac{\mu_1-\mu_2}{2}} \sum_{m\in\bZ} z^m t^{\frac{1}{2}(|m-\mu_1|+|m-\mu_2|)} y^{\frac{1}{2}(|m-\mu_1|-|m-\mu_2|-\mu_1+\mu_2)}\cdot\\
	&\cdot  g_1\left(B+\tfrac{1}{2}(|m-\mu_1|+|m-\mu_2|-\mu_1-\mu_2)\right)~.
\end{split}
\ee
We omitted the implicit dependence on fugacities where possible to avoid clutter. Here $g_1(B_{\rm eff}(m))$ of \eref{g1_conifold} is a function of the effective baryonic charge 
\be\label{Beff}
\begin{split}
	B_{\rm eff}(m)&=B+\tfrac{1}{2}(|m-\mu_1|+|m-\mu_2|-\mu_1-\mu_2)\\
	&=B+
	\begin{cases}
		m-\mu_1-\mu_2~, & \max(\mu_1,\mu_2)\leq m\\
		-\min(\mu_1,\mu_2)~, & \min(\mu_1,\mu_2)\le m\le \max(\mu_1,\mu_2)\\
		-m~, &m\le \min(\mu_1,\mu_2)
	\end{cases}
\end{split}
\ee
%$B_{\rm eff}=B+\tfrac{1}{2}(|m-\mu_1|+|m-\mu_2|-\mu_1-\mu_2)$, 
as well as of $t,x,y$. We inserted the prefactor $z^{-\frac{\mu_1+\mu_2}{2}}y^{\frac{\mu_1-\mu_2}{2}}$ in \eqref{Q111_FT} to compensate our asymmetric choice $\min(\mu_1,\mu_2)=0$ and make the Hilbert series \eref{Q111_FT} invariant under common shifts of $B$, $\mu_1$ and $\mu_2$. Similarly, $B_{\rm eff}(m)$ is invariant under common shifts of $m$, $B$, $\mu_1$ and $\mu_2$.

Setting $x=\alpha$, $y=\beta\gamma$, $z=\gamma/\beta$ and defining the generating function  
\be\label{f}
f(m_\alpha,m_\beta,m_\gamma):=t^{m_\alpha + \frac{1}{2}(m_\beta+m_\gamma)} \sum_{n=0}^\infty t^{2n}[n+m_\alpha;n+m_\beta;n+m_\gamma]_{\alpha,\beta,\gamma}~,
\ee
a straightforward though tedious computation reveals that the Hilbert series is
\be\label{Q111_baryonic_result}
\begin{split}
	H(B,\mu_1,\mu_2)&=
	%\begin{cases}
	%f(B,\mu_1,\mu_2) ~, & B\geq 0\\
	%f(0,\mu_1-B,\mu_2-B) ~, & B\leq 0
	%\end{cases}=
	f(B-\min(B,\mu_i),\mu_1-\min(B,\mu_i),\mu_2-\min(B,\mu_i))\\
	&= \begin{cases}
		f(0,\mu_1-B,\mu_2-B) ~, & \mu_1-B\geq 0 \wedge \mu_2-B\geq 0 \\
		f(B,0,\mu_2) ~, & B\geq 0 \wedge \mu_2\geq \mu_1=0\\
		f(B,\mu_1,0) ~, & B\geq 0 \wedge \mu_1\geq \mu_2=0
	\end{cases}
	~.
\end{split}
\ee

\begin{table}[t]
	\centering
	\begin{tabular}{cccccc}
		$t_1$ & $r_2$ & $s_2$ & $t_3$ & $r_4$ & $s_4$ \\ \hline  
		$1$ & $1$ & $1$ & $1$ & $1$ & $1$ \\ %& $T^2$\\
		$0$ & $1$ & $1$ & $1$ & $0$ & $0$ \\ %& $X$\\
		$0$ & $0$ & $0$ & $1$ & $1$ & $1$ \\ %& $Y$\\
		$0$ & $-1$& $0$ & $0$ & $0$ & $1$ \\ %& $Z$
	\end{tabular}
	\qquad
	\begin{tabular}{c|cccccc|c}
		& $t_1$ & $r_2$ & $s_2$ & $t_3$ & $r_4$ & $s_4$ & \\ \hline  
		$U(1)_1$ & $-1$ & $1$ & $0$ & $-1$ & $0$ & $1$ & $N_1$\\
		$U(1)_2$ & $-1$ & $0$ & $1$ & $-1$ & $1$ & $0$ & $N_2$
	\end{tabular}
	\caption{Toric data and GLSM charges for the cone over $Q^{1,1,1}$. }
	\label{tab:toric_Q111}
\end{table}

In the rest of this section we will reproduce this result from a geometric viewpoint. 
The cone over $Q^{1,1,1}$ is a toric variety: its toric data and the charge matrix of the $U(1)^2$ GLSM of which it is the vacuum moduli space are given in table \ref{tab:toric_Q111}. Introducing fugacities $w_1, w_2$ and baryonic charges $N_1$, $N_2$ for the $U(1)^2$ gauge symmetry of the GLSM, and fugacities $t$, $\alpha$, $\beta$, $\gamma$ for the $U(1)_R\times SU(2)^3$ toric global symmetry,  we have the Hilbert series
\be\label{Hilb_Q111_geom}
\begin{split}
	\hspace{-3pt}H_{\rm geom}(N_1,N_2) &= \oint \frac{dw_1}{2\pi i w_1} w_1^{-N_1} \oint \frac{dw_2}{2\pi i w_2} w_2^{-N_2} \\
	&\quad \PE[ t (\alpha+\alpha^{-1}) w_1^{-1}  w_2^{-1} + t^{1/2} (\beta+\beta^{-1})w_1+ t^{1/2} (\gamma+\gamma^{-1})w_2]
\end{split}
\ee
where the powers of $t$ have been chosen conveniently. This is computed to be
\be\label{Hilb_Q111_result}
\begin{split}
	\hspace{-3pt}H_{\rm geom}(N_1,N_2) &= \begin{cases}
		f(0,N_1,N_2)~, & N_1\ge 0\wedge N_2\ge 0\\
		f(-N_1,0,N_2-N_1)~, & -N_1 \ge 0 \wedge N_2-N_1\ge 0\\
		f(-N_2,N_1-N_2,0)~, & -N_2 \ge 0 \wedge N_1-N_2\ge 0~.
	\end{cases}
\end{split}
\ee
We therefore see that the field theory computation \eref{Q111_baryonic_result} matches the geometric computation \eref{Hilb_Q111_result} upon the identification $N_i=\mu_i-B$.

It is also possible to directly reproduce the monopole formula \eref{Q111_FT} geometrically, using the fact that the conifold $\cC$ is a K\"ahler quotient of the cone over $Q^{1,1,1}$, that is $\cC=C(Q^{1,1,1}//U(1)_M$. To do so, we consider the equivalent toric GLSM of charges 
\be
\begin{tabular}{c|cccccc|c}
	& $t_1$ & $r_2$ & $s_2$ & $t_3$ & $r_4$ & $s_4$ & \\ \hline  
	$w$ & $-1$ & $1$ & $0$ & $-1$ & $0$ & $1$ & $N_1=\mu_1-B$\\
	$u_1$ & $0$ & $1$ & $-1$ & $0$ & $0$ & $0$ & $N_1-N_2-M=\mu_1-m$ \\ \hline
	$u_2$ & $0$ & $0$ & $0$ & $0$ & $-1$ & $1$ & $M=m-\mu_2$ %\\ \hline
	%& $\frac{t\alpha}{w}$ & $t^{\xfrac{1}{2}} \beta w u_1$ & 
\end{tabular}
\ee
where we have included in the last line the charges under the $U(1)_M$ symmetry that was used in the construction of \cite{Benini2010} to reduce M-theory on the cone over $Q^{1,1,1}$ to type IIA on the conifold fibered over $\bR$. The first two lines of charges correspond to genuine gauge symmetries of the GLSM for $C(Q^{1,1,1})$, and the fugacities $w$ and $u_1$ will be integrated over. Instead we gauge and ungauge the global $U(1)_M$ symmetry, integrating over the fugacity $u_2$ and summing over the charge $M$, or equivalently $m$. Doing so, we can write the Hilbert series as 
\be\label{Q111_geom_alt}
\begin{split}
	H_{\rm geom} &=\sum_{m \in \bZ} \oint\frac{dw}{2\pi i w} w^{B-\mu_1} \oint\frac{du_1}{2\pi i u_1} u_1^{-(\mu_1-m)} \oint\frac{du_2}{2\pi i u_2} u_2^{-(m-\mu_2)} \\
	&\qquad \PE\left[ t \frac{\alpha}{w}+ t^{\frac{1}{2}} \beta w u_1 + t^{\frac{1}{2}} \frac{\gamma}{u_1}+ t \frac{1}{\alpha w}+ t^{\frac{1}{2}} \frac{1}{\gamma u_2}+ t^{\frac{1}{2}} \frac{w u_2}{\beta}\right]~.  
\end{split}
\ee
Using repeatedly the identity
\be
\begin{split}
	\oint\frac{du}{2\pi i u^{1+N}}
	\PE \left[ au +\frac{b}{u}\right] &= \PE[ab] ~a^{N\Theta(N)}b^{-N\Theta(-N)}=\PE[ab] ~a^{\frac{|N|+N}{2}}b^{\frac{|N|-N}{2}}~,
\end{split}
\ee
where $\Theta(x)$ denotes the Heaviside step function and we assumed that $|a|,|b|<1$ to Taylor expand, we can perform the integrals over $u_1$ and $u_2$ in \eref{Q111_geom_alt} and obtain 
\be\label{Q111_geom_alt_2}
\begin{split}
	&H_{\rm geom} =\sum_{m \in \bZ} t^{\frac{1}{2}(|m-\mu_1|+|m-\mu_2|)} 
	\left(\gamma\beta^{-1}\right)^{m-\frac{\mu_1+\mu_2}{2}} (\beta \gamma)^{\frac{1}{2}(|m-\mu_1|-|m-\mu_2|)}\cdot \\
	&\cdot \oint \frac{dw}{2\pi i w} w^{B+\frac{1}{2}\sum_i(-\mu_i+ |m-\mu_i|)} \PE\left[t w^{-1}(\alpha+\alpha^{-1})+ t w(\beta\gamma+(\beta\gamma)^{-1})\right]~.
\end{split}
\ee
Setting $x=\alpha$, $y=\beta\gamma$ and $z=\gamma\beta^{-1}$, the second line is   the baryonic Hilbert series of the conifold theory with the effective baryonic charge \eref{Beff}, hence \eref{Q111_geom_alt_2} precisely reproduces the field theory monopole formula \eref{Q111_FT}. The $U(1)_M$ symmetry maps to the topological symmetry of the field theory, and the sum over its background electric charges $M$ becomes the sum over the topological charges $m$ of monopole operators.

\bibliographystyle{ytphys}
\bibliography{ref}

\providecommand{\href}[2]{#2}\begingroup\raggedright\begin{thebibliography}{10}

\bibitem{BoerHoriOz1997}
J.~de~Boer, K.~Hori, and Y.~Oz, ``Dynamics of n=2 supersymmetric gauge theories
  in three dimensions,'' \href{http://arxiv.org/abs/hep-th/9703100}{{\ttfamily
  hep-th/9703100}}. \url{http://arxiv.org/abs/hep-th/9703100}.

\bibitem{AharonyHananyIntriligatorEtAl1997}
O.~Aharony, A.~Hanany, K.~Intriligator, N.~Seiberg, and M.~J. Strassler,
  ``Aspects of n=2 supersymmetric gauge theories in three dimensions,''
  \href{http://arxiv.org/abs/hep-th/9703110}{{\ttfamily hep-th/9703110}}.
  \url{http://arxiv.org/abs/hep-th/9703110}.

\bibitem{Tong2000}
D.~Tong, ``Dynamics of n=2 supersymmetric chern-simons theories,''
  \href{http://arxiv.org/abs/hep-th/0005186}{{\ttfamily hep-th/0005186}}.
  \url{http://arxiv.org/abs/hep-th/0005186}.

\bibitem{Intriligator:2013lca}
K.~Intriligator and N.~Seiberg, ``{Aspects of 3d N=2 Chern-Simons-Matter
  Theories},'' \href{http://dx.doi.org/10.1007/JHEP07(2013)079}{{\em JHEP}
  {\bfseries 1307} (2013) 079},
\href{http://arxiv.org/abs/1305.1633}{{\ttfamily arXiv:1305.1633 [hep-th]}}.
%%CITATION = ARXIV:1305.1633;%%.

\bibitem{'tHooft:1977hy}
G.~'t~Hooft, ``{On the Phase Transition Towards Permanent Quark Confinement},''
\href{http://dx.doi.org/10.1016/0550-3213(78)90153-0}{{\em Nucl.Phys.}
  {\bfseries B138} (1978) 1}.
%%CITATION = NUPHA,B138,1;%%.

\bibitem{Borokhov:2002ib}
V.~Borokhov, A.~Kapustin, and X.-k. Wu, ``{Topological disorder operators in
  three-dimensional conformal field theory},'' {\em JHEP} {\bfseries 0211}
  (2002) 049,
\href{http://arxiv.org/abs/hep-th/0206054}{{\ttfamily arXiv:hep-th/0206054
  [hep-th]}}.
%%CITATION = HEP-TH/0206054;%%.

\bibitem{Borokhov:2002cg}
V.~Borokhov, A.~Kapustin, and X.-k. Wu, ``{Monopole operators and mirror
  symmetry in three-dimensions},'' {\em JHEP} {\bfseries 0212} (2002) 044,
\href{http://arxiv.org/abs/hep-th/0207074}{{\ttfamily arXiv:hep-th/0207074
  [hep-th]}}.
%%CITATION = HEP-TH/0207074;%%.

\bibitem{Borokhov:2003yu}
V.~Borokhov, ``{Monopole operators in three-dimensional N=4 SYM and mirror
  symmetry},'' \href{http://dx.doi.org/10.1088/1126-6708/2004/03/008}{{\em
  JHEP} {\bfseries 0403} (2004) 008},
\href{http://arxiv.org/abs/hep-th/0310254}{{\ttfamily arXiv:hep-th/0310254
  [hep-th]}}.
%%CITATION = HEP-TH/0310254;%%.

\bibitem{Gaiotto:2008ak}
D.~Gaiotto and E.~Witten, ``{S-Duality of Boundary Conditions In N=4 Super
  Yang-Mills Theory},'' {\em Adv. Theor. Math. Phys.} {\bfseries 13} (2009)
  721,
\href{http://arxiv.org/abs/0807.3720}{{\ttfamily arXiv:0807.3720 [hep-th]}}.
%%CITATION = ARXIV:0807.3720;%%.

\bibitem{Bashkirov:2010kz}
D.~Bashkirov and A.~Kapustin, ``{Supersymmetry enhancement by monopole
  operators},'' \href{http://dx.doi.org/10.1007/JHEP05(2011)015}{{\em JHEP}
  {\bfseries 1105} (2011) 015},
\href{http://arxiv.org/abs/1007.4861}{{\ttfamily arXiv:1007.4861 [hep-th]}}.
%%CITATION = ARXIV:1007.4861;%%.

\bibitem{Bashkirov:2010hj}
D.~Bashkirov, ``{Examples of global symmetry enhancement by monopole
  operators},''
\href{http://arxiv.org/abs/1009.3477}{{\ttfamily arXiv:1009.3477 [hep-th]}}.
%%CITATION = ARXIV:1009.3477;%%.

\bibitem{Cremonesi:2013lqa}
S.~Cremonesi, A.~Hanany, and A.~Zaffaroni, ``{Monopole operators and Hilbert
  series of Coulomb branches of $3d$ $\mathcal{N} = 4$ gauge theories},''
  \href{http://dx.doi.org/10.1007/JHEP01(2014)005}{{\em JHEP} {\bfseries 1401}
  (2014) 005},
\href{http://arxiv.org/abs/1309.2657}{{\ttfamily arXiv:1309.2657 [hep-th]}}.
%%CITATION = ARXIV:1309.2657;%%.

\bibitem{Cremonesi:2014kwa}
S.~Cremonesi, A.~Hanany, N.~Mekareeya, and A.~Zaffaroni, ``{Coulomb branch
  Hilbert series and Hall-Littlewood polynomials},''
\href{http://arxiv.org/abs/1403.0585}{{\ttfamily arXiv:1403.0585 [hep-th]}}.
%%CITATION = ARXIV:1403.0585;%%.

\bibitem{Cremonesi:2014vla}
S.~Cremonesi, A.~Hanany, N.~Mekareeya, and A.~Zaffaroni, ``{Coulomb branch
  Hilbert series and Three Dimensional Sicilian Theories},''
\href{http://arxiv.org/abs/1403.2384}{{\ttfamily arXiv:1403.2384 [hep-th]}}.
%%CITATION = ARXIV:1403.2384;%%.

\bibitem{Cremonesi:2014xha}
S.~Cremonesi, G.~Ferlito, A.~Hanany, and N.~Mekareeya, ``{Coulomb Branch and
  The Moduli Space of Instantons},''
  \href{http://dx.doi.org/10.1007/JHEP12(2014)103}{{\em JHEP} {\bfseries 1412}
  (2014) 103},
\href{http://arxiv.org/abs/1408.6835}{{\ttfamily arXiv:1408.6835 [hep-th]}}.
%%CITATION = ARXIV:1408.6835;%%.

\bibitem{Cremonesi:2014uva}
S.~Cremonesi, A.~Hanany, N.~Mekareeya, and A.~Zaffaroni,
  ``{$T^{\sigma}_{\rho}(G)$ Theories and Their Hilbert Series},''
\href{http://arxiv.org/abs/1410.1548}{{\ttfamily arXiv:1410.1548 [hep-th]}}.
%%CITATION = ARXIV:1410.1548;%%.

\bibitem{Cremonesi2015a}
S.~Cremonesi, ``{The Hilbert series of 3d $\mathcal{N}=2$ Yang-Mills theories
  with vectorlike matter},''
  \href{http://dx.doi.org/10.1088/1751-8113/48/45/455401}{{\em J. Phys.}
  {\bfseries A48} no.~45, (2015) 455401},
\href{http://arxiv.org/abs/1505.02409}{{\ttfamily arXiv:1505.02409 [hep-th]}}.
%%CITATION = ARXIV:1505.02409;%%.

\bibitem{Hanany:2015via}
A.~Hanany, C.~Hwang, H.~Kim, J.~Park, and R.-K. Seong, ``{Hilbert Series for
  Theories with Aharony Duals},''
  \href{http://dx.doi.org/10.1007/JHEP11(2015)132,
  10.1007/JHEP04(2016)064}{{\em JHEP} {\bfseries 11} (2015) 132},
  \href{http://arxiv.org/abs/1505.02160}{{\ttfamily arXiv:1505.02160
  [hep-th]}}.
[Addendum: JHEP04,064(2016)].
%%CITATION = ARXIV:1505.02160;%%.

\bibitem{Mekareeya:2015bla}
N.~Mekareeya, ``{The moduli space of instantons on an ALE space from 3d
  $\mathcal{N}=4$ field theories},''
  \href{http://dx.doi.org/10.1007/JHEP12(2015)174}{{\em JHEP} {\bfseries 12}
  (2015) 174},
\href{http://arxiv.org/abs/1508.06813}{{\ttfamily arXiv:1508.06813 [hep-th]}}.
%%CITATION = ARXIV:1508.06813;%%.

\bibitem{Hanany:2011db}
A.~Hanany and N.~Mekareeya, ``{Complete Intersection Moduli Spaces in N=4 Gauge
  Theories in Three Dimensions},''
  \href{http://dx.doi.org/10.1007/JHEP01(2012)079}{{\em JHEP} {\bfseries 1201}
  (2012) 079},
\href{http://arxiv.org/abs/1110.6203}{{\ttfamily arXiv:1110.6203 [hep-th]}}.
%%CITATION = ARXIV:1110.6203;%%.

\bibitem{Razamat:2014pta}
S.~S. Razamat and B.~Willett, ``{Down the rabbit hole with theories of class
  S},''
\href{http://arxiv.org/abs/1403.6107}{{\ttfamily arXiv:1403.6107 [hep-th]}}.
%%CITATION = ARXIV:1403.6107;%%.

\bibitem{DoreyTong2000}
N.~Dorey and D.~Tong, ``Mirror symmetry and toric geometry in three-dimensional
  gauge theories,'' \href{http://arxiv.org/abs/hep-th/9911094}{{\ttfamily
  hep-th/9911094}}. \url{http://arxiv.org/abs/hep-th/9911094}.

\bibitem{Aharony:2008ug}
O.~Aharony, O.~Bergman, D.~L. Jafferis, and J.~Maldacena, ``{N=6 superconformal
  Chern-Simons-matter theories, M2-branes and their gravity duals},''
  \href{http://dx.doi.org/10.1088/1126-6708/2008/10/091}{{\em JHEP} {\bfseries
  0810} (2008) 091},
\href{http://arxiv.org/abs/0806.1218}{{\ttfamily arXiv:0806.1218 [hep-th]}}.
%%CITATION = ARXIV:0806.1218;%%.

\bibitem{kronheimer1990yang}
P.~B. Kronheimer and H.~Nakajima, ``Yang-mills instantons on ale gravitational
  instantons,'' {\em Mathematische Annalen} {\bfseries 288} no.~1, (1990)
  263--307.

\bibitem{Assel:2014awa}
B.~Assel, ``{Hanany-Witten effect and SL(2, $\mathbb{Z}$) dualities in matrix
  models},'' \href{http://dx.doi.org/10.1007/JHEP10(2014)117}{{\em JHEP}
  {\bfseries 1410} (2014) 117},
\href{http://arxiv.org/abs/1406.5194}{{\ttfamily arXiv:1406.5194 [hep-th]}}.
%%CITATION = ARXIV:1406.5194;%%.

\bibitem{Benini2010}
F.~Benini, C.~Closset, and S.~Cremonesi, ``{Chiral flavors and M2-branes at
  toric CY4 singularities},''
  \href{http://dx.doi.org/10.1007/JHEP02(2010)036}{{\em JHEP} {\bfseries 02}
  (2010) 036},
\href{http://arxiv.org/abs/0911.4127}{{\ttfamily arXiv:0911.4127 [hep-th]}}.
%%CITATION = ARXIV:0911.4127;%%.

\bibitem{Cremonesi2011}
S.~Cremonesi, ``{Type IIB construction of flavoured ABJ(M) and fractional M2
  branes},'' \href{http://dx.doi.org/10.1007/JHEP01(2011)076}{{\em JHEP}
  {\bfseries 01} (2011) 076},
\href{http://arxiv.org/abs/1007.4562}{{\ttfamily arXiv:1007.4562 [hep-th]}}.
%%CITATION = ARXIV:1007.4562;%%.

\bibitem{Jafferis2013}
D.~L. Jafferis, ``{Quantum corrections to $\mathcal{N} = 2$ Chern-Simons
  theories with flavor and their AdS$_{4}$ duals},''
  \href{http://dx.doi.org/10.1007/JHEP08(2013)046}{{\em JHEP} {\bfseries 08}
  (2013) 046},
\href{http://arxiv.org/abs/0911.4324}{{\ttfamily arXiv:0911.4324 [hep-th]}}.
%%CITATION = ARXIV:0911.4324;%%.

\bibitem{Benini2011}
F.~Benini, C.~Closset, and S.~Cremonesi, ``{Quantum moduli space of
  Chern-Simons quivers, wrapped D6-branes and AdS4/CFT3},''
  \href{http://dx.doi.org/10.1007/JHEP09(2011)005}{{\em JHEP} {\bfseries 09}
  (2011) 005},
\href{http://arxiv.org/abs/1105.2299}{{\ttfamily arXiv:1105.2299 [hep-th]}}.
%%CITATION = ARXIV:1105.2299;%%.

\bibitem{Closset2012}
C.~Closset and S.~Cremonesi, ``{Toric Fano varieties and Chern-Simons
  quivers},'' \href{http://dx.doi.org/10.1007/JHEP05(2012)060}{{\em JHEP}
  {\bfseries 05} (2012) 060},
\href{http://arxiv.org/abs/1201.2431}{{\ttfamily arXiv:1201.2431 [hep-th]}}.
%%CITATION = ARXIV:1201.2431;%%.

\bibitem{Martelli:2009ga}
D.~Martelli and J.~Sparks, ``{AdS(4) / CFT(3) duals from M2-branes at
  hypersurface singularities and their deformations},''
  \href{http://dx.doi.org/10.1088/1126-6708/2009/12/017}{{\em JHEP} {\bfseries
  12} (2009) 017},
\href{http://arxiv.org/abs/0909.2036}{{\ttfamily arXiv:0909.2036 [hep-th]}}.
%%CITATION = ARXIV:0909.2036;%%.

\bibitem{Jafferis:2009th}
D.~L. Jafferis, ``{Quantum corrections to $\mathcal{N} = 2$ Chern-Simons
  theories with flavor and their AdS$_{4}$ duals},''
  \href{http://dx.doi.org/10.1007/JHEP08(2013)046}{{\em JHEP} {\bfseries 08}
  (2013) 046},
\href{http://arxiv.org/abs/0911.4324}{{\ttfamily arXiv:0911.4324 [hep-th]}}.
%%CITATION = ARXIV:0911.4324;%%.

\bibitem{ImamuraYokoyama2011}
Y.~Imamura and S.~Yokoyama, ``Index for three dimensional superconformal field
  theories with general r-charge assignments,''
  \href{http://arxiv.org/abs/1101.0557}{{\ttfamily 1101.0557}}.
  \url{http://arxiv.org/abs/1101.0557}.

\bibitem{M2}
D.~R. Grayson and M.~E. Stillman, ``Macaulay2, a software system for research
  in algebraic geometry.'' Available at
  \href{http://www.math.uiuc.edu/Macaulay2/}%
  {http://www.math.uiuc.edu/Macaulay2/}.

\bibitem{SeibergTaylor2011}
N.~Seiberg and W.~Taylor, ``Charge lattices and consistency of 6d
  supergravity,'' \href{http://arxiv.org/abs/1103.0019}{{\ttfamily 1103.0019}}.
  \url{http://arxiv.org/abs/1103.0019}.

\bibitem{Forcella:2007wk}
D.~Forcella, A.~Hanany, and A.~Zaffaroni, ``{Baryonic Generating Functions},''
  \href{http://dx.doi.org/10.1088/1126-6708/2007/12/022}{{\em JHEP} {\bfseries
  0712} (2007) 022},
\href{http://arxiv.org/abs/hep-th/0701236}{{\ttfamily arXiv:hep-th/0701236
  [HEP-TH]}}.
%%CITATION = HEP-TH/0701236;%%.

\bibitem{Kapustin1999}
A.~Kapustin and M.~J. Strassler, ``{On mirror symmetry in three-dimensional
  Abelian gauge theories},''
  \href{http://dx.doi.org/10.1088/1126-6708/1999/04/021}{{\em JHEP} {\bfseries
  04} (1999) 021},
\href{http://arxiv.org/abs/hep-th/9902033}{{\ttfamily arXiv:hep-th/9902033
  [hep-th]}}.
%%CITATION = HEP-TH/9902033;%%.

\bibitem{Witten2003}
E.~Witten, ``{SL(2,Z) action on three-dimensional conformal field theories with
  Abelian symmetry},''
\href{http://arxiv.org/abs/hep-th/0307041}{{\ttfamily arXiv:hep-th/0307041
  [hep-th]}}.
%%CITATION = HEP-TH/0307041;%%.

\bibitem{Intriligator:1996ex}
K.~A. Intriligator and N.~Seiberg, ``{Mirror symmetry in three-dimensional
  gauge theories},'' \href{http://dx.doi.org/10.1016/0370-2693(96)01088-X}{{\em
  Phys.Lett.} {\bfseries B387} (1996) 513--519},
\href{http://arxiv.org/abs/hep-th/9607207}{{\ttfamily arXiv:hep-th/9607207
  [hep-th]}}.
%%CITATION = HEP-TH/9607207;%%.

\bibitem{Aharony:1997ju}
O.~Aharony and A.~Hanany, ``{Branes, superpotentials and superconformal fixed
  points},'' \href{http://dx.doi.org/10.1016/S0550-3213(97)00472-0}{{\em
  Nucl.Phys.} {\bfseries B504} (1997) 239--271},
\href{http://arxiv.org/abs/hep-th/9704170}{{\ttfamily arXiv:hep-th/9704170
  [hep-th]}}.
%%CITATION = HEP-TH/9704170;%%.

\bibitem{AganagicHoriKarchEtAl2001}
M.~Aganagic, K.~Hori, A.~Karch, and D.~Tong, ``Mirror symmetry in 2+1 and 1+1
  dimensions,'' \href{http://arxiv.org/abs/hep-th/0105075}{{\ttfamily
  hep-th/0105075}}. \url{http://arxiv.org/abs/hep-th/0105075}.

\bibitem{Goddard:1976qe}
P.~Goddard, J.~Nuyts, and D.~I. Olive, ``{Gauge Theories and Magnetic
  Charge},''
\href{http://dx.doi.org/10.1016/0550-3213(77)90221-8}{{\em Nucl.Phys.}
  {\bfseries B125} (1977) 1}.
%%CITATION = NUPHA,B125,1;%%.

\bibitem{JafferisYin2011}
D.~Jafferis and X.~Yin, ``{A Duality Appetizer},''
\href{http://arxiv.org/abs/1103.5700}{{\ttfamily arXiv:1103.5700 [hep-th]}}.
%%CITATION = ARXIV:1103.5700;%%.

\bibitem{Bergman1999}
O.~Bergman, A.~Hanany, A.~Karch, and B.~Kol, ``{Branes and supersymmetry
  breaking in three-dimensional gauge theories},''
  \href{http://dx.doi.org/10.1088/1126-6708/1999/10/036}{{\em JHEP} {\bfseries
  10} (1999) 036},
\href{http://arxiv.org/abs/hep-th/9908075}{{\ttfamily arXiv:hep-th/9908075
  [hep-th]}}.
%%CITATION = HEP-TH/9908075;%%.

\bibitem{Benini2011a}
F.~Benini, C.~Closset, and S.~Cremonesi, ``{Comments on 3d Seiberg-like
  dualities},'' \href{http://dx.doi.org/10.1007/JHEP10(2011)075}{{\em JHEP}
  {\bfseries 10} (2011) 075},
\href{http://arxiv.org/abs/1108.5373}{{\ttfamily arXiv:1108.5373 [hep-th]}}.
%%CITATION = ARXIV:1108.5373;%%.

\bibitem{Aharony:2008gk}
O.~Aharony, O.~Bergman, and D.~L. Jafferis, ``{Fractional M2-branes},''
  \href{http://dx.doi.org/10.1088/1126-6708/2008/11/043}{{\em JHEP} {\bfseries
  0811} (2008) 043},
\href{http://arxiv.org/abs/0807.4924}{{\ttfamily arXiv:0807.4924 [hep-th]}}.
%%CITATION = ARXIV:0807.4924;%%.

\bibitem{GaiottoTomasiello2010}
D.~Gaiotto and A.~Tomasiello, ``{The gauge dual of Romans mass},''
  \href{http://dx.doi.org/10.1007/JHEP01(2010)015}{{\em JHEP} {\bfseries 01}
  (2010) 015},
\href{http://arxiv.org/abs/0901.0969}{{\ttfamily arXiv:0901.0969 [hep-th]}}.
%%CITATION = ARXIV:0901.0969;%%.

\bibitem{Klebanov:1998hh}
I.~R. Klebanov and E.~Witten, ``{Superconformal field theory on three-branes at
  a Calabi-Yau singularity},''
  \href{http://dx.doi.org/10.1016/S0550-3213(98)00654-3}{{\em Nucl. Phys.}
  {\bfseries B536} (1998) 199--218},
\href{http://arxiv.org/abs/hep-th/9807080}{{\ttfamily arXiv:hep-th/9807080
  [hep-th]}}.
%%CITATION = HEP-TH/9807080;%%.

\bibitem{Forcella:2008bb}
D.~Forcella, A.~Hanany, Y.-H. He, and A.~Zaffaroni, ``{The Master Space of N=1
  Gauge Theories},''
  \href{http://dx.doi.org/10.1088/1126-6708/2008/08/012}{{\em JHEP} {\bfseries
  0808} (2008) 012},
\href{http://arxiv.org/abs/0801.1585}{{\ttfamily arXiv:0801.1585 [hep-th]}}.
%%CITATION = ARXIV:0801.1585;%%.

\bibitem{Hanany:2008cd}
A.~Hanany and A.~Zaffaroni, ``{Tilings, Chern-Simons Theories and M2 Branes},''
  \href{http://dx.doi.org/10.1088/1126-6708/2008/10/111}{{\em JHEP} {\bfseries
  0810} (2008) 111},
\href{http://arxiv.org/abs/0808.1244}{{\ttfamily arXiv:0808.1244 [hep-th]}}.
%%CITATION = ARXIV:0808.1244;%%.

\bibitem{Hanany:2008fj}
A.~Hanany, D.~Vegh, and A.~Zaffaroni, ``{Brane Tilings and M2 Branes},''
  \href{http://dx.doi.org/10.1088/1126-6708/2009/03/012}{{\em JHEP} {\bfseries
  0903} (2009) 012},
\href{http://arxiv.org/abs/0809.1440}{{\ttfamily arXiv:0809.1440 [hep-th]}}.
%%CITATION = ARXIV:0809.1440;%%.

\bibitem{Aharony:2000pp}
O.~Aharony, ``{A Note on the holographic interpretation of string theory
  backgrounds with varying flux},''
  \href{http://dx.doi.org/10.1088/1126-6708/2001/03/012}{{\em JHEP} {\bfseries
  0103} (2001) 012},
\href{http://arxiv.org/abs/hep-th/0101013}{{\ttfamily arXiv:hep-th/0101013
  [hep-th]}}.
%%CITATION = HEP-TH/0101013;%%.

\bibitem{Dymarsky:2005xt}
A.~Dymarsky, I.~R. Klebanov, and N.~Seiberg, ``{On the moduli space of the
  cascading SU(M+p) x SU(p) gauge theory},''
  \href{http://dx.doi.org/10.1088/1126-6708/2006/01/155}{{\em JHEP} {\bfseries
  0601} (2006) 155},
\href{http://arxiv.org/abs/hep-th/0511254}{{\ttfamily arXiv:hep-th/0511254
  [hep-th]}}.
%%CITATION = HEP-TH/0511254;%%.

\bibitem{Furuuchi:2010gu}
K.~Furuuchi and K.~Okuyama, ``{D-branes Wrapped on Fuzzy del Pezzo Surfaces},''
  \href{http://dx.doi.org/10.1007/JHEP01(2011)043}{{\em JHEP} {\bfseries 1101}
  (2011) 043},
\href{http://arxiv.org/abs/1008.5012}{{\ttfamily arXiv:1008.5012 [hep-th]}}.
%%CITATION = ARXIV:1008.5012;%%.

\bibitem{Jafferis:2008qz}
D.~L. Jafferis and A.~Tomasiello, ``{A Simple class of N=3 gauge/gravity
  duals},'' \href{http://dx.doi.org/10.1088/1126-6708/2008/10/101}{{\em JHEP}
  {\bfseries 0810} (2008) 101},
\href{http://arxiv.org/abs/0808.0864}{{\ttfamily arXiv:0808.0864 [hep-th]}}.
%%CITATION = ARXIV:0808.0864;%%.

\bibitem{Hosomichi:2008jd}
K.~Hosomichi, K.-M. Lee, S.~Lee, S.~Lee, and J.~Park, ``{N=4 Superconformal
  Chern-Simons Theories with Hyper and Twisted Hyper Multiplets},''
  \href{http://dx.doi.org/10.1088/1126-6708/2008/07/091}{{\em JHEP} {\bfseries
  0807} (2008) 091},
\href{http://arxiv.org/abs/0805.3662}{{\ttfamily arXiv:0805.3662 [hep-th]}}.
%%CITATION = ARXIV:0805.3662;%%.

\bibitem{Imamura:2008nn}
Y.~Imamura and K.~Kimura, ``{On the moduli space of elliptic
  Maxwell-Chern-Simons theories},''
  \href{http://dx.doi.org/10.1143/PTP.120.509}{{\em Prog.Theor.Phys.}
  {\bfseries 120} (2008) 509--523},
\href{http://arxiv.org/abs/0806.3727}{{\ttfamily arXiv:0806.3727 [hep-th]}}.
%%CITATION = ARXIV:0806.3727;%%.

\bibitem{deBoer:1996mp}
J.~de~Boer, K.~Hori, H.~Ooguri, and Y.~Oz, ``{Mirror symmetry in
  three-dimensional gauge theories, quivers and D-branes},''
  \href{http://dx.doi.org/10.1016/S0550-3213(97)00125-9}{{\em Nucl.Phys.}
  {\bfseries B493} (1997) 101--147},
\href{http://arxiv.org/abs/hep-th/9611063}{{\ttfamily arXiv:hep-th/9611063
  [hep-th]}}.
%%CITATION = HEP-TH/9611063;%%.

\bibitem{Porrati:1996xi}
M.~Porrati and A.~Zaffaroni, ``{M theory origin of mirror symmetry in
  three-dimensional gauge theories},''
  \href{http://dx.doi.org/10.1016/S0550-3213(97)00061-8}{{\em Nucl.Phys.}
  {\bfseries B490} (1997) 107--120},
\href{http://arxiv.org/abs/hep-th/9611201}{{\ttfamily arXiv:hep-th/9611201
  [hep-th]}}.
%%CITATION = HEP-TH/9611201;%%.

\bibitem{Dey:2013nf}
A.~Dey and J.~Distler, ``{Three Dimensional Mirror Symmetry and Partition
  Function on $S^3$},''
\href{http://arxiv.org/abs/1301.1731}{{\ttfamily arXiv:1301.1731 [hep-th]}}.
%%CITATION = ARXIV:1301.1731;%%.

\bibitem{Martelli2008d}
D.~Martelli and J.~Sparks, ``{Moduli spaces of Chern-Simons quiver gauge
  theories and AdS(4)/CFT(3)},''
  \href{http://dx.doi.org/10.1103/PhysRevD.78.126005}{{\em Phys. Rev.}
  {\bfseries D78} (2008) 126005},
\href{http://arxiv.org/abs/0808.0912}{{\ttfamily arXiv:0808.0912 [hep-th]}}.
%%CITATION = ARXIV:0808.0912;%%.

\bibitem{Hohenegger2009}
S.~Hohenegger and I.~Kirsch, ``{A Note on the holography of Chern-Simons matter
  theories with flavour},''
  \href{http://dx.doi.org/10.1088/1126-6708/2009/04/129}{{\em JHEP} {\bfseries
  04} (2009) 129},
\href{http://arxiv.org/abs/0903.1730}{{\ttfamily arXiv:0903.1730 [hep-th]}}.
%%CITATION = ARXIV:0903.1730;%%.

\bibitem{Gaiotto2012}
D.~Gaiotto and D.~L. Jafferis, ``{Notes on adding D6 branes wrapping RP**3 in
  AdS(4) x CP**3},'' \href{http://dx.doi.org/10.1007/JHEP11(2012)015}{{\em
  JHEP} {\bfseries 11} (2012) 015},
\href{http://arxiv.org/abs/0903.2175}{{\ttfamily arXiv:0903.2175 [hep-th]}}.
%%CITATION = ARXIV:0903.2175;%%.

\bibitem{Lee2007}
K.-M. Lee and H.-U. Yee, ``{New AdS(4) x X(7) Geometries with CN=6 in M
  Theory},'' \href{http://dx.doi.org/10.1088/1126-6708/2007/03/012}{{\em JHEP}
  {\bfseries 03} (2007) 012},
\href{http://arxiv.org/abs/hep-th/0605214}{{\ttfamily arXiv:hep-th/0605214
  [hep-th]}}.
%%CITATION = HEP-TH/0605214;%%.

\bibitem{Benvenuti:2010pq}
S.~Benvenuti, A.~Hanany, and N.~Mekareeya, ``{The Hilbert Series of the One
  Instanton Moduli Space},''
  \href{http://dx.doi.org/10.1007/JHEP06(2010)100}{{\em JHEP} {\bfseries 06}
  (2010) 100},
\href{http://arxiv.org/abs/1005.3026}{{\ttfamily arXiv:1005.3026 [hep-th]}}.
%%CITATION = 1005.3026;%%.

\bibitem{Martelli2006}
D.~Martelli, J.~Sparks, and S.-T. Yau, ``{The Geometric dual of a-maximisation
  for Toric Sasaki-Einstein manifolds},''
  \href{http://dx.doi.org/10.1007/s00220-006-0087-0}{{\em Commun. Math. Phys.}
  {\bfseries 268} (2006) 39--65},
\href{http://arxiv.org/abs/hep-th/0503183}{{\ttfamily arXiv:hep-th/0503183
  [hep-th]}}.
%%CITATION = HEP-TH/0503183;%%.

\bibitem{Gauntlett2004b}
J.~P. Gauntlett, D.~Martelli, J.~F. Sparks, and D.~Waldram, ``{A New infinite
  class of Sasaki-Einstein manifolds},''
  \href{http://dx.doi.org/10.4310/ATMP.2004.v8.n6.a3}{{\em Adv. Theor. Math.
  Phys.} {\bfseries 8} no.~6, (2004) 987--1000},
\href{http://arxiv.org/abs/hep-th/0403038}{{\ttfamily arXiv:hep-th/0403038
  [hep-th]}}.
%%CITATION = HEP-TH/0403038;%%.

\bibitem{Gauntlett2005}
J.~P. Gauntlett, D.~Martelli, J.~Sparks, and D.~Waldram, ``{Supersymmetric AdS
  backgrounds in string and M-theory},'' {\em IRMA Lect. Math. Theor. Phys.}
  {\bfseries 8} (2005) 217--252,
\href{http://arxiv.org/abs/hep-th/0411194}{{\ttfamily arXiv:hep-th/0411194
  [hep-th]}}.
%%CITATION = HEP-TH/0411194;%%.

\bibitem{Martelli2008c}
D.~Martelli and J.~Sparks, ``{Notes on toric Sasaki-Einstein seven-manifolds
  and AdS(4) / CFT(3)},''
  \href{http://dx.doi.org/10.1088/1126-6708/2008/11/016}{{\em JHEP} {\bfseries
  11} (2008) 016},
\href{http://arxiv.org/abs/0808.0904}{{\ttfamily arXiv:0808.0904 [hep-th]}}.
%%CITATION = ARXIV:0808.0904;%%.

\bibitem{Hanany:2014dia}
A.~Hanany and R.~Kalveks, ``{Highest Weight Generating Functions for Hilbert
  Series},'' \href{http://dx.doi.org/10.1007/JHEP10(2014)152}{{\em JHEP}
  {\bfseries 1410} (2014) 152},
\href{http://arxiv.org/abs/1408.4690}{{\ttfamily arXiv:1408.4690 [hep-th]}}.
%%CITATION = ARXIV:1408.4690;%%.

\bibitem{Fabbri:1999hw}
D.~Fabbri, P.~Fre', L.~Gualtieri, C.~Reina, A.~Tomasiello, A.~Zaffaroni, and
  A.~Zampa, ``{3-D superconformal theories from Sasakian seven manifolds: New
  nontrivial evidences for AdS(4) / CFT(3)},''
  \href{http://dx.doi.org/10.1016/S0550-3213(00)00098-5}{{\em Nucl. Phys.}
  {\bfseries B577} (2000) 547--608},
\href{http://arxiv.org/abs/hep-th/9907219}{{\ttfamily arXiv:hep-th/9907219
  [hep-th]}}.
%%CITATION = HEP-TH/9907219;%%.

\bibitem{ChesterIliesiuPufuEtAl2015}
S.~M. Chester, L.~V. Iliesiu, S.~S. Pufu, and R.~Yacoby, ``{Bootstrapping
  $O(N)$ Vector Models with Four Supercharges in $3 \leq d \leq4$},''
\href{http://arxiv.org/abs/1511.07552}{{\ttfamily arXiv:1511.07552 [hep-th]}}.
%%CITATION = ARXIV:1511.07552;%%.

\bibitem{Nakajima2015}
H.~Nakajima, ``{Towards a mathematical definition of Coulomb branches of
  $3$-dimensional $\mathcal N=4$ gauge theories, I},''
\href{http://arxiv.org/abs/1503.03676}{{\ttfamily arXiv:1503.03676 [math-ph]}}.
%%CITATION = ARXIV:1503.03676;%%.

\bibitem{MorrisonPlesser1999}
D.~R. Morrison and M.~R. Plesser, ``{Nonspherical horizons. 1.},'' {\em Adv.
  Theor. Math. Phys.} {\bfseries 3} (1999) 1--81,
\href{http://arxiv.org/abs/hep-th/9810201}{{\ttfamily arXiv:hep-th/9810201
  [hep-th]}}.
%%CITATION = HEP-TH/9810201;%%.

\bibitem{Gray:2008yu}
J.~Gray, A.~Hanany, Y.-H. He, V.~Jejjala, and N.~Mekareeya, ``{SQCD: A
  Geometric Apercu},''
  \href{http://dx.doi.org/10.1088/1126-6708/2008/05/099}{{\em JHEP} {\bfseries
  0805} (2008) 099},
\href{http://arxiv.org/abs/0803.4257}{{\ttfamily arXiv:0803.4257 [hep-th]}}.
%%CITATION = ARXIV:0803.4257;%%.

\bibitem{Hanany:2008kn}
A.~Hanany and N.~Mekareeya, ``{Counting Gauge Invariant Operators in SQCD with
  Classical Gauge Groups},''
  \href{http://dx.doi.org/10.1088/1126-6708/2008/10/012}{{\em JHEP} {\bfseries
  0810} (2008) 012},
\href{http://arxiv.org/abs/0805.3728}{{\ttfamily arXiv:0805.3728 [hep-th]}}.
%%CITATION = ARXIV:0805.3728;%%.

\bibitem{Witten:2009xu}
E.~Witten, ``{Branes, Instantons, And Taub-NUT Spaces},''
  \href{http://dx.doi.org/10.1088/1126-6708/2009/06/067}{{\em JHEP} {\bfseries
  0906} (2009) 067},
\href{http://arxiv.org/abs/0902.0948}{{\ttfamily arXiv:0902.0948 [hep-th]}}.
%%CITATION = ARXIV:0902.0948;%%.

\bibitem{Hanany:1996ie}
A.~Hanany and E.~Witten, ``{Type IIB superstrings, BPS monopoles, and
  three-dimensional gauge dynamics},''
  \href{http://dx.doi.org/10.1016/S0550-3213(97)00157-0}{{\em Nucl.Phys.}
  {\bfseries B492} (1997) 152--190},
\href{http://arxiv.org/abs/hep-th/9611230}{{\ttfamily arXiv:hep-th/9611230
  [hep-th]}}.
%%CITATION = HEP-TH/9611230;%%.

\bibitem{kronheimer1990}
P.~B. Kronheimer, ``Instantons and the geometry of the nilpotent variety,''
  \href{http://projecteuclid.org/euclid.jdg/1214445316}{{\em J. Differential
  Geom.} {\bfseries 32} no.~2, (1990) 473--490}.

\bibitem{Br}
R.~Brylinski, ``{Instantons and K\"ahler geometry of nilpotent orbits},'' in
  {\em {Representation theories and algebraic geometry}}, vol.~514 of {\em NATO
  Adv. Sci. Inst. Ser. C Math. Phys. Sci.}, pp.~85--125.
\newblock Kluwer, 1998.
\newblock \href{http://arxiv.org/abs/math.SG/9811032}{{\ttfamily
  math.SG/9811032}}.

\bibitem{KS1}
P.~Kobak and A.~Swann, ``{The hyperk\"ahler geometry associated to Wolf
  spaces},'' {\em Boll. Unione Mat. Ital. Serie 8, Sez. B Artic. Ric. Mat.}
  {\bfseries 4} (2001) 587,
  \href{http://arxiv.org/abs/math.DG/0001025}{{\ttfamily math.DG/0001025}}.

\bibitem{Gaiotto:2008nz}
D.~Gaiotto, A.~Neitzke, and Y.~Tachikawa, ``{Argyres-Seiberg duality and the
  Higgs branch},'' \href{http://dx.doi.org/10.1007/s00220-009-0938-6}{{\em
  Commun.Math.Phys.} {\bfseries 294} (2010) 389--410},
\href{http://arxiv.org/abs/0810.4541}{{\ttfamily arXiv:0810.4541 [hep-th]}}.
%%CITATION = ARXIV:0810.4541;%%.

\bibitem{Dey:2013fea}
A.~Dey, A.~Hanany, N.~Mekareeya, D.~Rodriguez-Gomez, and R.-K. Seong,
  ``{Hilbert Series for Moduli Spaces of Instantons on
  $\mathbb{C}$$^{2}$/$\mathbb{Z}$$_{n}$},''
  \href{http://dx.doi.org/10.1007/JHEP01(2014)182}{{\em JHEP} {\bfseries 1401}
  (2014) 182},
\href{http://arxiv.org/abs/1309.0812}{{\ttfamily arXiv:1309.0812 [hep-th]}}.
%%CITATION = ARXIV:1309.0812;%%.

\bibitem{Butti:2006au}
A.~Butti, D.~Forcella, and A.~Zaffaroni, ``{Counting BPS baryonic operators in
  CFTs with Sasaki-Einstein duals},''
  \href{http://dx.doi.org/10.1088/1126-6708/2007/06/069}{{\em JHEP} {\bfseries
  0706} (2007) 069},
\href{http://arxiv.org/abs/hep-th/0611229}{{\ttfamily arXiv:hep-th/0611229
  [hep-th]}}.
%%CITATION = HEP-TH/0611229;%%.

\end{thebibliography}\endgroup

\end{document}